\pdfoutput=1
\documentclass[aoas]{imsart}

\RequirePackage[OT1]{fontenc}
\RequirePackage{amsthm,amsmath}
\usepackage{amsfonts}
\RequirePackage{natbib}
\usepackage{graphicx}
\usepackage{pdflscape}
\usepackage{mathabx}
\RequirePackage[colorlinks,citecolor=blue,urlcolor=blue]{hyperref}
\usepackage[figuresright]{rotating}

\startlocaldefs
\numberwithin{equation}{section}
\theoremstyle{plain}

\endlocaldefs

\begin{document}

\begin{frontmatter}
\title{Critical Window Variable Selection for Mixtures: Estimating the Impact of Multiple Air Pollutants on Stillbirth}
\runtitle{Critical Window Variable Selection for Mixtures}

\begin{aug}
  \author[A]{\fnms{Joshua L.}  \snm{Warren}\ead[label=e1]{joshua.warren@yale.edu}},
  \author[B]{\fnms{Howard H.} \snm{Chang}\ead[label=e2,mark]{howard.chang@emory.edu}},
  \author[C]{\fnms{Lauren K.} \snm{Warren}\ead[label=e3,mark]{lklein@rti.org}},
  \author[D]{\fnms{Matthew J.} \snm{Strickland}\ead[label=e4,mark]{mstrickland@unr.edu, ldarrow@unr.edu}},
    \author[D]{\fnms{Lyndsey A.} \snm{Darrow}},
  \and
  \author[E]{\fnms{James A.}  \snm{Mulholland}%
  \ead[label=e5,mark]{james.mulholland@ce.gatech.edu}}

  \runauthor{J.L.\ Warren et al.}

\address[A]{Yale University,
\printead{e1}}

\address[B]{Emory University,
\printead{e2}}

\address[C]{RTI International,
\printead{e3}}

\address[D]{University of Nevada,
\printead{e4}}

\address[E]{Georgia Institute of Technology,
\printead{e5}}
\end{aug}

\begin{abstract}
Understanding the role of time-varying pollution mixtures on human health is critical as people are simultaneously exposed to multiple pollutants during their lives.  For vulnerable sub-populations who have well-defined exposure periods (e.g., pregnant women), questions regarding critical windows of exposure to these mixtures are important for mitigating harm.  We extend Critical Window Variable Selection (CWVS) to the multipollutant setting by introducing CWVS for Mixtures (CWVSmix), a hierarchical Bayesian method that combines smoothed variable selection and temporally correlated weight parameters to (i) identify critical windows of exposure to mixtures of time-varying pollutants, (ii) estimate the time-varying relative importance of each individual pollutant and their first order interactions within the mixture, and (iii) quantify the impact of the mixtures on health.  Through simulation, we show that CWVSmix offers the best balance of performance in each of these categories in comparison to competing methods.  Using these approaches, we investigate the impact of exposure to multiple ambient air pollutants on the risk of stillbirth in New Jersey, 2005-2014.  We find consistent elevated risk in gestational weeks 2, 16-17, and 20 for non-Hispanic Black mothers, with pollution mixtures dominated by ammonium (weeks 2, 17, 20), nitrate (weeks 2, 17), nitrogen oxides (weeks 2, 16), PM$_{2.5}$ (week 2), and sulfate (week 20).  The method is available in the R package \texttt{CWVSmix}.
\end{abstract}


\end{frontmatter}


\section{Introduction}
Throughout their lives, humans are simultaneously exposed to multiple contaminants that may adversely impact their health.  Ambient air pollutants represent a major source of potentially hazardous exposures.  Numerous past studies have estimated associations between ambient exposures and multiple adverse health outcomes, with some findings consistent enough to suggest a causal relationship \citep{brunekreef2002air, kampa2008human, stieb2012ambient}.  However, the majority of these studies relied on statistical models that examined the role of a single pollutant on a health outcome.  Therefore, the findings may represent an incomplete understanding of the impact that pollution mixtures have on health.  New statistical methods are needed to address this issue, though a number of challenges in their development have been previously identified, particularly high correlation between exposures due to shared emission sources and meteorological drivers \citep{dominici2010protecting}.

Recently, there has been a shift away from single pollutant approaches towards the development of multipollutant methods (e.g., \cite{davalos2017current, ferrari2019identifying, ferrari2020bayesian, antonelli2020estimating, reich2020integrative}).  However, the majority of these methods were not designed for the analysis of time-varying mixtures, a feature that introduces additional modeling complications such as strong temporal correlation between exposures.  Methods that examine the role of time-varying pollutants on health are particularly important for identifying critical periods of exposure as is often of interest when analyzing pregnancy outcomes, though other health outcomes have also been explored in this context \citep{warren2012bayesian, warren2012spatial, warren2016bayesian, warren2019critical, warren2020spatially, chang2015assessment, wilson2017bayesian}.  

Methods that seek to identify critical windows of exposure most often rely on distributed lag models (DLMs) in some form to quantify the relationship between exposure over time and health, with the majority of methods developed for a single or small number of pollutants.  Using a Gaussian process, \cite{warren2012spatial} introduced a model which allowed for two pollutants to be analyzed jointly while accounting for their cross-covariance.  \cite{warren2012bayesian} extended this model using a semiparametric Bayesian approach which more generally considered four pollutants.  However, both frameworks became computationally demanding as the number of pollutants increased.  \cite{chen2019distributed} proposed a method to explore the impact of two time-varying exposures, including interactions, within a DLM framework.  \cite{liu2018lagged} extended Bayesian kernel machine regression (BKMR) to accommodate time-varying mixtures and further introduced a computationally feasible framework using mean field variational Bayesian inference \citep{liu2018modeling}.  However, the authors noted that this method is better suited for a smaller number of shared exposure time periods.  More recently, \cite{wilson2019kernel} introduced BKMR-DLM, a method designed to more flexibly and generally handle multiple exposure time periods while exploring nonlinear and interaction associations between pollutants and the outcome.     

Weighted quantile sum (WQS) regression \citep{carrico2015characterization} was developed to investigate the impact of multiple non-time-varying exposures on an outcome.  It introduces regularization via a set of pollutant-specific weights that are forced to sum to one to stabilize model fitting when considering a large number of contaminants.  Extending WQS to the setting of multiple exposure periods, \cite{bello2017extending} proposed an algorithm that relied on a DLM, though \cite{gennings2020lagged} later suggested that it may be more appropriate for a smaller number of pollutants.  As a result, \cite{gennings2020lagged} introduced lagged WQS (LWQS) regression which proposed to fit WQS regression separately at each exposure period to estimate the time-varying weight parameters and obtain the weighted average exposures across time.  It then used a DLM to estimate the association between the estimated weighted average exposures and the outcome.  However, this two-stage algorithm may be ignoring sources of uncertainty during model fitting (e.g., weight parameters are assumed known in the second stage DLM), may be inefficient since weights are estimated separately at each exposure period (i.e., ignoring potential correlation between weights in nearby exposure periods), and ignores the impact of interactions between exposures.  

In this work, we extend the recently developed method for more accurately identifying critical exposure windows in a single pollutant setting (Critical Window Variable Selection (CWVS)) \citep{warren2019critical} to the multipollutant setting by introducing CWVS for Mixtures (CWVSmix).  Similar to LWQS, CWVSmix introduces sum-to-one constraints on time-varying mixture weights in order to introduce regularization and accommodate a large number of pollutants.  However, unlike LWQS, CWVSmix is designed to jointly estimate these weights and the estimated mixtures' impacts on health while employing hierarchical Bayesian variable selection techniques to more formally identify critical windows of exposure and the important pollutants involved in the time-varying mixtures.  Additionally, first order interactions between pollutants on a given exposure period are also considered within the framework.  

We design a simulation study to compare the performances of CWVSmix and other competing options (including LWQS) with respect to (i) accuracy of the identified critical exposure periods, (ii) estimation of the time-varying pollutant-specific weights, and (iii) risk parameter estimation.  We then apply the methodology to investigate the impact that multiple gas and particulate matter ambient air pollutants, estimated using a novel data fusion model \citep{senthilkumar2019application}, have on stillbirth risk in New Jersey (NJ), 2005-2014, with a focus on investigating differences between various race/ethnicity groups.  

\section{Data}
We analyze live birth and fetal death records from the Division of Family Health Services in the NJ Department of Health in 2005-2014.  All births/deaths included in our study are singletons with a clinically estimated gestational age of at least 20 weeks, no birth defects, and a conception date in 2005-2013 (based on gestational age and date of birth/death) to avoid the fixed-cohort bias \citep{strand2011methodological}.  A stillbirth was defined as a fetal death occurring on or after 20 weeks of gestation (e.g., \cite{faiz2012ambient, green2015association}).  From these records, we extracted information on maternal educational attainment ($<$ high school, high school, $>$ high school), maternal age category ($< 25$, $[25, 30)$, $[30, 35)$, $> 35$), maternal race/ethnicity (non-Hispanic Black, Hispanic, non-Hispanic White), maternal tobacco use during pregnancy (yes, no), maternal residence at delivery (latitude, longitude), sex of the fetus (male, female), and the season and year of conception.  

Due to the large number of live births in each year of analysis and the low prevalence of stillbirth, we opt for a case-control study design similar to \cite{warren2019critical} who investigated the impact of air pollution on very preterm birth.  Additionally, we analyze each maternal race/ethnicity group separately to better understand the unique associations corresponding to each group instead of assuming common patterns of risk.  Therefore, for each eligible stillbirth in the dataset, we select five controls (i.e., live births) while matching only on race/ethnicity.  In Tables S1-S3 of the Supplement, we display summary information for each of the three race/ethnicity groups by stillbirth status.

We obtained daily estimates of ambient air pollution concentrations from 12 pollutants that covered a 12 by 12 kilometer grid across NJ from 2005-2014 with no spatial or temporal missingness.  The pollutants include 24-hour average particulate matter with an aerodynamic diameter less than or equal to 2.5 and 10 microns (PM$_{2.5}$ and PM$_{10}$, respectively) and PM$_{2.5}$ constituents nitrate (NO$_3^-$), ammonium (NH$_4^+$), sulfate (SO$_4^{2-}$), elemental carbon (EC), and organic carbon (OC); 1-hour maximum carbon monoxide (CO), nitrogen dioxide (NO$_2$), nitrogen oxides (NO$_\text{x}$), and sulfur dioxide (SO$_2$); and 8-hour maximum ozone (O$_3$).  These estimates come from a recently introduced spatiotemporal data fusion model that combines measured ambient concentrations with deterministic output from the Community Multiscale Air Quality model \citep{senthilkumar2019application}.  The latitude/longitude of the residence at delivery for each birth in the study was linked with the closest 12 by 12 kilometer grid cell centroid and weekly exposures for each pollutant (through gestational week 20) were calculated based on the date of birth and gestational age.  This study was approved by the institutional review boards at Rowan and Yale Universities.  

\section{Critical Window Variable Selection for Mixtures}
We introduce CWVSmix, a model for (i) identifying critical windows of exposure to mixtures of time-varying pollutants, (ii) estimating the time-varying relative importance of each individual pollutant and their first order interactions within the mixture, and (iii) quantifying the impact that the identified mixtures have on a health outcome.  CWVSmix incorporates hierarchical Bayesian variable selection techniques for defining critical windows and identifying important pollutants within a time-varying mixture, and uses Gaussian processes to allow for smoothness/regularization when estimating a potentially high-dimensional set of parameters that vary across time.     

Specifically, we extend CWVS \citep{warren2019critical} from the single pollutant setting to accommodate multiple exposures and their interactions such that \begin{align}\begin{split}
    &Y_i|p_i \stackrel{\text{ind}}{\sim}\text{Bernoulli}\left(p_i\right),\ i = 1, \hdots, n,\ \text{and }
    \ln\left(\frac{p_i}{1 - p_i}\right) = \\ &\textbf{x}_i^{\text{T}}\boldsymbol{\beta} + \sum_{t=1}^m \left[\sum_{j=1}^q \lambda_j\left(t\right) \text{z}_{ij}\left(t\right) + \sum_{j=1}^{q-1} \sum_{k=j+1}^q \widetilde{\lambda}_{jk}\left(t\right) \text{z}_{ij}\left(t\right)\text{z}_{ik}\left(t\right)\right]\alpha\left(t\right) \end{split}
\end{align} where $Y_i$ is the binary adverse health outcome of interest observed for study participant $i$; $n$ is the total number of people in the study; $p_i$ is the probability of experiencing the adverse health outcome; $\textbf{x}_i$ is a vector of person-specific covariates/confounders with accompanying regression parameters $\boldsymbol{\beta}$; $\text{z}_{ij}\left(t\right)$ represents exposure to the $j^{\text{th}}$ pollutant ($j=1,\hdots,q$), corresponding to person $i$'s spatial location, averaged over exposure period $t$ ($t=1,\hdots,m$) corresponding to the relevant calendar date range specific to person $i$'s timing of exposure (e.g., calendar dates that align with the selected gestational week); $\lambda_{j}\left(t\right) \in \left[0,1\right]$ are time-varying weight parameters that describe the main effect of pollutant $j$ in the mixture corresponding to exposure period $t$; $\widetilde{\lambda}_{jk}\left(t\right) \in \left[0,1\right)$ describes the interaction effect between pollutants $j$ and $k$ on exposure period $t$; and $\alpha\left(t\right)$ describes the impact that the mixture of pollutants defined by \begin{equation}\sum_{j=1}^q \lambda_{j}\left(t\right)\text{z}_{ij}\left(t\right) + \sum_{j=1}^{q-1} \sum_{k=j+1}^q \widetilde{\lambda}_{jk}\left(t\right) \text{z}_{ij}\left(t\right) \text{z}_{ik}\left(t\right)\end{equation} has on the probability of developing the adverse health outcome.

\subsection{Time-Varying Weights}
The weight parameters, \begin{equation}\boldsymbol{\lambda}\left(t\right) = \left\{\lambda_{1}\left(t\right), \hdots, \lambda_{q}\left(t\right), \widetilde{\lambda}_{12}\left(t\right), \hdots, \widetilde{\lambda}_{q-1,q}\left(t\right)\right\}^{\text{T}},\end{equation} define the mixture on exposure period $t$ and vary across pollutant, interaction, and exposure period, allowing for the possibility that potentially important mixtures are shifting over time.  These weights are defined to sum to one during each exposure period such that $$\sum_{j=1}^q \lambda_{j}\left(t\right) + \sum_{j=1}^{q-1} \sum_{k=j+1}^q \widetilde{\lambda}_{jk}\left(t\right) = 1$$ for all $t$, allowing the user to determine the relative importance of each term within the mixture.  In order to enforce this constraint while simultaneously accounting for the possibility of temporal correlation between these parameters across time (e.g., similar mixture profiles may adversely impact health during nearby periods of pregnancy) and carrying out variable selection for the individual components, we model the weights using multivariate latent variables as \begin{align}\begin{split}&\lambda_j\left(t\right) = \frac{\max\left\{\lambda^*_j\left(t\right), 0\right\}}{d\left(t\right)},\\ &\widetilde{\lambda}_{jk}\left(t\right) = \frac{\max\left\{\widetilde{\lambda}^*_{jk}\left(t\right) , 0\right\} 1\left\{\lambda^*_j\left(t\right) > 0\right\} 1\left\{\lambda^*_k\left(t\right) > 0\right\}}{d\left(t\right)}\end{split}\end{align} where the denominator is defined as the sum of all individual numerator terms corresponding to exposure period $t$ such as \begin{align}\begin{split}d\left(t\right) = &\sum_{j=1}^q \max\left\{\lambda^*_j\left(t\right), 0\right\} + \\ &\sum_{j=1}^{q-1} \sum_{k=j+1}^q \max\left\{\widetilde{\lambda}^*_{jk}\left(t\right), 0\right\} 1\left\{\lambda^*_j\left(t\right) > 0\right\} 1\left\{\lambda^*_k\left(t\right) > 0\right\}\end{split}
\end{align} and $1\left(.\right)$ represents the indicator function which is equal to one when the input statement is true and is equal to zero otherwise.    

The weight parameters in (3.3-3.5) are defined by latent variables $\boldsymbol{\lambda}^* = \left\{\boldsymbol{\lambda}^*\left(1\right)^{\text{T}}, \hdots, \boldsymbol{\lambda}^*\left(m\right)^{\text{T}}\right\}^{\text{T}},$ where $\boldsymbol{\lambda}^*\left(t\right)$ has the same form as $\boldsymbol{\lambda}\left(t\right)$ in (3.3), and are modeled jointly as \begin{align}\begin{split}&\boldsymbol{\lambda}^*|\phi_{\lambda} \sim \text{MVN}\left\{\boldsymbol{0}_{mq(q + 1)/2},\ \Sigma\left(\phi_{\lambda}\right) \otimes I_{q(q + 1)/2}\right\},\\ &\Sigma\left(\phi_{\lambda}\right)_{tt'} = \exp\left\{-\phi_{\lambda}|t-t'|\right\}\end{split}\end{align} where $\boldsymbol{0}_r$ is a length $r$ column vector of zeros; $I_r$ is an $r$ by $r$ identity matrix; $q(q+1)/2$ represents the total number of main and interaction effect parameters at a single exposure period; $\otimes$ represents the Kronecker product; and $\Sigma\left(\phi_{\lambda}\right)$ describes the correlation between the parameter vectors from different exposure periods with $\phi_{\lambda} > 0$ defining the level of temporal correlation.  Large values of $\phi_{\lambda}$ suggest that the mixtures are largely independent across time while small values indicate smoothness in the estimated mixtures.  The latent variables are assumed to have a marginal variance equal to one due to identifiability issues caused by the truncation and summation of individual parameters shown in (3.4) and (3.5).   

As a result of this specification, the entries of each $\boldsymbol{\lambda}\left(t\right)$ vector individually sum to one, we allow for the possibility that the mixture weights are similar across exposure time, and variable selection for these parameters is performed simultaneously.  Specifically, when a latent variable is $\leq 0$, the corresponding weight parameter will be exactly zero based on the truncation seen in (3.4).  This formulation also allows for the scenario that only a single pollutant is impacting the risk, as a main effect weight can exactly equal one when its latent variable is $> 0$ and all others are $\leq 0$.  Additionally an interaction weight can only be non-zero when its corresponding latent parameters is $> 0$ \textbf{and} the associated main effects are both larger than zero; satisfying the strong hierarchy interaction condition \citep{lim2015learning}.  

\subsection{Quantifying Mixture Risk}
Because the weight parameters sum to one at each exposure period, the expression in (3.2) can be interpreted as a weighted average of exposures and interactions whose total impact on the outcome is described by $\alpha\left(t\right)$.  These parameters represent the global variable selection process which describes whether any of the pollutants or interactions have an impact on risk at a specific exposure period.  The model for these risk magnitude parameters are specified using the original CWVS framework such that \begin{align*} &\alpha\left(t\right) = \theta\left(t\right)\gamma\left(t\right),\\ &\gamma\left(t\right) | \pi\left(t\right) \stackrel{\text{ind}}{\sim} \text{Bernoulli}\left\{\pi\left(t\right)\right\} \text{ where } \Phi^{-1}\left\{\pi\left(t\right)\right\} = \eta\left(t\right),\\ &\left[\begin{array}{c}
\theta\left(t\right)  \\
\eta\left(t\right) \end{array}\right]=A\left[\begin{array}{c}
\delta_1\left(t\right)  \\
\delta_2\left(t\right) \end{array}\right], \text{ and } A=\left[\begin{array}{cc}
A_{11} & 0  \\
A_{21} & A_{22} \end{array}\right].
\end{align*}  These parameters are decomposed into continuous, $\theta\left(t\right)$, and binary, $\gamma\left(t\right)$, time-varying components for the purposes of conducting temporally smoothed variable selection, where $\gamma\left(t\right)$ is controlled by the latent time-varying parameters $\eta\left(t\right)$ through a probit regression (i.e., $\Phi^{-1}\left(.\right)$ is the inverse cumulative distribution function of the standard normal distribution).  In order to account for temporal correlation and cross-covariance between these parameters, $\theta\left(t\right)$ and $\eta\left(t\right)$ are jointly modeled using the linear model of coregionalization \citep{wackernagel2013multivariate} where \begin{align*} &\boldsymbol{\delta}_j = \left\{\delta_j\left(1\right),\hdots,\delta_j\left(m\right)\right\}^\text{T}|\phi_j \stackrel{\text{ind}}{\sim} \text{MVN}\left\{\textbf{0}_m, \Sigma\left(\phi_j\right)\right\},\ j=1,2, \text{ and} \\ &\text{Corr}\left\{\delta_j\left(t\right), \delta_j\left(t'\right)\right\} = \Sigma\left(\phi_j\right)_{tt'} = \exp\left\{-\phi_j |t - t'|\right\},\ \phi_j > 0\end{align*} defines the smoothness of both processes over time. Finally, the variability and cross-covariance of the parameters are defined by $A_{jj} > 0$ for $j=1,2$ and $A_{21} \in \mathbb{R}$.  More information regarding the induced covariance between these parameters can be found in \cite{warren2019critical}.

\subsection{Interpreting Parameters}
The impact on the log-odds of developing the adverse outcome for a one unit increase in each pollutant (i.e, moving from $\text{z}_{ij}\left(t\right)$ to $\text{z}_{ij}\left(t\right) + 1$ for all $j$) during exposure period $t$ is given as $$\left[1 + \sum_{j=1}^{q-1}\sum_{k=j+1}^q \widetilde{\lambda}_{jk}\left(t\right)\left\{\text{z}_{ij}\left(t\right) + \text{z}_{ik}\left(t\right)\right\}\right]\alpha\left(t\right),$$ where the baseline pollution levels are involved in the expression due to the inclusion of interaction terms.  Note that the main effect parameters are not involved in this expression because all of the weights on a given exposure period sum to one.  Similarly, for individual pollutant $j$, $$\left\{\lambda_{j}\left(t\right) + \sum_{k=1}^{j-1} \widetilde{\lambda}_{kj}\left(t\right)\text{z}_{ik}\left(t\right) + \sum_{k=j+1}^q \widetilde{\lambda}_{jk}\left(t\right)\text{z}_{ik}\left(t\right)\right\}\alpha\left(t\right)$$ represents the increase in the log-odds for a one unit increase in pollutant $j$ during exposure period $t$ (i.e., moving from $\text{z}_{ij}\left(t\right)$ to $\text{z}_{ij}\left(t\right) + 1$ for pollutant $j$, with $\text{z}_{ik}\left(t\right)$ remaining the same for all $k \neq j$).  Therefore, $\alpha\left(t\right)$ and $\lambda_{j}\left(t\right)\alpha\left(t\right)$ represent the expected change in the log-odds when all exposures start at zero and the relevant exposure(s) are increased by one unit.  

\subsection{Prior Specifications}
We finalize the model by selecting prior distributions for the remaining parameters.  When possible, we favor weakly informative priors such that $\beta_j \stackrel{\text{iid}}{\sim} \text{N}\left(0, \sigma^2_{\boldsymbol{\beta}}\right)$, $j=1,\hdots,p$, where $\sigma^2_{\boldsymbol{\beta}}$ is fixed at a large value and $p$ is the length of the $\textbf{x}_i$ vector (including an intercept); $\phi_{\lambda}, \phi_j \stackrel{\text{iid}}{\sim} \text{Gamma}\left(\alpha_{\phi}, \beta_{\phi}\right),\ j=1,2$, where $\alpha_{\phi}$ and $\beta_{\phi}$ are fixed at small values; $\ln\left(A_{jj}\right) \sim \text{Normal}\left(0, \sigma^2_{A}\right),\ j=1,2$, and $A_{21}\sim \text{Normal}\left(0, \sigma^2_{A}\right)$, where $\sigma^2_{A}$ is fixed at a large value.  

\section{Simulation Study}
We design a simulation study to determine how CWVSmix performs in comparison to competing approaches with respect to (i) identification of the true critical window set, (ii) estimation of the important pollutants and interactions that comprise the time-varying mixtures, and (iii) estimation of time-varying mixture risk parameters. 

For the study, we simulate data from the model in (3.1) where $\textbf{x}_i^{\text{T}} \boldsymbol{\beta} \equiv 0$ for all $i$ since estimation of the covariates/confounders is not the primary interest.  The intercept is also set to zero so that approximately half of the simulated outcomes result in the adverse event, similar to the case-control design of our stillbirth application in NJ (Section 5).  To investigate realistic settings, we simulate data that closely resemble data from the stillbirth analysis.  Specifically, from (3.1) we set the number of exposure periods to 20 ($m = 20$) and the number of pollutants to 5 ($q=5$).  The vector of exposures specific to person $i$ across all exposure periods and pollutants, $$\textbf{z}_i = \left[\text{z}_{i1}\left(1\right), \hdots, \text{z}_{iq}\left(1\right), \hdots, \text{z}_{iq}\left(m\right)\right]^{\text{T}},$$ is randomly selected from the complete NJ dataset (i.e., including all race/ ethnicities).  An entire exposure profile is selected from a real individual and assigned to a simulated individual in order to maintain the temporal correlation in exposures across time and the cross-covariance among pollutants; both of which would be difficult to mimic otherwise.  For each simulated dataset, we use this process to create full exposure profiles for $n=2,534$ individuals, where around half of these people ($1,267$) develop the adverse outcome in each dataset.  This matches the number of stillbirths seen in the non-Hispanic Black stillbirth analysis from the real data application.

From (3.1), we modify $\lambda_{j}\left(t\right)$, $\widetilde{\lambda}_{jk}\left(t\right)$, and $\alpha\left(t\right)$ to simulate data from different scenarios, with an interest in determining how the competing methods perform as the complexity of the time-varying mixtures increases.  In Settings 1, 2, 3, 4, and 5, we assume that 1, 2, 3, 4, and 5 pollutants, respectively, are important drivers of the risk due to mixtures (i.e., those pollutants where $\lambda_{j}\left(t\right) > 0$).  We include Setting 1 (a single important pollutant) to determine how the methods perform if mixtures are not important features of the analysis and a single pollutant approach is optimal.  Within each of these settings, we include sub-settings that determine the exact composition of the time-varying mixture (i.e., choice of $\lambda_{j}\left(t\right)$ and $\widetilde{\lambda}_{jk}\left(t\right)$) and control the level of smoothness of these parameters across exposure period.  

In Sub-Setting A, we assume that the same pollutants and interactions are impacting the outcome at each exposure period and at the same magnitude during each period (i.e., $\boldsymbol{\lambda}\left(t\right)$ is the same for each $t$ included in the critical window set).  This represents complete smoothness of the weights across time.  In Sub-Setting B, we allow for an intermediate level of smoothness by once again assuming that the same set of pollutants and interactions are important at each exposure period but allowing their relative contributions to the mixtures to change across time (i.e., \textbf{only} the non-zero entries of $\boldsymbol{\lambda}\left(t\right)$ change for each $t$ in the critical window set).  In Sub-Setting C, we assume that the set of important pollutants/interactions and their relative contributions to the mixtures, can entirely vary across exposure periods, with only the total number of important main effects consistent across time (i.e., the zero \textbf{and} non-zero entries of $\boldsymbol{\lambda}\left(t\right)$ can change for each $t$ in the critical set, with only the total number of non-zero $\lambda_j\left(t\right)$ parameters consistent across time).  This results in a complete lack of smoothness in these parameters across time.  

The interaction pattern changes with each simulated dataset and is determined by the specific simulation setting.  For example, in Setting 2 where only two main effect pollutants are non-zero, only the interaction between those two pollutants is eligible to be non-zero (i.e., strict hierarchy for interactions).  Additionally, each eligible interaction term has a 50\% chance of being active in a given simulated dataset.  This allows there to be variability in the interaction pattern across the simulated datasets, resulting in a more realistic setup where only a subset of potential interactions are active.  In Sub-Setting C, this selection process happens separately at each critical week so that the important interactions are varying across time.  Across all settings, we simulate the non-zero weight parameters using a $\text{Dirichlet}\left(1,\hdots,1\right)$ distribution to ensure that they sum to one.    

One simulated realization of the weight parameters across all simulation study settings is shown in Figure 1 where the true critical window set includes five exposure periods that we arbitrarily define as the first five weeks of exposure for plotting purposes.  We note that simulating data from Setting 1B is not possible with only one important pollutant at each critical week since we can not vary the weights among this single pollutant (i.e., the non-zero weight must always be equal to one).  From Figure 1, it is clear how the complexity of the mixtures increase as we move from Settings 1 to 5 and how the weights become less smooth as we move from Sub-Settings A to C. 

\begin{sidewaysfigure}
\begin{center}
\includegraphics[scale=0.21]{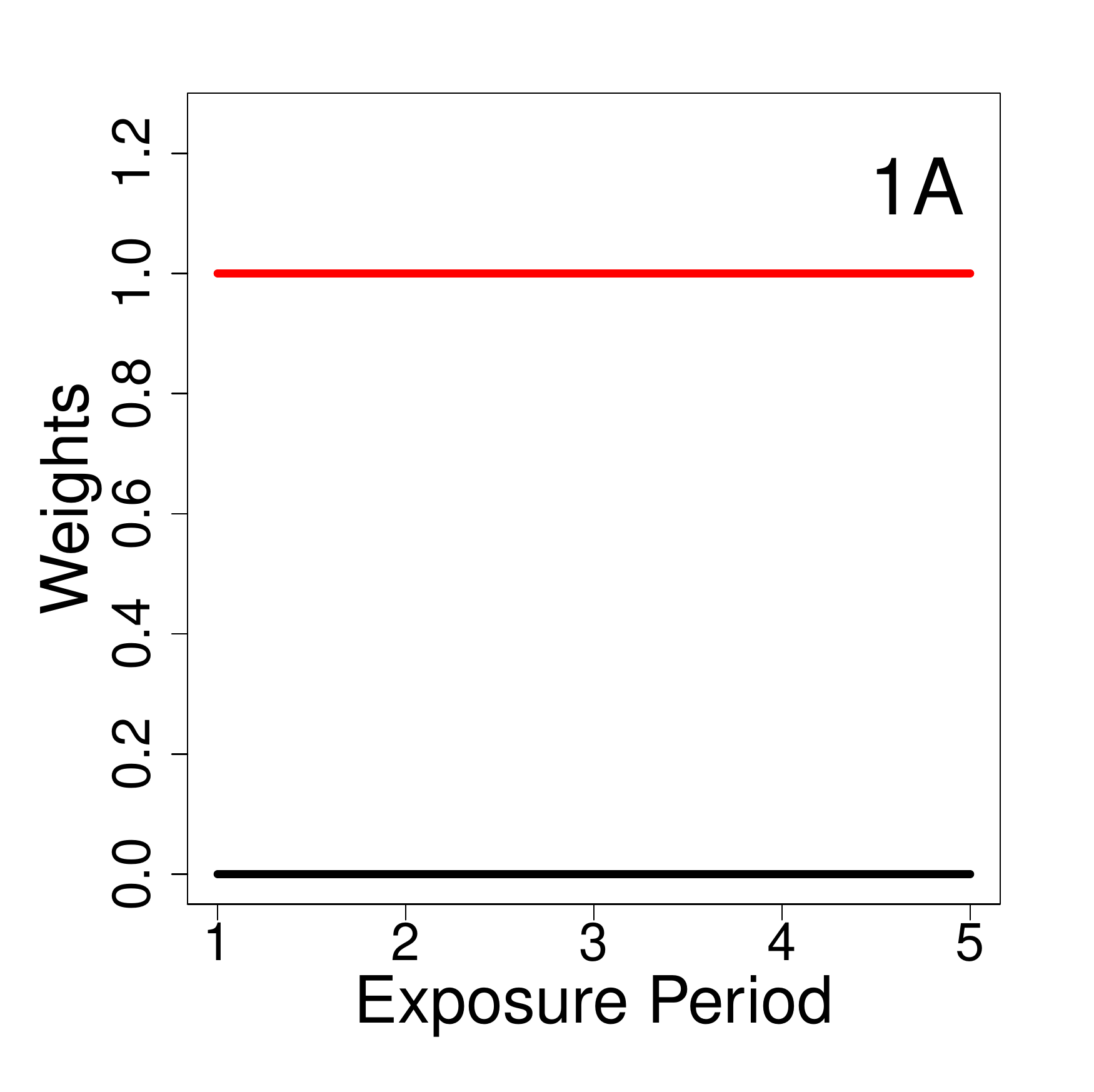}
\includegraphics[scale=0.21]{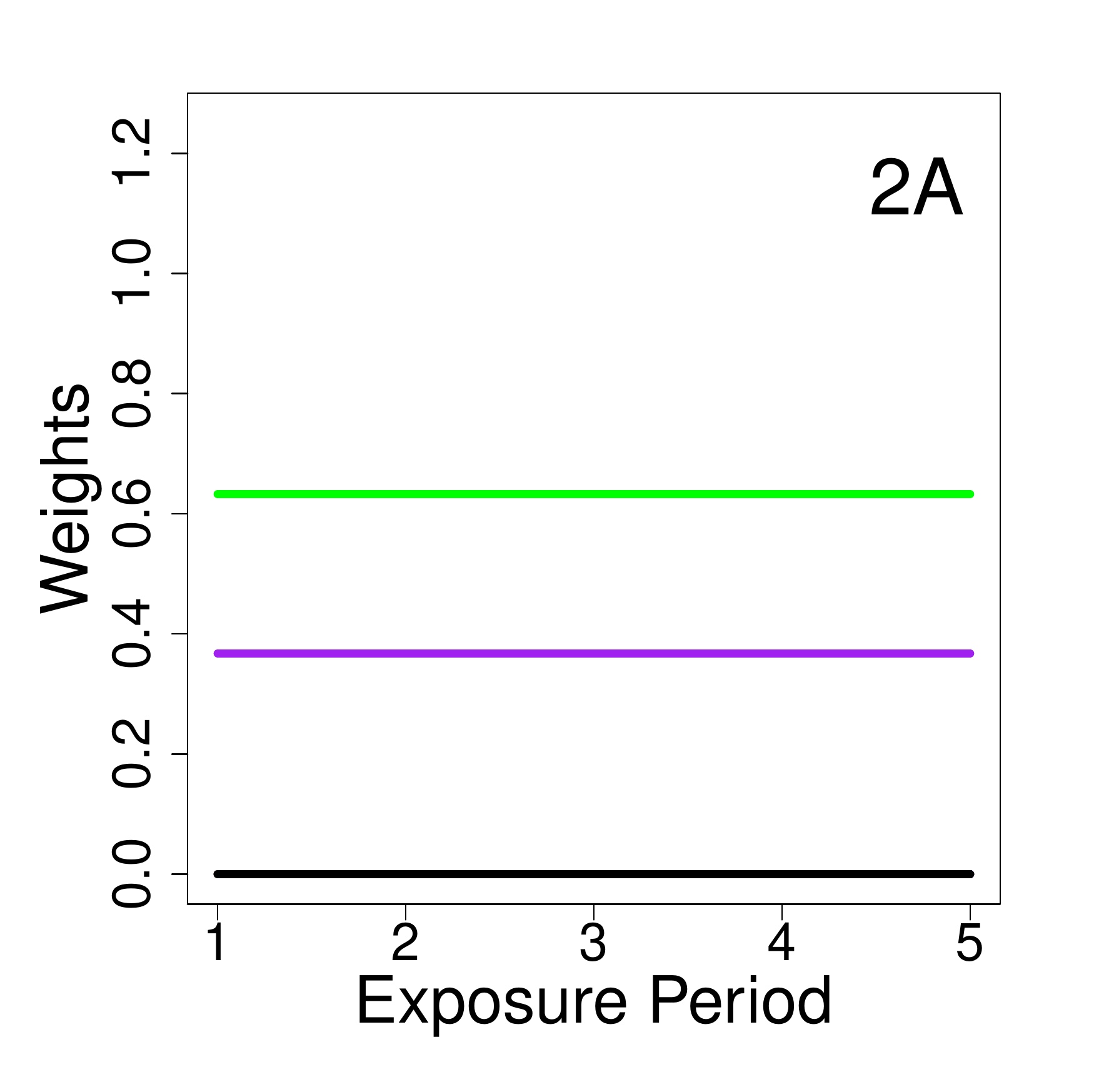}
\includegraphics[scale=0.21]{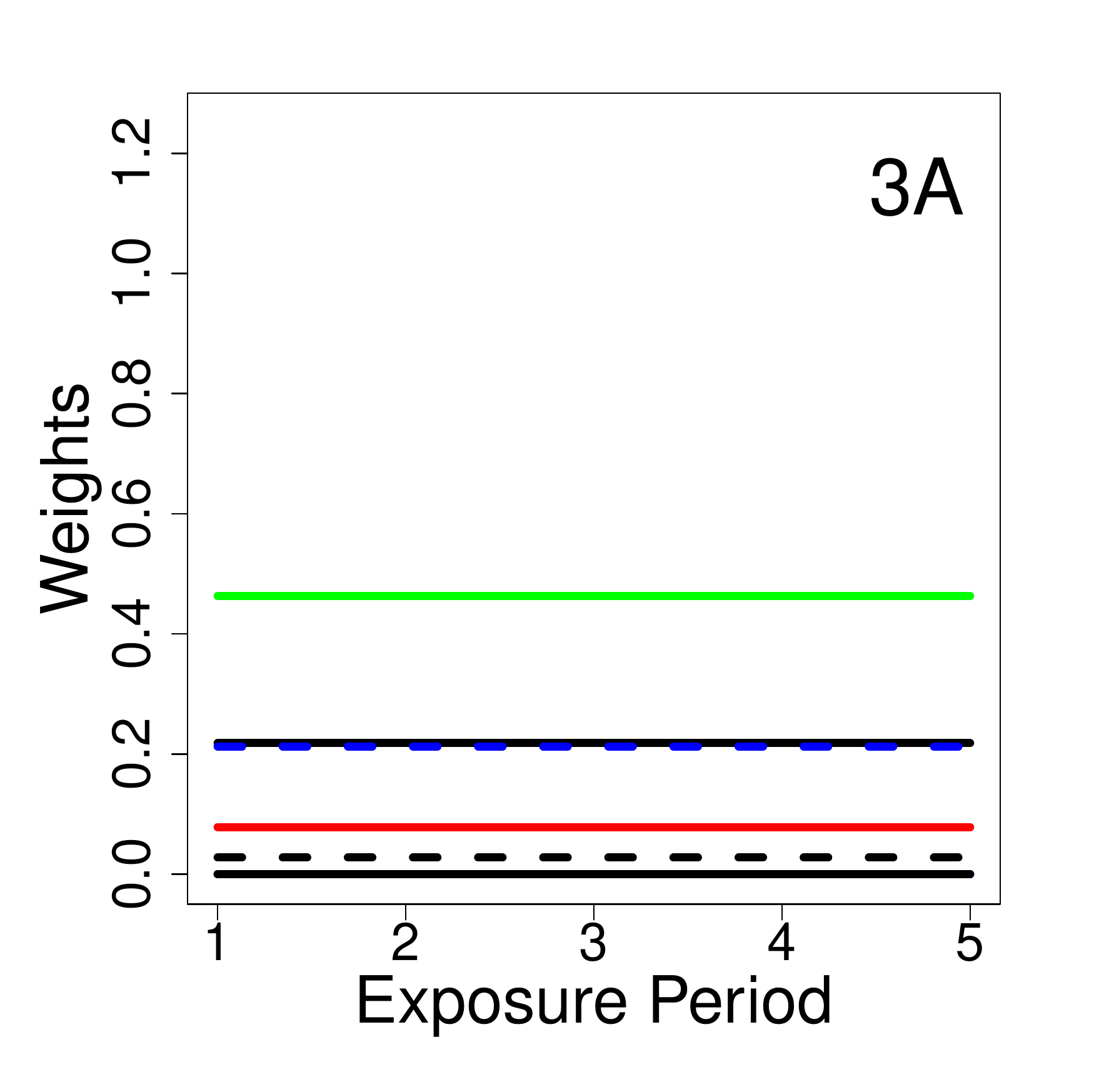}
\includegraphics[scale=0.21]{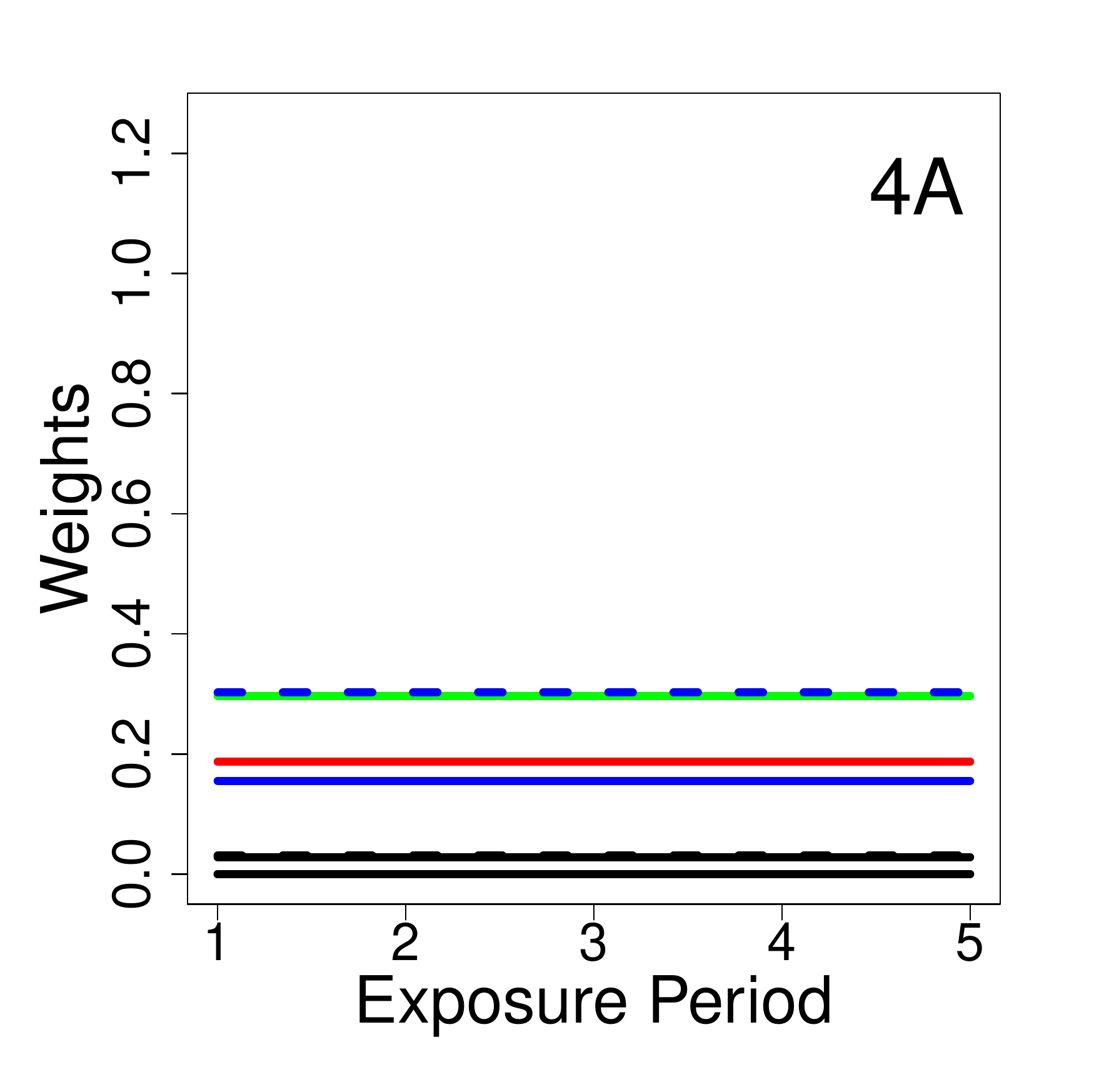}
\includegraphics[scale=0.21]{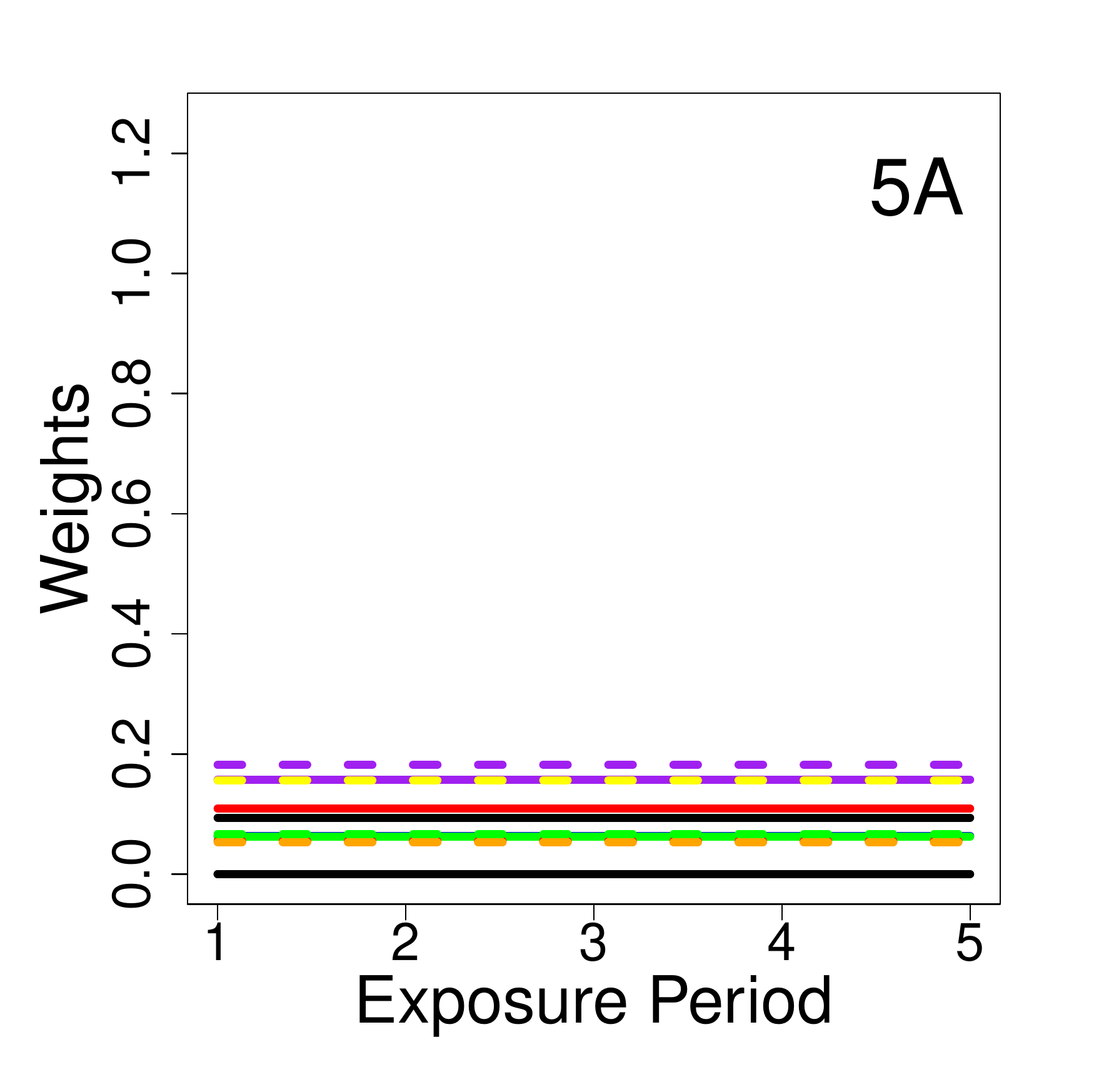}\\
\includegraphics[scale=0.21]{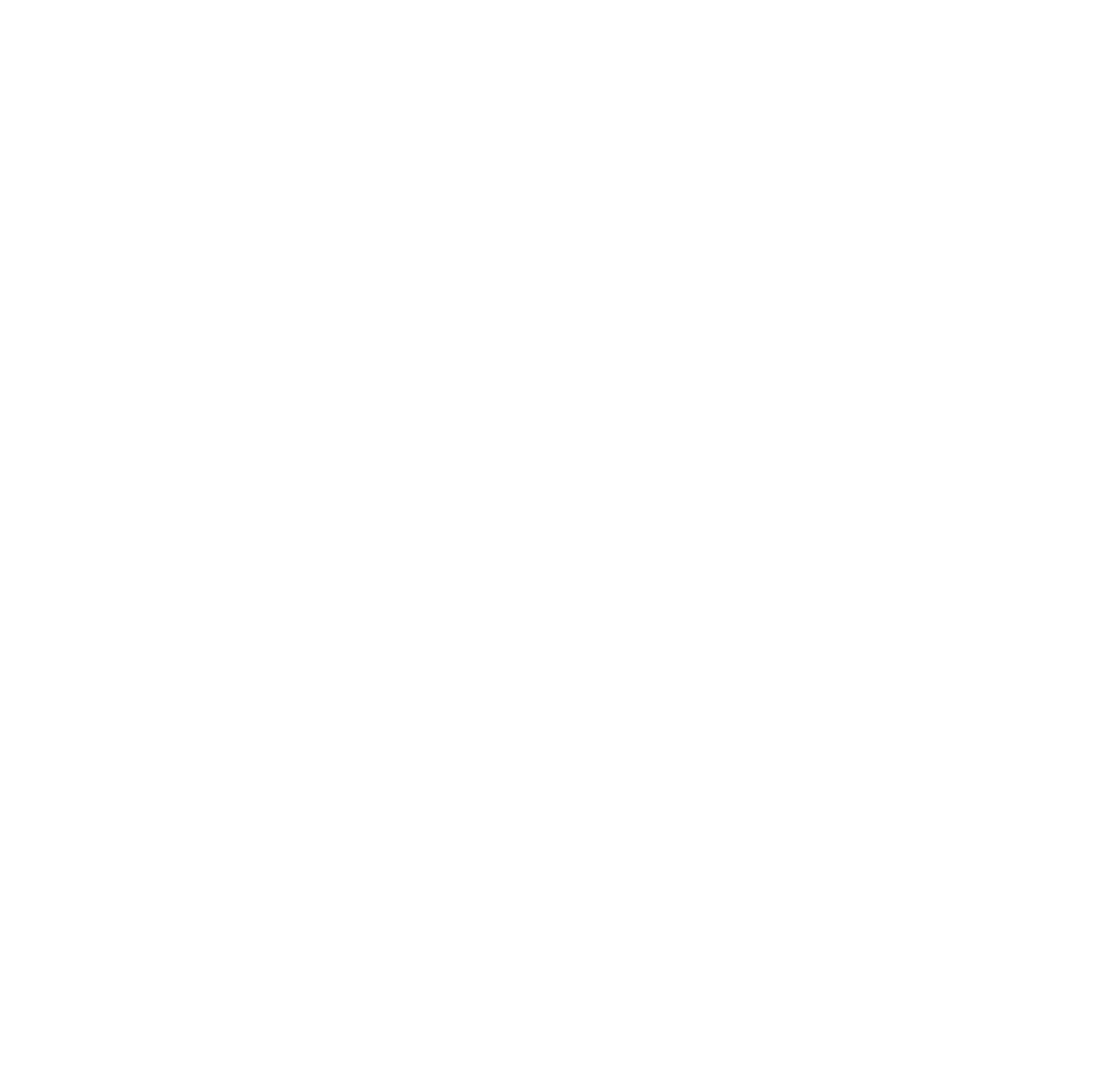}
\includegraphics[scale=0.21]{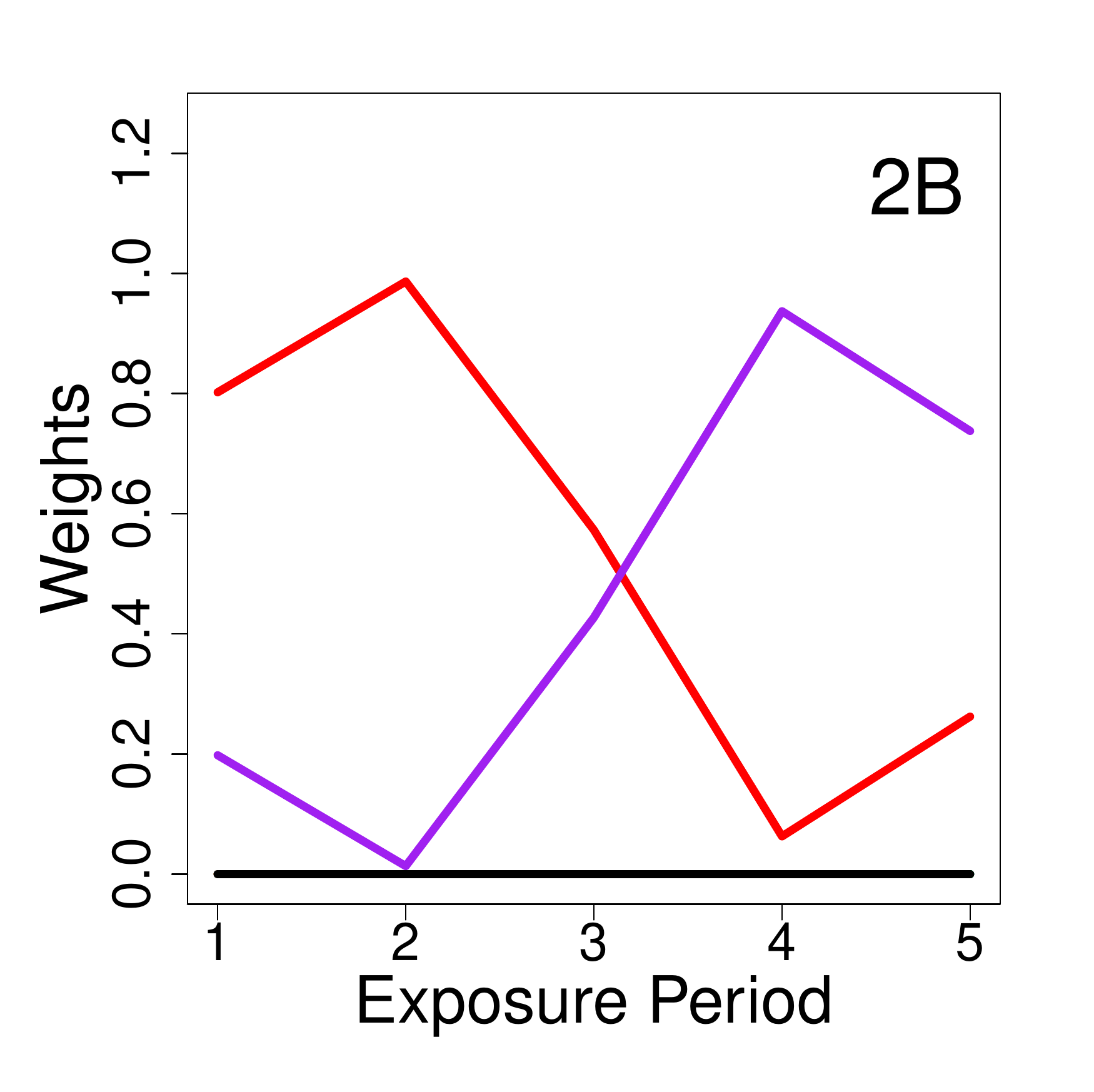}
\includegraphics[scale=0.21]{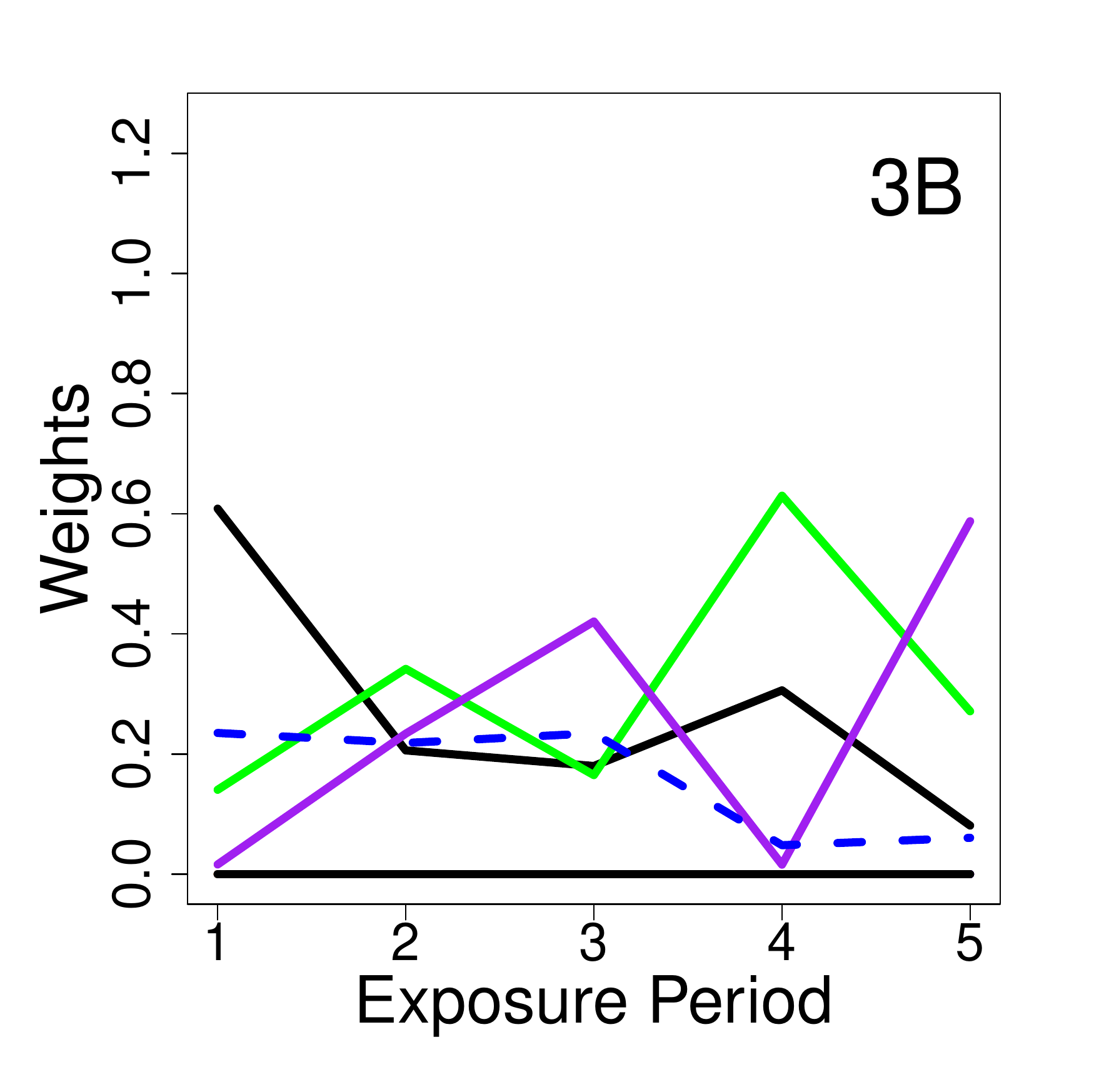}
\includegraphics[scale=0.21]{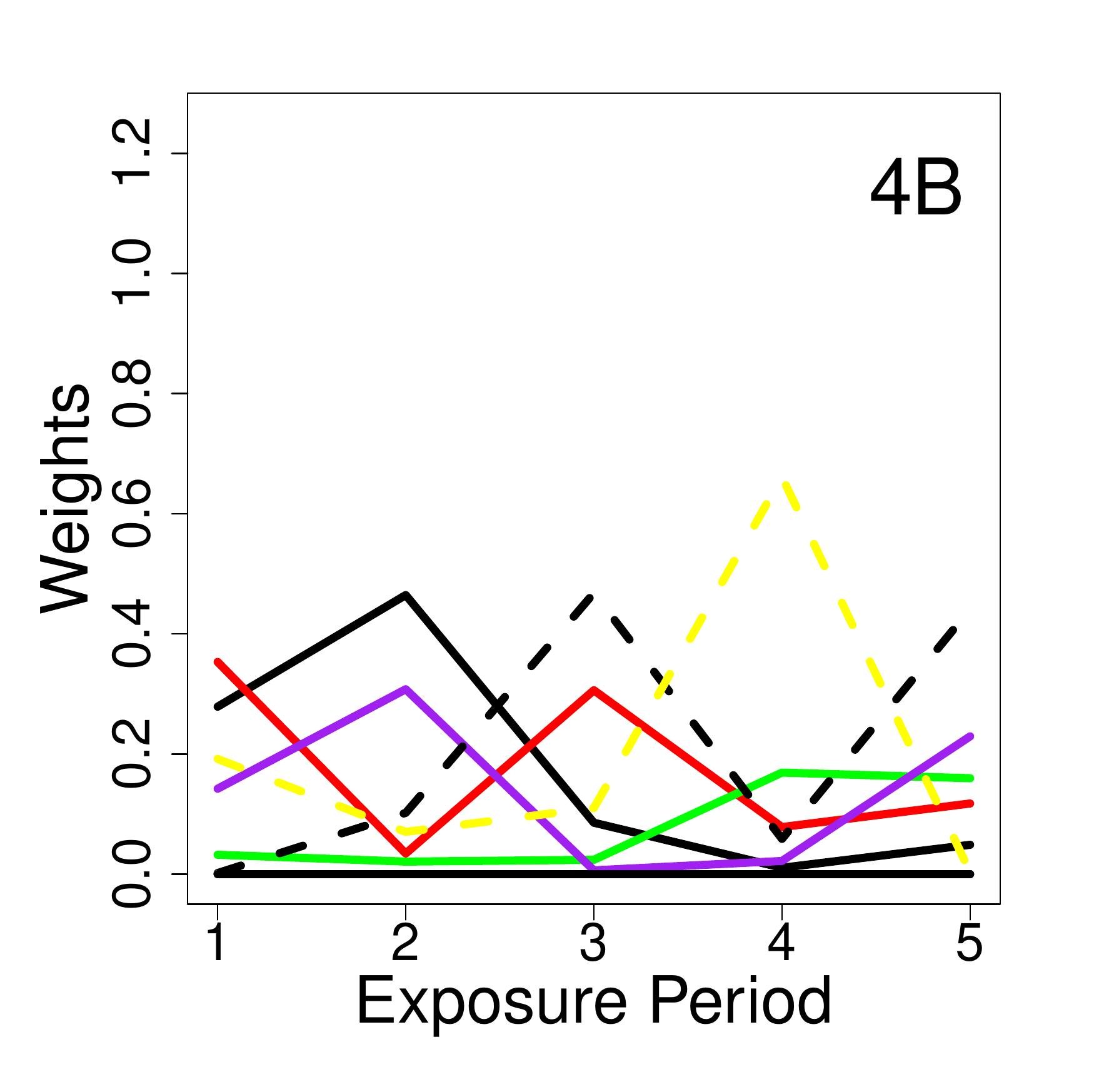}
\includegraphics[scale=0.21]{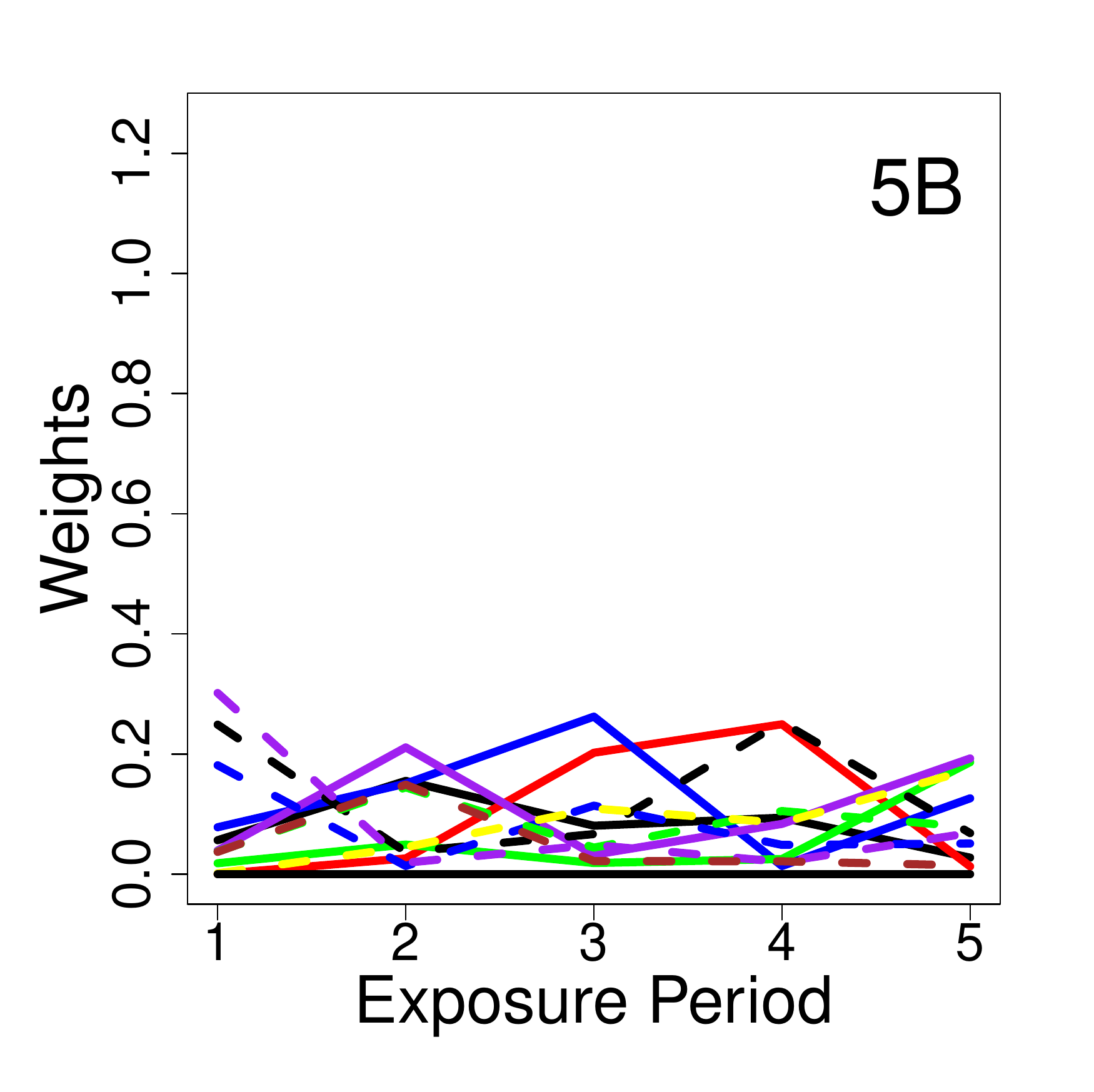}\\
\includegraphics[scale=0.21]{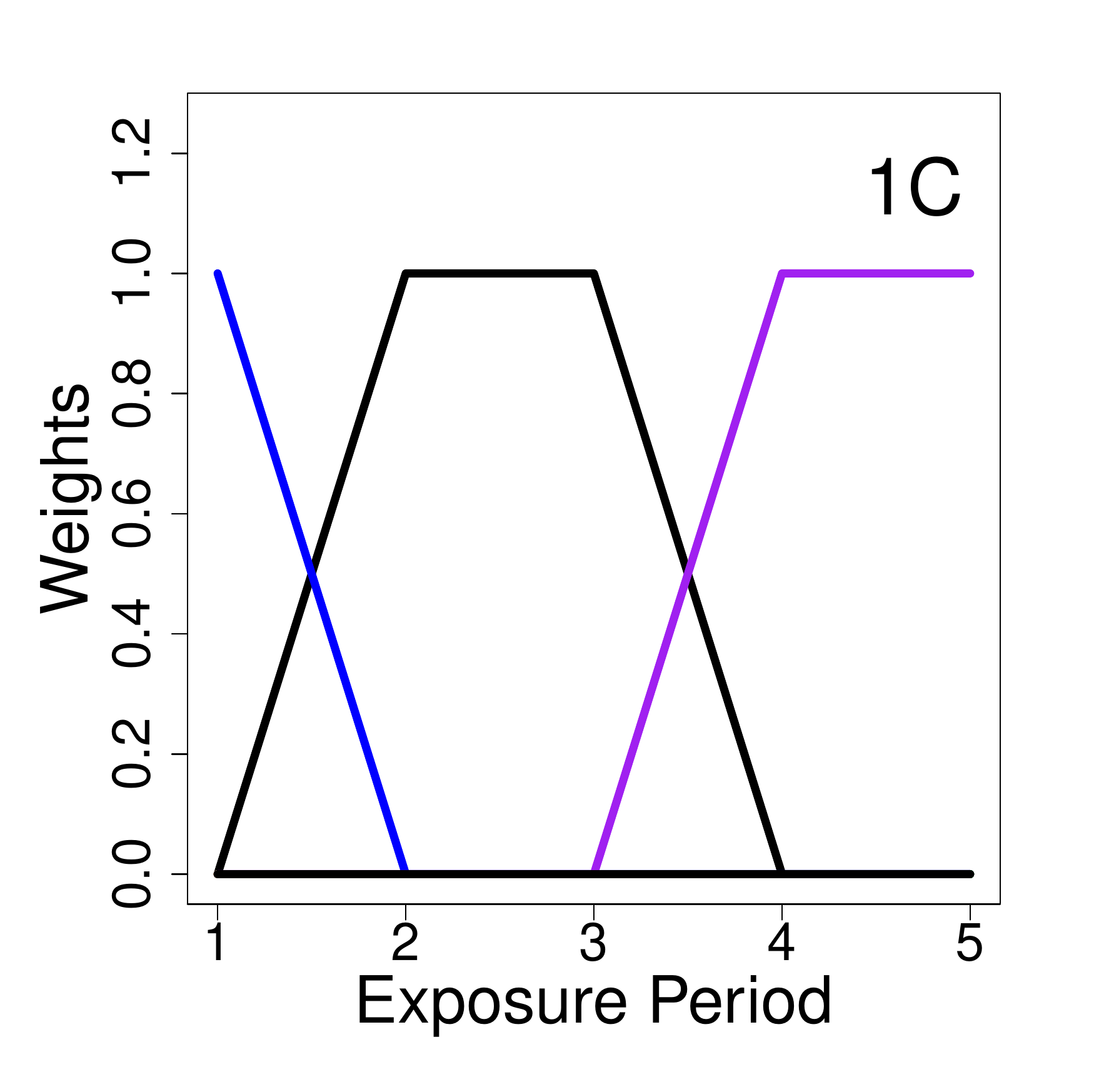}
\includegraphics[scale=0.21]{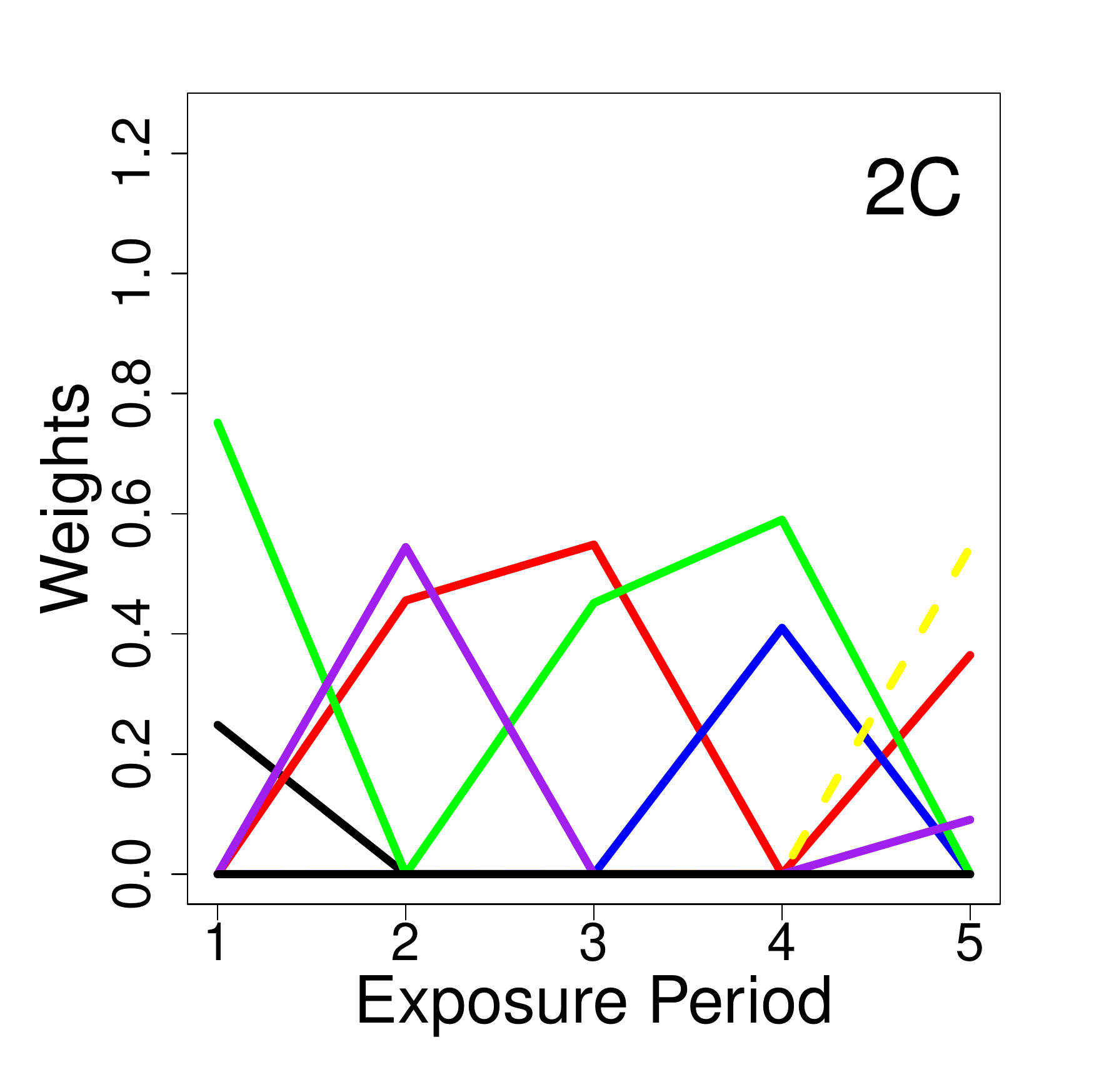}
\includegraphics[scale=0.21]{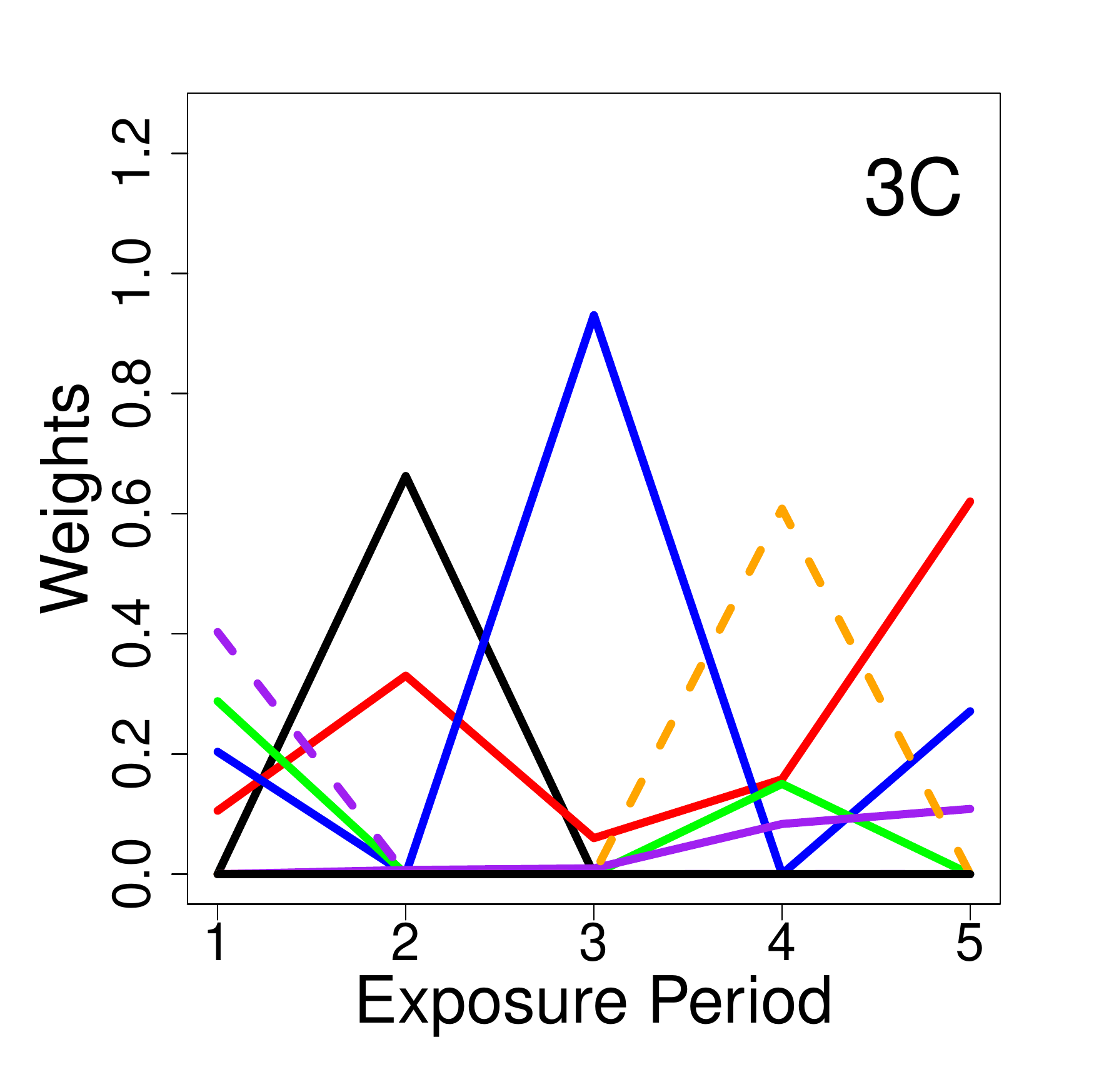}
\includegraphics[scale=0.21]{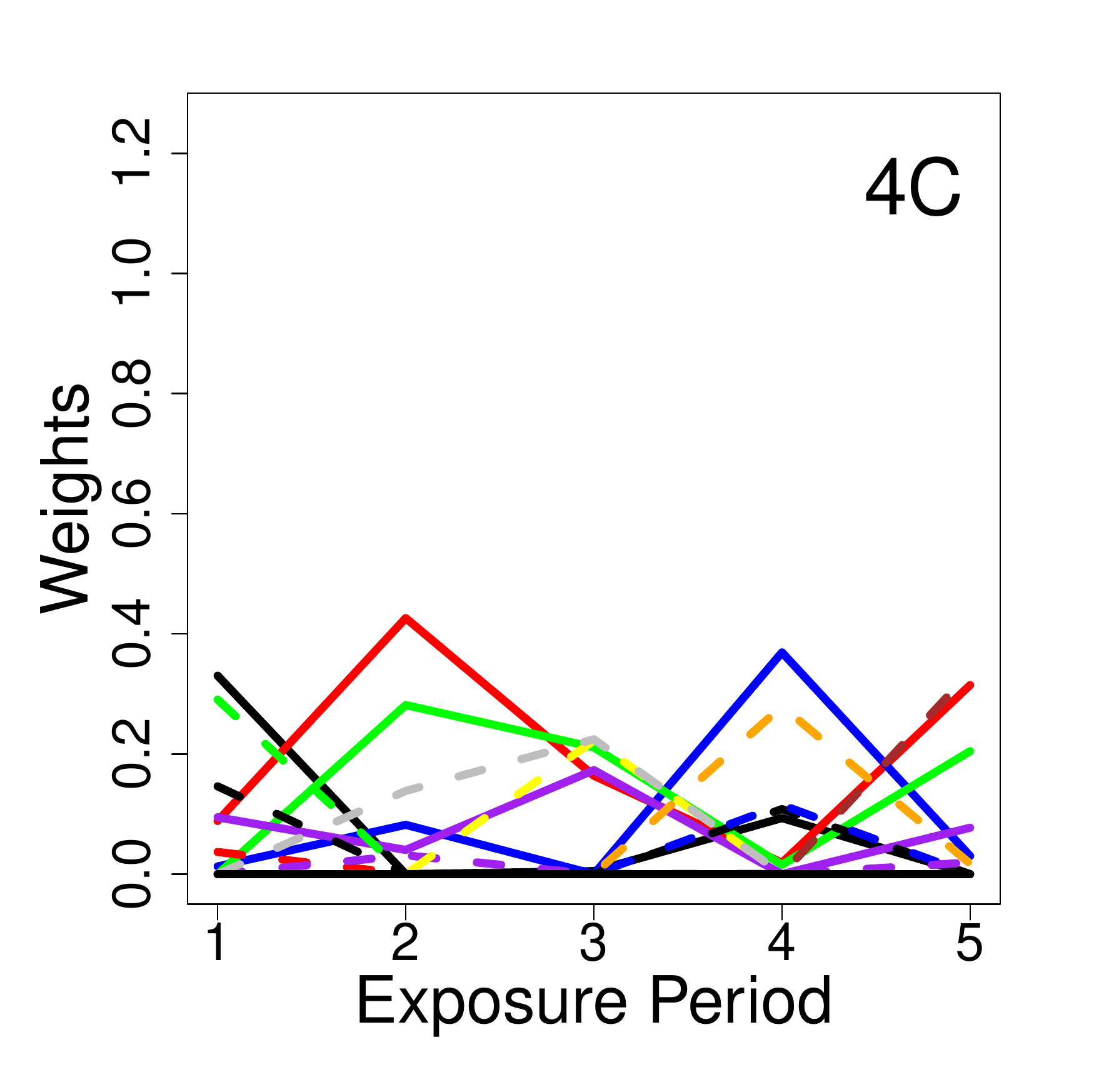}
\includegraphics[scale=0.21]{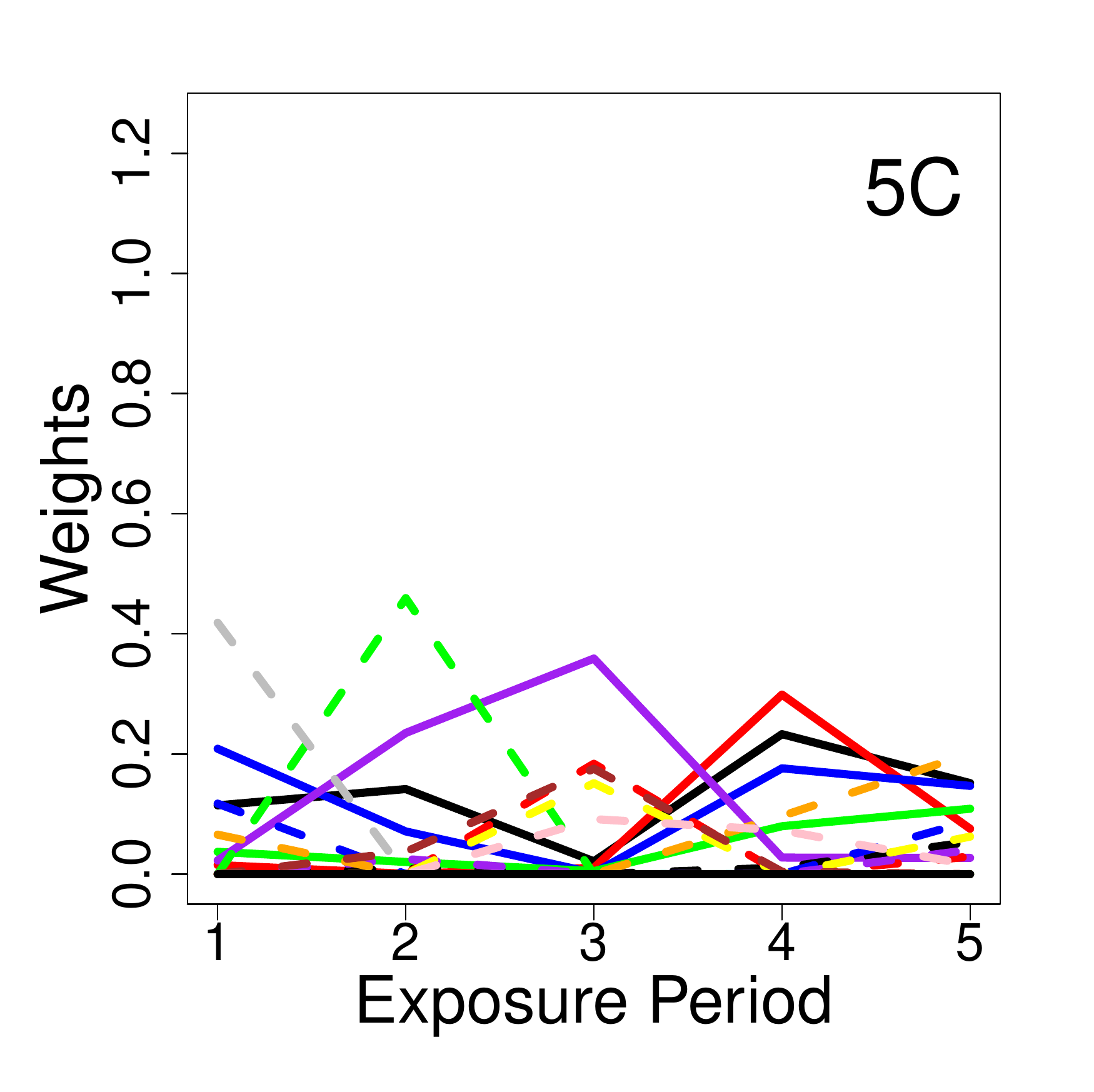}
\caption{A single simulated set of weight parameters from each simulation study setting where a line color corresponds to a unique pollutant, solid lines represent main effects, and dashed lines represent interaction effects.}
\end{center}
\end{sidewaysfigure}

For each simulated dataset, we randomly select how many exposure periods will be part of the critical window set (i.e., between one and seven; e.g., $m_0$), the first exposure period where the risk becomes elevated (i.e., between one and $m$ when possible; e.g., $t_0$), and define the critical window set as the consecutive $m_0$ exposure periods beginning on $t_0$.  For the exposure periods selected in the critical window set, the corresponding $\alpha\left(t\right)$ parameters are fixed at $0.23$, which is the sum of the posterior means of $\alpha\left(t\right)|\gamma\left(t\right) = 1$ for all exposure periods identified in the critical window set from the non-Hispanic Black stillbirth NJ CWVSmix data analysis; providing realistic effect sizes for simulation.  The remaining $\alpha\left(t\right)$ parameters are set to zero.  The result is simulated datasets where the true critical window sets differ by length and location during the exposure period, with consecutively defined non-zero effect sizes during this period.  A similar simulation framework was used in \cite{warren2019critical}.

In addition to fitting CWVSmix to each simulated dataset, we also explore two comparable alternative approaches for identifying critical exposure windows when multiple time-varying exposures are available.  Both methods represent modifications of (3.1) and are given as: \begin{itemize}
\item Equal Weight CWVS (EW):  All pollutants and interactions are given the same weight at each exposure period (i.e., $\lambda_{j}\left(t\right) = \widetilde{\lambda}_{jk}\left(t\right) = 2/q(q + 1)$ for all $t$, $j$, and $k$) and the original CWVS method is used to analyze the averaged time-varying exposures.  Therefore, (3.1) becomes \begin{equation*} \sum_{t=1}^m \left[\frac{2}{q(q + 1)} \left\{\sum_{j=1}^q \text{z}_{ij}\left(t\right) + \sum_{j=1}^{q-1} \sum_{k = j+1}^q \text{z}_{ij}\left(t\right)\text{z}_{ik}\left(t\right)\right\}\right]\alpha\left(t\right).\end{equation*}
\item LWQS CWVS:  At each individual exposure period, WQS regression is used to estimate $\boldsymbol{\lambda}\left(t\right)$ as $\widecheck{\boldsymbol{\lambda}}\left(t\right)$ and the original CWVS method is used to analyze the resulting weighted time-varying exposures.  Therefore, (3.1) becomes \begin{equation*}
    \sum_{t=1}^m \left\{\sum_{j=1}^q \widecheck{\lambda}_{j}\left(t\right) \text{z}_{ij}\left(t\right) + \sum_{j=1}^{q-1} \sum_{k=j+1}^q \widecheck{\widetilde{\lambda}}_{jk}\left(t\right) \text{z}_{ij}\left(t\right) \text{z}_{ik}\left(t\right)\right\}\alpha\left(t\right). \end{equation*}     
\end{itemize}  EW serves as a baseline method to determine the impact that ignoring the time-varying weighting has on various aspects of analysis.  In our version of LWQS, we opt for CWVS instead of a traditional DLM that does not include a variable selection component in order to more fairly compare findings across methods.  Comparing CWVSmix with LWQS in this study allows us to determine the benefits of jointly estimating the weights and risk parameters within a single framework that accounts for temporal correlation between parameters and variable selection of the weights (CWVSmix) as opposed to a more ad hoc two-stage model fitting approach (LWQS).  

\subsection{Simulation Study Results}  
For the CWVSmix prior distributions from Section 3.4, we select $\sigma_{\boldsymbol{\beta}} = 100$, $\alpha_{\phi} = \beta_{\phi} = 1.00$, and $\sigma^2_A = 1.00$.  CWVSmix is fit using the CWVSmix R package (\texttt{https://github.com/warrenjl/CWVSmix}) with model fitting details provided in Section S1 of the Supplement.  For CWVS, we select prior distributions as described in \cite{warren2019critical} and use the CWVS R package for model fitting (\texttt{https://github.com/warrenjl/CWVS}).  For LWQS, we use the gWQS R package \citep{gWQS} for fitting the WQS regressions based on 100 bootstrap samples where we modify the original algorithm so that all samples are used in the final estimation instead of choosing a risk direction \textit{a priori}.  This adjustment was made based on the failure of the original algorithm across several simulated datasets to produce effects in the prespecified direction and is in line more recent WQS work which also avoids this directionality assumption \citep{colicino2020per}.  

From each simulation setting, we generate 100 datasets, resulting in a total of 1,400 for analysis.  For each dataset, we fit each competing method and collect 1,000 samples from the joint posterior distribution after removing the first 10,000 iterations during a burn-in period and thinning the next 10,000 samples by a factor of 10 to reduce posterior autocorrelation.  

For each method and setting, we estimate the accuracy of a method to identify the true critical weeks as CW Accuracy = $$\frac{1}{100m} \sum_{s=1}^{100}\sum_{t=1}^m \left[1\left\{\alpha^{(s)}\left(t\right) > 0,\ t \in \widehat{\text{CW}}_s\right\} + 1\left\{\alpha_s\left(t\right) = 0, t \notin \widehat{\text{CW}}_s\right\} \right]$$ where $\alpha^{(s)}\left(t\right)$ is the true global risk parameter at exposure period $t$ in simulated dataset $s$ and $\widehat{\text{CW}}_s$ is the set of estimated critical window exposure periods.  We define an exposure period as being in the critical window set if the posterior inclusion probability is larger than 0.50 and its exponentiated quantile-based 90\% credible interval, excluding exact ones (i.e., conditional on $\gamma\left(t\right) = 1$), is entirely above or below one.  
This critical window definition was shown to outperform competing options with respect to accuracy in \cite{warren2019critical}. 

We also calculate the average mean squared error (AMSE) of the estimators of $\lambda_{j}\left(t\right)$ and $\widetilde{\lambda}_{jk}\left(t\right)$, averaged over all exposure time periods in the critical window set and pollutants/interactions, such as $\text{AMSE}\left(\boldsymbol{\widehat{\lambda}}_{\text{cw}}\right) = $ \begin{align*}\begin{split} &\frac{1}{100} \sum_{s=1}^{100} \frac{2}{q(q + 1)|\text{CW}_s|} \sum_{t \in \text{CW}_s} \left(\sum_{j=1}^q \left[\lambda_{j}^{\left(s\right)}\left(t\right) - \text{E}\left\{\lambda_{j}\left(t\right)|\gamma\left(t\right) = 1, \boldsymbol{Y}_s\right\}\right]^2 + \right. \\ & \left. \sum_{j=1}^{q-1} \sum_{k=j+1}^q \left[\widetilde{\lambda}_{jk}^{(s)}\left(t\right) - \text{E}\left\{\widetilde{\lambda}_{jk}\left(t\right)|\gamma\left(t\right) = 1, \boldsymbol{Y}_s\right\}\right]^2\right)\end{split}\end{align*} where $\text{CW}_s$ is the true set of exposure periods in the critical window set for simulated dataset $s$; $|\text{CW}_s|$ represents the number of exposure periods in the critical window set; $\boldsymbol{Y}_s$ is the vector of simulated outcomes for dataset $s$; $\text{E}\left\{\lambda_{j}\left(t\right)|\gamma\left(t\right) = 1, \boldsymbol{Y}_s\right\}$ is the posterior mean of the main effect weight parameters when the corresponding risk magnitude parameter is non-zero (similar definition holds for the interaction parameters); and $\lambda_j^{(s)}\left(t\right)$ is the true main effect weight parameter (similar definition holds for the interaction parameters).  When calculating AMSE$\left(\widehat{\boldsymbol{\lambda}}_{\text{cw}}\right)$ for EW and LWQS, $\text{E}\left\{\lambda_{j}\left(t\right)|\gamma\left(t\right) = 1, \boldsymbol{Y}_s\right\}$ and $\text{E}\left\{\widetilde{\lambda}_{jk}\left(t\right)|\gamma\left(t\right) = 1, \boldsymbol{Y}_s\right\}$ are both replaced by $2/q(q + t)$ for EW, and by $\widecheck{\lambda}_{j}\left(t\right)$ and $\widecheck{\widetilde{\lambda}}_{jk}\left(t\right)$, respectively, for LWQS. 

Next, we estimate the AMSE of the estimators of $\exp\left\{\alpha\left(t\right)\right\}$ (odds ratio scale), averaged over all exposure time periods, such as $\text{AMSE}\left(\widehat{\exp\left\{\boldsymbol{\alpha}\right\}}\right) =$ $$\frac{1}{100m} \sum_{s=1}^{100} \sum_{t=1}^m \left(\exp\left\{\alpha^{(s)}\left(t\right)\right\} - \text{E}\left[\exp\left\{\alpha\left(t\right)\right\}|\gamma\left(t\right) = 1, \boldsymbol{Y}_s\right]\right)^2$$ where $\text{E}\left[\exp\left\{\alpha\left(t\right)\right\}|\gamma\left(t\right) = 1, \boldsymbol{Y}_s\right]$ is the posterior mean of the non-zero component of the risk parameter.  

We also estimate the accuracy of CWVSmix to identify the important weight parameters across the critical window time periods.  A main effect weight parameter is ``selected'' in the model if its posterior inclusion probability is larger than 0.50 while an interaction weight parameter is selected if this probability is larger than 0.125 \textbf{and} the corresponding main effect parameters are both selected.  The threshold of 0.125 is based on the weight definition in (3.4), where $\lambda^*_j\left(t\right)$, $\lambda^*_k\left(t\right)$, and $\widetilde{\lambda}^*_{jk}\left(t\right)$ must each be greater than zero for $\widetilde{\lambda}_{jk}\left(t\right)$ to be non-zero; and the prior probability of each of these events is 0.50.  For each analyzed dataset, we use this procedure to classify each weight parameter during the true critical weeks as selected or not and then compare these estimates with the true classifications to calculate accuracy.  EW and LWQS do not include a variable selection component for the weights, so their performances can not be compared.

\begin{landscape}
\begin{table}[h!]
\centering
\caption{Simulation study model comparison results.  Bold entries indicate optimal values across methods.  All values are multiplied by 100 (i.e., AMSE($\widehat{\boldsymbol{\lambda}}_{\text{cw}}$), CW Accuracy) or 1,000 (i.e., AMSE($\widehat{\exp\left\{\boldsymbol{\alpha}\right\}}$)) for presentation purposes.  AMSE:  average mean squared error; Max.\ SE:  maximum standard error across all estimates in a column.}
\begin{tabular}{lrrrrrrrrrrr}
\hline
 & \multicolumn{3}{c}{CW Accuracy}  & & \multicolumn{3}{c}{AMSE($\widehat{\boldsymbol{\lambda}}_{\text{cw}}$)} & & \multicolumn{3}{c}{AMSE($\widehat{\exp\left\{\boldsymbol{\alpha}\right\}}$)}\\
 \cline{2-4} 
 \cline{6-8}
 \cline{10-12}
Setting         & EW    & LWQS    & CWVSMix &           & EW              & LWQS    & CWVSMix           & & EW    & LWQS    & CWVSmix   \\
\hline
1A              &  90.45 & 94.65   & \textbf{96.70} & & 6.22            & 0.76    & \textbf{0.71} & &  3.84 &  2.85   & \textbf{2.59}           \\
1C              &  94.45 & 94.05   & \textbf{95.90} & & 6.22            & 3.64    & \textbf{2.58} & &  3.02 &  3.03   & \textbf{2.69}           \\
\hline
2A              &  92.65 & 94.50   & \textbf{96.45} & & 3.42            & 1.21    & \textbf{0.95} & &  3.19 &  3.00   & \textbf{2.68}           \\
2B              &  92.25 & 92.65   & \textbf{95.45} & & 3.49            & 1.85    & \textbf{1.42} & &  3.07 &  3.19   & \textbf{2.77}           \\
2C              &  94.60 & 95.05   & \textbf{96.00} & & 3.30            & 2.43    & \textbf{1.83} & &  2.65 &  2.88   & \textbf{2.58}           \\
\hline
3A              &  94.25 & 94.50   & \textbf{96.25} & & 2.19            & 1.39    & \textbf{1.12} & &  \textbf{2.61} &  3.10   & 2.70           \\
3B              &  94.20 & 94.25   & \textbf{95.85} & & 2.07            & 1.73    & \textbf{1.27} & &  \textbf{2.71} &  3.33   & 2.81           \\
3C              &  95.80 & 94.95   & \textbf{96.20} & & 2.05            & 1.99    & \textbf{1.40} & &  \textbf{2.51} &  2.96   & 2.65           \\
\hline
4A              &  95.00 & 95.00   & \textbf{96.10} & & 1.23            & 1.37    & \textbf{1.01} & &  \textbf{2.33} &  3.02   & 2.70           \\
4B              &  95.70 & 94.20   & \textbf{96.45} & & 1.20            & 1.54    & \textbf{1.09} & &  \textbf{2.42} &  3.25   & 2.75           \\
4C              &  95.75 & 95.55   & \textbf{96.45} & & 1.29            & 1.69    & \textbf{1.16} & &  \textbf{2.29} &  3.08   & 2.58           \\
\hline
5A              &  95.45 & 95.50   & \textbf{96.05} & & \textbf{0.77}   & 1.35    & 0.85  & &  \textbf{2.34} &  3.08   & 2.78                  \\
5B              &  96.60 & 94.95   & \textbf{96.85} & & \textbf{0.77}   & 1.60    & 0.97  & &  \textbf{2.13} &  2.94   & 2.50                    \\
5C              &  95.45 & 94.60   & \textbf{95.80} & & \textbf{0.79}   & 1.51    & 0.93  & &  \textbf{2.18} &  3.09   & 2.65                  \\
\hline
Max.\ SE        &  1.13  & 0.70    &  0.54          & & 0.11            & 0.15    & 0.12  & &  0.22 &  0.20   & 0.16                    \\
\hline
\end{tabular}
\end{table}
\end{landscape}

In Table 1 we display the results from the study across all methods and simulation settings.  CWVSmix outperforms both competing methods in terms of accuracy of the identified critical windows across all settings.  In terms of estimating the mixture weight parameters, use of CWVSmix results in reduced AMSE in Settings 1-4 (across all sub-settings).  In Setting 5, where the proportion of all pollutants/interactions that are important becomes large, EW results in improved performance.  This finding is sensible because as this proportion grows, it means that most of the weight parameters will be non-zero as shown in Figure 1.  Because so many weights are non-zero, it becomes less likely that one or two pollutants will dominate the mixture.  Therefore, a model that assumes all pollutants are contributing equally is likely a reasonable option.  LWQS more often outperforms EW in terms of estimating the weight parameters while the accuracy results between the two methods are similar.  CWVSmix results in improved estimation of the global risk parameters in Settings 1 and 2, and is outperformed in Settings 3-5 by EW.  As the proportion of important pollutants/interactions grows, EW becomes a more efficient option for understanding the global impact of all pollutants.  However, EW does not provide insights on the important drivers of this global risk like CWVSmix and LWQS. 

\begin{table}[ht]
\centering
\caption{Simulation study variable selection results.  Accuracy of CWVSmix in correctly classifying zero and non-zero critical week weight parameters is displayed along with standard errors in parentheses.  All values are multiplied by 100 for presentation purposes.}
\begin{tabular}{lrr}
\hline
 & \multicolumn{2}{c}{Effects}\\
 \cline{2-3}
Setting         & Main    & Interaction    \\
\hline
1A              &  94.26 (1.14)  &  96.12 (1.00)  \\
1C              &  78.20 (1.47) &  86.19 (1.22)   \\
\hline
2A              &  86.45 (1.27) &  91.22 (0.98)   \\
2B              &  80.52 (1.68) &  85.09 (1.57)   \\
2C              &  69.91 (1.35) &  78.92 (1.34)   \\
\hline
3A              &  75.03 (1.56) &  78.56 (1.85)   \\
3B              &  78.71 (1.61) &  75.52 (2.00)   \\
3C              &  69.15 (1.11) &  70.99 (1.45)   \\
\hline
4A              &  72.64 (1.76) &  65.29 (1.51)   \\
4B              &  73.94 (1.40) &  64.65 (1.57)   \\
4C              &  67.35 (1.45) &  58.60 (1.43)   \\
\hline
5A              &  67.55 (1.40) &  57.49 (1.30)   \\
5B              &  70.01 (1.53) &  55.13 (1.09)   \\
5C              &  72.59 (1.61) &  55.05 (1.02)   \\
\hline
\end{tabular}
\end{table}

In Table 2, the accuracy of the variable selection procedure for the critical week weight parameters is displayed across each simulation setting.  Generally, the performance improves when the proportion of important pollutants/interactions is small.  However, across all settings the accuracy is above 0.50, suggesting that the method is differentiating important and unimportant features with some success.  Overall, the simulation study results suggest that CWVSmix offers an improved balance between critical window accuracy, mixture estimation, and risk estimation than the other methods and is relatively robust to the data simulation setting.  Additionally, it consistently outperforms LWQS across all considered settings.  

\section{Stillbirth in New Jersey, 2005-2014}
We analyze the impact that exposure to multiple ambient air pollutants has on the risk of stillbirth using the data from NJ, 2005-2014, previously described in Section 2.  Stillbirth is defined as the death of a baby before or during a delivery occurring after 20 completed weeks of gestation and impacts around 1 in 160 births in the United States \citep{hoyert2016cause}.  Some known maternal risk factors include Black race, 35 years of age and older, low socioeconomic status, and smoking.  Past studies have investigated the link between stillbirth and exposure to various indoor and ambient air pollution with somewhat mixed findings, though the evidence suggests further studies are warranted; see \cite{pope2010risk}, \cite{siddika2016prenatal}, and \cite{bekkar2020association} for recent reviews.  

First, we conduct single pollutant analyses using CWVS, where each of the 12 pollutants in the study are analyzed individually and critical windows of exposure are identified.  These analyses are conducted for each pollutant and race/ethnicity (i.e., non-Hispanic Black, Hispanic, and non-Hispanic White) separately.  Important individual pollutants identified by the single pollutant analyses are then used in the multipollutant/mixture approaches for critical window identification by applying CWVSmix and the competing methods from Section 4 to the data.  

We also include an additional competing method which uses principal component analyses (PCAs) to reduce the dimensionality of the exposures prior to analysis.  Specifically, at a selected gestational week, we apply PCA to the standardized exposures from all pollutants and interactions (across all individuals in the study) and extract factor loadings from the first principal component (i.e., pollutant/interaction- and gestational week-specific weights).  Using these factor loadings, we construct the factor score (i.e., weighted exposure) for each individual.  This process is repeated separately at each gestational week, resulting in 20 weighted exposures for each individual.  We then use CWVS to analyze the impact that these standardized PCA weighted exposures have on stillbirth risk.  This method differs from CWVSmix and LWQS because the weights are not forced to sum to one, and the weight values are determined by correlation between pollutants and do not consider associations with the stillbirth outcome.  

For all models, the outcome $Y_i$ is an indicator of a stillbirth for pregnancy $i$ and we select the first $m=20$ exposure weeks since each included pregnancy had a gestational age of at least 20 weeks.  In total, we consider average weekly exposures to the $q=12$ pollutants described in Section 2 (i.e., CO, EC, NH$_4^-$, NO$_2$, NO$_3^-$, NO$_\text{x}$, O$_3$, OC, PM$_{10}$, PM$_{2.5}$, SO$_2$, and SO$_4^{2-}$).  The covariates included in $\textbf{x}_i$ include categorical variables for the year and season of conception, an indicator of maternal tobacco use during pregnancy (yes vs.\ no), maternal age category ([25, 30) vs.\ $<$25, [30, 35) vs.\ $<$ 25, $\geq$ 35 vs.\ $<$25), maternal educational attainment (high school vs.\ $<$ high school, $>$ high school vs.\ $<$ high school), sex of the fetus (male vs.\ female), and the latitude/longitude of the maternal residence at delivery to account for unexplained spatial variation in stillbirth risk.

Average weekly air pollution exposures were scaled by subtracting off the median value and dividing by the interquartile range (IQR) on a specific gestational week of pregnancy, resulting in an IQR interpretation for the risk parameters.  Specifically, an IQR increase from the median exposure level of \textbf{each} pollutant during gestational week $t$ corresponds to an odds ratio of $\exp\left\{\alpha\left(t\right)\right\}$ with respect to stillbirth risk for EW, LWQS, and CWVSmix (see Section 3.3 for more information).  For PCA, because the weight parameters do not sum to one, the corresponding effect is given as $$\exp\left[\alpha\left(t\right)\left\{\sum_{j=1}^q \lambda_{j}^{(\text{pca})}\left(t\right) + \sum_{j=1}^{q-1} \sum_{k=j+1}^q \widetilde{\lambda}_{jk}^{(\text{pca})}\left(t\right)\right\}\right]$$ where $\lambda_{j}^{(\text{pca})}$ are the PCA main effect weights (similar definition holds for the interaction parameters).  In Figures S1-S3 of the Supplement, we display the IQRs for each pollutant and race/ethnicity across every gestational week for the purposes of interpretation.    

For CWVSmix and each application of CWVS, we collect 10,000 posterior samples from the joint posterior distribution after discarding the first 10,000 during a burn-in period and thinning the next 100,000 by a factor of 10 to reduce posterior autocorrelation.  Trace plots of individual parameters and the Geweke diagnostic \citep{geweke1991evaluating} were used to assess convergence of each model, with no obvious signs of non-convergence observed.  Prior distributions and model fitting details match those from Section 4.1.   

\vspace{-0.35cm}
\subsection{Data Application Results}
In Figures S4-S9 of the Supplement, we display the single pollutant results for all race/ethnicity analyses and individual pollutants.  Similar to \cite{warren2019critical}, we present posterior means and quantile-based 95\% credible intervals for $\exp\left\{\alpha\left(t\right)\right\}|\gamma\left(t\right) = 1$, along with posterior inclusion probabilities at each gestational week (i.e., $\text{P}\left\{\gamma\left(t\right) = 1|\boldsymbol{Y}\right\}$).  For non-Hispanic Black mothers, increased exposure during weeks 2 (PM$_{2.5}$), 16 (NO$_{\text{x}}$), 17 (NH$_4^+$, NO$_3^-$), and 20 (NH$_4^+$, SO$_4^{2-}$), is associated with elevated risk of stillbirth.  The results for Hispanic  and non-Hispanic White mothers are null across all pollutants and gestational weeks.  Based on this initial screening, we use these five pollutants in the subsequent multipollutant modeling. 

\begin{figure}
\begin{center}
\includegraphics[trim={0.5cm 0.5cm 1cm 0.5cm}, clip, scale=0.18]{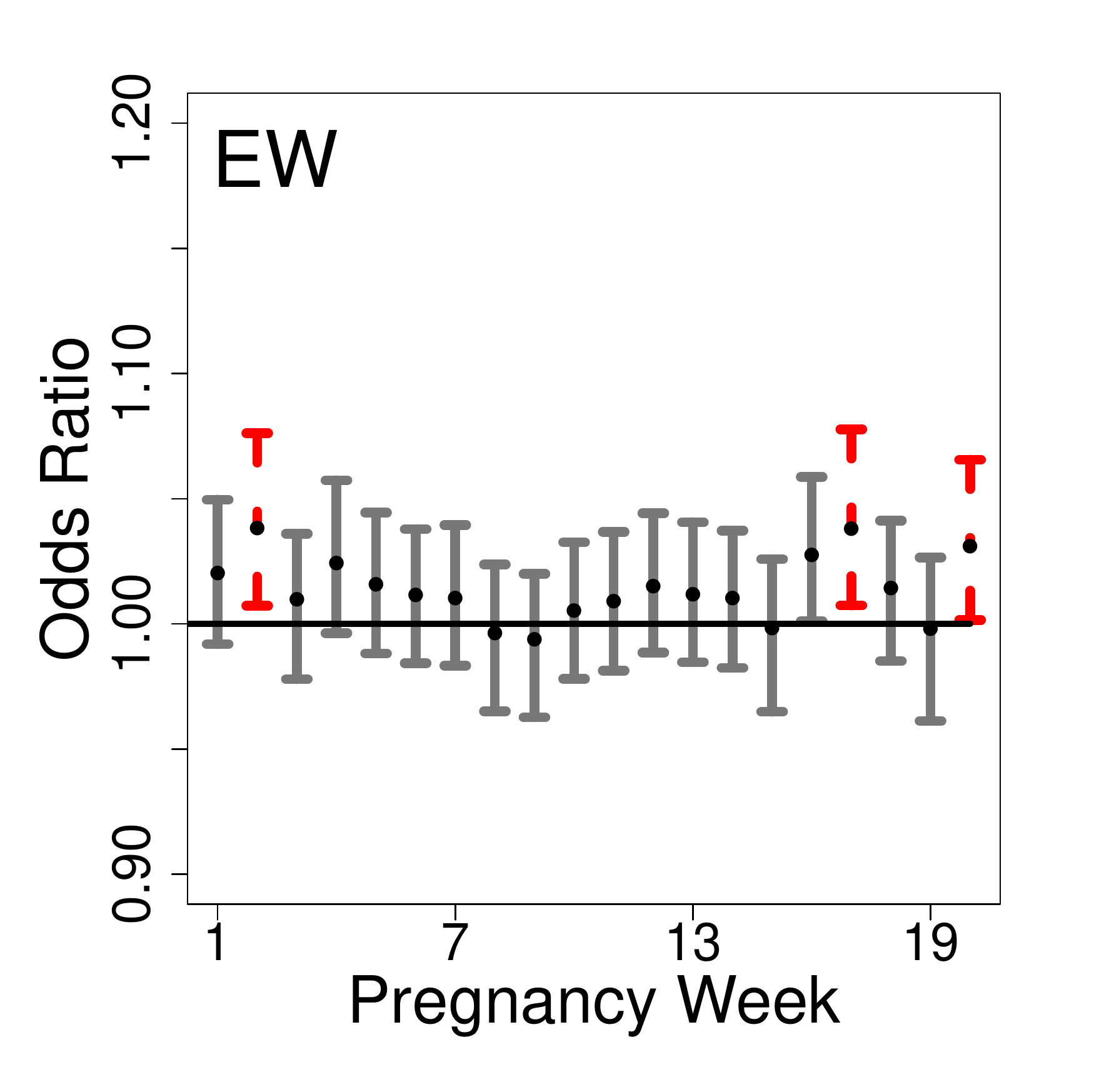}
\includegraphics[trim={0.5cm 0.5cm 1cm 0.5cm}, clip, scale=0.18]{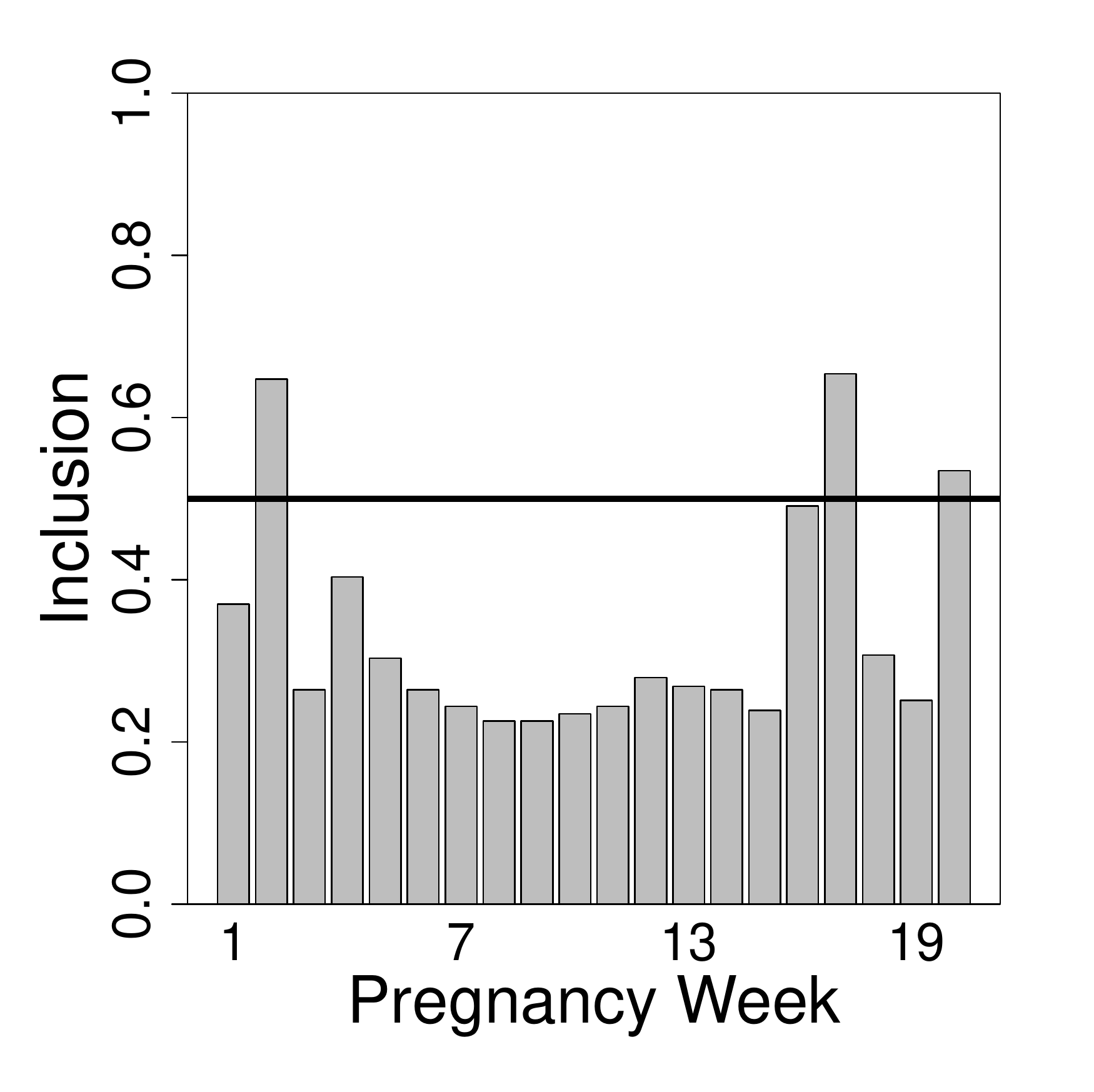}
\includegraphics[trim={0.25cm 0.5cm 1cm 0.5cm}, clip, scale=0.18]{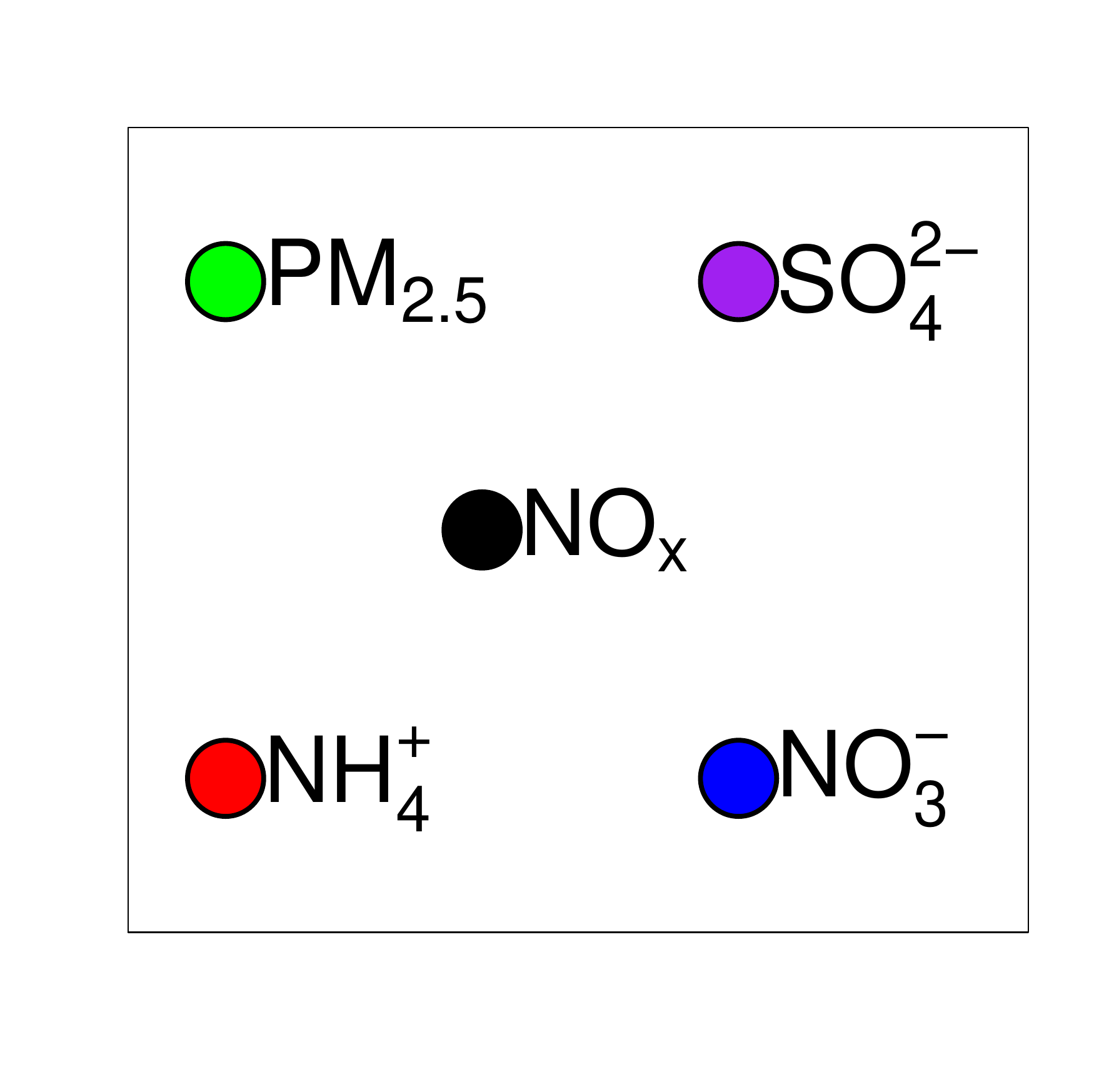}
\includegraphics[trim={0cm 0.5cm 1cm 0.5cm}, clip, scale=0.18]{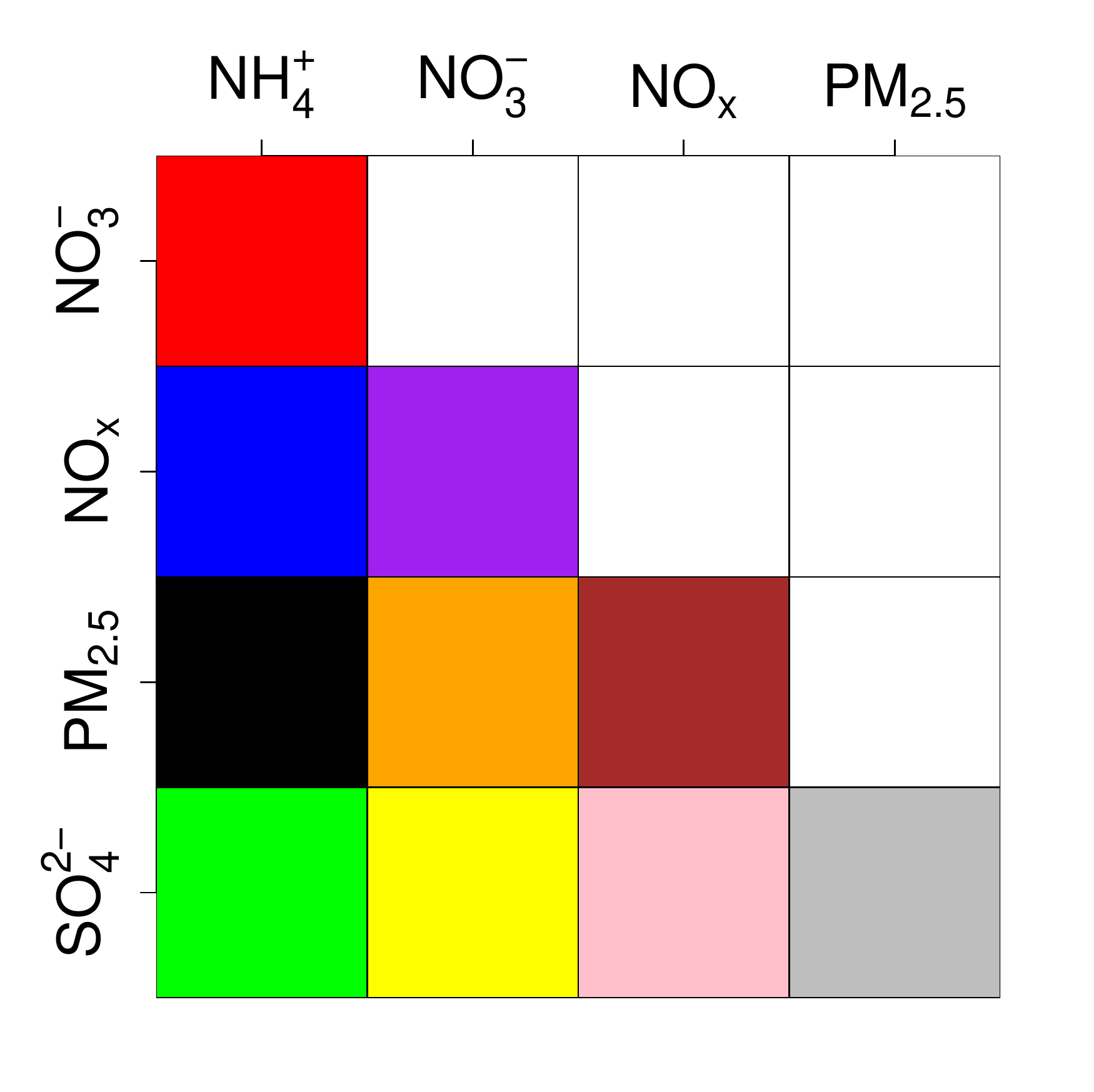}\\
\includegraphics[trim={0.5cm 0.5cm 1cm 0.5cm}, clip, scale=0.18]{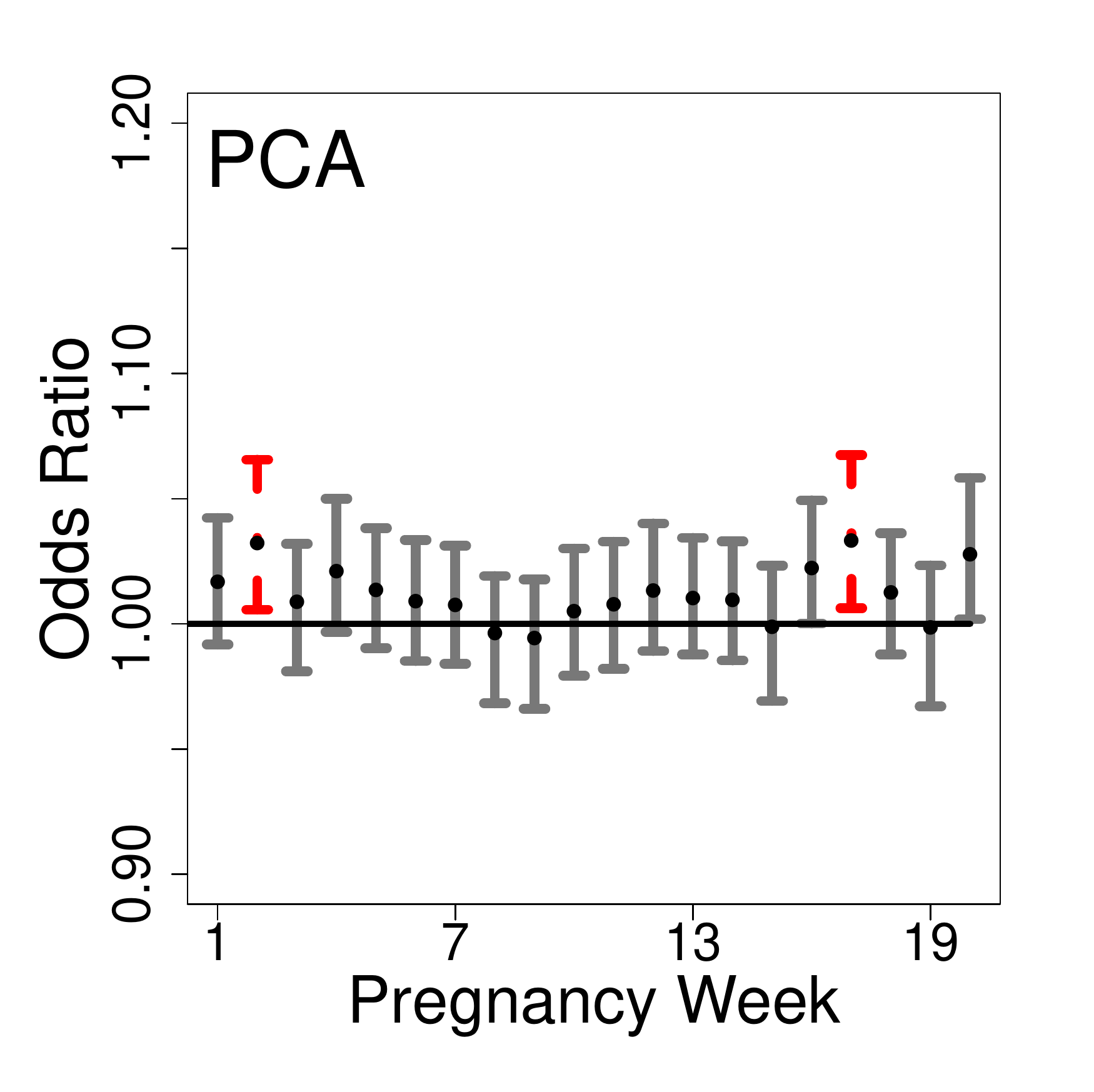}
\includegraphics[trim={0.5cm 0.5cm 1cm 0.5cm}, clip, scale=0.18]{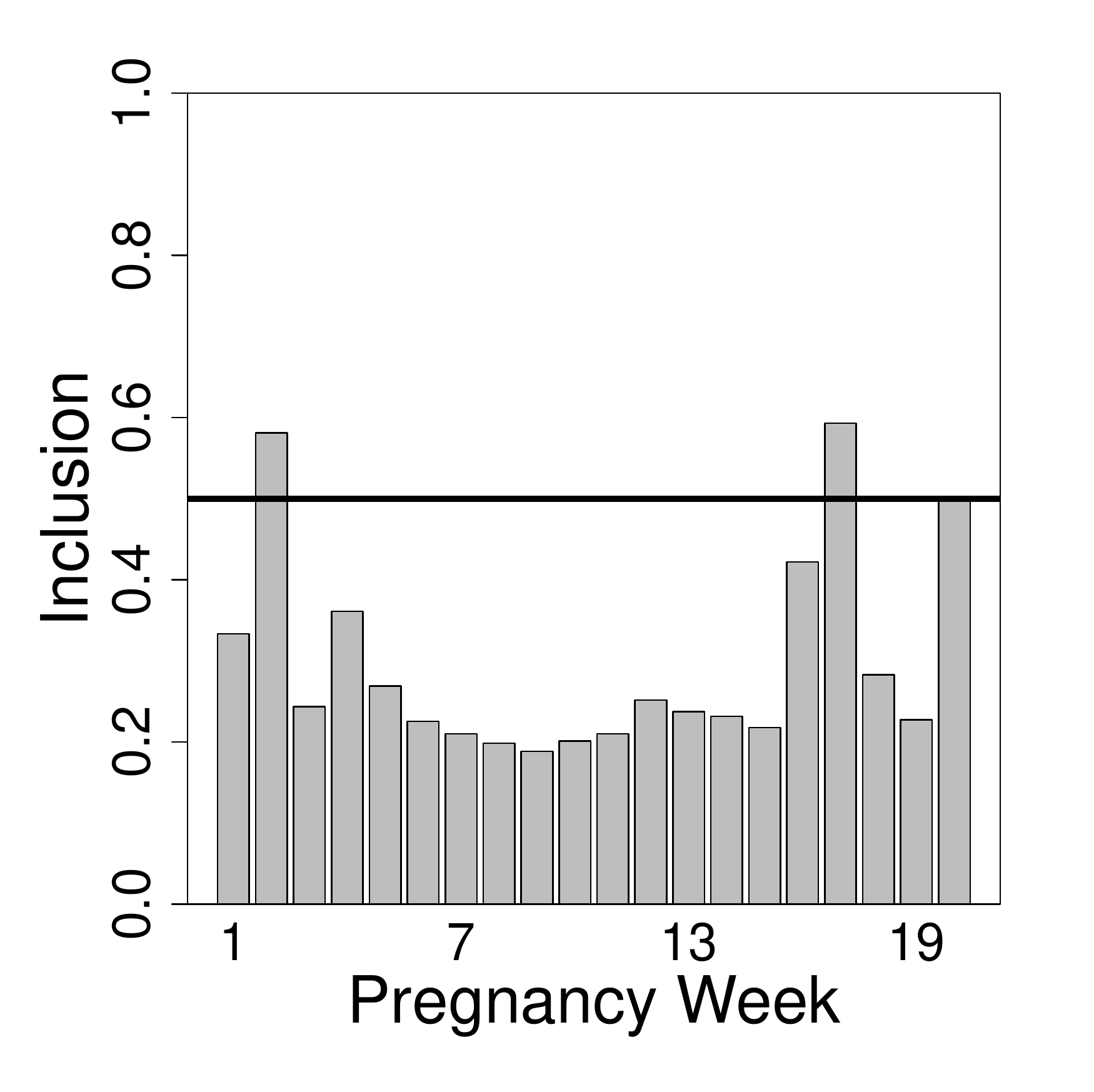}
\includegraphics[trim={0.25cm 0.5cm 1cm 0.5cm}, clip, scale=0.18]{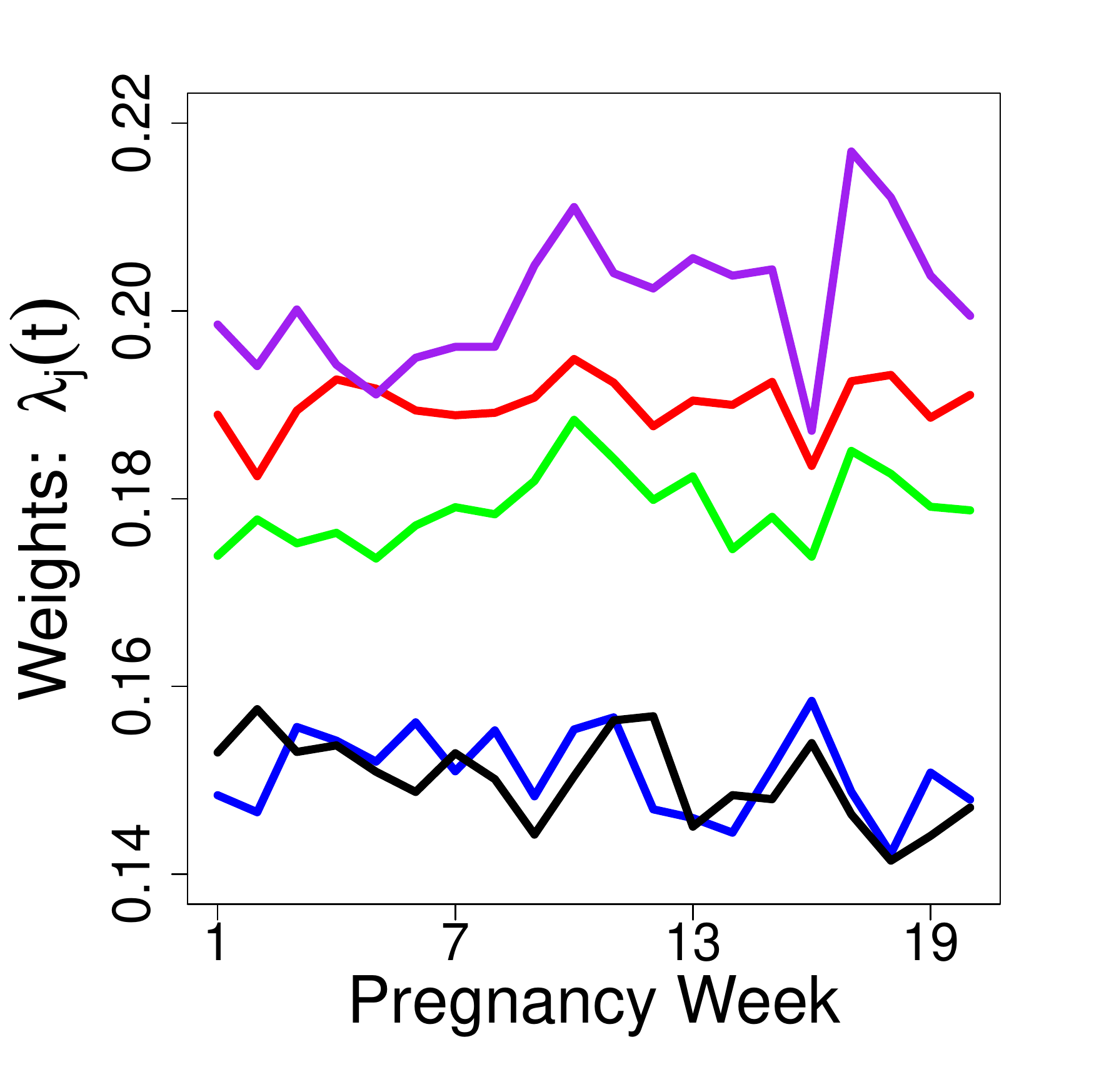}
\includegraphics[trim={0cm 0.5cm 1cm 0.5cm}, clip, scale=0.18]{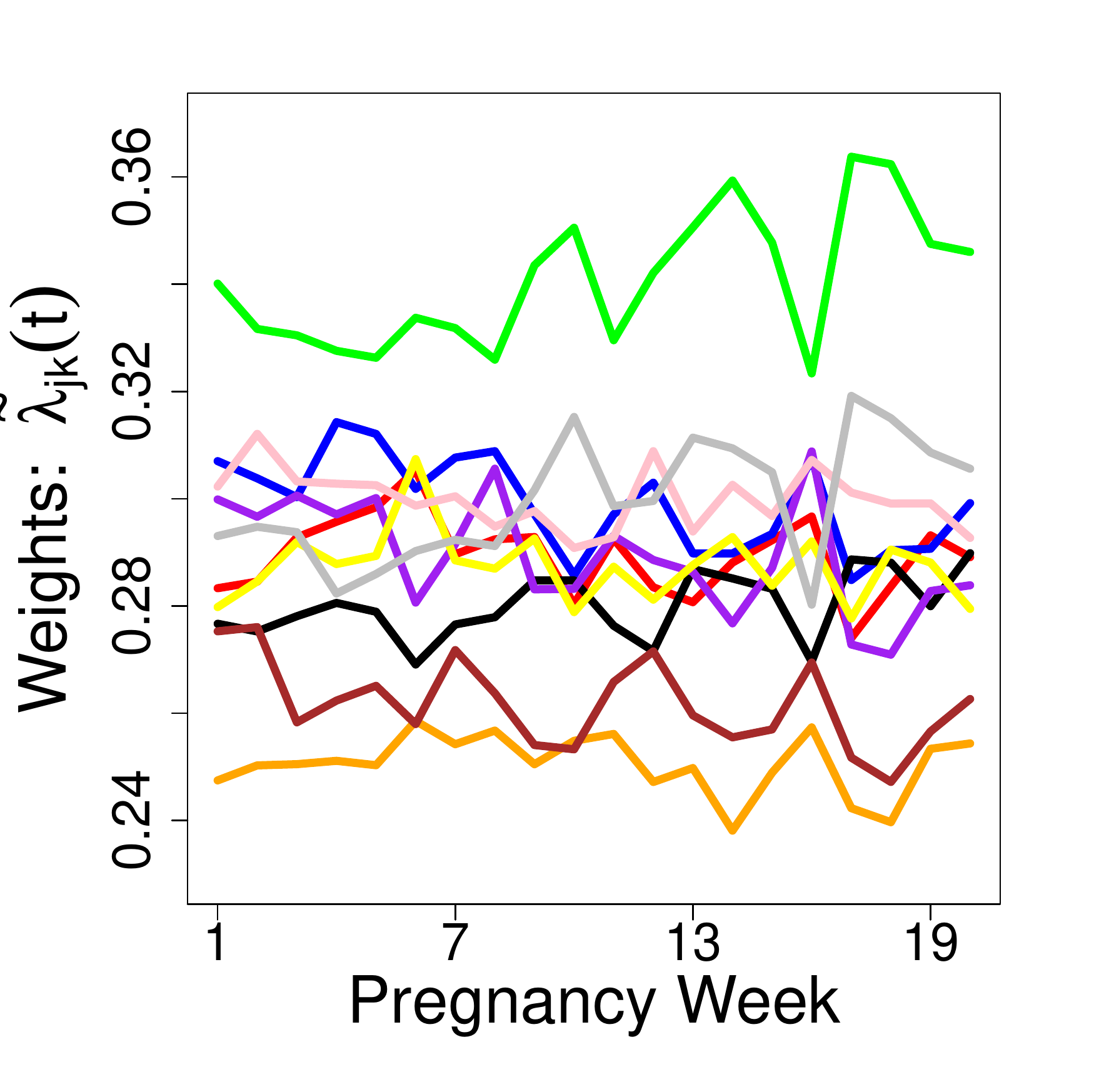}\\
\includegraphics[trim={0.5cm 0.5cm 1cm 0.5cm}, clip, scale=0.18]{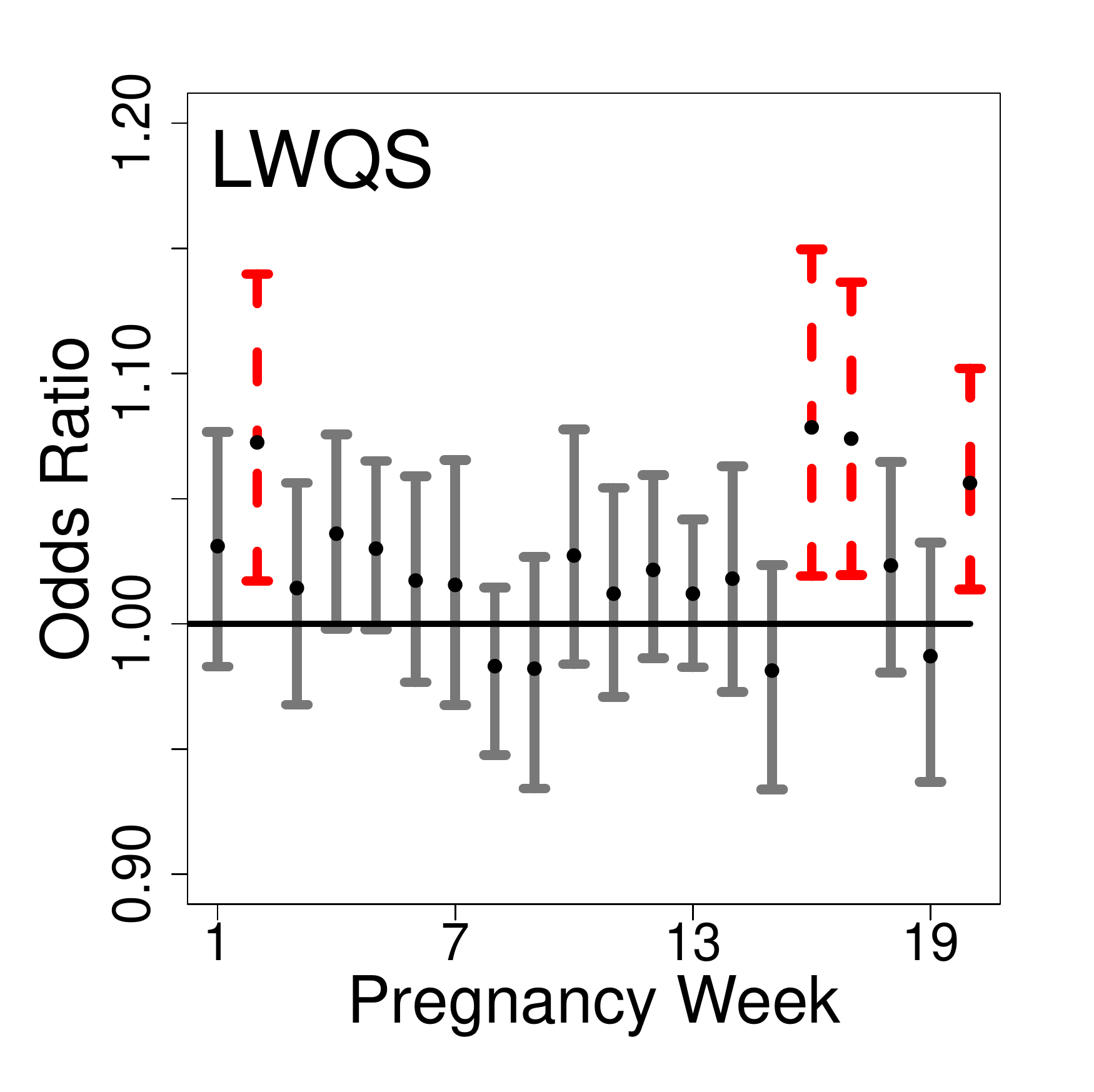}
\includegraphics[trim={0.5cm 0.5cm 1cm 0.5cm}, clip, scale=0.18]{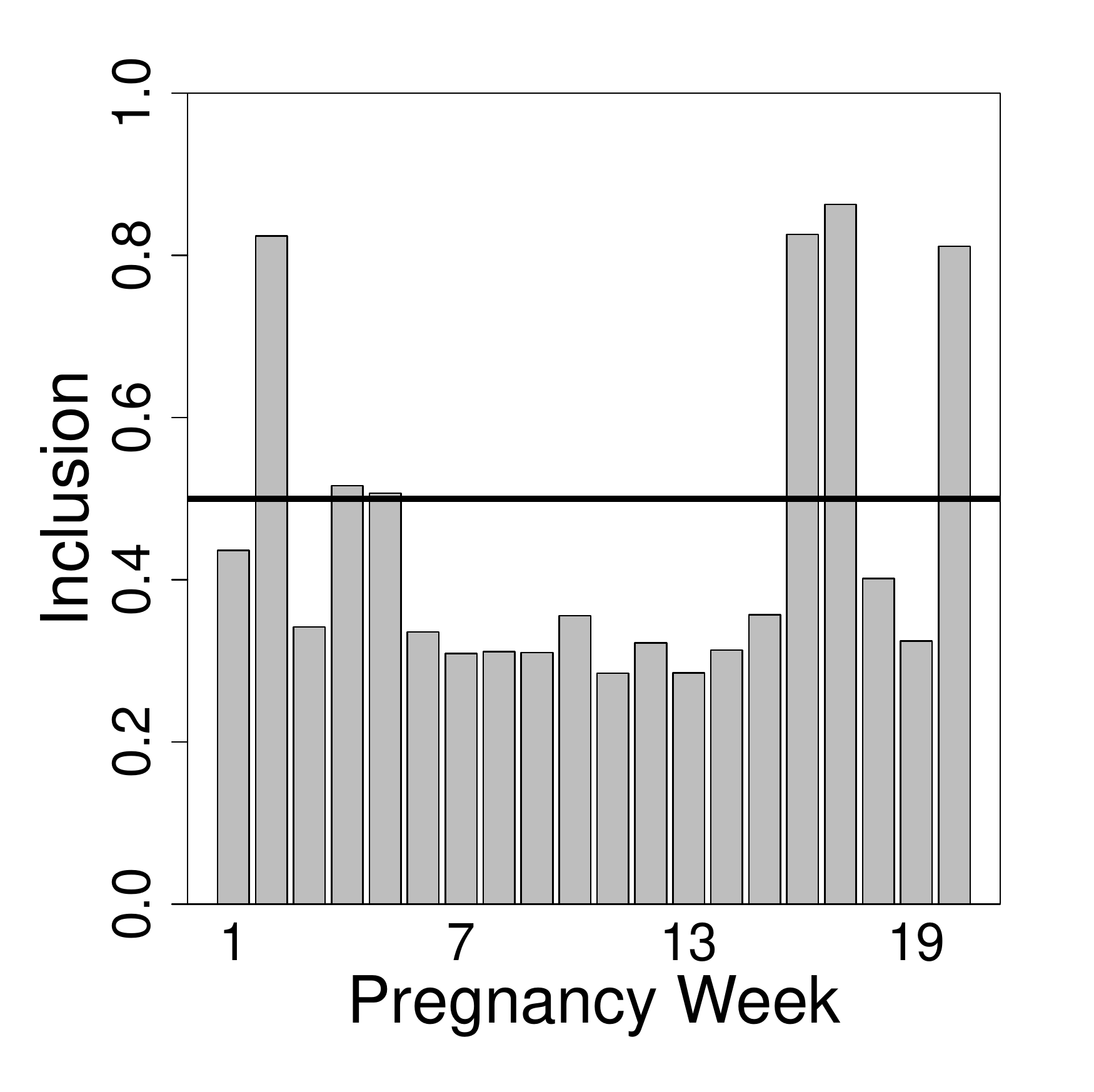}
\includegraphics[trim={0.25cm 0.5cm 1cm 0.5cm}, clip, scale=0.18]{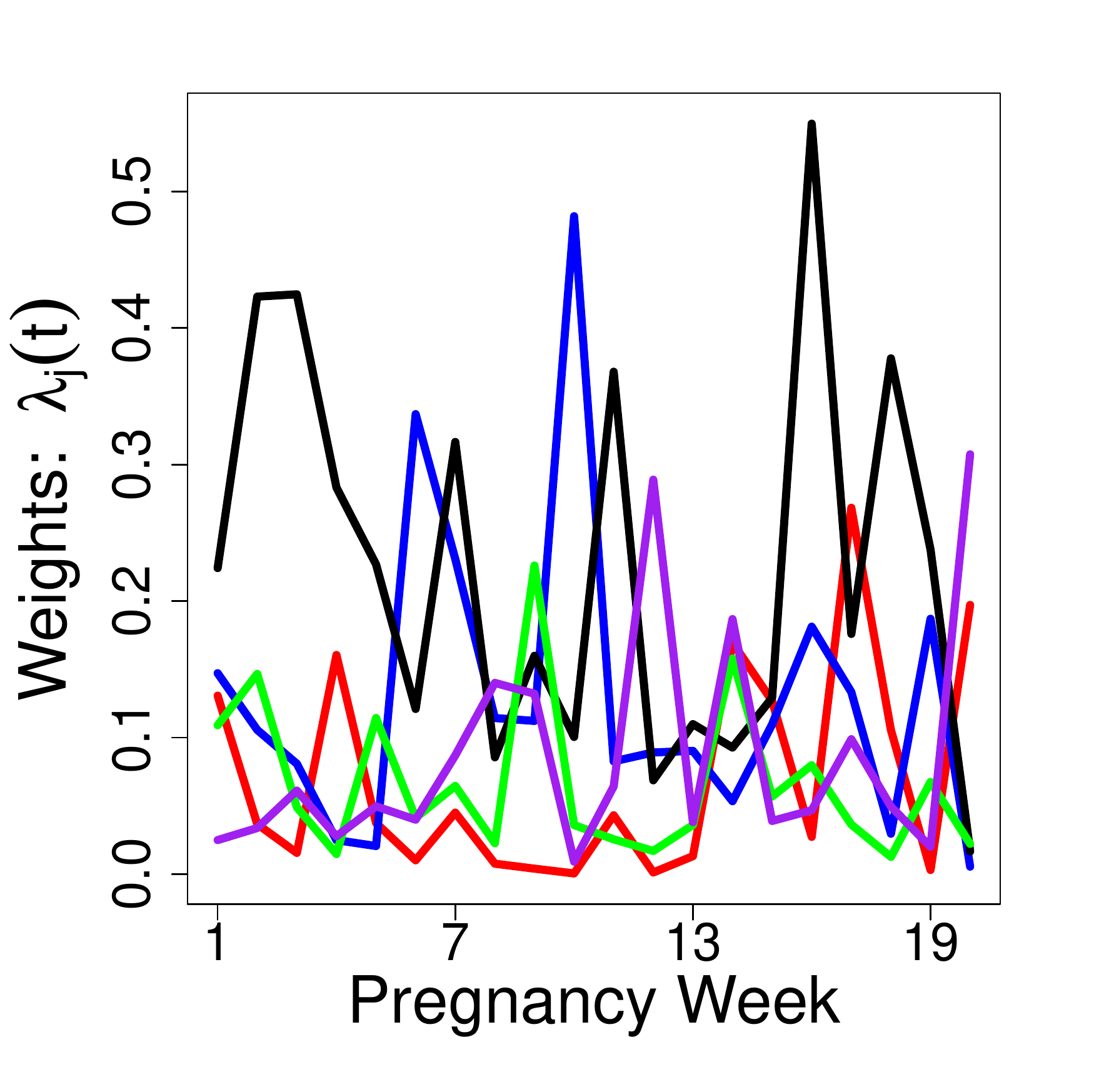}
\includegraphics[trim={0cm 0.5cm 1cm 0.5cm}, clip, scale=0.18]{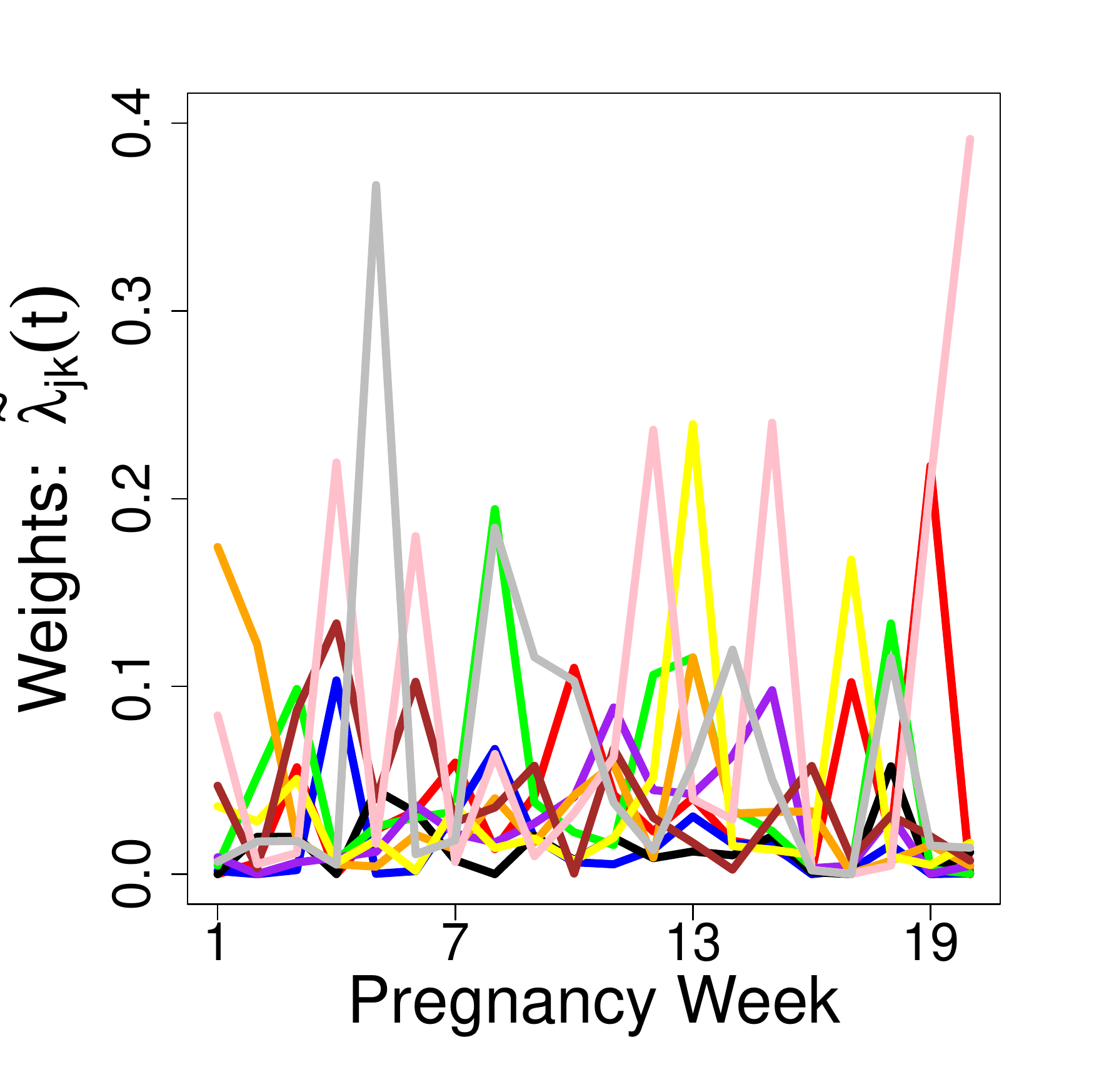}\\
\includegraphics[trim={0.5cm 0.5cm 1cm 0.5cm}, clip, scale=0.18]{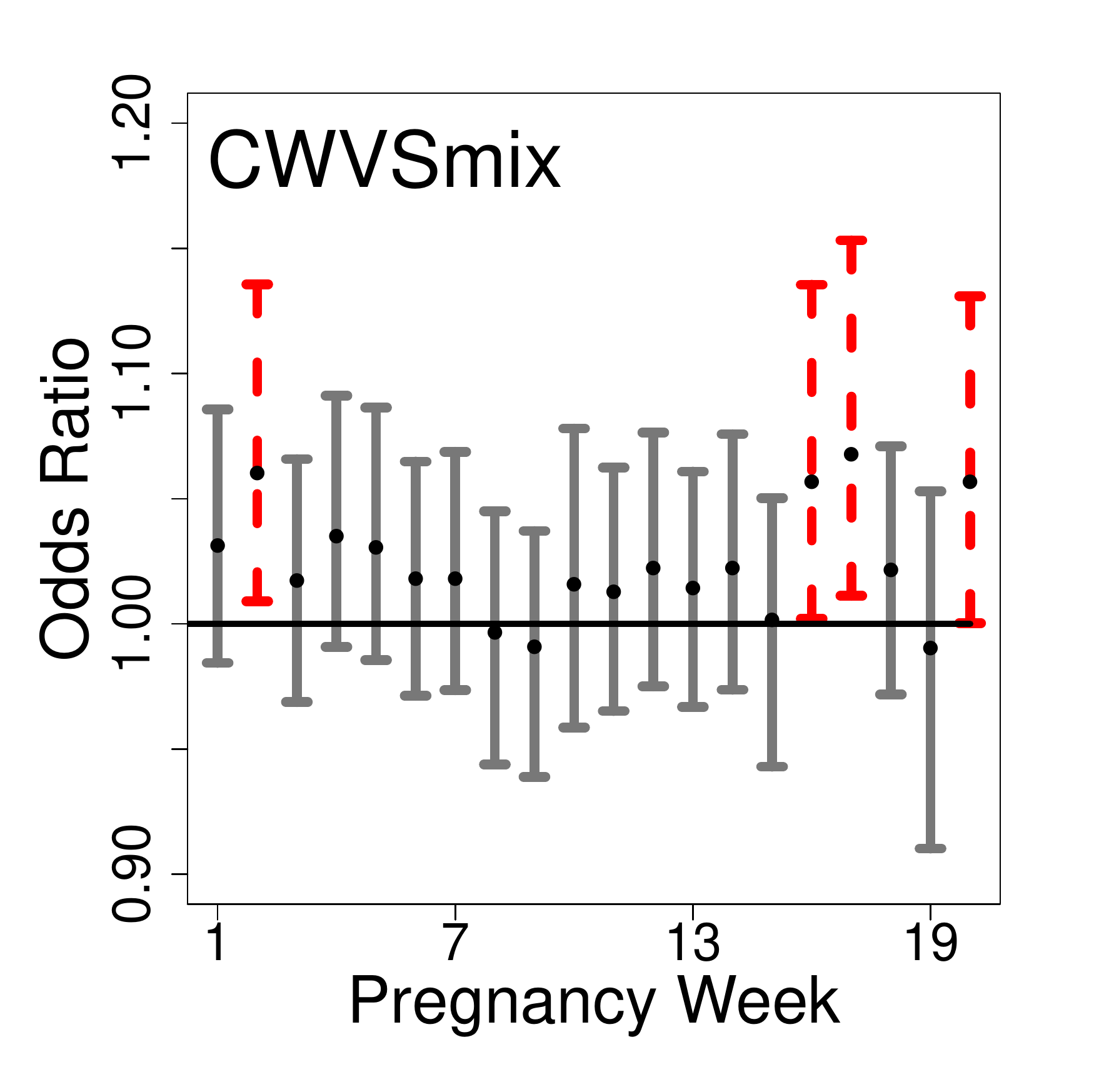}
\includegraphics[trim={0.5cm 0.5cm 1cm 0.5cm}, clip, scale=0.18]{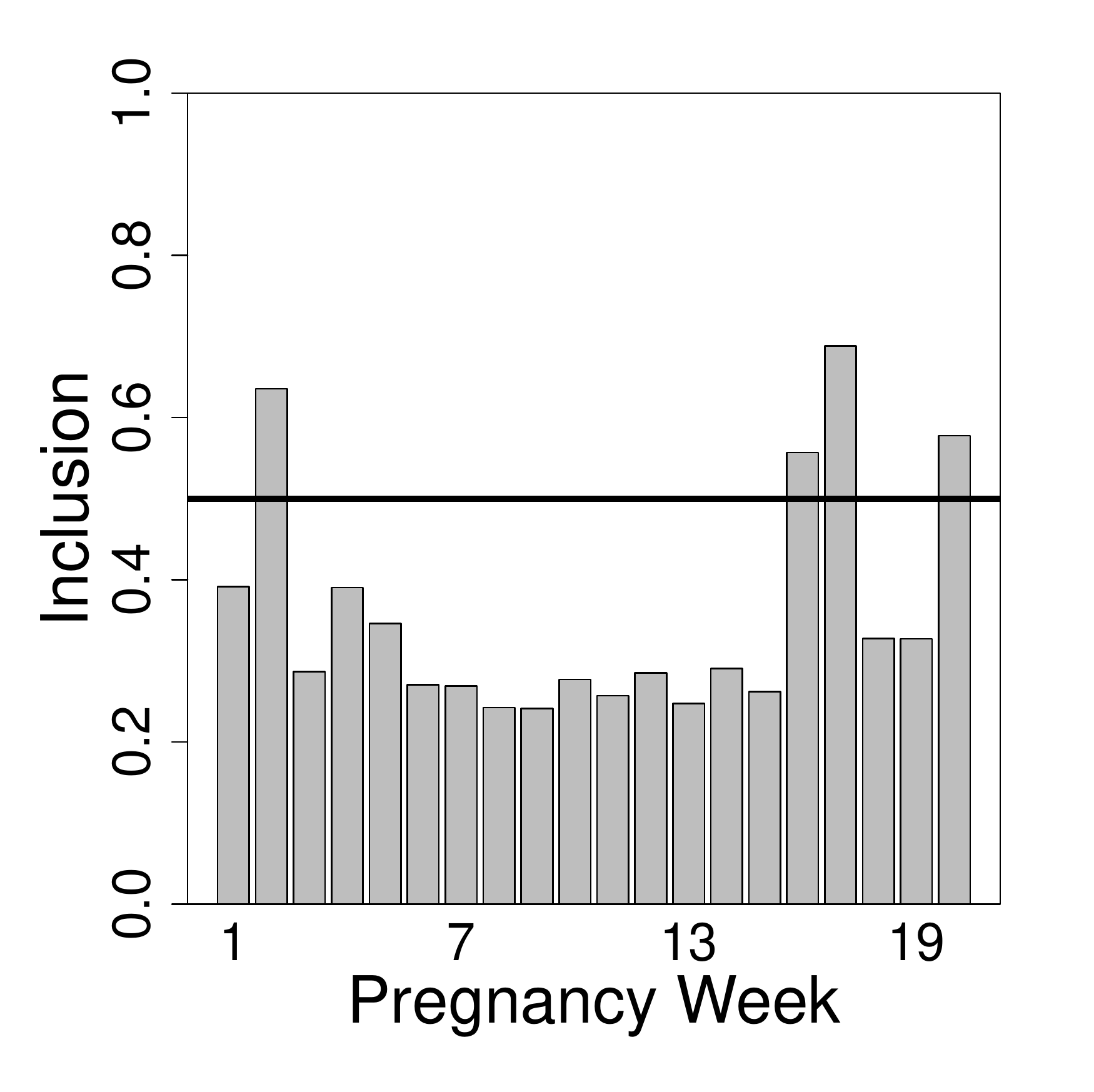}
\includegraphics[trim={0.25cm 0.5cm 1cm 0.5cm}, clip, scale=0.18]{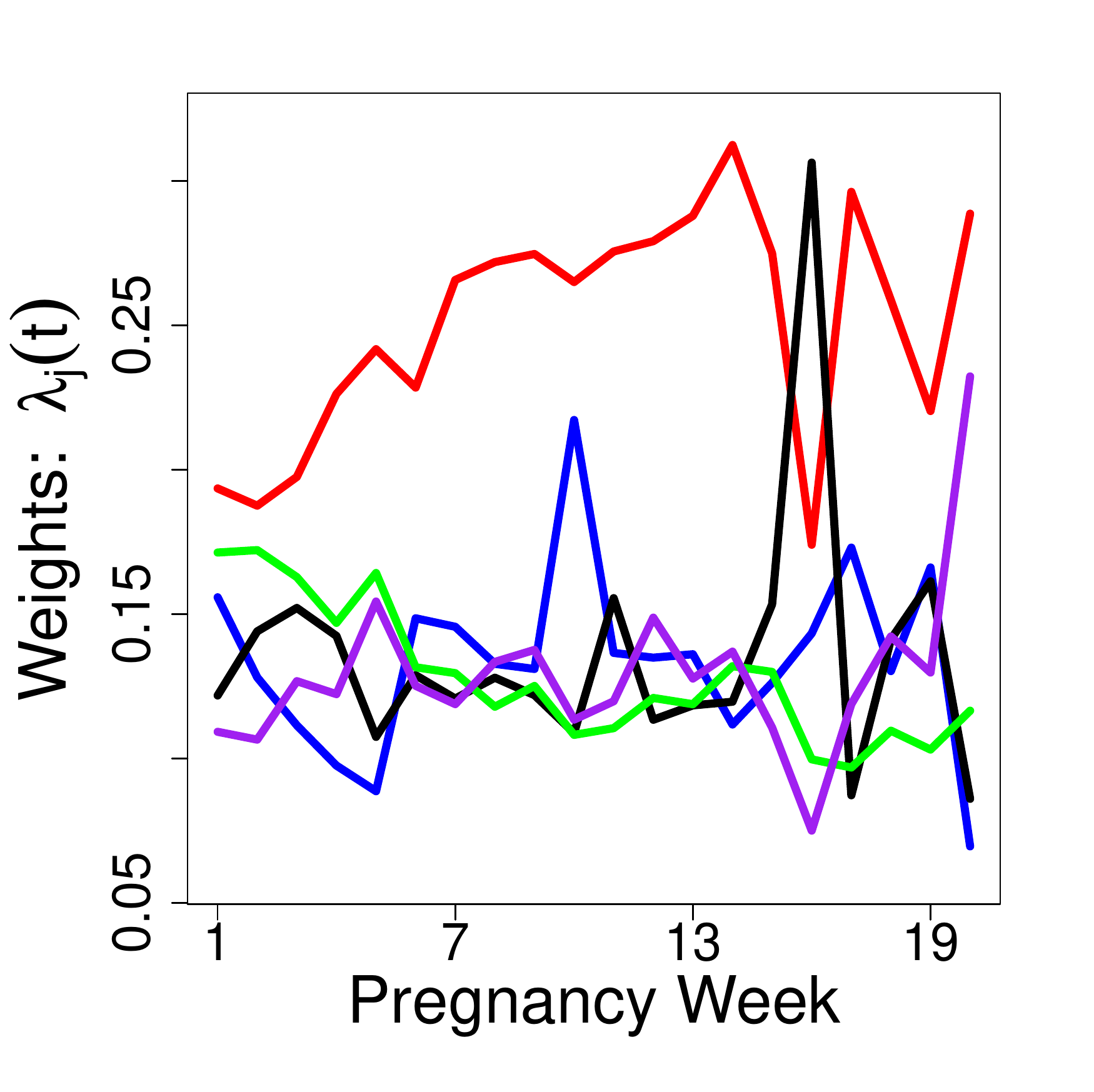}
\includegraphics[trim={0cm 0.5cm 1cm 0.5cm}, clip, scale=0.18]{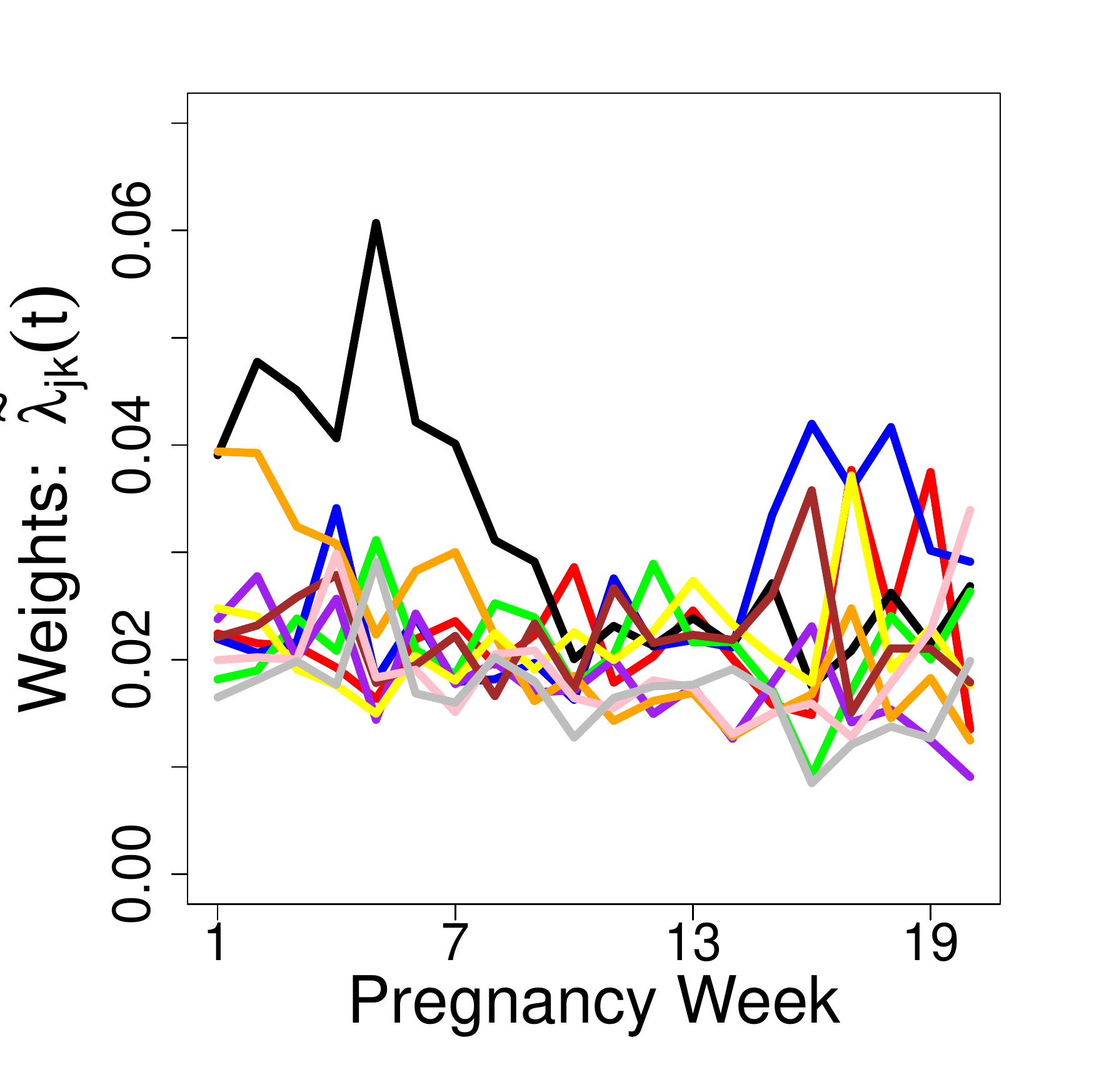}\\
\caption{Posterior means and 90\% credible intervals for the risk parameters (first column), posterior inclusion probabilities (second column), posterior means for the main effect weight parameters (third column), and posterior means for the interaction effect weight parameters (fourth column) for the \textbf{non-Hispanic Black} stillbirth and multiple exposures analyses in New Jersey, 2005-2014. Results based on an interquartile range increase in weekly exposure. Weeks identified as part of the critical window set are shown in red/dashed (harmful) and blue/dashed (protective).  The average posterior standard deviation for the weight parameters from CWVSmix is 0.13 (range: 0.04-0.37).}
\end{center}
\end{figure}

In Tables S4-S6 of the Supplement, we display posterior inference for the regression parameters across each race/ethnicity analysis using CWVSmix.  Consistent findings across the groups suggest that older mothers with less educational attainment are more likely to have a stillbirth; male babies are also more likely to result in a stillbirth.  While tobacco use during pregnancy had an odds ratio greater than one in each analysis, only for non-Hispanic White mothers did its credible interval exclude one.  

\begin{table}[ht]
\centering
\caption{Weight parameter variable selection results for critical weeks identified by CWVSmix in the non-Hispanic Black stillbirth and multiple exposures analyses in New Jersey, 2005-2014.  Posterior inclusion probabilities are presented with bolded entries indicating the pollutant/interaction was selected based on the procedure detailed in Section 4.1.}
\begin{tabular}{lrrrr}
\hline
 & \multicolumn{4}{c}{Pregnancy Week}\\
 \cline{2-5}
Effect                      & 2              & 16             & 17             & 20  \\
\hline
NH$^+_4$                    & \textbf{61.24} & 48.24          & \textbf{69.09} & \textbf{64.81} \\
NO$^-_3$                    & \textbf{57.37} & 47.90          & \textbf{58.01} & 32.66 \\
NO$_{\text{x}}$             & \textbf{52.79} & \textbf{69.46} & 40.38          & 39.19 \\
PM$_{2.5}$                  & \textbf{62.17} & 40.79          & 42.15          & 42.70 \\
SO$^{2-}_4$                 & 44.36          & 34.51          & 46.79          & \textbf{61.45} \\
\hline
NH$^+_4$ $\times$ NO$^-_3$           & \textbf{14.30} & 10.18          & \textbf{18.79} &  9.61 \\
NH$^+_4$ $\times$ NO$_{\text{x}}$    & \textbf{15.00} & 16.12          & 14.89          & 13.76 \\
NH$^+_4$ $\times$ PM$_{2.5}$         & \textbf{24.39} & 10.65          & 13.41          & 15.20 \\
NH$^+_4$ $\times$ SO$^{2-}_4$        & 13.30          &  6.91          & 11.90          & \textbf{15.79} \\
\hline
NO$^-_3$ $\times$ NO$_{\text{x}}$    & \textbf{16.88} & 13.45          & 10.46          &  7.53 \\
NO$^-_3$ $\times$ PM$_{2.5}$         & \textbf{21.23} & 10.56          & 14.12          &  8.69 \\
NO$^-_3$ $\times$ SO$^{2-}_4$        & 15.55          & 10.18          & 17.57          & 10.80 \\
\hline
NO$_{\text{x}}$ $\times$ PM$_{2.5}$  & \textbf{16.38} & 16.95          & 9.50           & 11.29 \\
NO$_{\text{x}}$ $\times$ SO$^{2-}_4$ & 14.29          &  9.62          & 9.65           & 15.27 \\
\hline
PM$_{2.5}$ $\times$ SO$^{2-}_4$      & 13.22          &  6.62          & 8.94           & 12.41 \\
\hline
\end{tabular}
\end{table}

In Figures 2 and 3, and Figure S10 in the Supplement, we display the results by race/ethnicity for each of the different multipollutant approaches (i.e., estimated risk parameters, posterior inclusion probabilities, and estimated weight parameters where applicable).  As in the simulation study and \cite{warren2019critical}, we present inference for the risk and weight parameters on a specific exposure period given that it was included in the model (i.e., $\gamma\left(t\right) = 1$).  For non-Hispanic Black mothers, all methods indicate that elevated exposures during gestational weeks 2 and 17 are associated with increased risk of stillbirth.  All methods other than PCA include week 20 in the critical set, while CWVSmix and LWQS additionally include week 16.  

To better understand the important drivers of this risk, we present the posterior inclusion probabilities (see Section 4.1) for the weights during each of the estimated critical weeks from CWVSmix in Table 3.  Recall that no other method performs variable selection at this level of the model.  During gestational week 2, the estimated mixture profile is mainly driven by NH$_4^+$, NO$_3^-$, NO$_{\text{x}}$, PM$_{2.5}$, and many of the first order interactions.  In week 16, the risk is driven by NO$_{\text{x}}$, while in 17 it is due to NH$_4^+$, NO$_3^-$, and their interaction.  NH$_4^+$, SO$^{2-}_4$, and their interaction dominate the risk on week 20.  The estimated weights from LWQS often align with those from CWVSmix, though without the ability to carry out variable selection it becomes more difficult to interpret the findings.  The PCA method gives similar weights to all pollutants and is consistent across all race/ethnicity groups as it does not consider the outcome when defining the weights.

\begin{figure}
\begin{center}
\includegraphics[trim={0.5cm 0.5cm 1cm 0.5cm}, clip, scale=0.18]{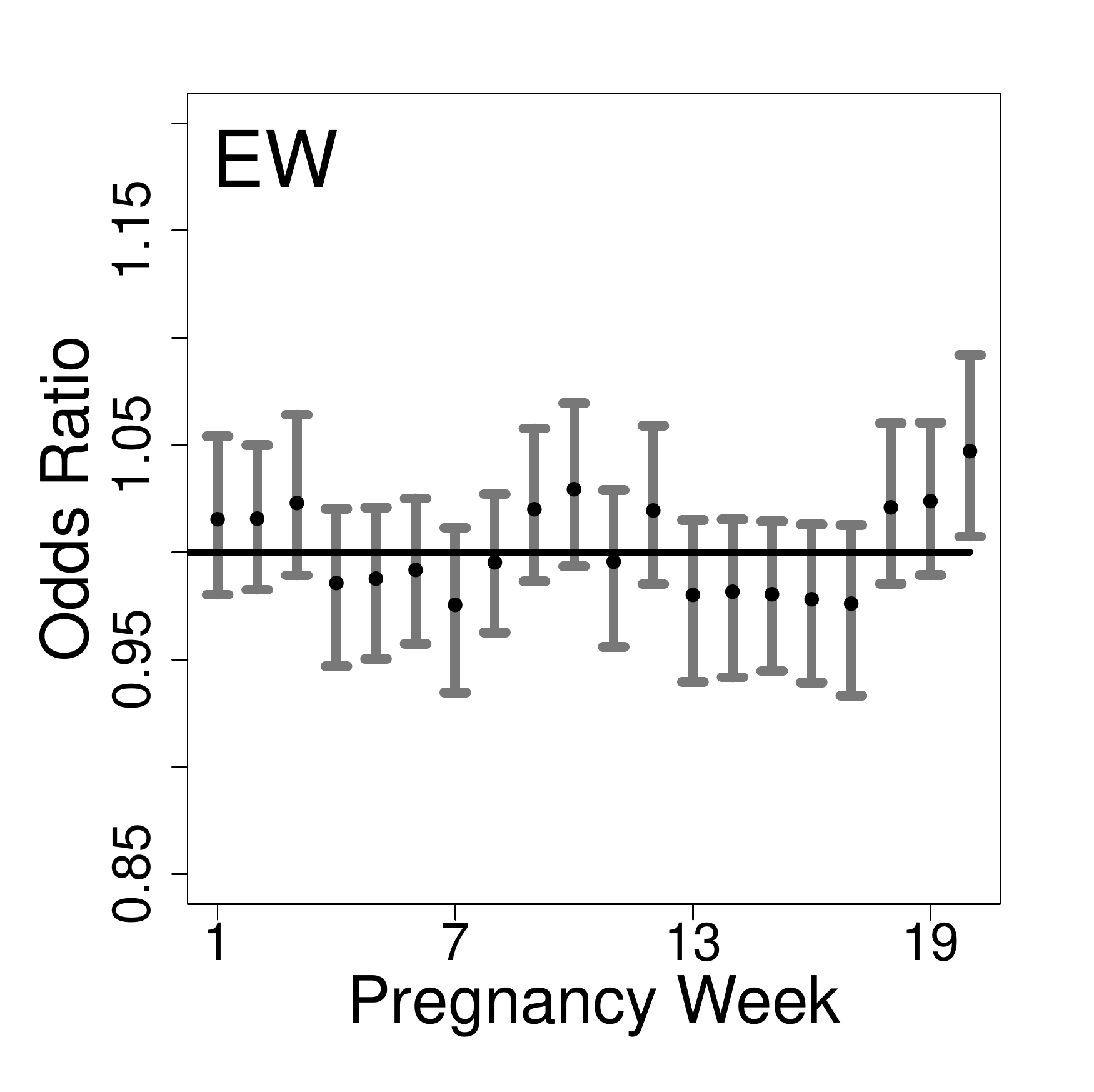}
\includegraphics[trim={0.5cm 0.5cm 1cm 0.5cm}, clip, scale=0.18]{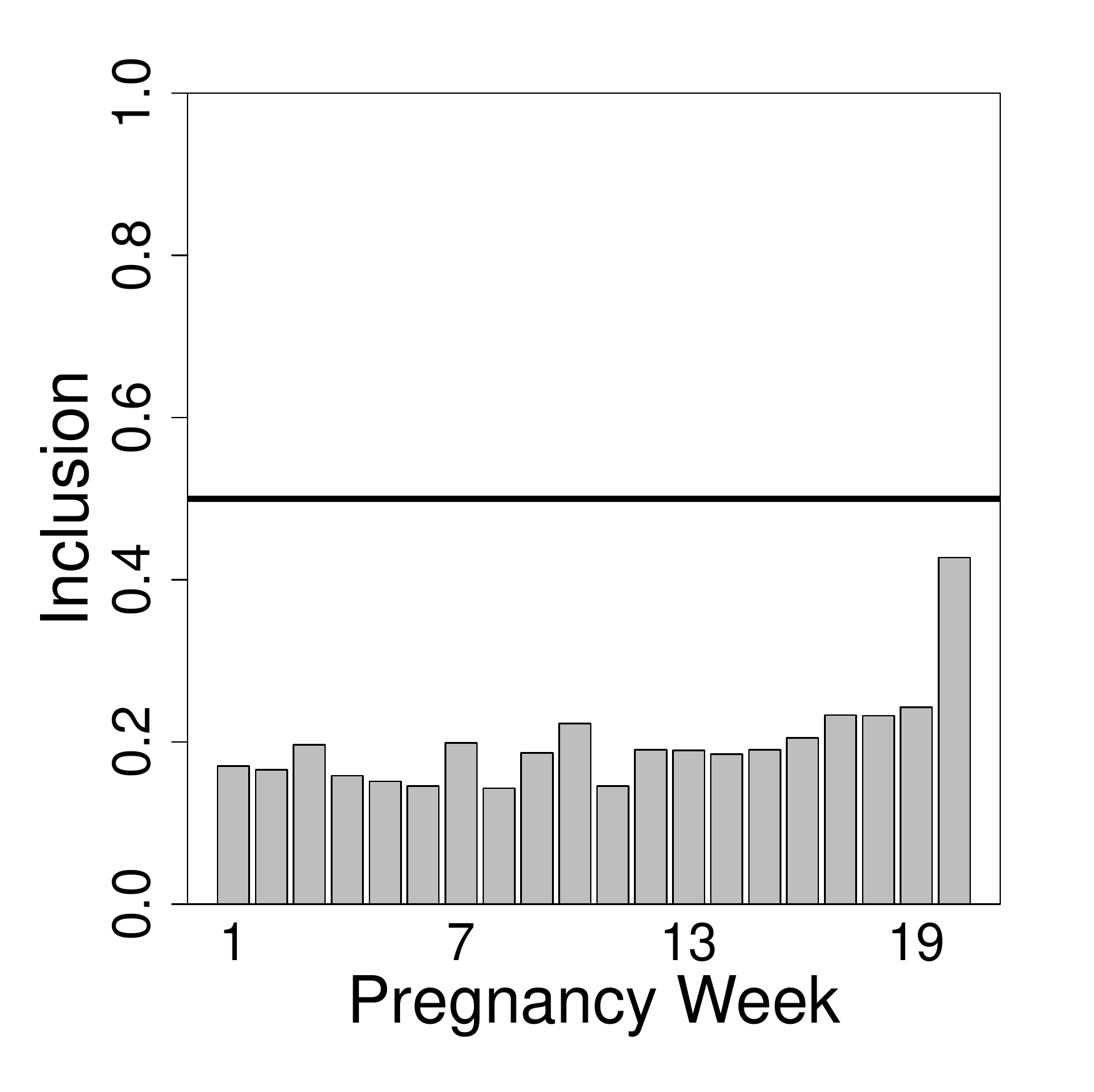}
\includegraphics[trim={0.25cm 0.5cm 1cm 0.5cm}, clip, scale=0.18]{Figures/color_key.pdf}
\includegraphics[trim={0cm 0.5cm 1cm 0.5cm}, clip, scale=0.18]{Figures/color_key-interactions.pdf}\\
\includegraphics[trim={0.5cm 0.5cm 1cm 0.5cm}, clip, scale=0.18]{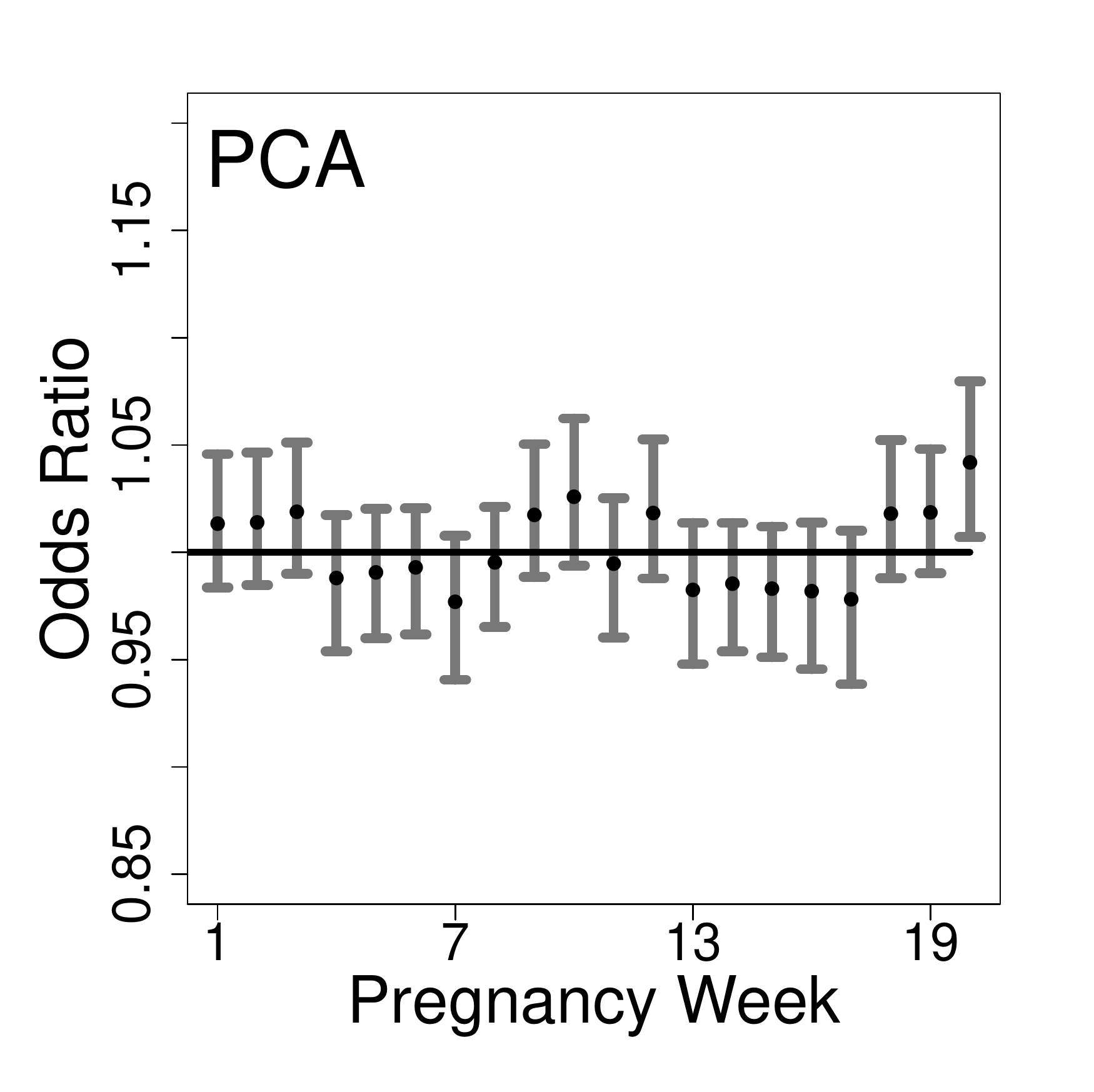}
\includegraphics[trim={0.5cm 0.5cm 1cm 0.5cm}, clip, scale=0.18]{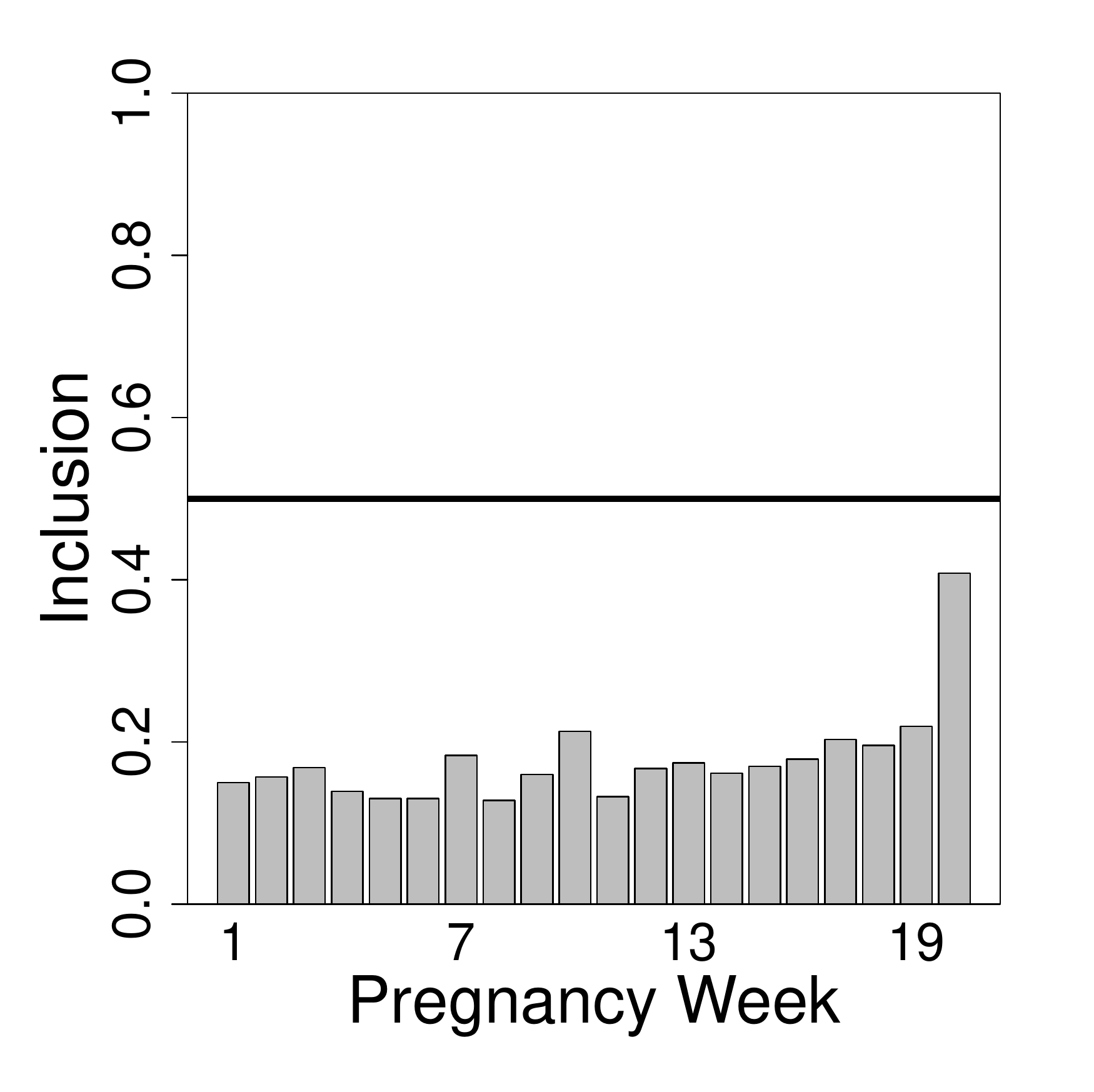}
\includegraphics[trim={0.25cm 0.5cm 1cm 0.5cm}, clip, scale=0.18]{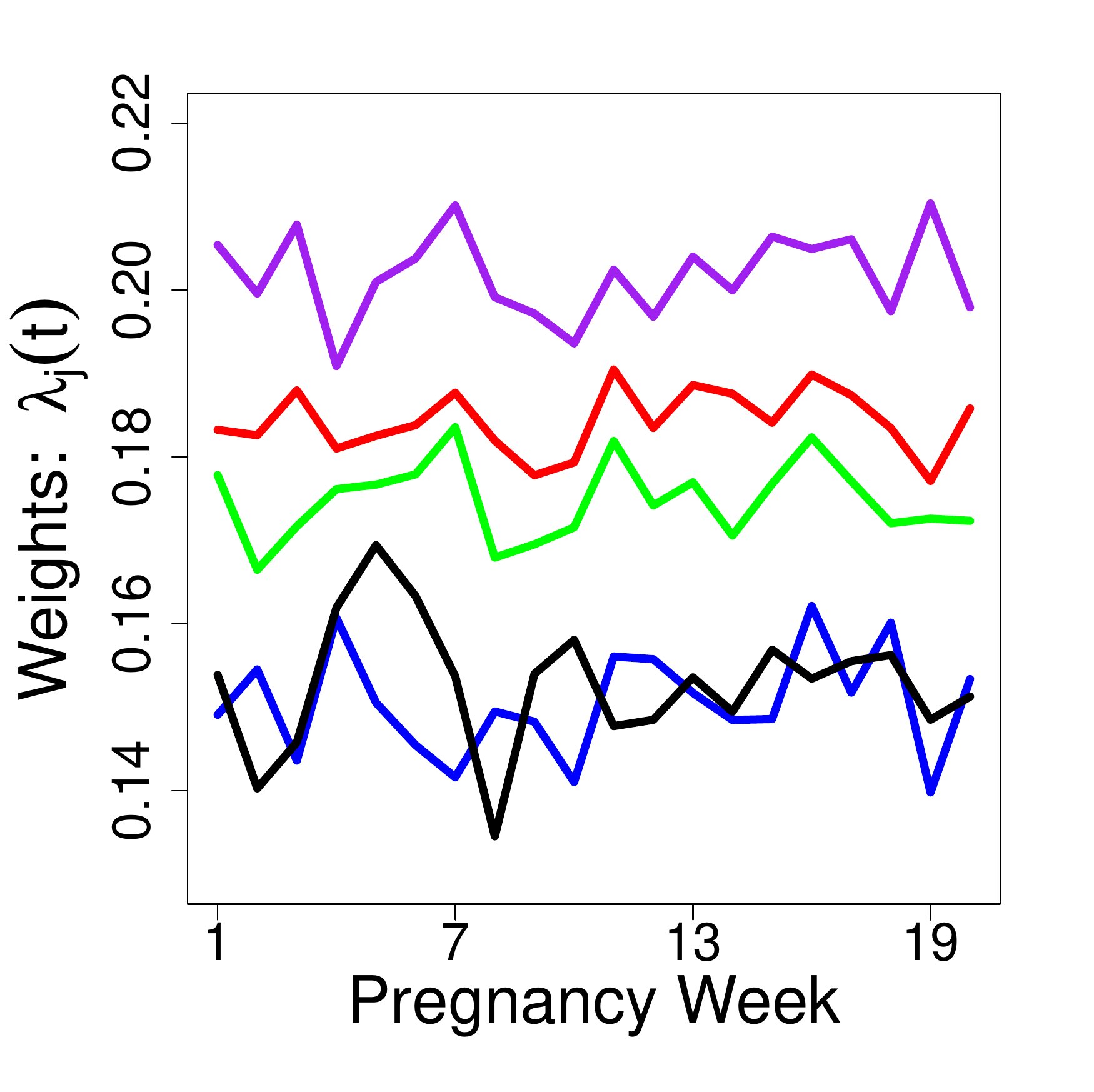}
\includegraphics[trim={0cm 0.5cm 1cm 0.5cm}, clip, scale=0.18]{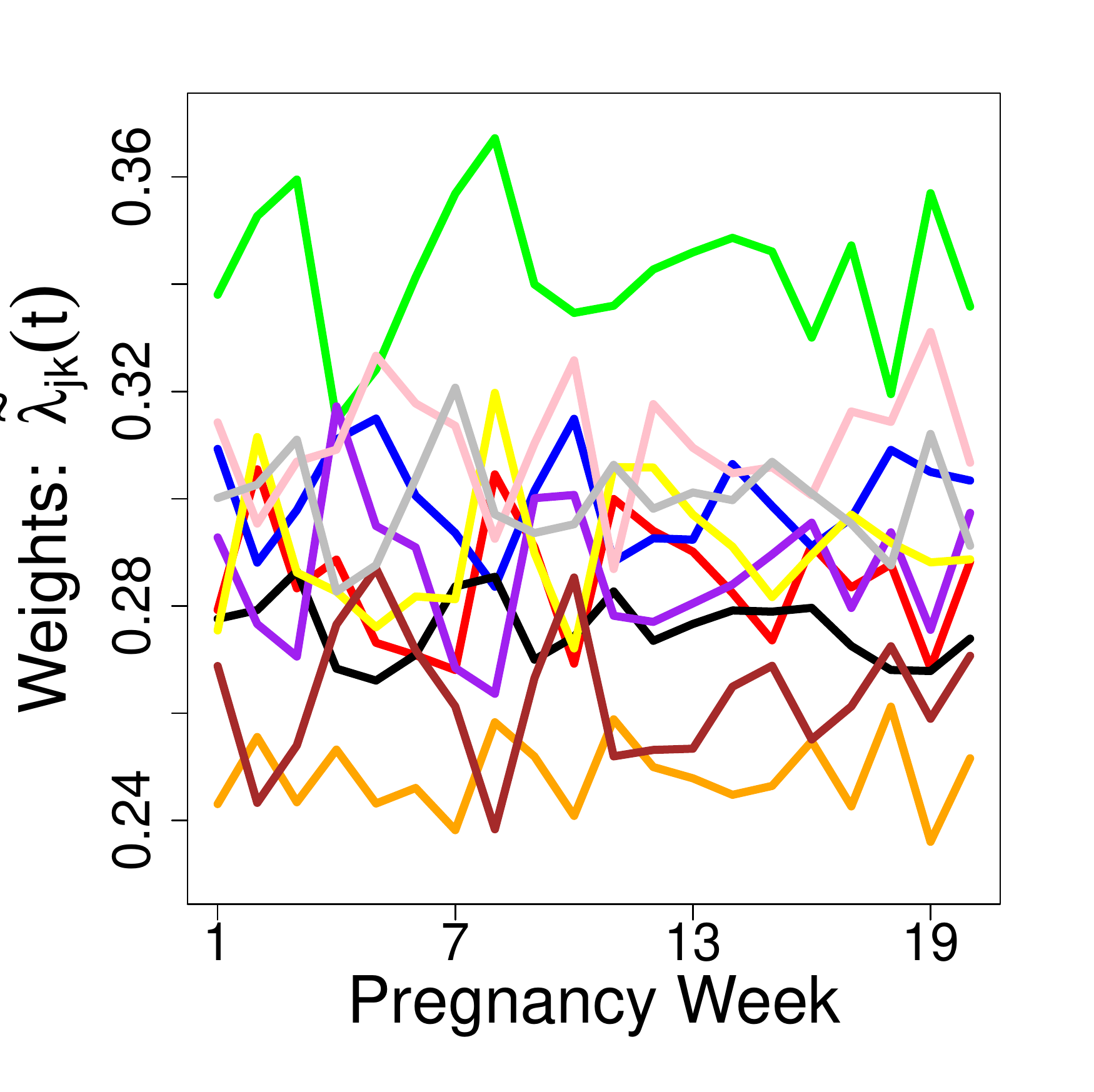}\\
\includegraphics[trim={0.5cm 0.5cm 1cm 0.5cm}, clip, scale=0.18]{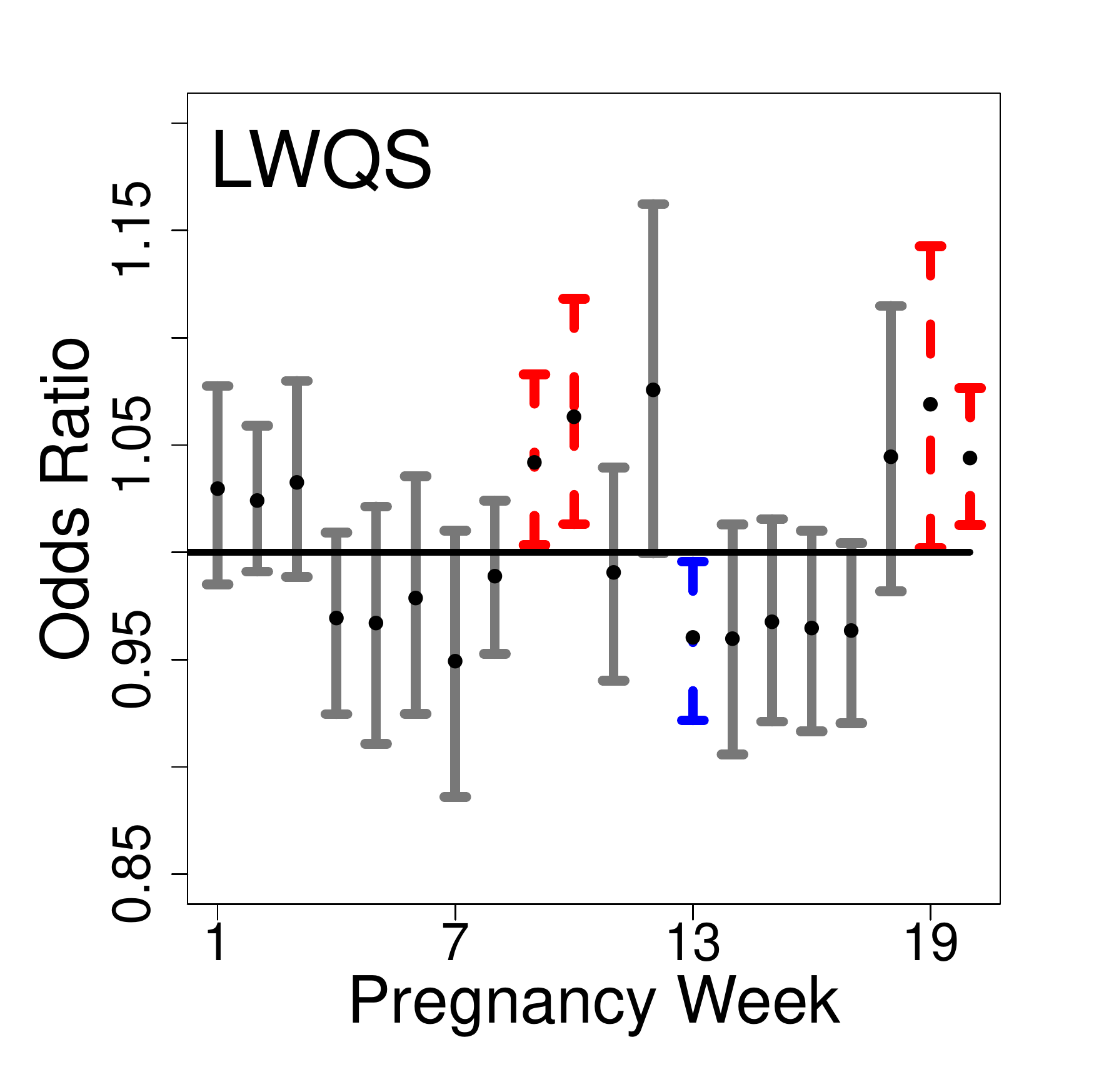}
\includegraphics[trim={0.5cm 0.5cm 1cm 0.5cm}, clip, scale=0.18]{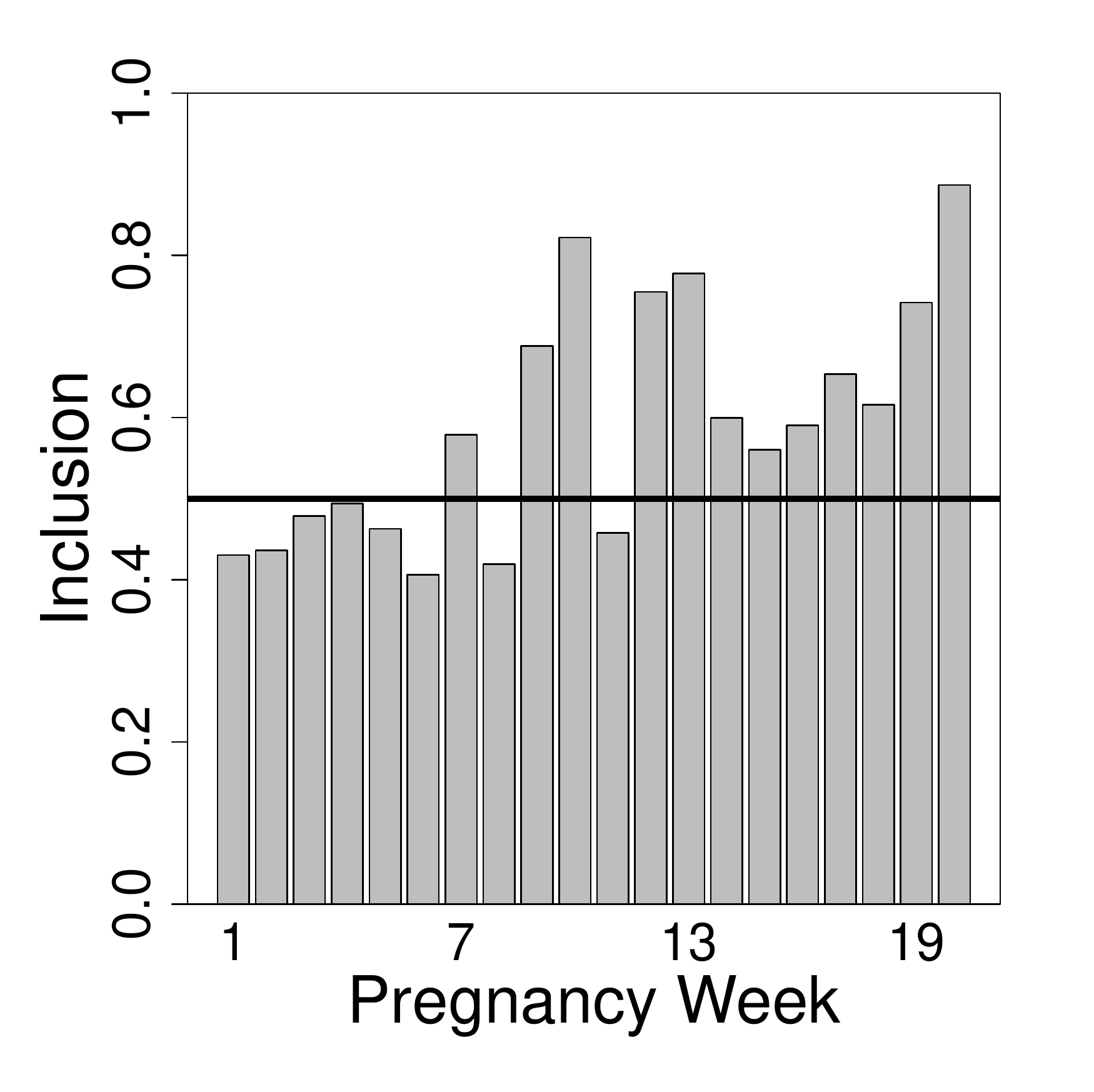}
\includegraphics[trim={0.25cm 0.5cm 1cm 0.5cm}, clip, scale=0.18]{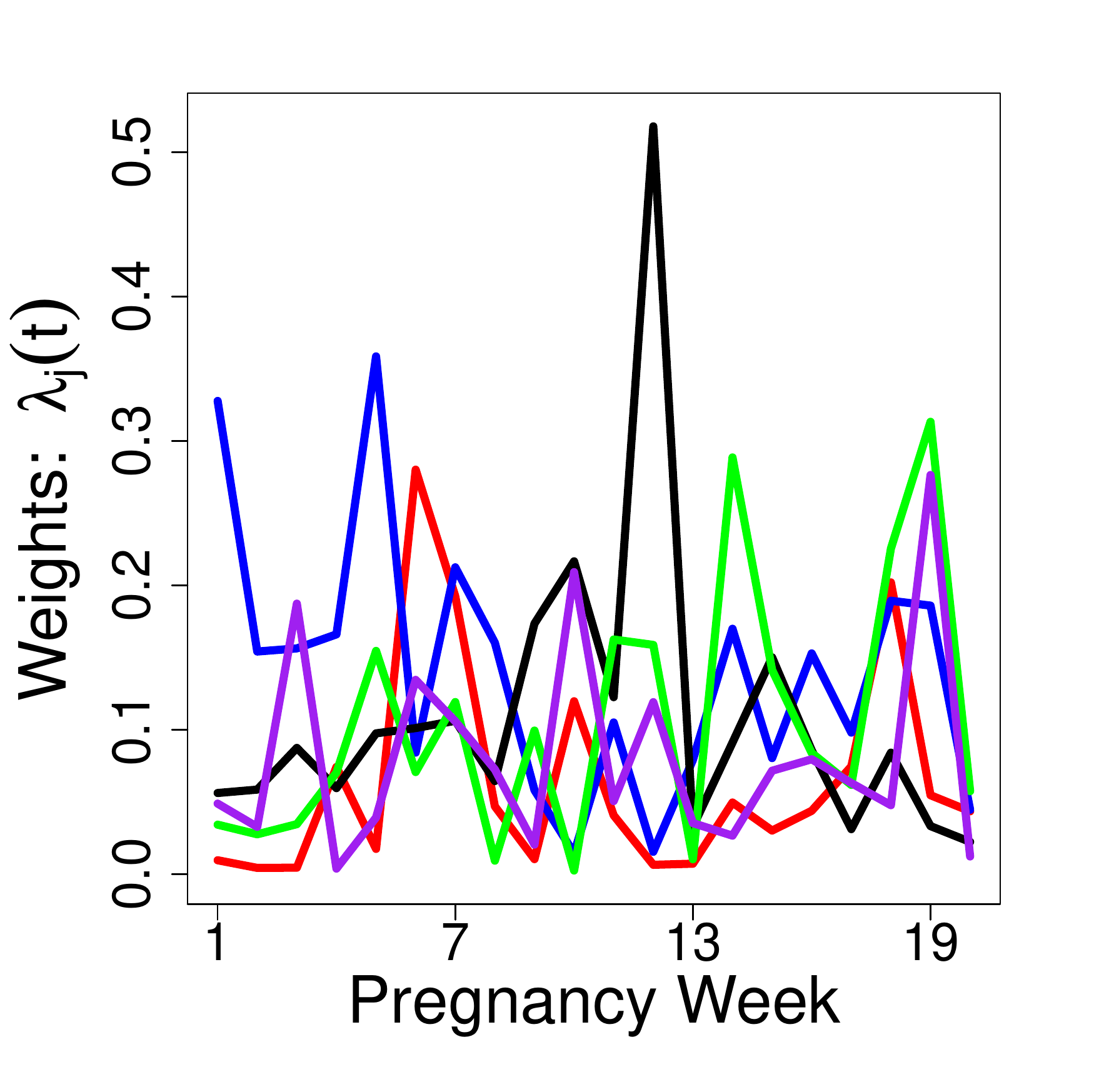}
\includegraphics[trim={0cm 0.5cm 1cm 0.5cm}, clip, scale=0.18]{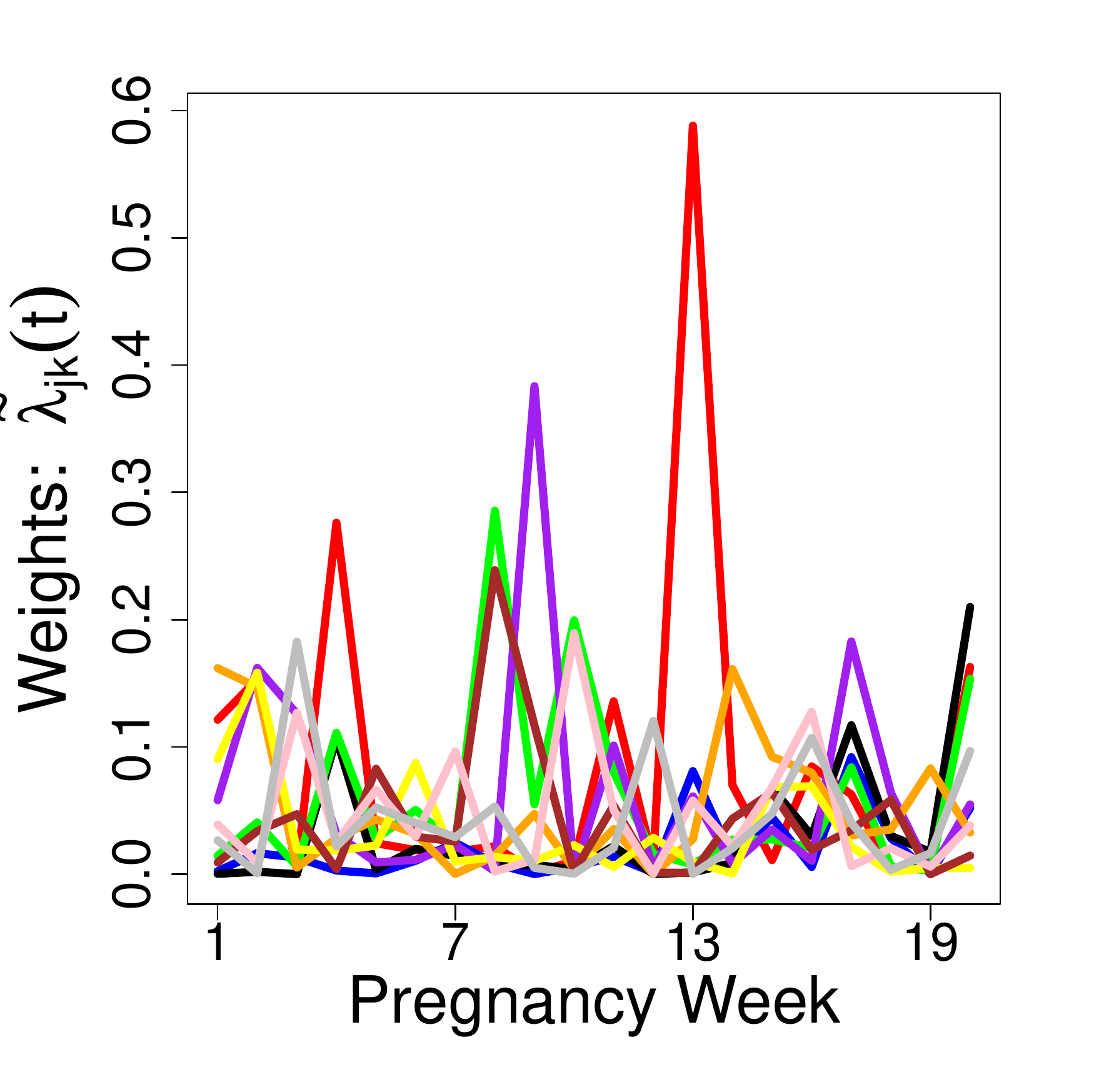}\\
\includegraphics[trim={0.5cm 0.5cm 1cm 0.5cm}, clip, scale=0.18]{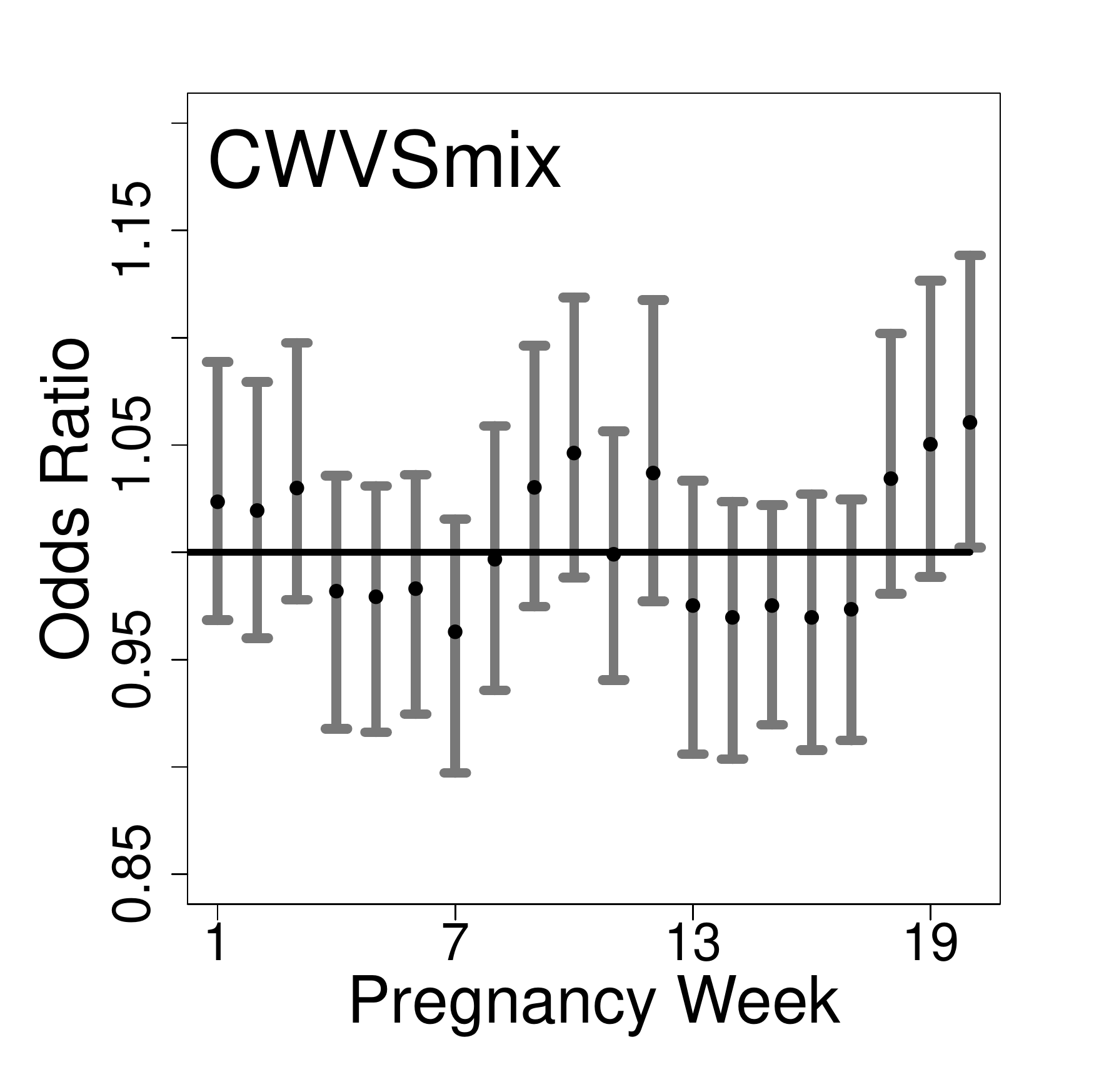}
\includegraphics[trim={0.5cm 0.5cm 1cm 0.5cm}, clip, scale=0.18]{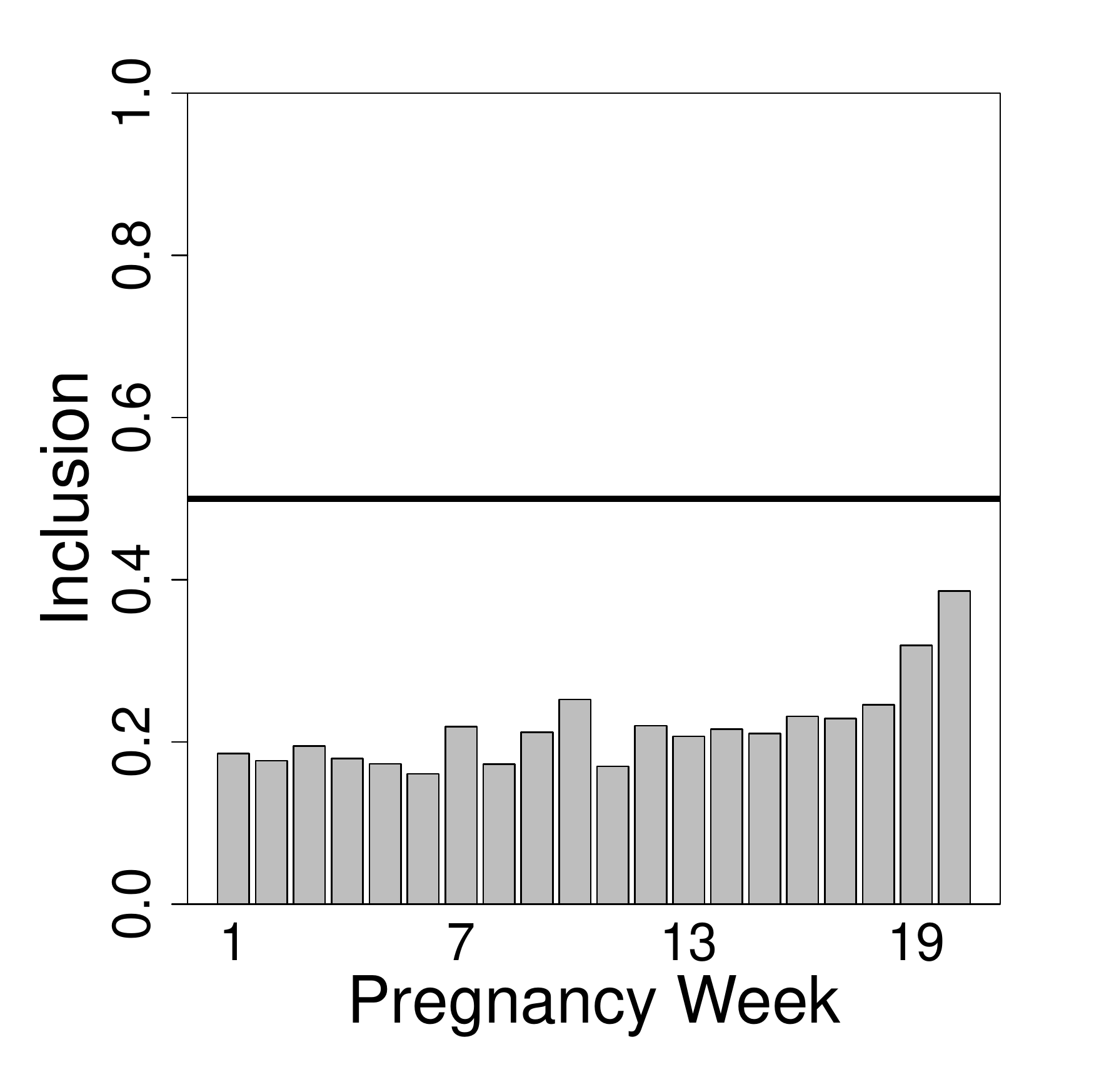}
\includegraphics[trim={0.25cm 0.5cm 1cm 0.5cm}, clip, scale=0.18]{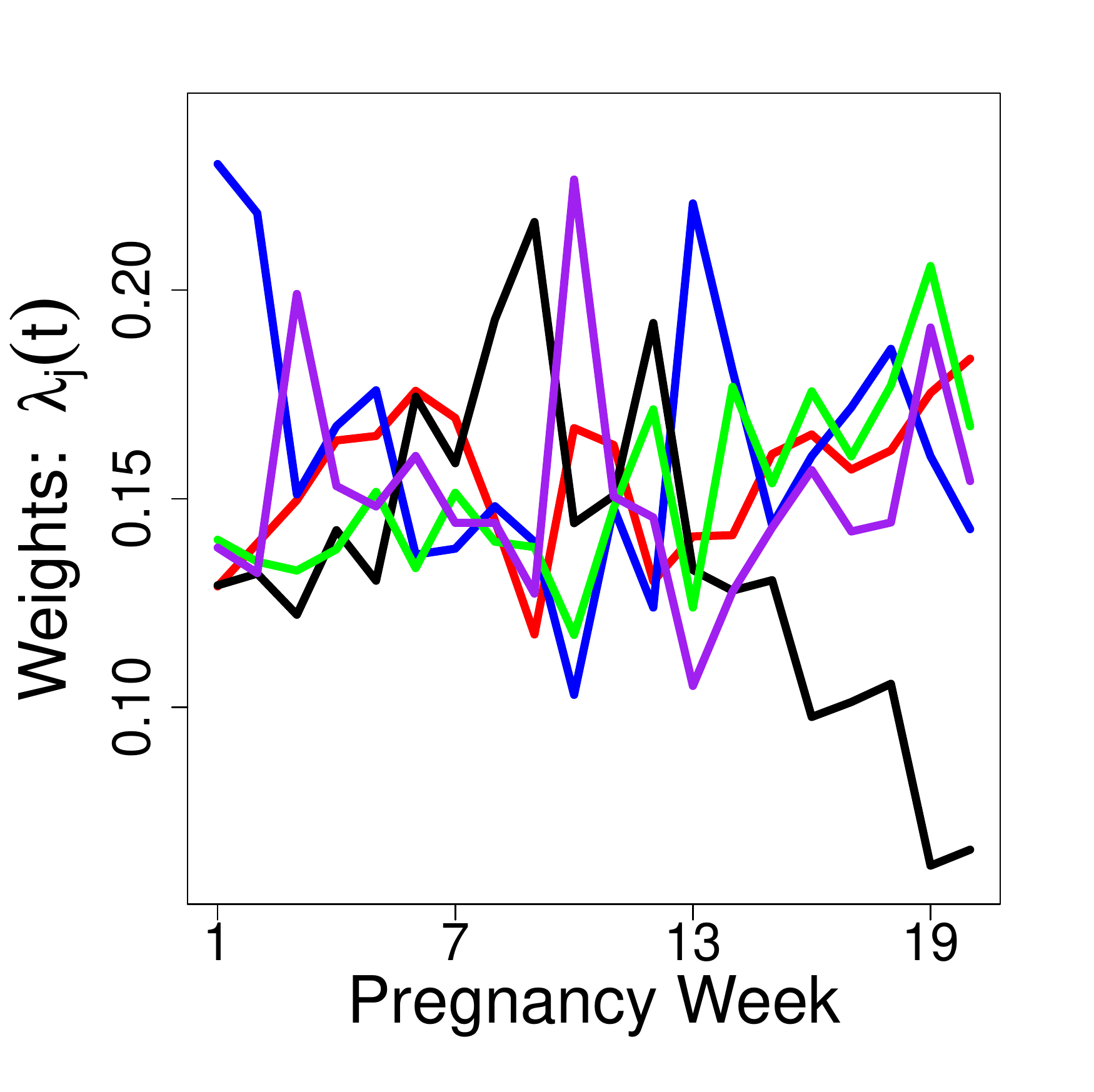}
\includegraphics[trim={0cm 0.5cm 1cm 0.5cm}, clip, scale=0.18]{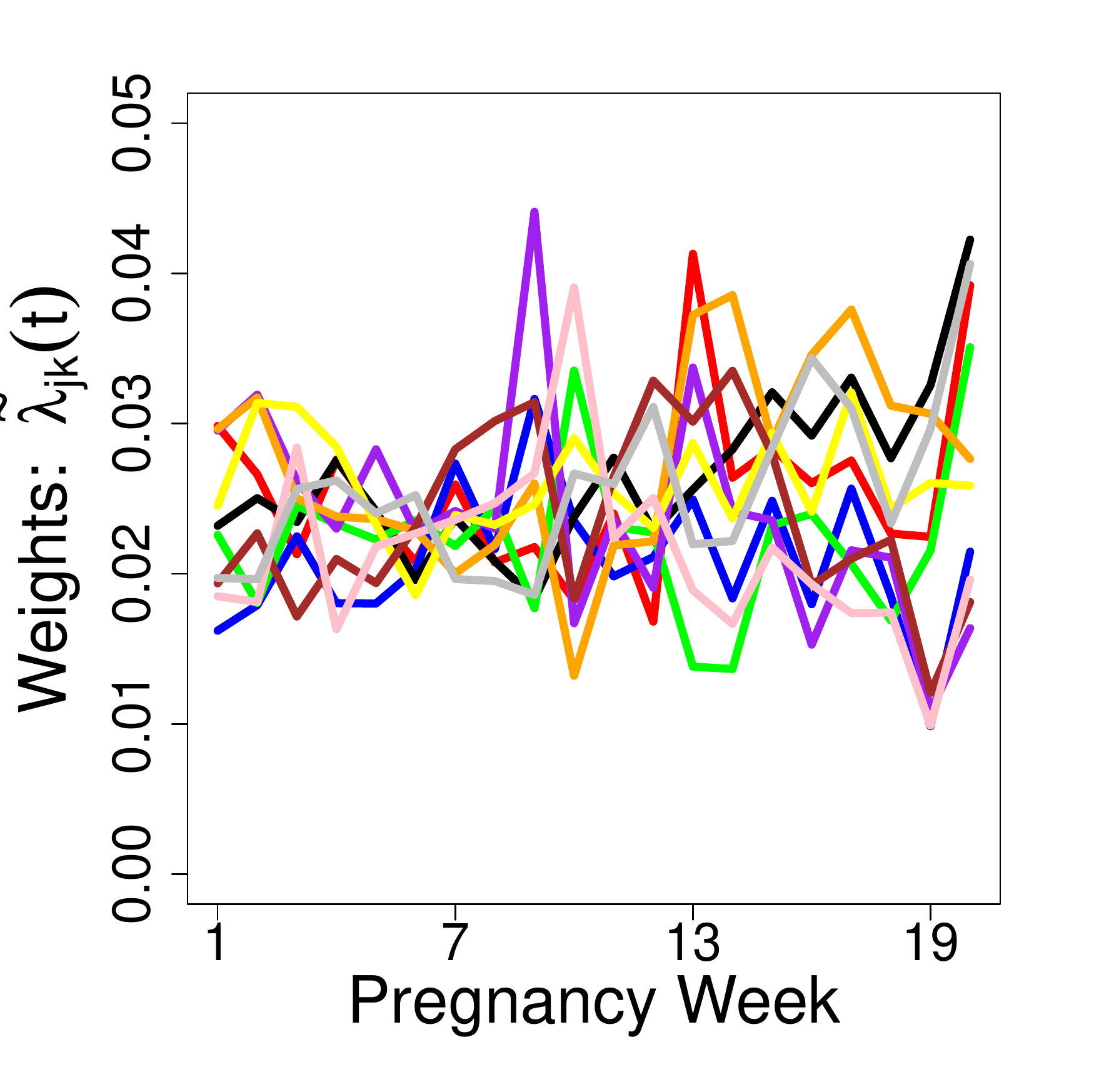}\\
\caption{Posterior means and 90\% credible intervals for the risk parameters (first column), posterior inclusion probabilities (second column), posterior means for the main effect weight parameters (third column), and posterior means for the interaction effect weight parameters (fourth column) for the \textbf{Hispanic} stillbirth and multiple exposures analyses in New Jersey, 2005-2014. Results based on an interquartile range increase in weekly exposure. Weeks identified as part of the critical window set are shown in red/dashed (harmful) and blue/dashed (protective).  The average posterior standard deviation for the weight parameters from CWVSmix is 0.13 (range: 0.05-0.30).}
\end{center}
\end{figure}

For Hispanic mothers (Figure 2), EW, PCA, and CWVSmix all suggest no significant associations, while LWQS indicates a protective association on gestational week 13, and an adverse association during weeks 9, 10, 19, and 20.  No weeks are identified as part of a critical window set for non-Hispanic White mothers using any of the methods (Figure S10 of the Supplement).

\section{Discussion}
In this work, we introduced CWVSmix, a method to simultaneously identify critical windows of exposure to multiple time-varying exposures, determine the time-varying relative importance of individual pollutants/interactions within the mixture, and estimate risk associations for exposure mixtures.  Through a simulation study, we showed that CWVSmix offers the best balance of performance across each of these areas with respect to the competing methods.  In particular, LWQS, which attempts to estimate these mixture weights and associations at each exposure time period separately, is shown to produce critical window identification accuracy, estimates of mixture weights, and estimates of risk that are always less optimal than CWVSmix.  This suggests that jointly estimating the weights and risk parameters and accounting for temporal correlation in exposures while identifying critical windows may be important features of an analysis.  Additionally, the variable selection procedure introduced for the weight parameters allows for a more intuitive explanation of the drivers of risk during a specific exposure period.  

By introducing weights that sum to one and are between zero and one, we assume that (i) pollutants impact the risk in the same direction on a given exposure period and (ii) main and interaction effects are in the same direction.  For (i), we do not enforce that the global risk parameter must be positive or negative, which avoids bias in these estimates when that assumption is violated.  For (ii), we anticipate that if two main effects have adverse (or protective) impacts on risk, then their interaction will either intensify this relationship or be negligible.  The benefit of this formulation is the interpretability of the findings when describing the composition of a harmful mixture, similar to source apportionment methods, and that it can effectively collapse to a single pollutant model given that a main effect weight can be exactly equal to one.       

In our application to stillbirth risk due to multiple ambient air pollution exposures, we found elevated risk of stillbirth for non-Hispanic Black mothers who experience higher concentrations of NH$_4^+$, NO$_3^-$, NO$_{\text{x}}$, PM$_{2.5}$, and SO$_4^{2-}$.  Previous work in this area has led to mixed findings regarding the harmful pollutants and exposure periods, with several works relying on single pollutant approaches fit across multiple exposure periods.  Using CWVSmix, we can now more accurately identify the critical exposure periods while accounting for exposure mixtures.  Additionally, use of the fused air pollution estimates gives us the opportunity to investigate less commonly studied pollutants.    

The differences observed between the competing methods in the data application must be interpreted in the context of the simulation study conclusions when determining which results are likely most reliable.  Findings in Table 1 from the simulation study, which was designed to resemble the NJ stillbirth data application, indicate that CWVSmix is more likely to accurately identify the true critical window set, yield improved estimates of the weight parameters, and yield improved estimates of the risk parameters when the proportion of important pollutants/interactions is relatively small.  The variable selection findings from the NJ stillbirth analysis in Table 3 suggest that the data application most closely resembles Settings 1 and 2 from the simulation study, where CWVSmix outperformed the competing methods across all metrics (Table 1).  Additionally, the protective association estimated using LWQS in the Hispanic data application (but not CWVSmix) is likely due to the lack of smoothness in the estimated weights, and provides some additional support to suggest that CWVSmix may yield more reliable results in this case.  Because WQS is fit separately at each exposure period in the LWQS algorithm, the model remains unadjusted for the other time-varying exposures when the weight parameters are estimated.  This may also contribute to the findings observed in the simulation study and NJ data applications. 

The ambient air pollution estimates from the data fusion model represent a major strength of this work.  Typically, only a handful of regularly monitored pollutants are able to be included in a particular analysis, depending on the timeframe and spatial location of the study.  In our case, we had excellent space-time coverage of our estimates, allowing us to investigate associations that are less common, with increased sample sizes.  However, for some of the pollutants (e.g., PM$_{2.5}$ constituents) the measurements in space-time are limited and the model-derived estimates further away from the measurements may be more biased than those near the measurements.  In addition, the SO$_2$ estimates have substantial uncertainty due to the plume nature of these pollutant fields that is difficult to capture with either monitoring or modeling.  

As is typically the case when linking environmental exposures to birth records data, we used residence at delivery to define the spatial location of data linkage.  This may introduce measurement error for some participants in the study, though \cite{warren2017investigating} showed that critical window identification is relatively robust to this type of error.  In (3.6), we assume independence \textit{a priori} between the weight parameters within an exposure time period due to the potentially large dimension of this matrix (i.e., $q(q+1)/2$ by $q(q+1)/2$) as the number of pollutants increases.  Future methods work that could reduce this dimensionality, possibly through incorporation of  auxiliary information about the chemicals in a mixture, may be able to accommodate this correlation \citep{reich2020integrative}.  Future work may also want to consider interactions between pollutants on different exposure periods, which will require further methodological extensions.  

When determining which method to use in future multipollutant applications, we recommend designing a simulation study based on important features of the analysis dataset and comparing the competing approaches.  Additionally, computing times may be important to consider based on the size of the dataset.  In Table S7 of the Supporting Information, we show the average computing times in minutes for each of the methods in the simulation study.  EW, which relies on CWVS, is clearly the fastest algorithm, followed by CWVSmix and LWQS.  The LWQS run times are impacted by the number of bootstrap samples requested and the number of exposure periods where the WQS algorithm must be applied.  Overall, CWVSmix offers an intuitive and interpretable method for analyzing the impact of multiple time-varying exposures on an adverse health outcome.       



\bibliographystyle{imsart-nameyear}
\bibliography{Main_Text}

\end{document}


\begin{frontmatter}
\title{Supplement to ``Critical Window Variable Selection for Mixtures: Estimating the Impact of Multiple Air Pollutants on Stillbirth''}
\runtitle{Supplement}

\begin{aug}
  \author[A]{\fnms{Joshua L.}  \snm{Warren}\ead[label=e1]{joshua.warren@yale.edu}},
  \author[B]{\fnms{Howard H.} \snm{Chang}\ead[label=e2,mark]{howard.chang@emory.edu}},
  \author[C]{\fnms{Lauren K.} \snm{Warren}\ead[label=e3,mark]{lklein@rti.org}},
  \author[D]{\fnms{Matthew J.} \snm{Strickland}\ead[label=e4,mark]{mstrickland@unr.edu, ldarrow@unr.edu}},
    \author[D]{\fnms{Lyndsey A.} \snm{Darrow}},
  \and
  \author[E]{\fnms{James A.}  \snm{Mulholland}%
  \ead[label=e5,mark]{james.mulholland@ce.gatech.edu}}

  \runauthor{J.L.\ Warren et al.}

\address[A]{Yale University,
\printead{e1}}

\address[B]{Emory University,
\printead{e2}}

\address[C]{RTI International,
\printead{e3}}

\address[D]{University of Nevada,
\printead{e4}}

\address[E]{Georgia Institute of Technology,
\printead{e5}}
\end{aug}

\end{frontmatter}

\renewcommand{\thesection}{S1}
\section{Model Fitting Details}
We fit critical window variable selection for mixtures (CWVSmix) using Markov chain Monte Carlo sampling techniques, including Gibbs and Metropolis (within Gibbs) sampling algorithms \citep{metropolis1953equation,geman1984stochastic,gelfand1990sampling}.  
All models are fit using R statistical software \citep{Rsoftware}.  We rely on the latent variable approach of \cite{polson2013bayesian} to allow for closed form full conditionals within our logistic regression framework for the included regression parameters.  Specifically, we introduce $$w_i|\boldsymbol{\beta}, \boldsymbol{\lambda}^*, \boldsymbol{\alpha} \stackrel{\text{ind}}{\sim}\text{P\'olya-Gamma}\left\{1, \textbf{x}_i^{\text{T}}\boldsymbol\beta + \boldsymbol{g}_i\left(\boldsymbol{\lambda}^*\right)^{\text{T}} \boldsymbol{\alpha}\right\},$$ $i=1,\hdots,n$; one latent variable for each observed $Y_i$ binary response variable.  The length $m$ vector of weighted exposures specific to subject $i$ is given as $\boldsymbol{g}_i\left(\boldsymbol{\lambda}^*\right)$ where $\boldsymbol{\lambda}^* = \left\{\boldsymbol{\lambda}^* \left(1\right)^{\text{T}}, \hdots, \boldsymbol{\lambda}^* \left(m\right)^{\text{T}}\right\}^{\text{T}}$ is the complete set of latent weight parameters across all exposure time periods; the $t^{th}$ entry of $\boldsymbol{g}_i\left(\boldsymbol{\lambda}^*\right)$ is $g_i\left(\boldsymbol{\lambda}^*, t\right) = \sum_{j=1}^{q} \lambda_j\left(t\right) \text{z}_{ij}\left(t\right) + \sum_{j=1}^{q-1} \sum_{k=j+1}^q \widetilde{\lambda}_{jk}\left(t\right)\text{z}_{ij}\left(t\right)\text{z}_{ik}\left(t\right)$; and $\boldsymbol{\alpha}=\left\{\alpha\left(1\right),\hdots,\alpha\left(m\right)\right\}^{\text{T}}$. Using these latent variables, we derive the full conditional densities needed for posterior sampling.

From \cite{polson2013bayesian}, the full conditional distributions of the $w_i$ latent variables are given as $$w_i|\boldsymbol{Y}, \Theta_{-w_i} \stackrel{\text{ind}}{\sim}\text{P\'olya-Gamma}\left\{1, \textbf{x}_i^{\text{T}}\boldsymbol\beta + \boldsymbol{g}_i\left(\boldsymbol{\lambda}^*\right)^{\text{T}} \boldsymbol{\alpha}\right\},\ i=1,\hdots,n$$ where $\boldsymbol{Y}=\left(Y_1, \hdots, Y_n\right)^{\text{T}}$ and $\Theta_{-w_i}$ represents all latent variables and model parameters after removing $w_i$.  Note that the full conditional and prior distributions are identical for these parameters given the lack of dependence between $w_i$ and $Y_i$.  We sample from this distribution using the \texttt{pgdraw} package in R \citep{pgdraw}.  

The vector of regression parameters associated with the covariates and confounders also has a closed form full conditional density such that $\boldsymbol{\beta}|\boldsymbol{Y},\Theta_{-\boldsymbol{\beta}}\sim\text{MVN}\left(\boldsymbol{\mu}_{\boldsymbol{\beta}}, \Sigma_{\boldsymbol{\beta}}\right)$ with $$\Sigma_{\beta}=\left(\text{X}^{\text{T}}\Omega \text{X} + \frac{1}{\sigma^2_{\boldsymbol{\beta}}}I_p\right)^{-1},\ \boldsymbol{\mu}_{\beta}=\Sigma_{\beta}\left\{\text{X}^{\text{T}}\Omega\left(\boldsymbol{\zeta} - G \boldsymbol{\alpha}\right)\right\}$$ where $\Omega_{ii}=w_i$ for $i=1,\hdots,n$ and $\Omega_{ij}=0$ for $i \neq j$; $\boldsymbol{\zeta}=\left\{\left(Y_1-0.50\right)/w_1, \hdots, \left(Y_n-0.50\right)/w_n\right\}^{\text{T}}$; $\text{X}$ is an $n$ by $p$ matrix with $i^{\text{th}}$ row equal to $\textbf{x}_i^{\text{T}}$; and $G$ is an $n$ by $m$ matrix with $i^{\text{th}}$ row equal to $\boldsymbol{g}_i\left(\boldsymbol{\lambda}^*\right)^{\text{T}}$. 

The variable selection parameters have independent Bernoulli full conditional distributions such that $\gamma\left(t\right)|\boldsymbol{Y}, \Theta_{-\gamma\left(t\right)}\stackrel{\text{ind}}{\sim}\text{Bernoulli}\left\{\kappa\left(t\right)\right\}$ where $\kappa\left(t\right)=$ $$\frac{\exp\left\{-\frac{1}{2}\left(\boldsymbol{\zeta} - \text{X}\boldsymbol{\beta} - G \boldsymbol{\alpha}_{\left\{\gamma\left(t\right)=1\right\}}\right)^{\text{T}} \Omega \left(\boldsymbol{\zeta} - \text{X}\boldsymbol{\beta} - G \boldsymbol{\alpha}_{\left\{\gamma\left(t\right)=1\right\}}\right)\right\}\pi\left(t\right)}{\sum_{j=0}^1 \exp\left\{-\frac{1}{2}\left(\boldsymbol{\zeta} - \text{X}\boldsymbol{\beta} - G \boldsymbol{\alpha}_{\left\{\gamma\left(t\right)=j\right\}}\right)^{\text{T}} \Omega \left(\boldsymbol{\zeta} - \text{X}\boldsymbol{\beta} - G \boldsymbol{\alpha}_{\left\{\gamma\left(t\right)=j\right\}}\right)\right\}\pi\left(t\right)^{I\left(j=1\right)} \left\{1 - \pi\left(t\right)\right\}^{I\left(j=0\right)}}$$ for $t=1,\hdots,m$ where $\boldsymbol{\alpha}_{\left\{\gamma\left(t\right)=j\right\}}$ is the vector of $\alpha\left(t\right)$ parameters defined by $\gamma\left(t\right)=j$ and $I\left(.\right)$ is the indicator function taking a value of one if the input statement is true and zero otherwise.

We introduce independent, normally distributed latent variables which define the underlying probability of a one or zero for $\gamma\left(t\right)$ such that $\gamma^*\left(t\right)|\eta\left(t\right)\stackrel{\text{ind}}{\sim}\text{N}\left\{\eta\left(t\right), 1\right\}$, $t=1,\hdots, m$ and $\pi\left(t\right)=\text{P}\left\{\gamma^*\left(t\right) > 0\right\}$.  The full conditional density of one of these latent parameters is given as $\gamma^*\left(t\right)|\boldsymbol{Y},\Theta_{-\gamma^*\left(t\right)}\stackrel{\text{ind}}{\sim}\text{Truncated Normal}\left\{\eta\left(t\right), 1\right\},\ t=1,\hdots,m$ where the truncation is $\leq 0$ if $\gamma\left(t\right)=0$ and is $> 0$ if $\gamma\left(t\right)=1$.  
Introduction of these latent parameters allows us to obtain a closed-form full conditional for $\boldsymbol{\delta}_2$ such that $\boldsymbol{\delta}_2|\boldsymbol{Y},\Theta_{-\boldsymbol{\delta}_2}\sim \text{MVN}\left(\boldsymbol{\mu}_{\boldsymbol{\delta}_2}, \Sigma_{\boldsymbol{\delta}_2}\right)$ where $$\Sigma_{\boldsymbol{\delta}_2}=\left\{A_{22}^2 I_m + \Sigma\left(\phi_2\right)^{-1}\right\}^{-1},\ \boldsymbol{\mu}_{\boldsymbol{\delta}_2}=\Sigma_{\boldsymbol{\delta}_2}\left\{A_{22}\left(\boldsymbol{\gamma}^*-A_{21}\boldsymbol{\delta}_1\right)\right\}$$ where $\boldsymbol{\gamma}^*=\left\{\gamma^*\left(1\right),\hdots,\gamma^*\left(m\right)\right\}^{\text{T}}$.  Similarly, $\boldsymbol{\delta}_1$ has a $\text{MVN}\left(\boldsymbol{\mu}_{\boldsymbol{\delta}_1}, \Sigma_{\boldsymbol{\delta}_1}\right)$ full conditional distribution with $$\Sigma_{\boldsymbol{\delta}_1}=\left\{A_{11}^2 G^{*\text{T}} \Omega G^* + A_{21}^2 I_m + \Sigma\left(\phi_1\right)^{-1}\right\}^{-1},$$ $$\boldsymbol{\mu}_{\boldsymbol{\delta}_1}=\Sigma_{\boldsymbol{\delta}_1}\left\{A_{11} G^{*\text{T}} \Omega \left(\boldsymbol{\zeta} - \text{X}\boldsymbol{\beta}\right) + A_{21}\left(\boldsymbol{\gamma}^* - A_{22}\boldsymbol{\delta}_2\right)\right\}$$ where $G^*$ is an $n$ by $m$ matrix with $\left(i,t\right)^{th}$ entry $G^*_{it}=g_i\left(\boldsymbol{\lambda}^*, t\right)\gamma\left(t\right)$.

Metropolis sampling is required for the entries of the $A$ matrix.  For $A_{21}$, the full conditional density is proportional to $$\exp\left\{-\frac{1}{2}\left(\boldsymbol{\gamma}^* - A_{21}\boldsymbol{\delta}_1 - A_{22}\boldsymbol{\delta}_2\right)^{\text{T}}\left(\boldsymbol{\gamma}^* - A_{21}\boldsymbol{\delta}_1 - A_{22}\boldsymbol{\delta}_2\right) -\frac{1}{2\sigma^2_{A}}A_{21}^2\right\}.$$  The full conditional distribution for $\ln\left(A_{22}\right)$ has the same form with $-\frac{1}{2\sigma^2_{A}}A_{21}^2$ replaced by $-\frac{1}{2\sigma^2_{A}}\ln\left(A_{22}\right)^2$.  For $\ln\left(A_{11}\right)$, the full conditional density if proportional to $$\exp\left\{-\frac{1}{2}\left(\boldsymbol{\zeta} - \text{X}\boldsymbol{\beta} - A_{11}G^*\boldsymbol{\delta}_1\right)^{\text{T}} \Omega \left(\boldsymbol{\zeta} - \text{X}\boldsymbol{\beta} - A_{11}G^*\boldsymbol{\delta}_1\right) -\frac{1}{2\sigma^2_{A}}\ln\left(A_{11}\right)^2\right\}.$$

The latent weight parameters are updated using blocked Metropolis sampling where the full conditional distribution is proportional to $$\exp\left[-\frac{1}{2}\left(\boldsymbol{\zeta} - \text{X}\boldsymbol{\beta} - A_{11}G^*\boldsymbol{\delta}_1\right)^{\text{T}} \Omega \left(\boldsymbol{\zeta} - \text{X}\boldsymbol{\beta} - A_{11}G^*\boldsymbol{\delta}_1\right) - \frac{1}{2} \boldsymbol{\lambda}^{*\text{T}} \left\{\Sigma\left(\phi_{\lambda}\right)^{-1} \otimes I_{q\left(q+1\right)/2}\right\} \boldsymbol{\lambda}^*\right];$$ $G^*$ is defined by the latent weight parameters; and we propose/evaluate new $\boldsymbol{\lambda}^*\left(t\right)$ vectors during model fitting. 

Finally, the three parameters that control the smoothness of the processes over time also require a Metropolis step with their full conditional densities proportional to $$\frac{1}{|\Sigma\left(\phi_j\right)|^{1/2}}\exp\left\{-\frac{1}{2}\boldsymbol{\delta}_j^{\text{T}} \Sigma\left(\phi_j\right)^{-1}\boldsymbol{\delta}_j\right\}\exp\left\{\alpha_{\phi} \psi_j - \beta_{\phi}\exp\left\{\psi_j\right\} \right\}$$ for $j=0,1$, and $$\frac{1}{|\Sigma\left(\phi_{\lambda}\right)|^{q(q+1)/4}}\exp\left[-\frac{1}{2}\boldsymbol{\boldsymbol{\lambda}^*}^{\text{T}} \left\{\Sigma\left(\phi_{\lambda}\right\}^{-1} \otimes I_{q(q+1)/2}\right)\boldsymbol{\lambda}^*\right]\exp\left\{\alpha_{\phi} \psi_{\lambda} - \beta_{\phi}\exp\left\{\psi_{\lambda}\right\} \right\}.$$  We transform each of these parameters to have support on the real line (working with the induced prior distribution) in order to improve properties of the posterior sampling such that $\psi_j=\ln\left(\phi_j\right)$ and $\psi_{\lambda} = \ln\left(\phi_{\lambda}\right)$ are all $\in \mathbb{R}$.
\clearpage

\renewcommand{\thesection}{S2}
\section{Supplemental Tables}
\hspace{3mm}

\begin{table}[h]
\centering
\caption{Background information for the \textbf{non-Hispanic Black} study population in New Jersey, 2005-2014.}
\begin{tabular}{lrr}
\hline                                                
Characteristic                  & Cases (Stillbirth)      & Controls (Live Birth) \\
\hline
Total                           & 1,267      & 6,335   \\
Tobacco Use (\% Yes)            & 11.37      & 8.97     \\
Age Category:                   &            &      \\
\ \ $<$25                     & 33.31      & 37.57 \\
\ \ [25, 30)                     & 23.91      & 25.11 \\
\ \ [30, 35)                     & 23.05      & 21.99 \\
\ \ $\geq$                     & 19.73      & 15.33 \\ 
Education:                      &            &      \\
\ \ $<$ High School             & 15.15      & 13.64 \\
\ \ High School                 & 46.88      & 41.10 \\
\ \ $>$ High School             & 37.96      & 45.26 \\
Sex (\% Male)                   & 56.20      & 51.84 \\
\hline
\end{tabular}
\end{table}
\clearpage

\begin{table}[ht]
\centering
\caption{Background information for the \textbf{Hispanic} study population in New Jersey, 2005-2014.}
\begin{tabular}{lrr}
\hline                                                
Characteristic                  & Cases (Stillbirth)      & Controls (Live Birth) \\
\hline
Total                           &   928      & 4,640   \\
Tobacco Use (\% Yes)            & 4.09       & 3.90     \\
Age Category:                   &            &      \\
\ \ $<$25                     & 30.82      & 35.67 \\
\ \ [25, 30)                     & 25.65      & 27.93 \\
\ \ [30, 35)                     & 24.89      & 21.94 \\
\ \ $\geq$35                     & 18.64      & 14.46 \\ 
Education:                      &            &      \\
\ \ $<$ High School             & 31.79      & 31.96 \\
\ \ High School                 & 40.73      & 35.78 \\
\ \ $>$ High School             & 27.48      & 32.26 \\
Sex (\% Male)                   & 54.85      & 51.14 \\
\hline
\end{tabular}
\end{table}
\clearpage

\begin{table}[ht]
\centering
\caption{Background information for the \textbf{non-Hispanic White} study population in New Jersey, 2005-2014.}
\begin{tabular}{lrr}
\hline                                                
Characteristic                  & Cases (Stillbirth)      & Controls (Live Birth) \\
\hline
Total                           & 1,129      & 5,645   \\
Tobacco Use (\% Yes)            & 14.79      & 8.06     \\
Age Category:                   &            &      \\
\ \ $<$25                     & 15.94      & 14.14 \\
\ \ [25, 30)                     & 23.91      & 24.66 \\
\ \ [30, 35)                     & 31.53      & 35.68 \\
\ \ $\geq$35                     & 28.61      & 25.53 \\ 
Education:                      &            &      \\
\ \ $<$ High School             &  6.64      &  3.53 \\
\ \ High School                 & 33.30      & 21.72 \\
\ \ $>$ High School             & 60.05      & 74.76 \\
Sex (\% Male)                   & 55.45      & 50.54 \\
\hline
\end{tabular}
\end{table}
\clearpage

\begin{table}
\centering
\caption{Covariate results from the \textbf{non-Hispanic Black} stillbirth and multiple exposure Critical Window Variable Selection for Mixtures (CWVSmix) analysis in New Jersey, 2005-2014.  Posterior inference on the odds ratio scale is presented.}
\begin{tabular}{lcccc}
  \hline
          &      &    & \multicolumn{2}{c}{Quantiles} \\ \cline{4-5}
Parameter & Mean & SD &          0.025 & 0.975        \\ 
\hline
Tobacco Use During Pregnancy: &&&&\\
\ \ \ Yes vs.\ No                          & 1.22 & 0.13 & 1.00 & 1.49\\ 
Age (Years): &&&&\\
\ \ \ [25, 30) vs.\ $<$25                  & 1.16 & 0.10 & 0.98 & 1.37\\ 
\ \ \ [30, 35) vs.\ $<$25                  & 1.33 & 0.12 & 1.12 & 1.57\\ 
\ \ \ $\geq$35 vs.\ $<$25                  & 1.64 & 0.15 & 1.36 & 1.95\\ 
Education: &&&&\\
\ \ \ High school vs.\ $<$ high school     & 1.00 & 0.10 & 0.83 & 1.20\\ 
\ \ \ $>$ High school vs.\ $<$ high school & 0.69 & 0.07 & 0.57 & 0.84\\ 
Sex of Child: &&&&\\
\ \ \ Male vs.\ Female                     & 1.20 & 0.08 & 1.06 & 1.36\\ 
Season of Conception: &&&&\\
\ \ \ Summer vs.\ Spring                   & 0.91 & 0.09 & 0.74 & 1.10\\ 
\ \ \ Fall vs.\ Spring                     & 0.86 & 0.09 & 0.69 & 1.04\\
\ \ \ Winter vs.\ Spring                   & 1.02 & 0.10 & 0.84 & 1.22\\
Year of Conception: &&&&\\
\ \ \ 2006 vs.\ 2005                       & 0.94 & 0.12 & 0.73 & 1.19\\ 
\ \ \ 2007 vs.\ 2005                       & 0.93 & 0.12 & 0.71 & 1.18\\
\ \ \ 2008 vs.\ 2005                       & 0.88 & 0.12 & 0.66 & 1.14\\
\ \ \ 2009 vs.\ 2005                       & 0.94 & 0.15 & 0.68 & 1.27\\ 
\ \ \ 2010 vs.\ 2005                       & 1.10 & 0.18 & 0.80 & 1.48\\
\ \ \ 2011 vs.\ 2005                       & 0.99 & 0.17 & 0.70 & 1.37\\
\ \ \ 2012 vs.\ 2005                       & 1.23 & 0.22 & 0.84 & 1.73\\ 
\ \ \ 2013 vs.\ 2005                       & 1.14 & 0.21 & 0.80 & 1.61\\
Latitude                                   & 0.90 & 0.05 & 0.81 & 1.00\\
Longitude                                  & 1.13 & 0.06 & 1.03 & 1.25\\
\hline
\end{tabular}
\end{table}
\clearpage

\begin{table}
\centering
\caption{Covariate results from the \textbf{Hispanic} stillbirth and multiple exposure Critical Window Variable Selection for Mixtures (CWVSmix) analysis in New Jersey, 2005-2014.  Posterior inference on the odds ratio scale is presented.}
\begin{tabular}{lcccc}
  \hline
          &      &    & \multicolumn{2}{c}{Quantiles} \\ \cline{4-5}
Parameter & Mean & SD &          0.025 & 0.975        \\ 
\hline
Tobacco Use During Pregnancy: &&&&\\
\ \ \ Yes vs.\ No                          & 1.09 & 0.20 & 0.73 & 1.53\\ 
Age (Years): &&&&\\
\ \ \ [25, 30) vs.\ $<$25                  & 1.11 & 0.11 & 0.91 & 1.34\\ 
\ \ \ [30, 35) vs.\ $<$25                  & 1.40 & 0.14 & 1.15 & 1.70\\ 
\ \ \ $\geq$35 vs.\ $<$25                  & 1.63 & 0.18 & 1.31 & 2.01\\ 
Education: &&&&\\
\ \ \ High school vs.\ $<$ high school     & 1.14 & 0.10 & 0.95 & 1.35\\ 
\ \ \ $>$ High school vs.\ $<$ high school & 0.79 & 0.08 & 0.65 & 0.95\\ 
Sex of Child: &&&&\\
\ \ \ Male vs.\ Female                     & 1.18 & 0.09 & 1.02 & 1.36\\ 
Season of Conception: &&&&\\
\ \ \ Summer vs.\ Spring                   & 0.94 & 0.10 & 0.75 & 1.16\\ 
\ \ \ Fall vs.\ Spring                     & 0.79 & 0.09 & 0.63 & 0.98\\
\ \ \ Winter vs.\ Spring                   & 1.08 & 0.12 & 0.86 & 1.32\\
Year of Conception: &&&&\\
\ \ \ 2006 vs.\ 2005                       & 1.00 & 0.16 & 0.73 & 1.35\\ 
\ \ \ 2007 vs.\ 2005                       & 1.18 & 0.18 & 0.86 & 1.58\\
\ \ \ 2008 vs.\ 2005                       & 0.82 & 0.14 & 0.59 & 1.13\\
\ \ \ 2009 vs.\ 2005                       & 0.80 & 0.14 & 0.56 & 1.12\\ 
\ \ \ 2010 vs.\ 2005                       & 0.82 & 0.15 & 0.57 & 1.14\\
\ \ \ 2011 vs.\ 2005                       & 0.98 & 0.17 & 0.69 & 1.36\\
\ \ \ 2012 vs.\ 2005                       & 1.01 & 0.19 & 0.70 & 1.43\\ 
\ \ \ 2013 vs.\ 2005                       & 1.15 & 0.21 & 0.81 & 1.62\\
Latitude                                   & 0.86 & 0.05 & 0.77 & 0.95\\
Longitude                                  & 1.20 & 0.06 & 1.08 & 1.33\\
\hline
\end{tabular}
\end{table}
\clearpage

\begin{table}
\centering
\caption{Covariate results from the \textbf{non-Hispanic White} stillbirth and multiple exposure Critical Window Variable Selection for Mixtures (CWVSmix) analysis in New Jersey, 2005-2014.  Posterior inference on the odds ratio scale is presented.}
\begin{tabular}{lcccc}
  \hline
          &      &    & \multicolumn{2}{c}{Quantiles} \\ \cline{4-5}
Parameter & Mean & SD &          0.025 & 0.975        \\ 
\hline
Tobacco Use During Pregnancy: &&&&\\
\ \ \ Yes vs.\ No                          & 1.55 & 0.16 & 1.25 & 1.89\\ 
Age (Years): &&&&\\
\ \ \ [25, 30) vs.\ $<$25                  & 1.14 & 0.13 & 0.91 & 1.40\\ 
\ \ \ [30, 35) vs.\ $<$25                  & 1.15 & 0.13 & 0.92 & 1.42\\ 
\ \ \ $\geq$35 vs.\ $<$25                  & 1.48 & 0.17 & 1.18 & 1.84\\ 
Education: &&&&\\
\ \ \ High school vs.\ $<$ high school     & 0.83 & 0.13 & 0.61 & 1.12\\ 
\ \ \ $>$ High school vs.\ $<$ high school & 0.45 & 0.07 & 0.33 & 0.60\\ 
Sex of Child: &&&&\\
\ \ \ Male vs.\ Female                     & 1.23 & 0.08 & 1.08 & 1.40\\ 
Season of Conception: &&&&\\
\ \ \ Summer vs.\ Spring                   & 0.88 & 0.08 & 0.73 & 1.05\\ 
\ \ \ Fall vs.\ Spring                     & 0.85 & 0.08 & 0.71 & 1.02\\
\ \ \ Winter vs.\ Spring                   & 0.99 & 0.09 & 0.82 & 1.18\\
Year of Conception: &&&&\\
\ \ \ 2006 vs.\ 2005                       & 0.85 & 0.11 & 0.65 & 1.09\\ 
\ \ \ 2007 vs.\ 2005                       & 0.68 & 0.10 & 0.51 & 0.89\\
\ \ \ 2008 vs.\ 2005                       & 0.76 & 0.11 & 0.57 & 0.98\\
\ \ \ 2009 vs.\ 2005                       & 0.97 & 0.14 & 0.73 & 1.27\\ 
\ \ \ 2010 vs.\ 2005                       & 0.72 & 0.11 & 0.54 & 0.95\\
\ \ \ 2011 vs.\ 2005                       & 0.89 & 0.13 & 0.66 & 1.16\\
\ \ \ 2012 vs.\ 2005                       & 1.11 & 0.15 & 0.84 & 1.43\\ 
\ \ \ 2013 vs.\ 2005                       & 1.01 & 0.14 & 0.77 & 1.31\\
Latitude                                   & 1.02 & 0.04 & 0.94 & 1.10\\
Longitude                                  & 0.93 & 0.04 & 0.86 & 1.00\\
\hline
\end{tabular}
\end{table}
\clearpage

\begin{table}[ht]
\centering
\caption{Average single dataset run times (minutes) from the simulation study analyses.}
\begin{tabular}{lrrr}
\hline                                                
Setting      & EW            & LWQS         & CWVSmix \\
\hline
1A           &  1.45 (0.00)  & 32.92 (0.38) &  9.57 (0.07) \\
1C           &  1.45 (0.00)  & 32.97 (0.48) & 10.31 (0.04) \\
\hline 
2A           &  1.43 (0.00)  & 32.52 (0.43) &  8.69 (0.05) \\
2B           &  1.45 (0.00)  & 32.25 (0.47) &  8.41 (0.05) \\
2C           &  1.35 (0.00)  & 34.95 (0.49) &  8.75 (0.05)\\
\hline
3A           &  1.44 (0.00)  & 33.98 (0.53) &  8.54 (0.03) \\
3B           &  1.45 (0.00)  & 36.56 (0.51) &  8.60 (0.03) \\
3C           &  1.35 (0.00)  & 34.62 (0.46) & 13.03 (0.09) \\
\hline
4A           &  1.36 (0.00)  & 34.25 (0.45) &  9.20 (0.09) \\
4B           &  1.36 (0.00)  & 34.14 (0.47) &  8.41 (0.05) \\
4C           &  1.42 (0.00)  & 34.33 (0.44) &  8.59 (0.06) \\
\hline
5A           &  1.45 (0.00)  & 33.18 (0.44) & 11.70 (0.07) \\
5B           &  1.45 (0.00)  & 34.97 (0.44) &  9.49 (0.05)\\
5C           &  1.40 (0.00)  & 33.96 (0.52) & 11.04 (0.09) \\
\hline
\end{tabular}
\end{table}
\clearpage

\renewcommand{\thesection}{S3}
\section{Supplemental Figures}
\hspace{3mm}

\begin{figure}[!hbt]
\begin{center}
\includegraphics[scale=0.26]{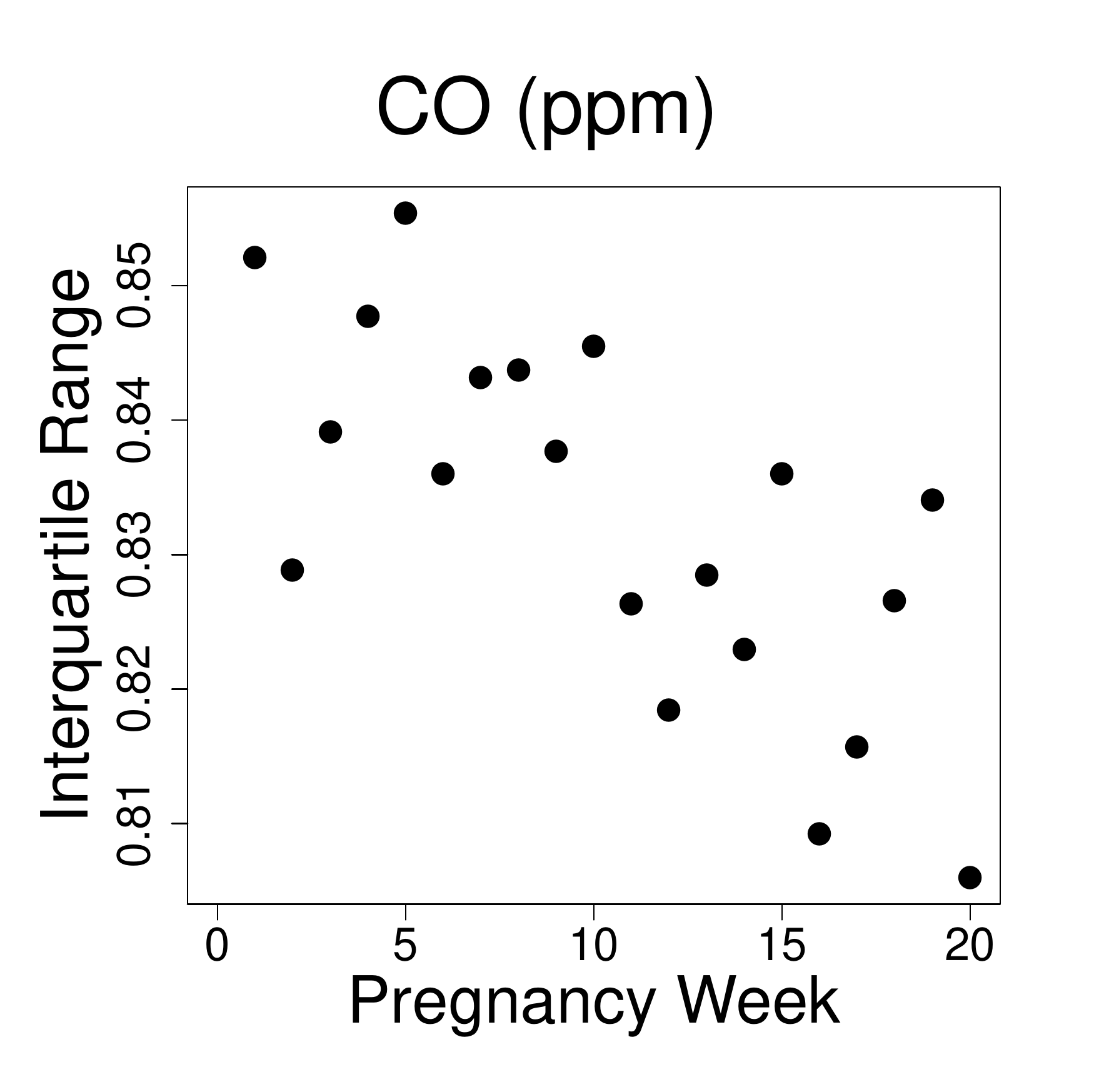}
\includegraphics[scale=0.26]{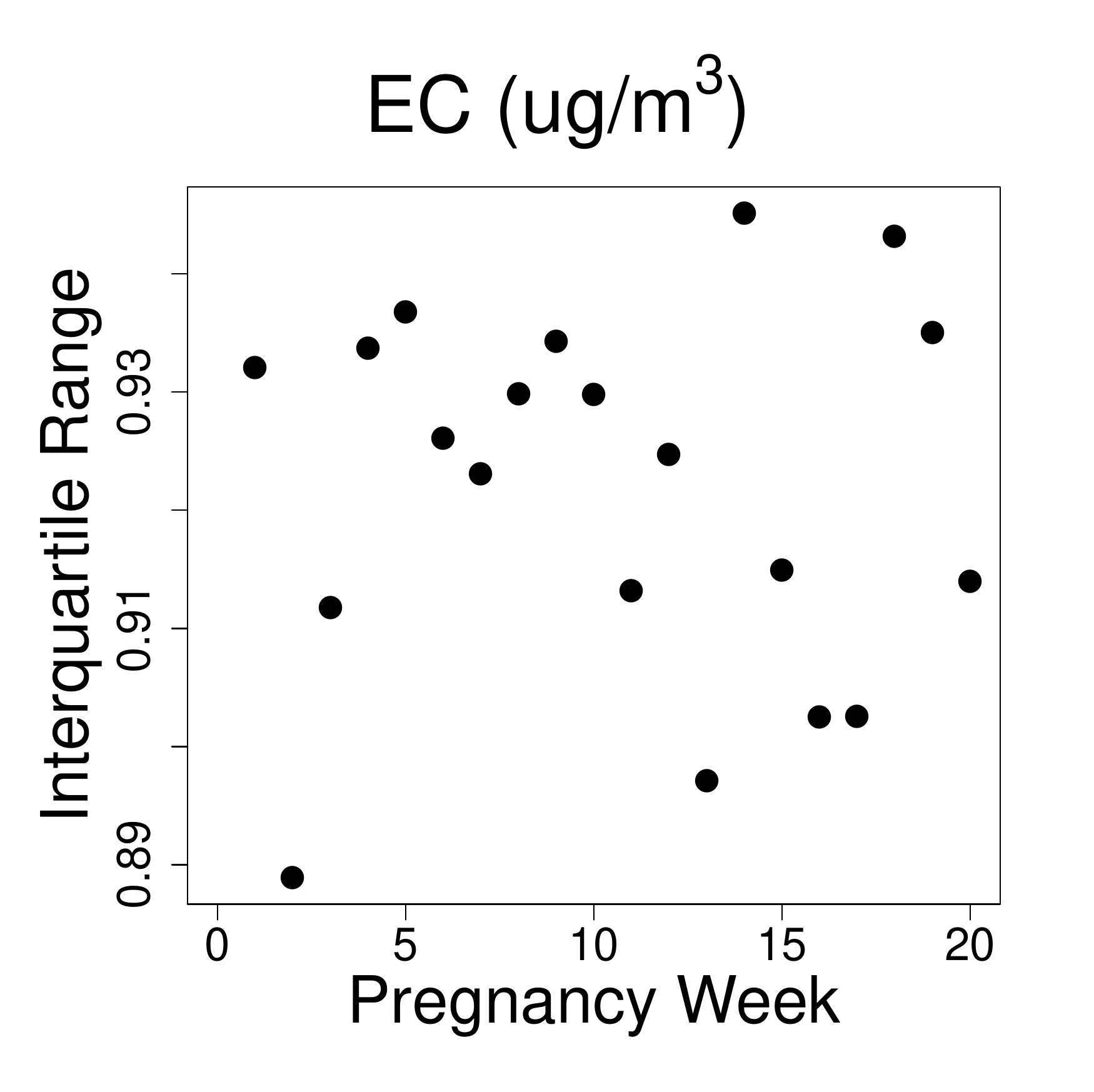}
\includegraphics[scale=0.26]{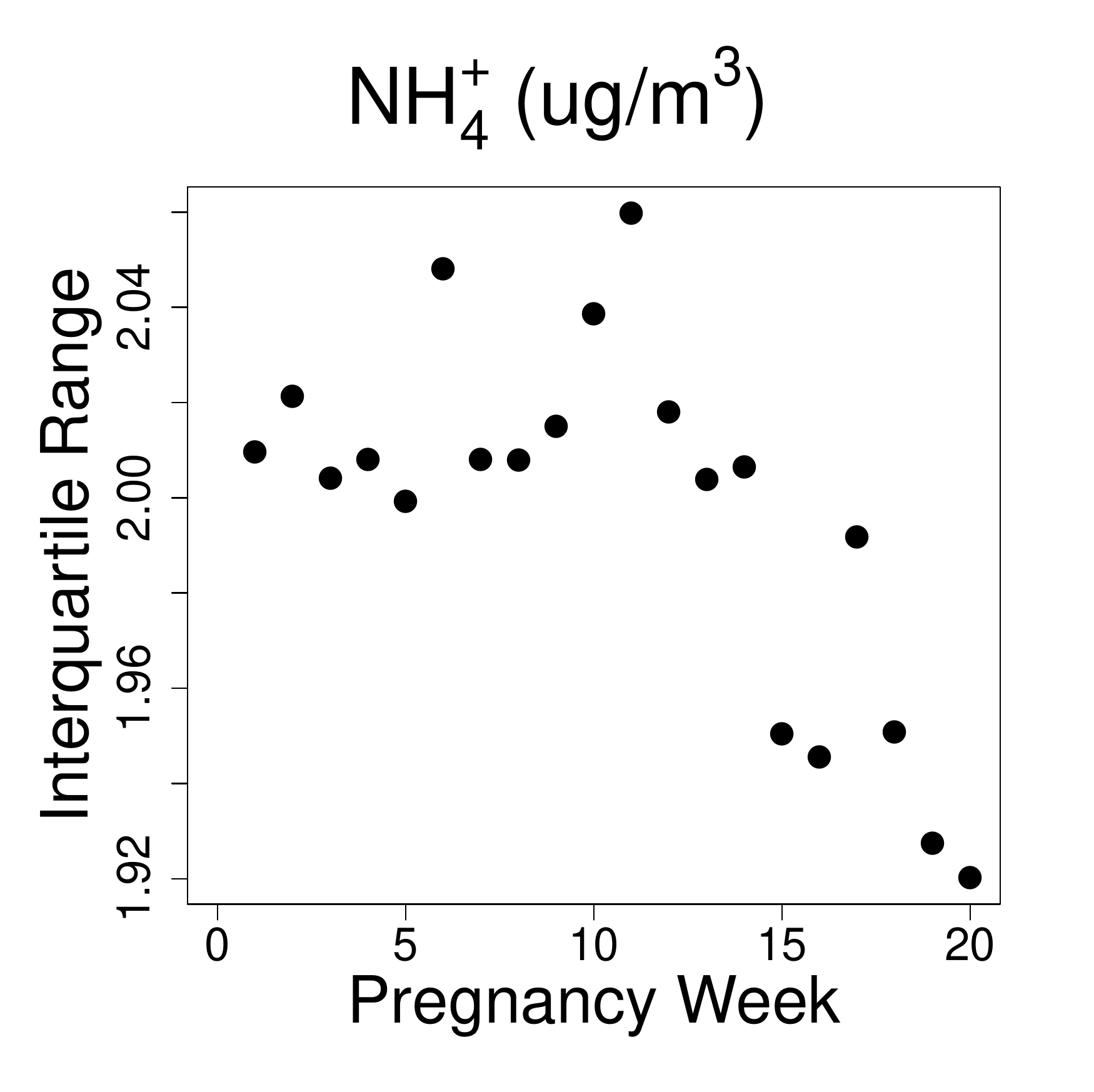}\\
\includegraphics[scale=0.26]{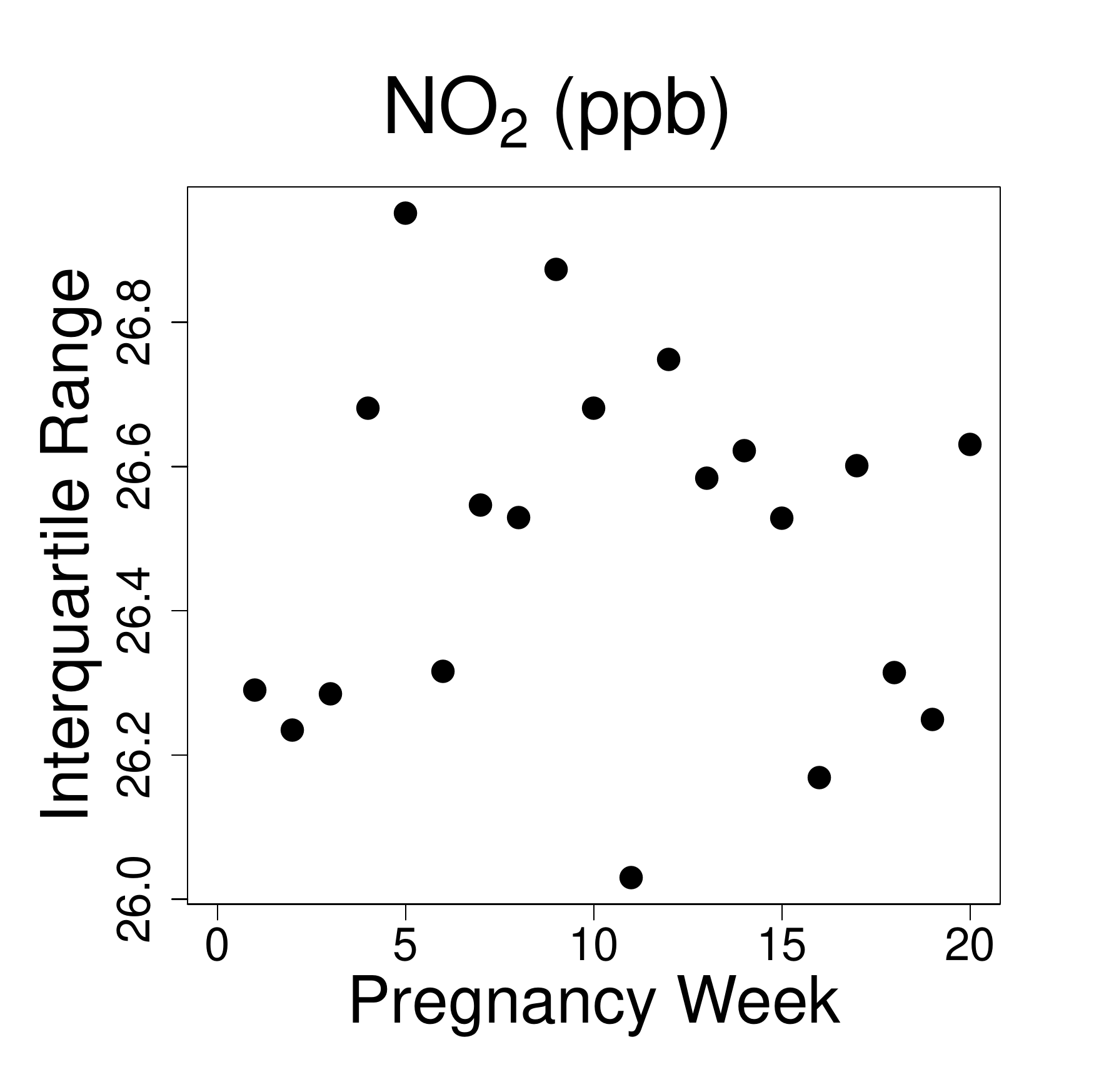}
\includegraphics[scale=0.26]{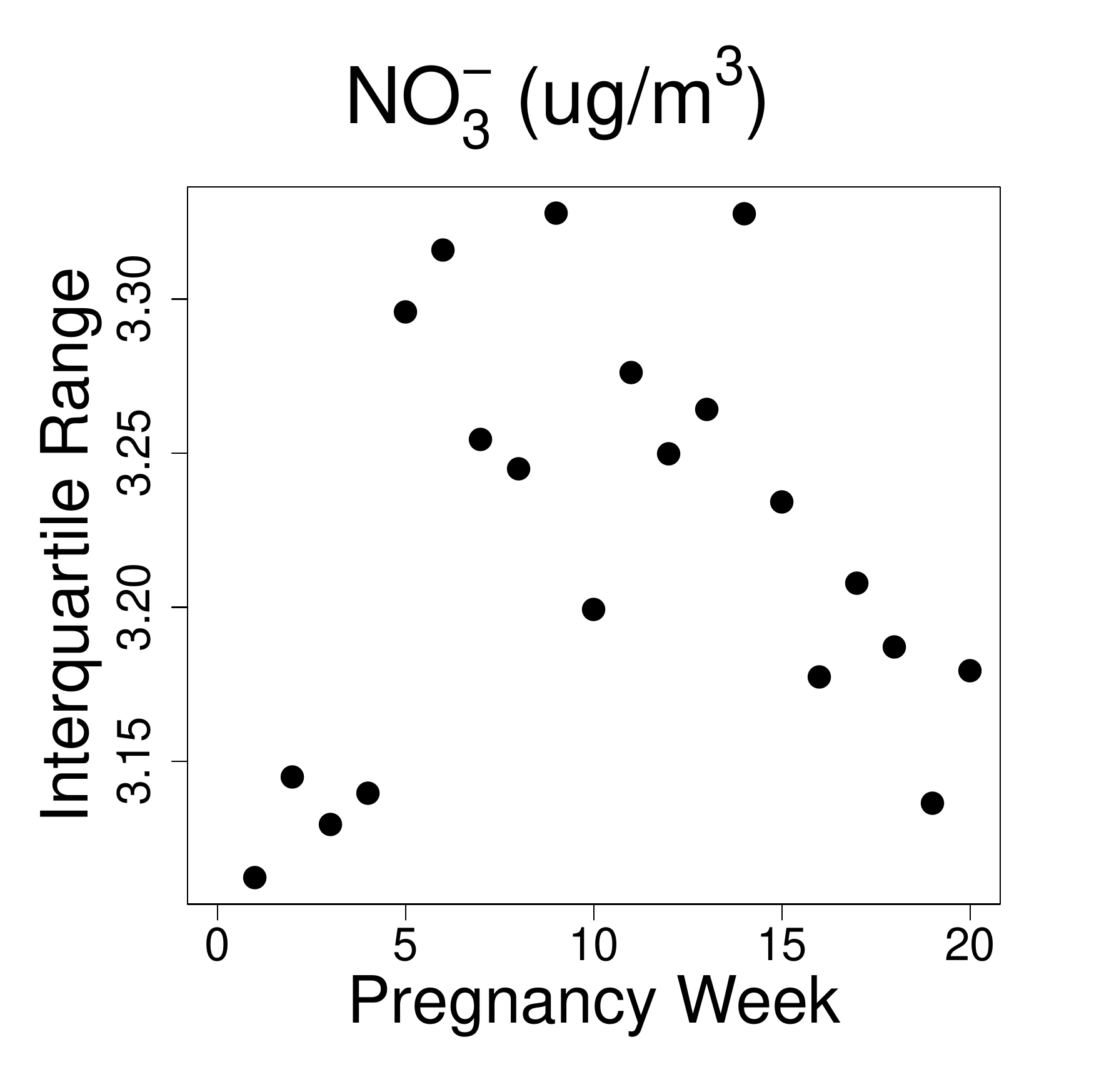}
\includegraphics[scale=0.26]{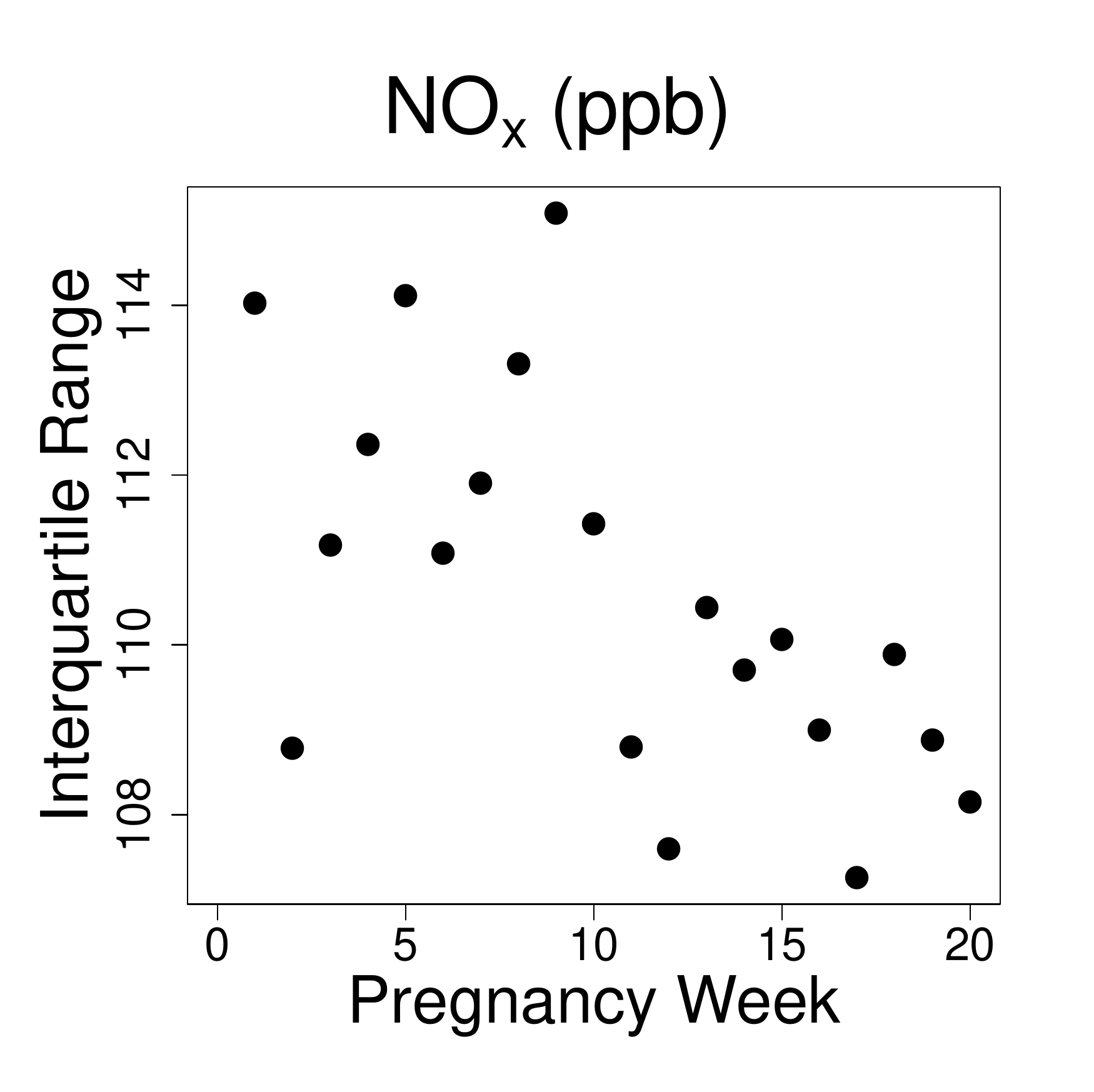}\\
\includegraphics[scale=0.26]{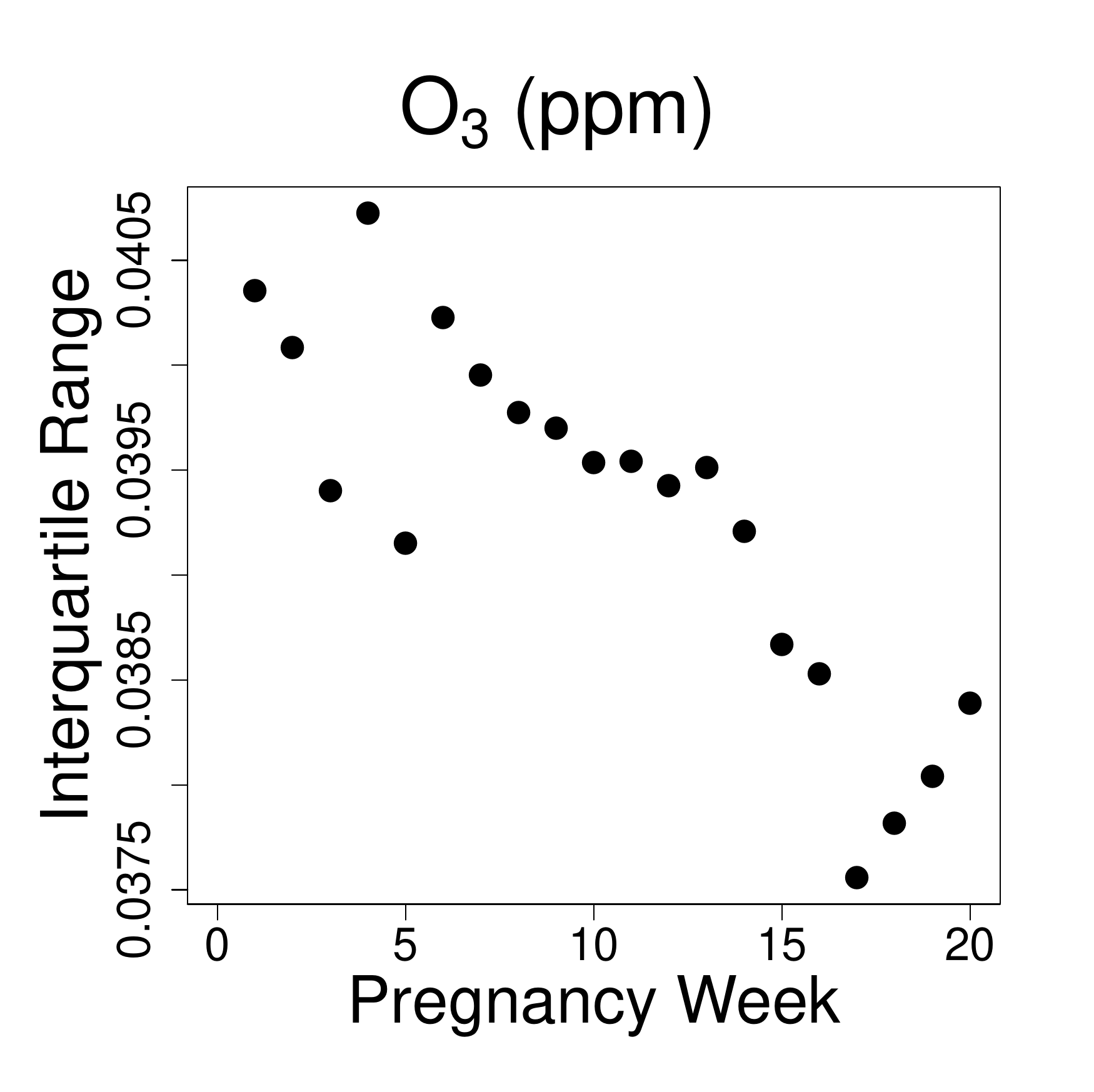}
\includegraphics[scale=0.26]{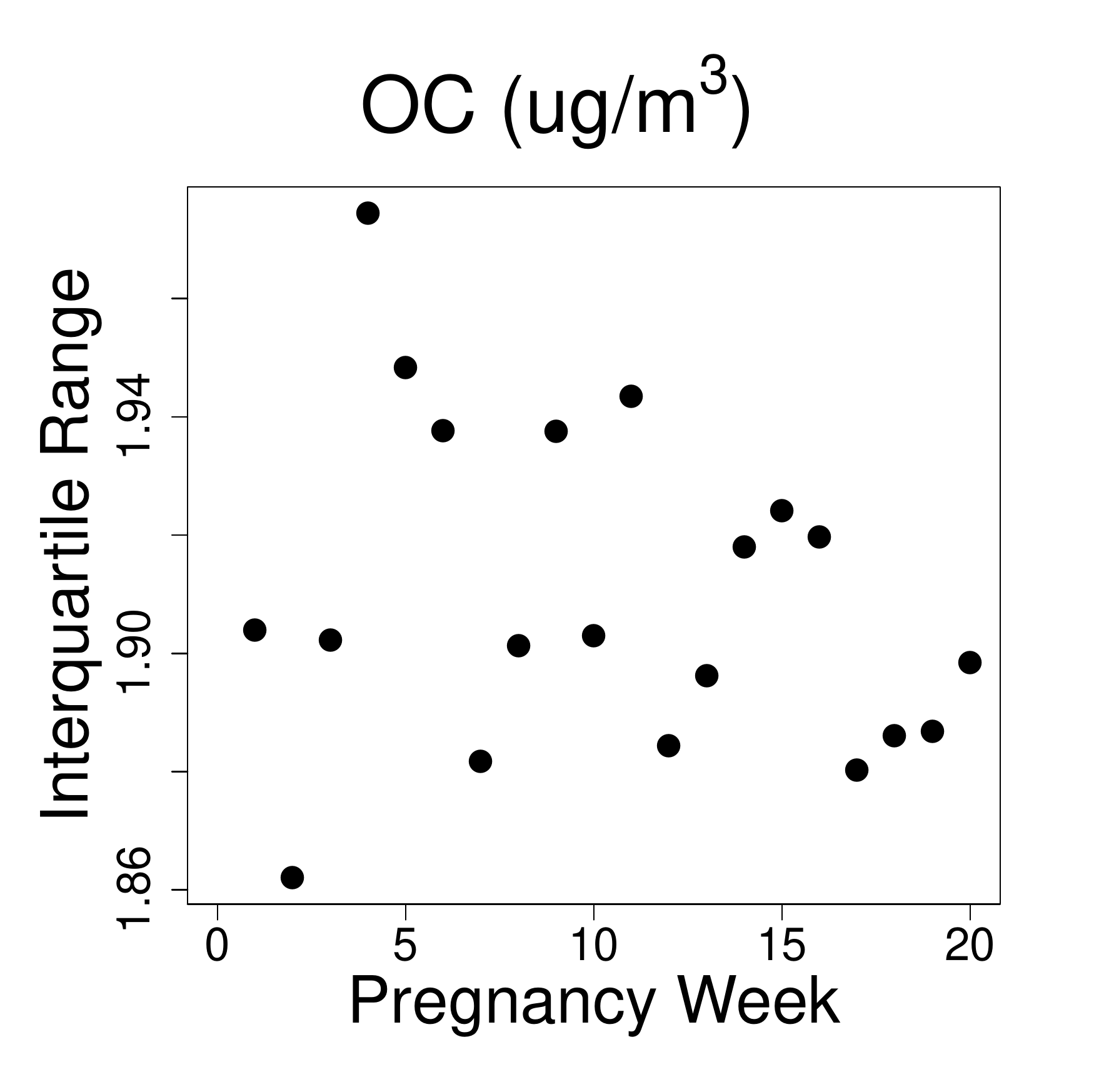}
\includegraphics[scale=0.26]{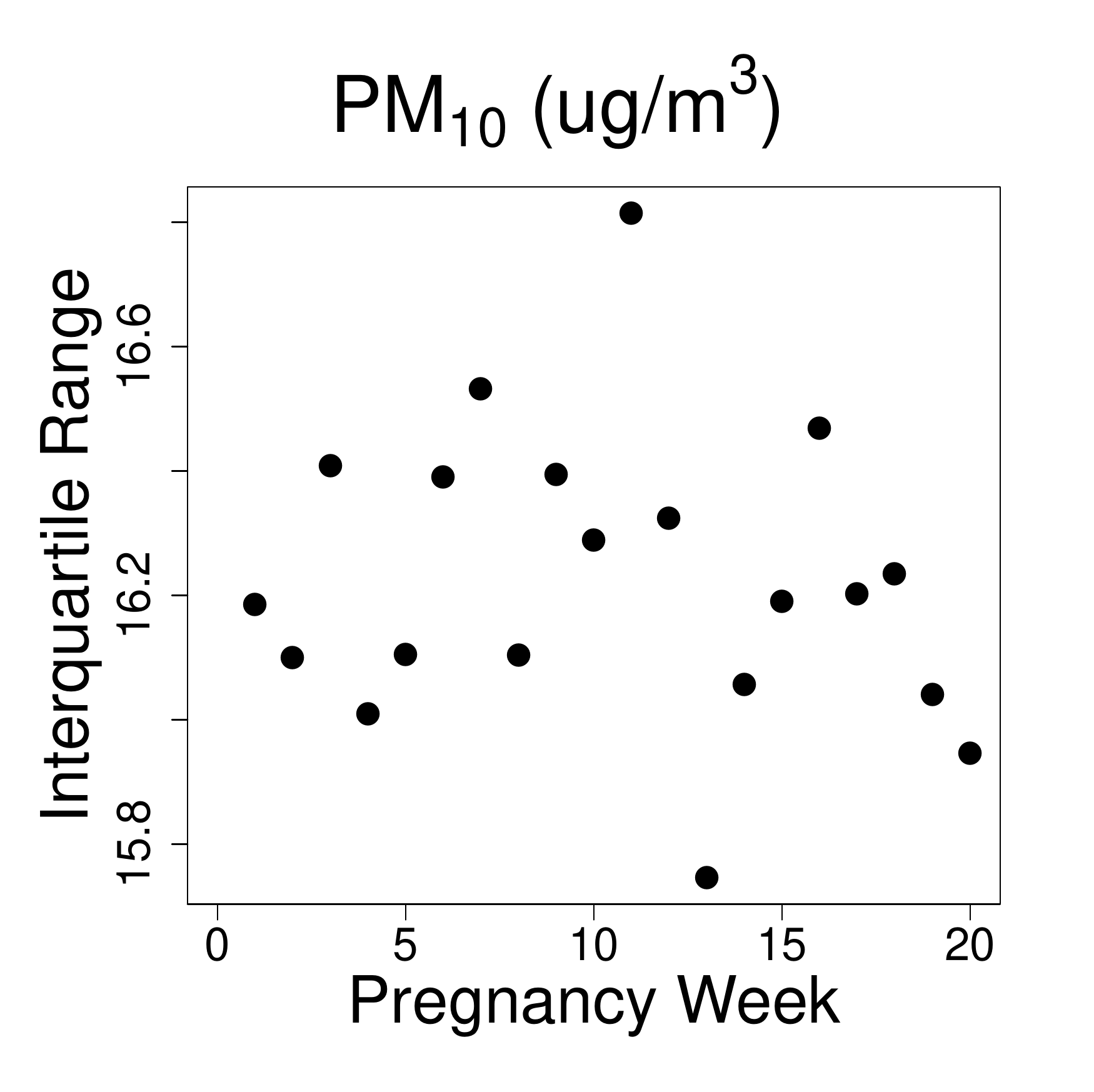}\\
\includegraphics[scale=0.26]{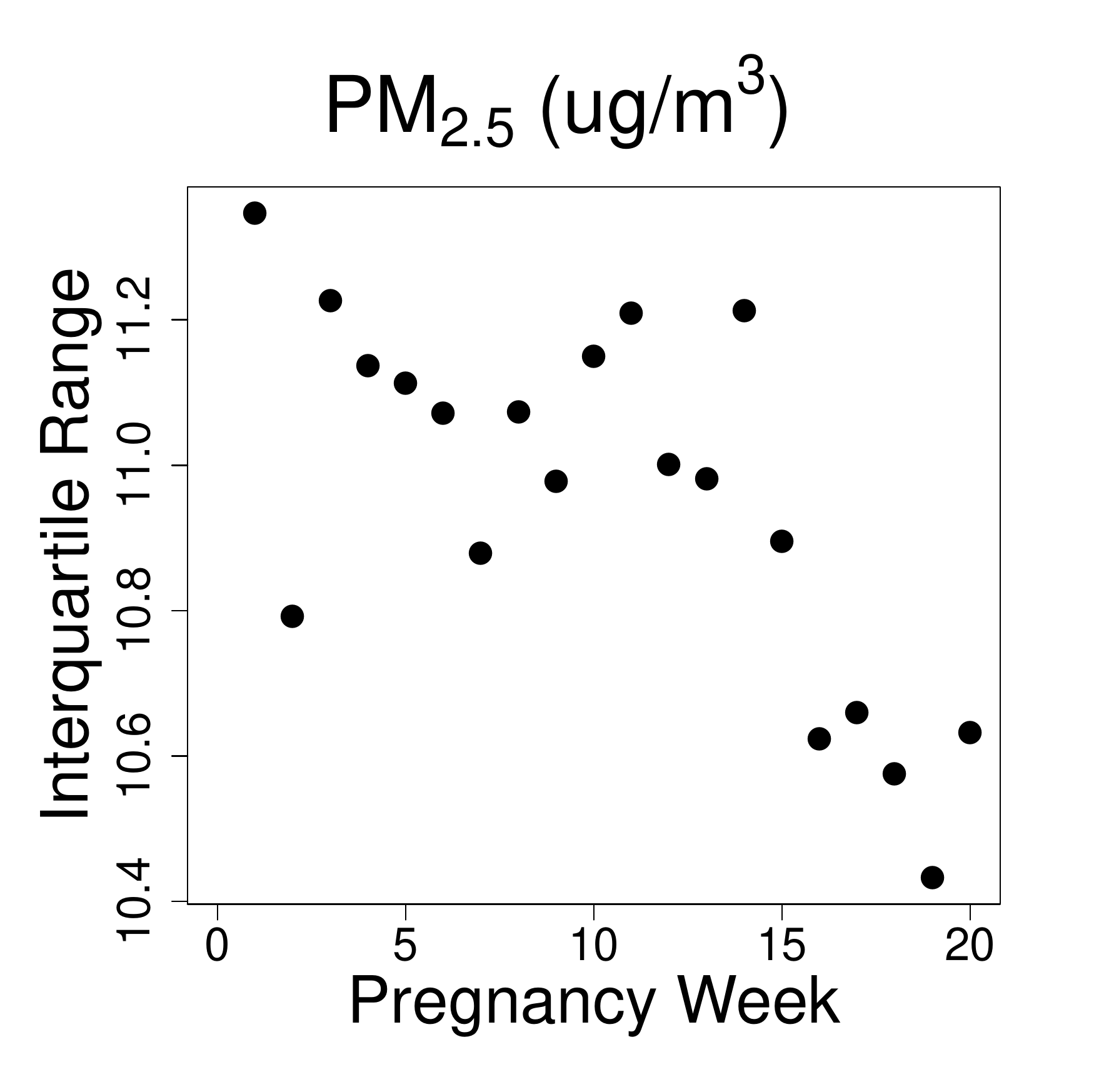}
\includegraphics[scale=0.26]{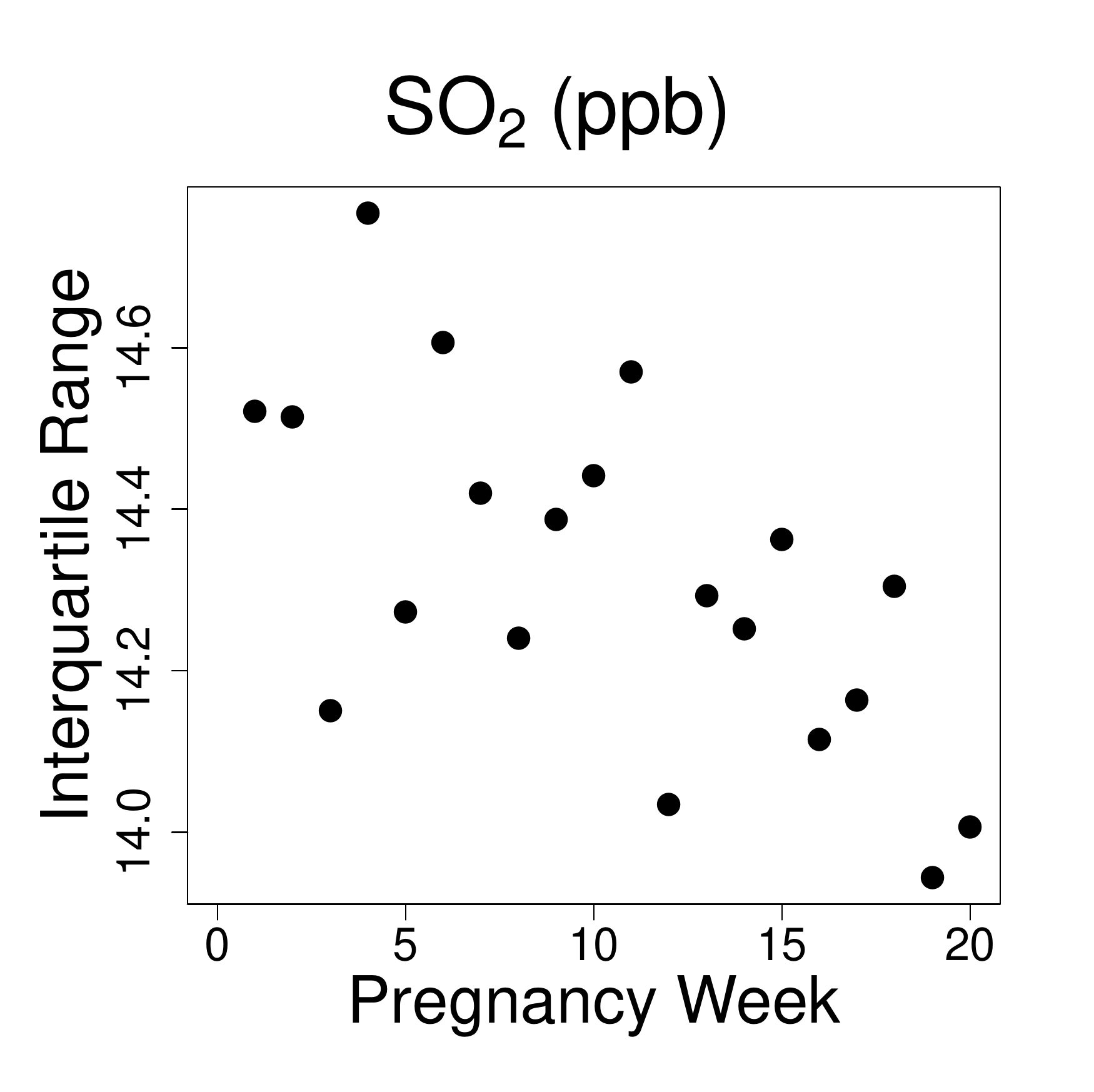}
\includegraphics[scale=0.26]{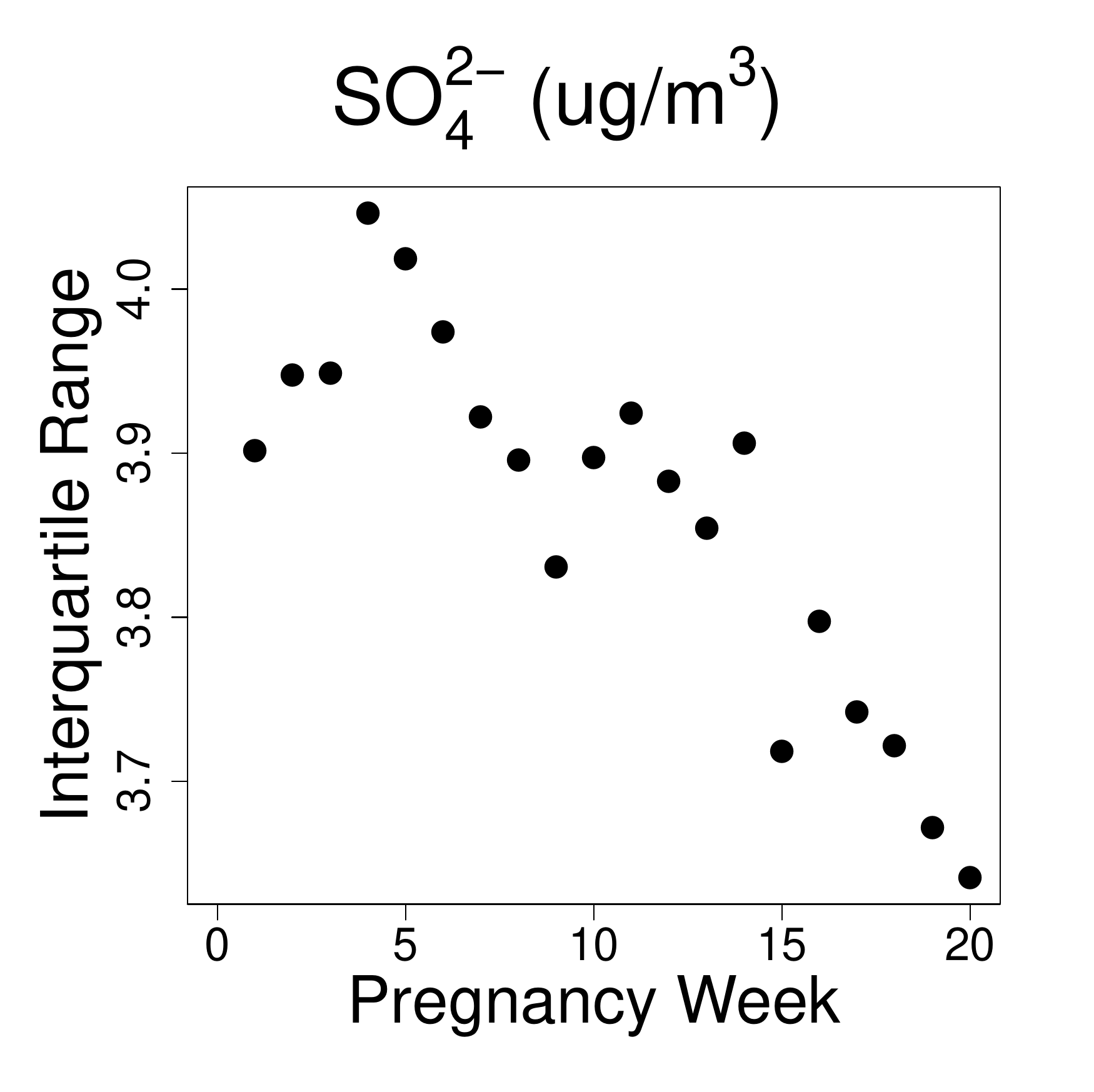}
\caption{Interquartile range for each pollutant across all gestational weeks for the \textbf{non-Hispanic Black} study population in New Jersey, 2005-2014.}
\end{center}
\end{figure}
\clearpage 

\begin{figure}[h]
\begin{center}
\includegraphics[scale=0.26]{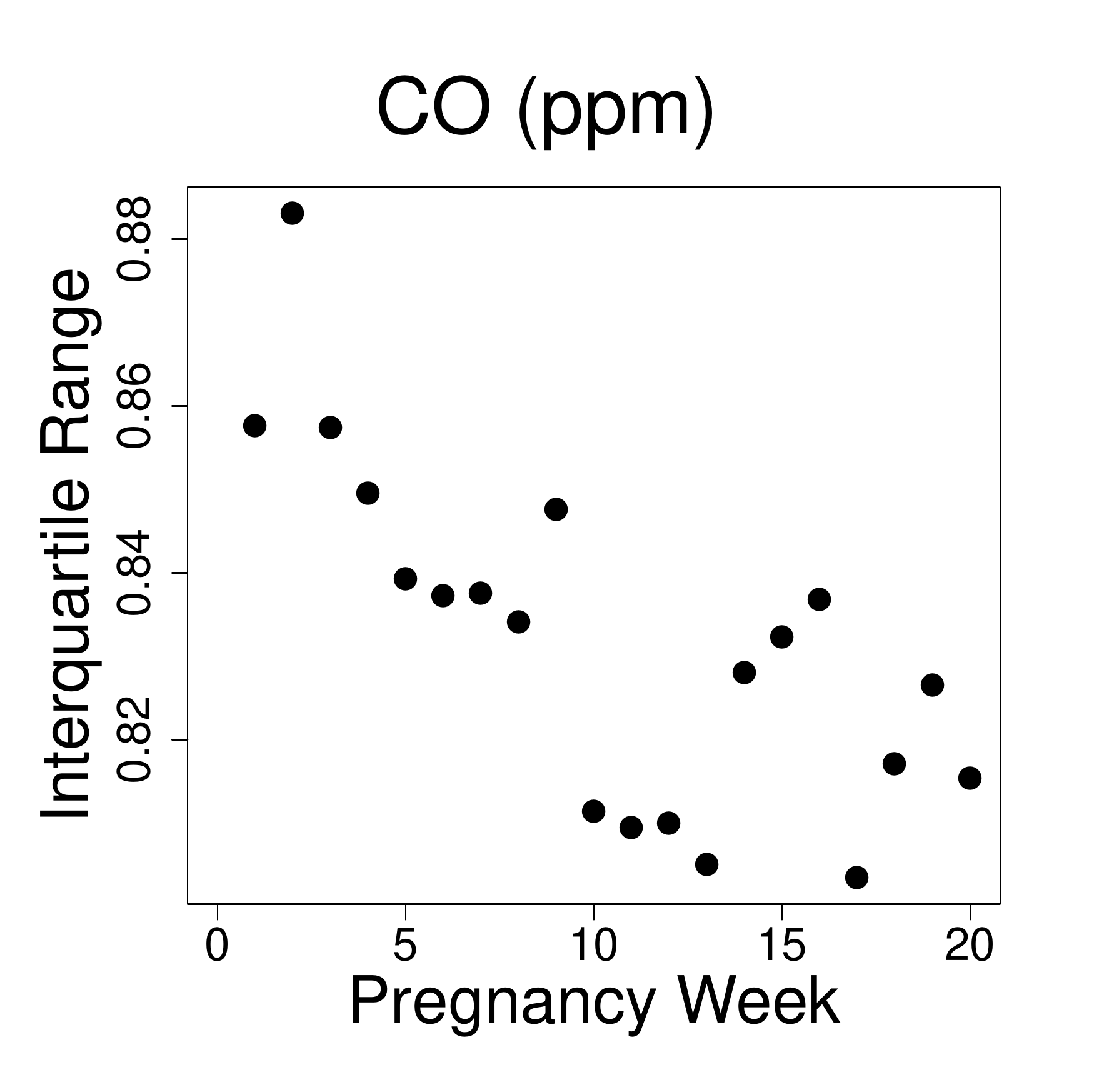}
\includegraphics[scale=0.26]{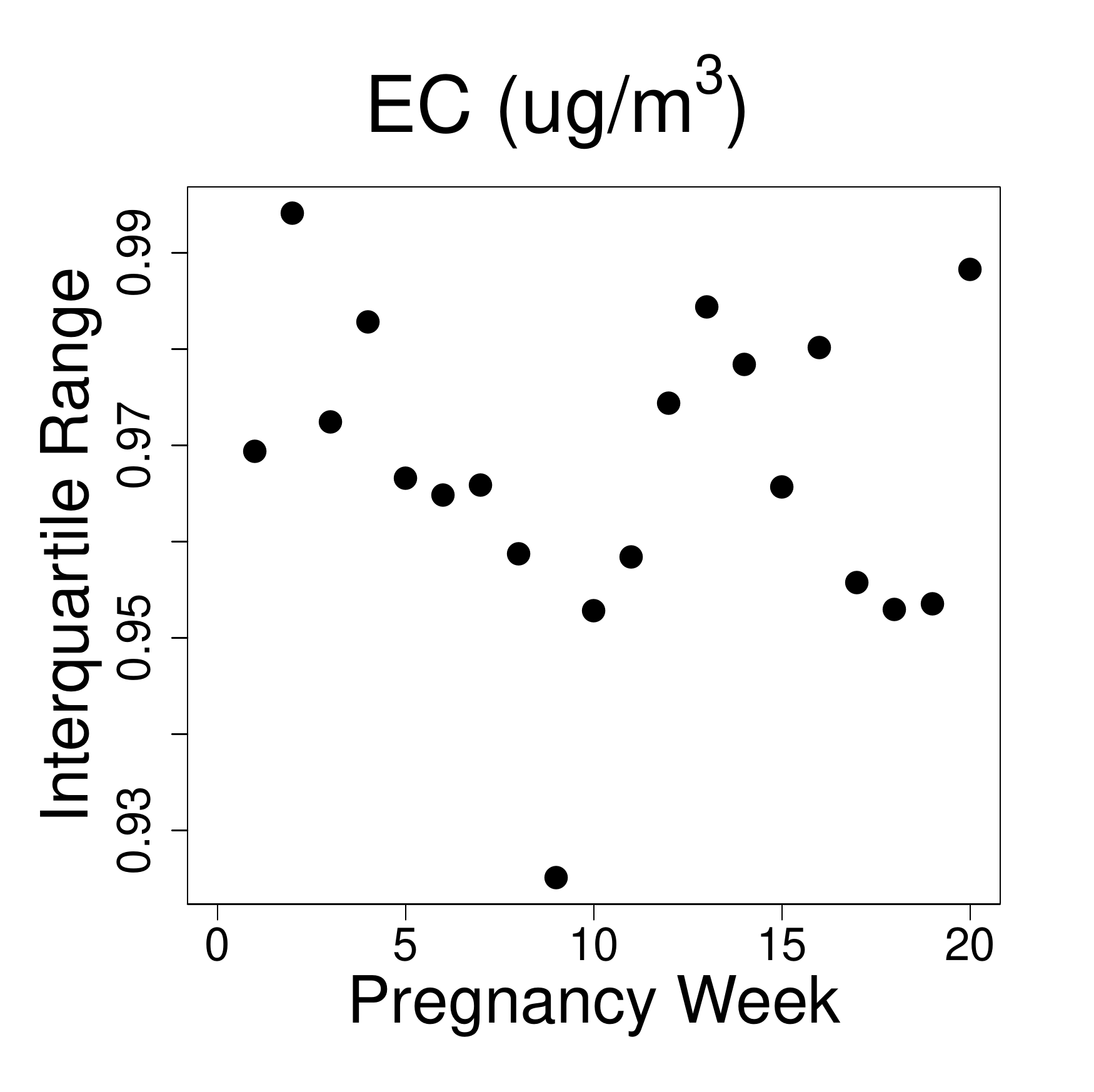}
\includegraphics[scale=0.26]{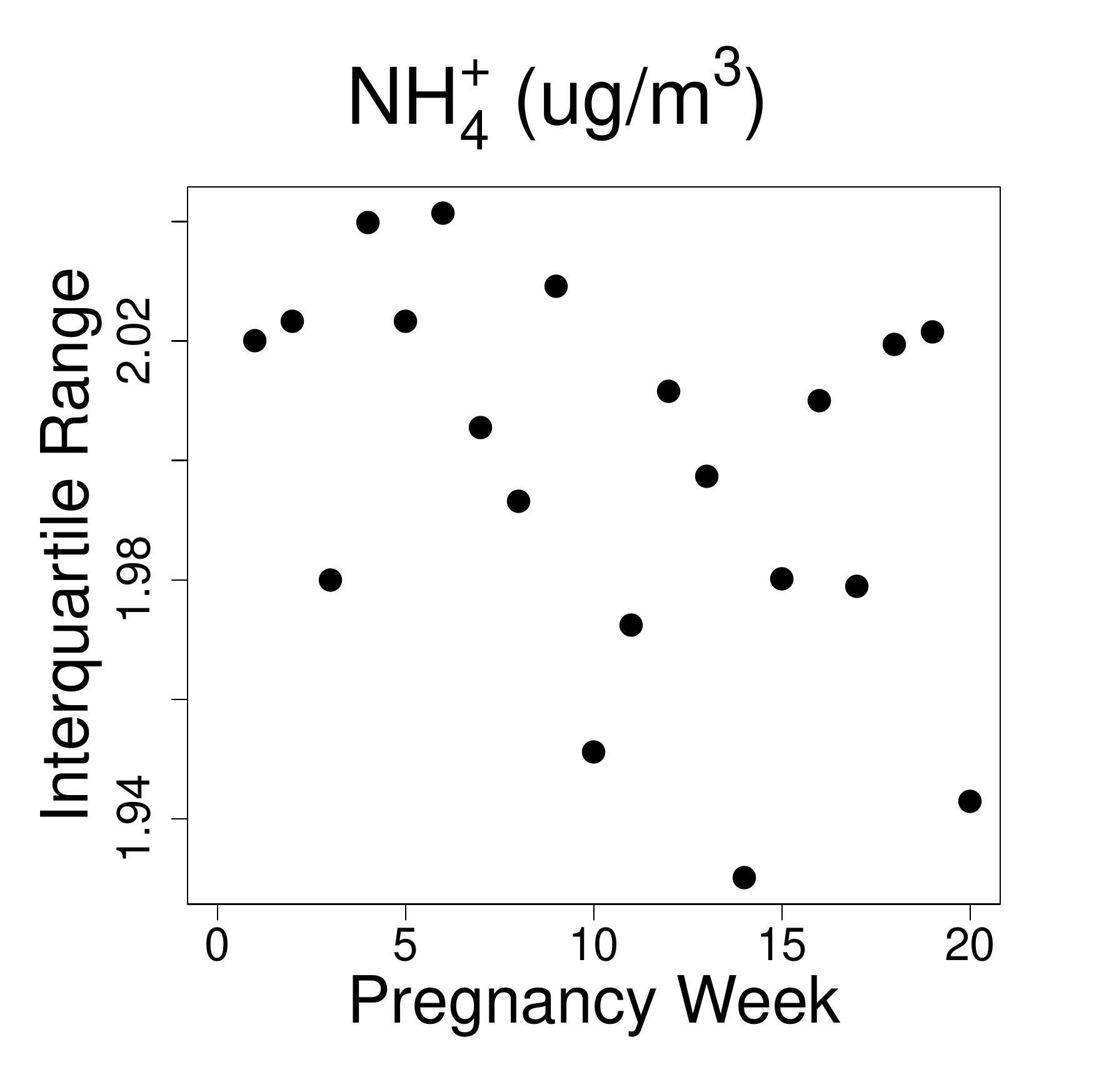}\\
\includegraphics[scale=0.26]{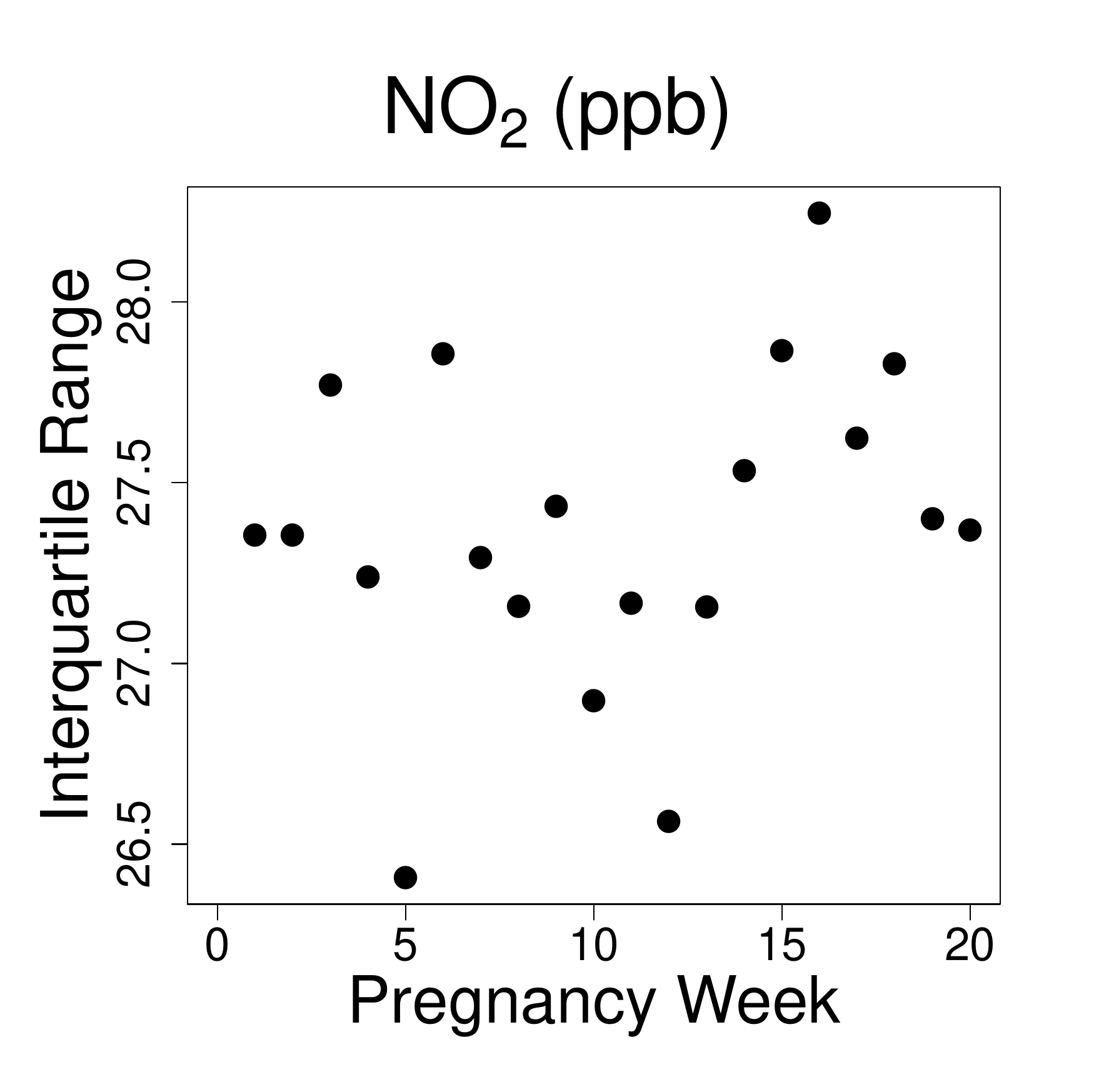}
\includegraphics[scale=0.26]{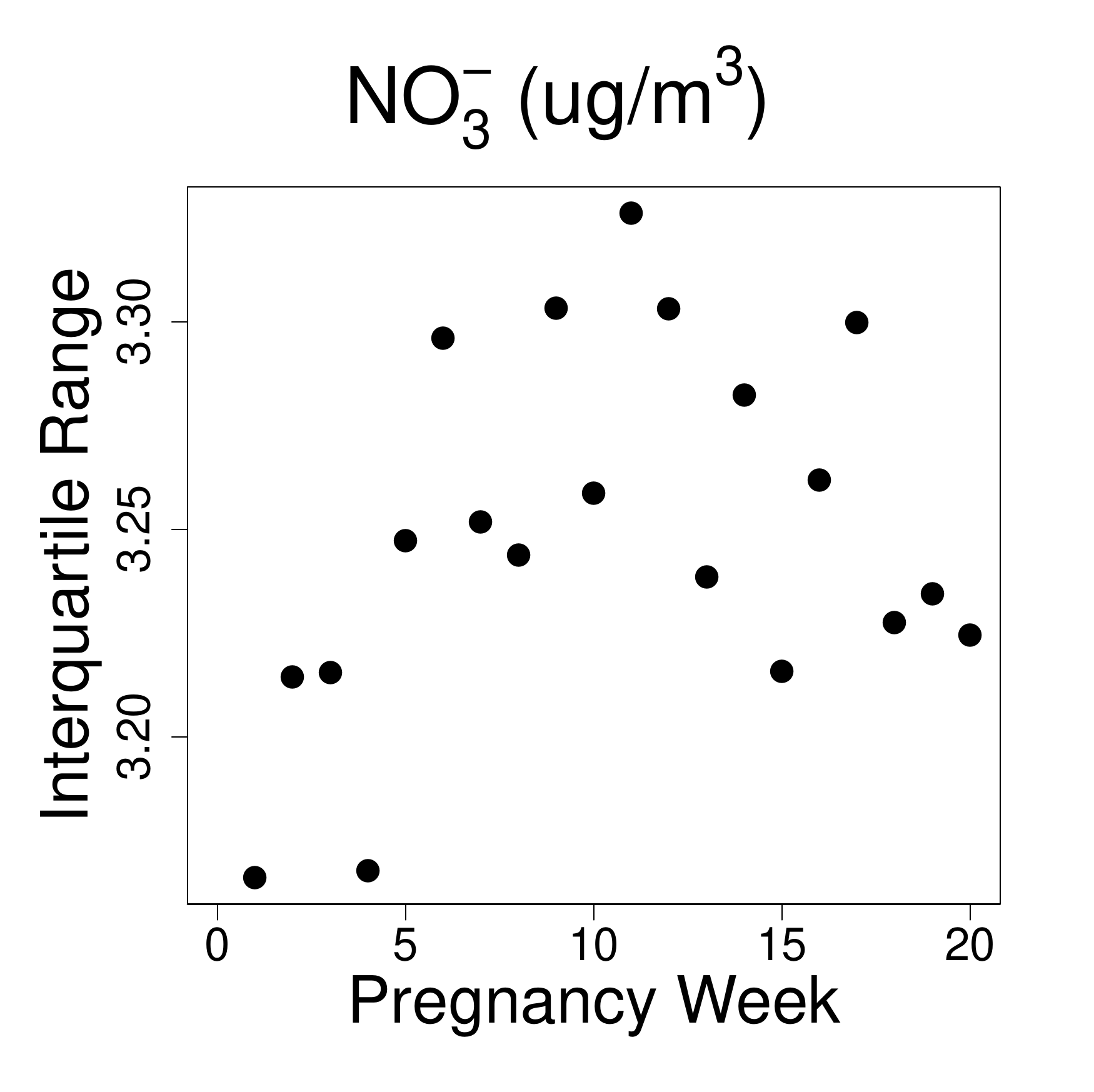}
\includegraphics[scale=0.26]{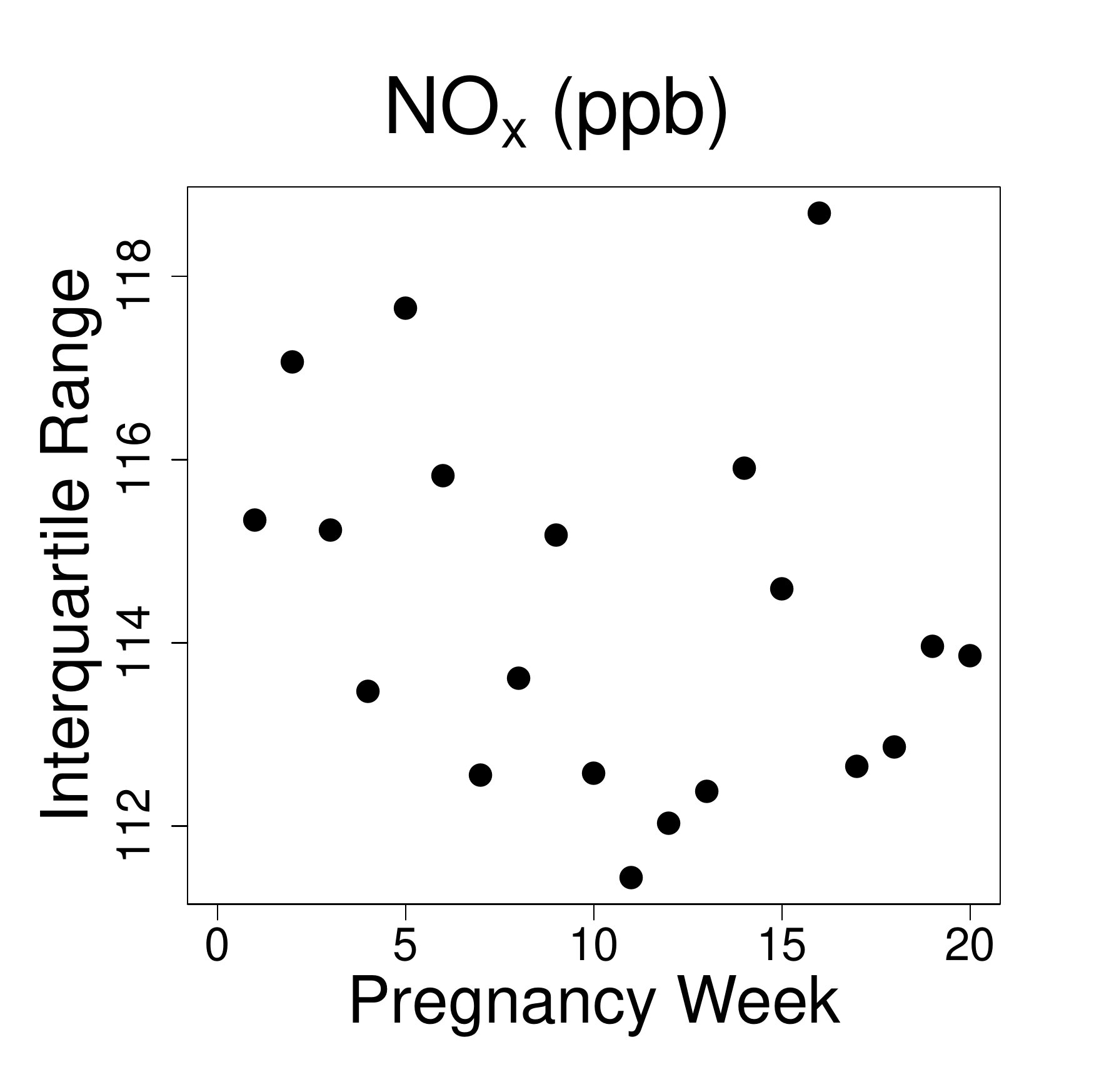}\\
\includegraphics[scale=0.26]{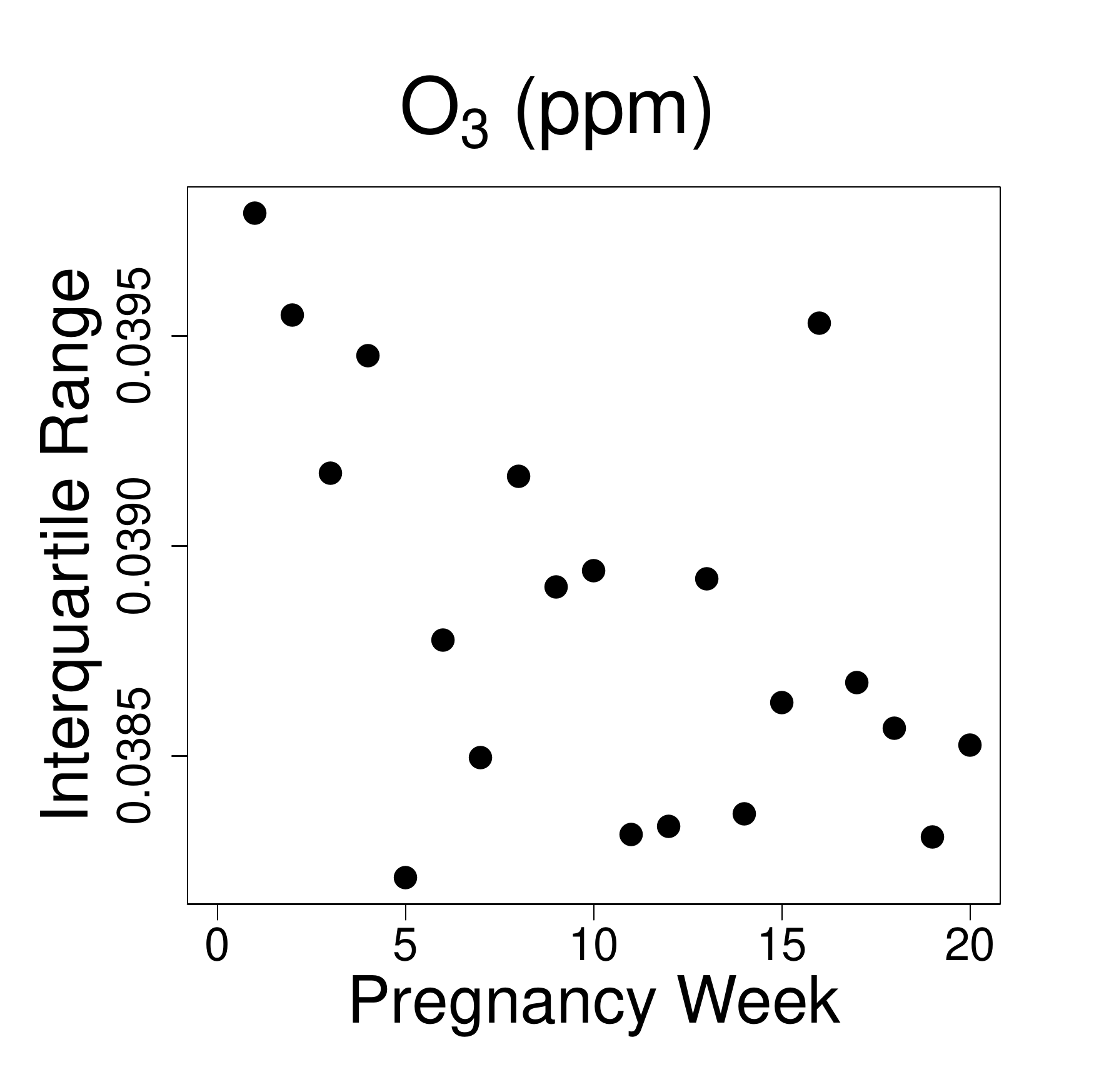}
\includegraphics[scale=0.26]{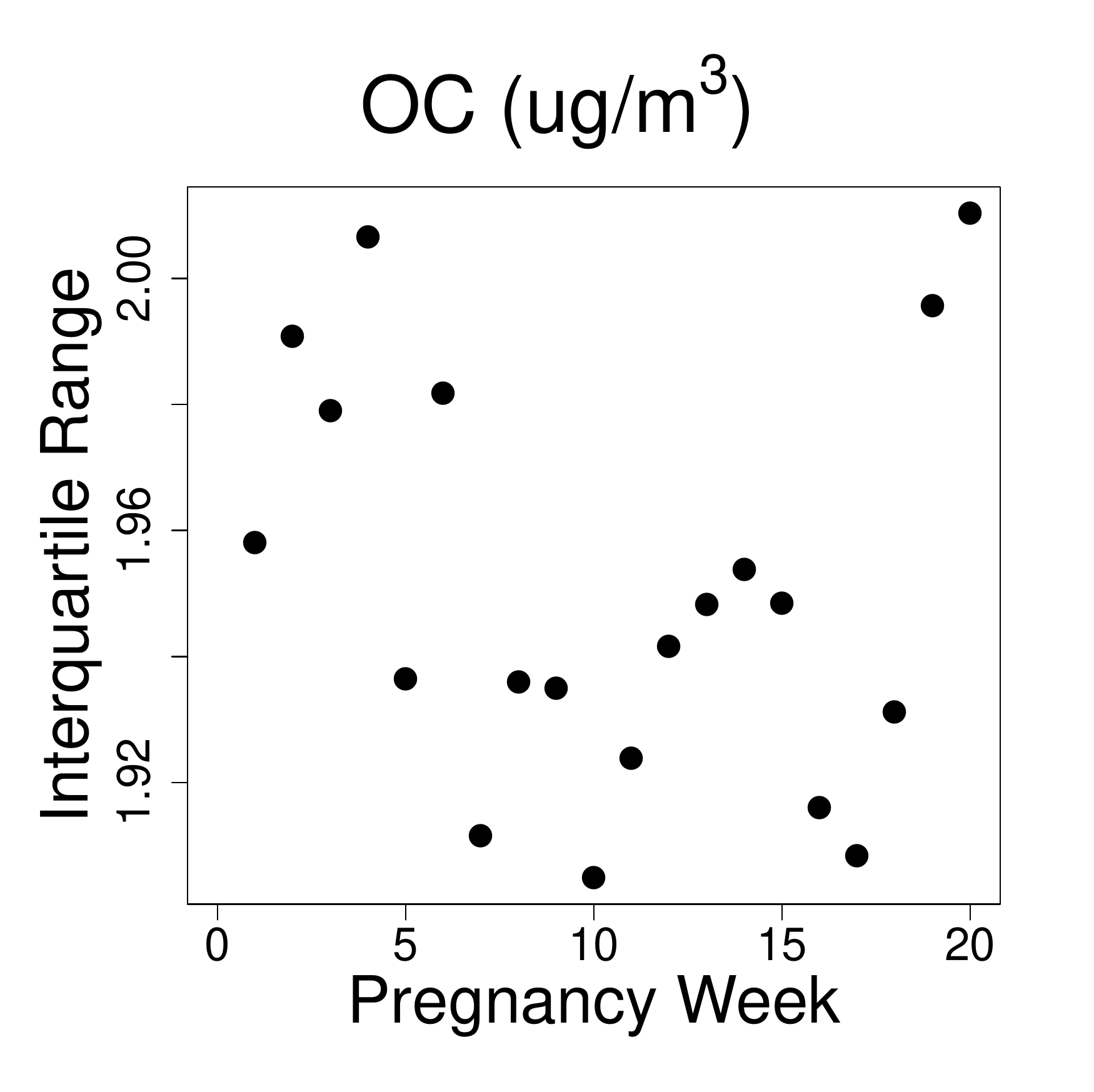}
\includegraphics[scale=0.26]{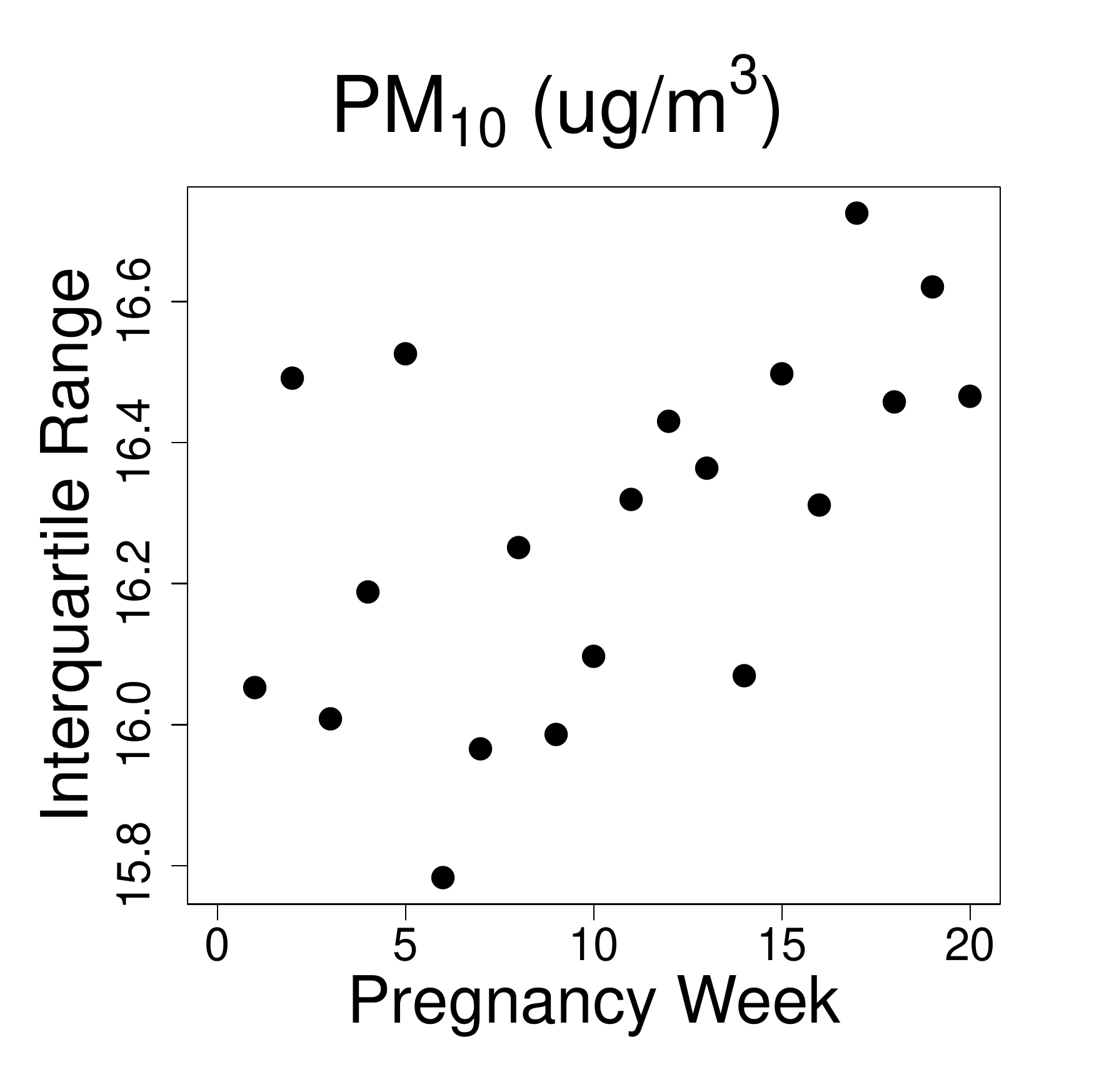}\\
\includegraphics[scale=0.26]{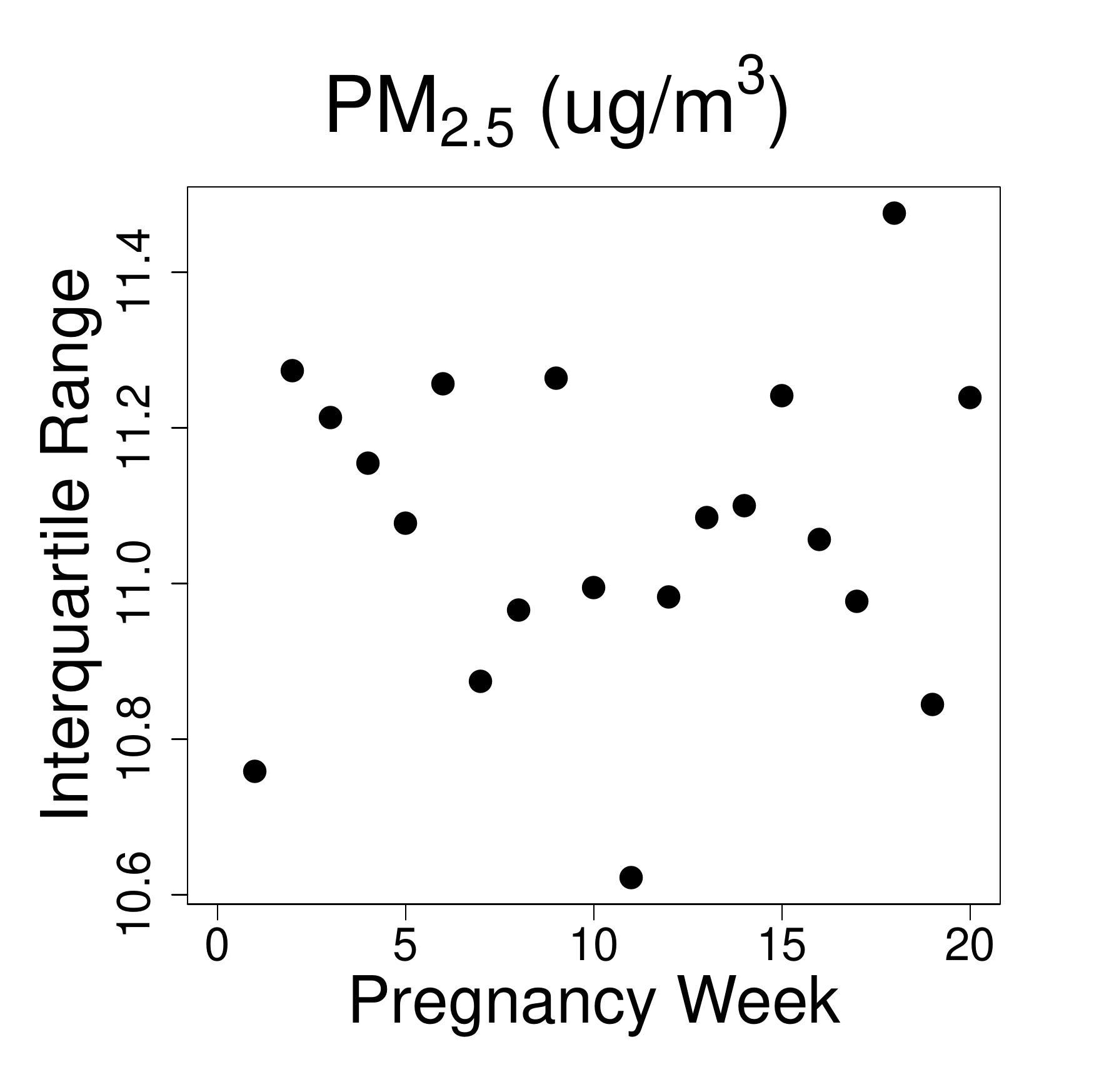}
\includegraphics[scale=0.26]{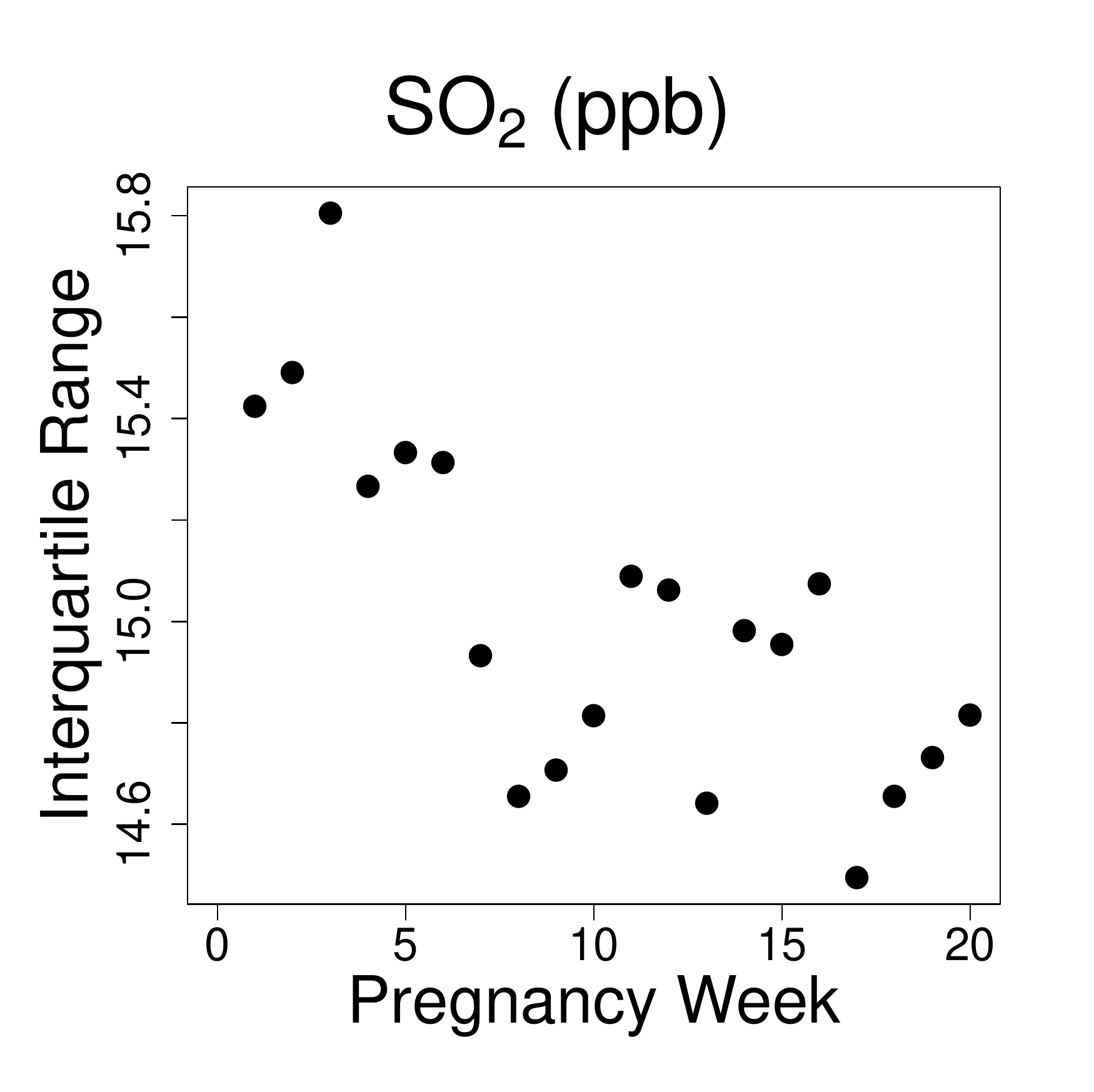}
\includegraphics[scale=0.26]{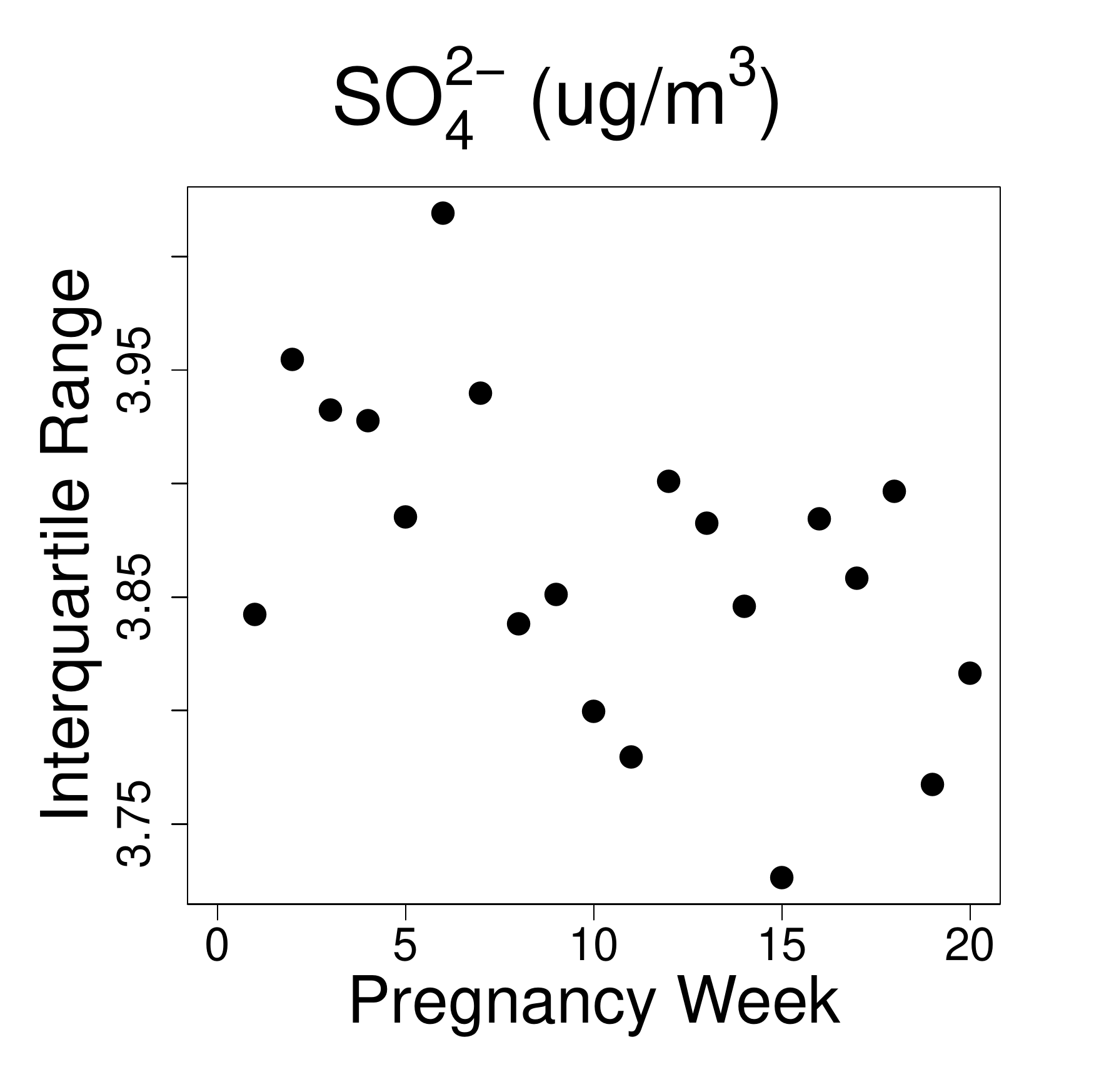}
\caption{Interquartile range for each pollutant across all gestational weeks for the \textbf{Hispanic} study population in New Jersey, 2005-2014.}
\end{center}
\end{figure}
\clearpage

\begin{figure}[h]
\begin{center}
\includegraphics[scale=0.26]{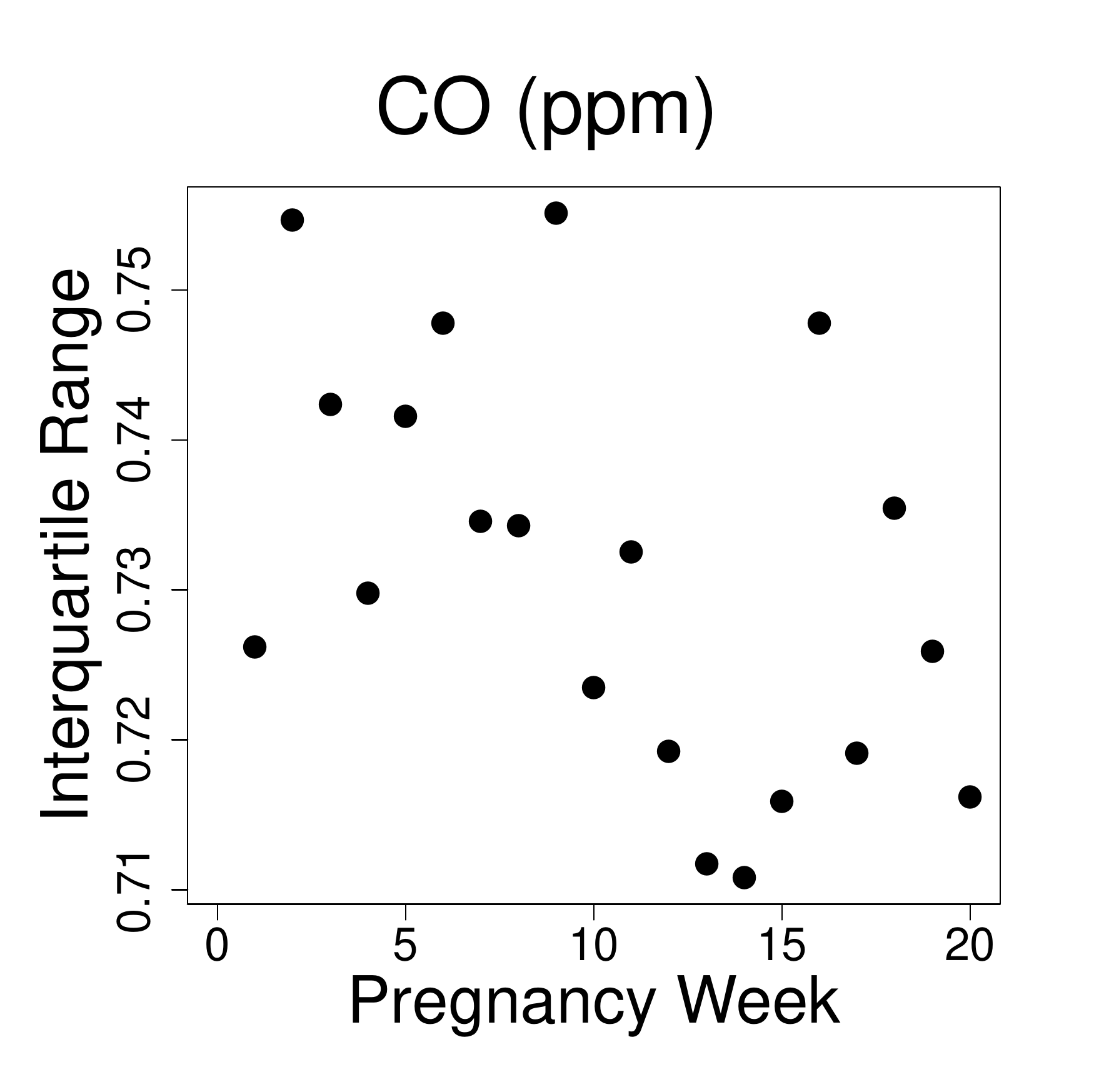}
\includegraphics[scale=0.26]{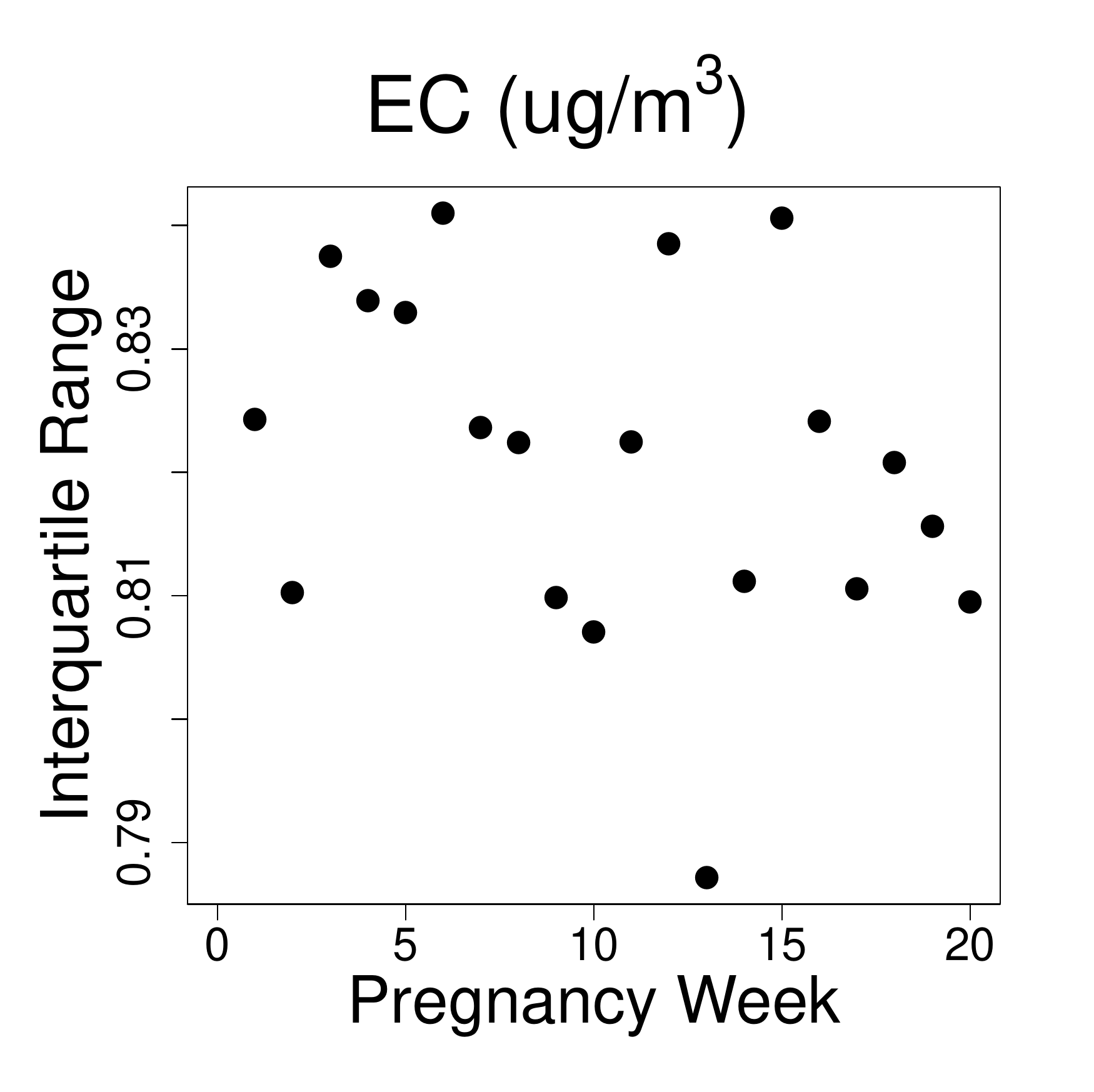}
\includegraphics[scale=0.26]{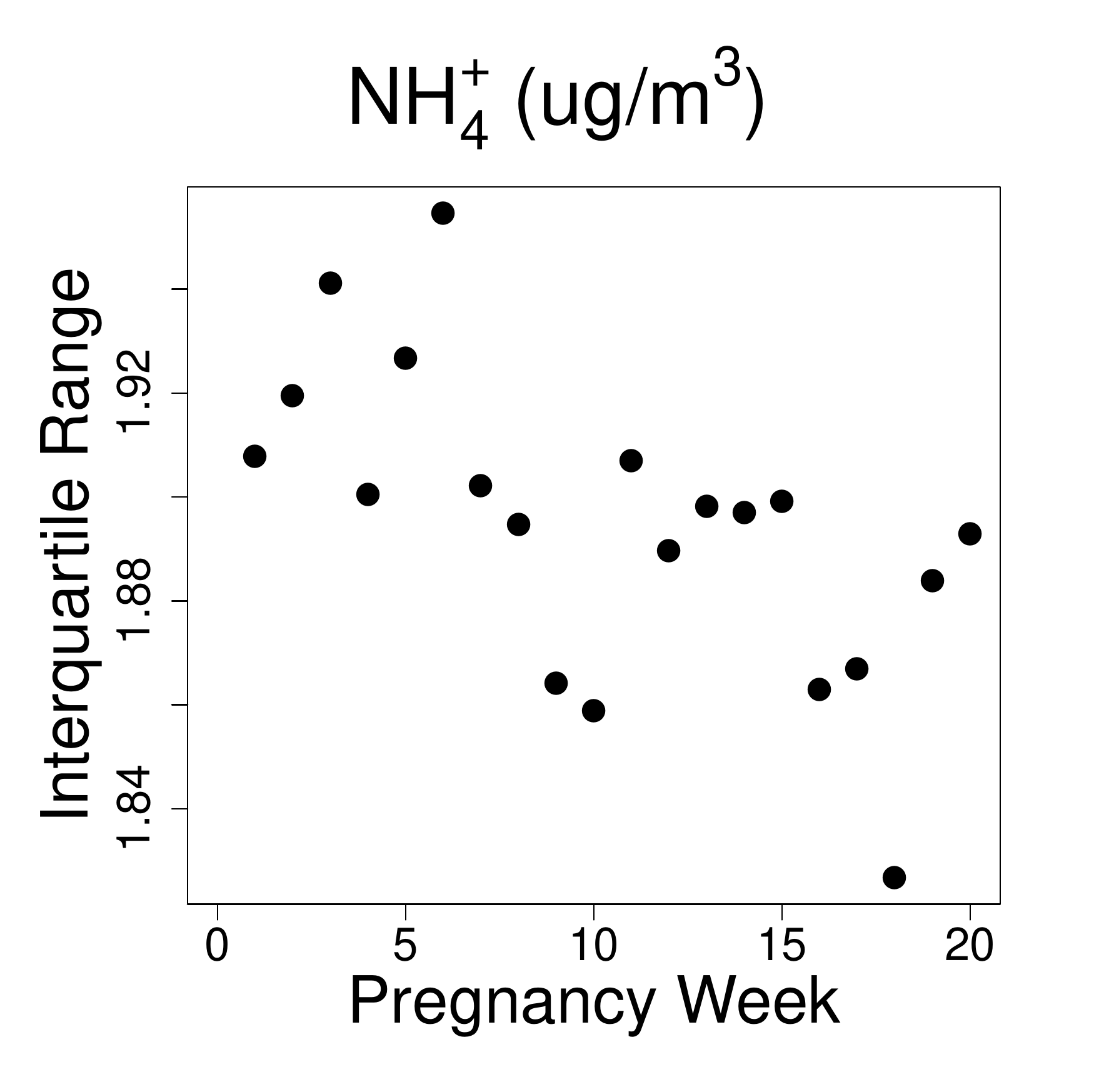}\\
\includegraphics[scale=0.26]{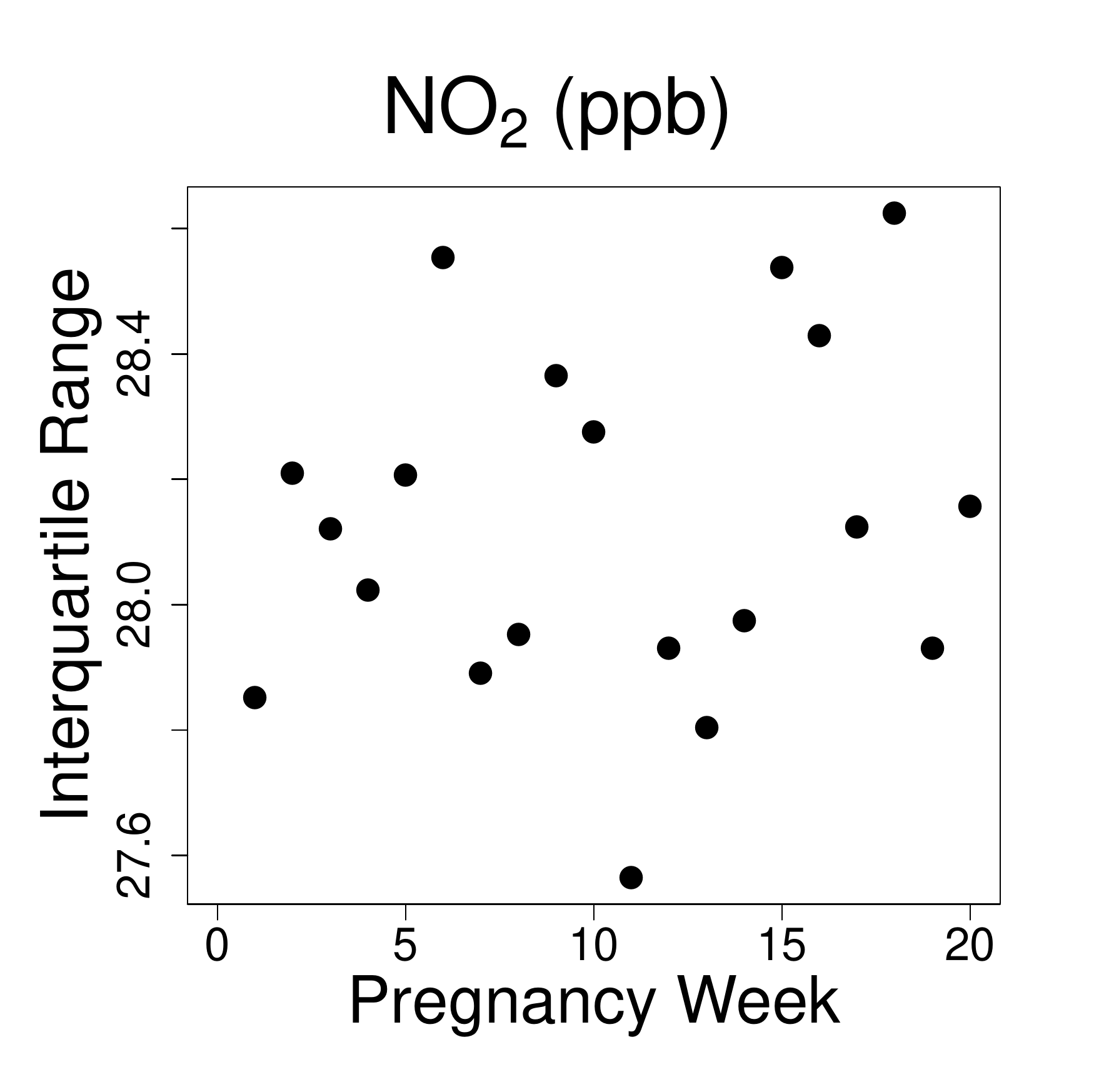}
\includegraphics[scale=0.26]{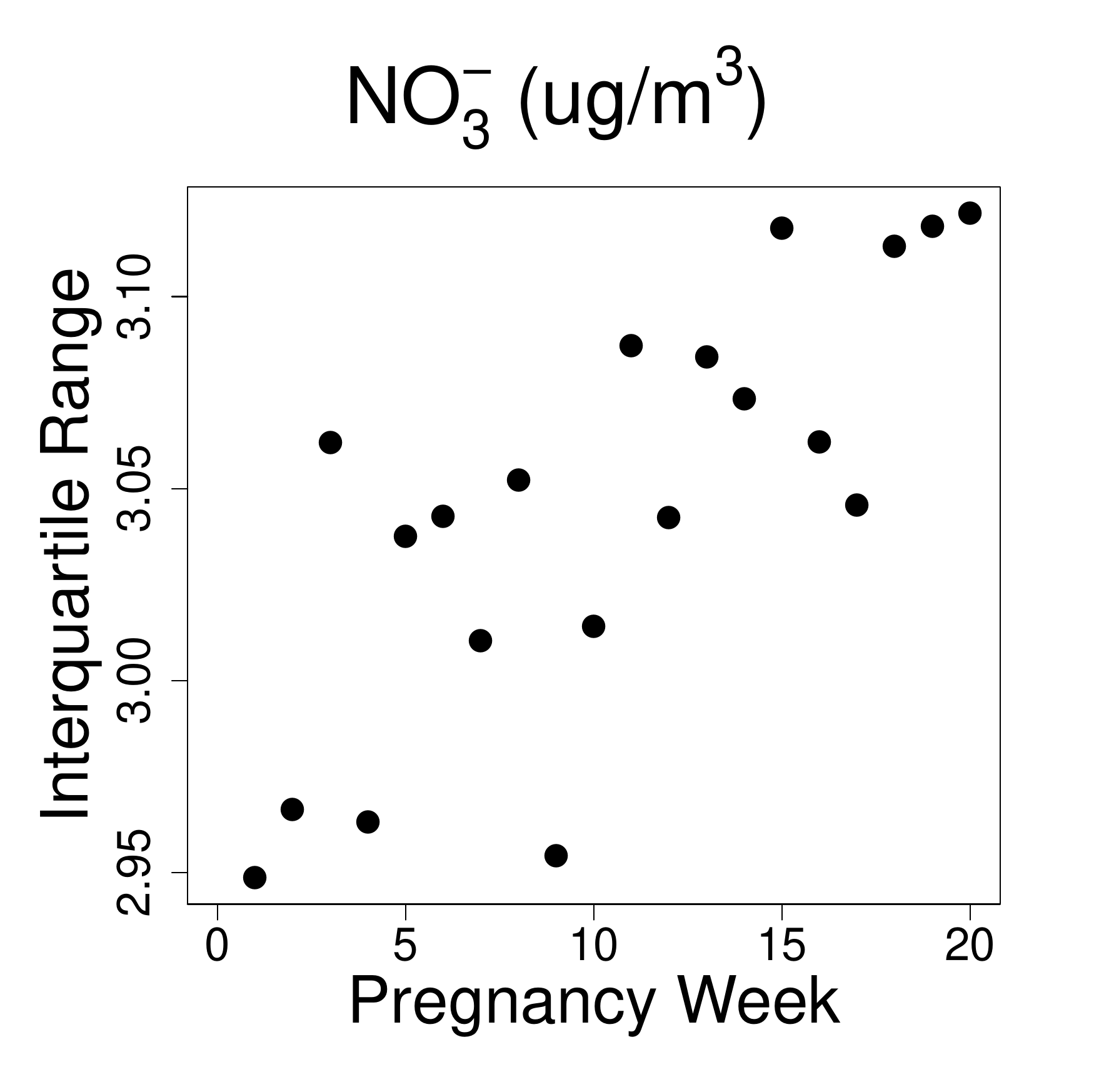}
\includegraphics[scale=0.26]{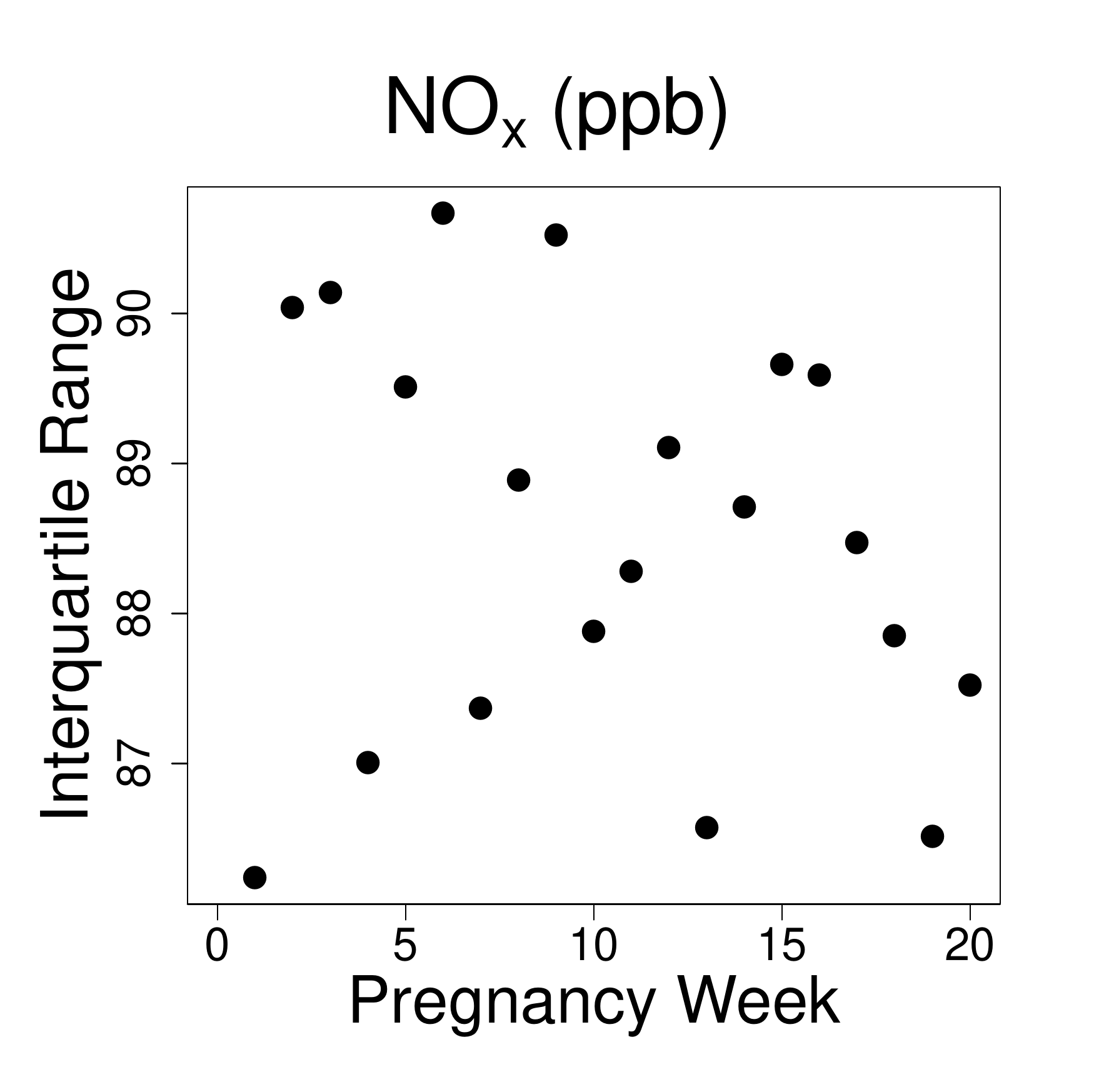}\\
\includegraphics[scale=0.26]{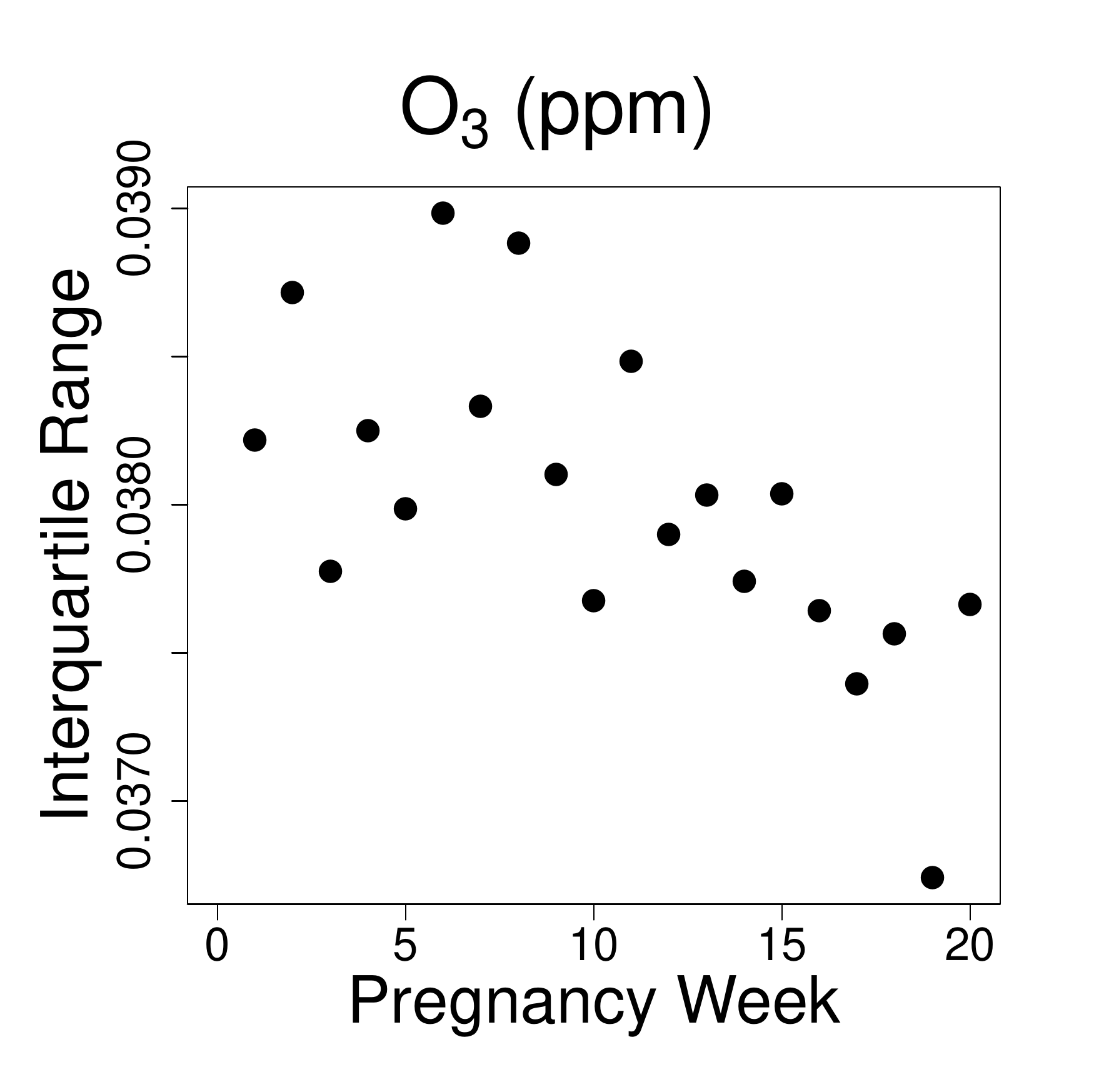}
\includegraphics[scale=0.26]{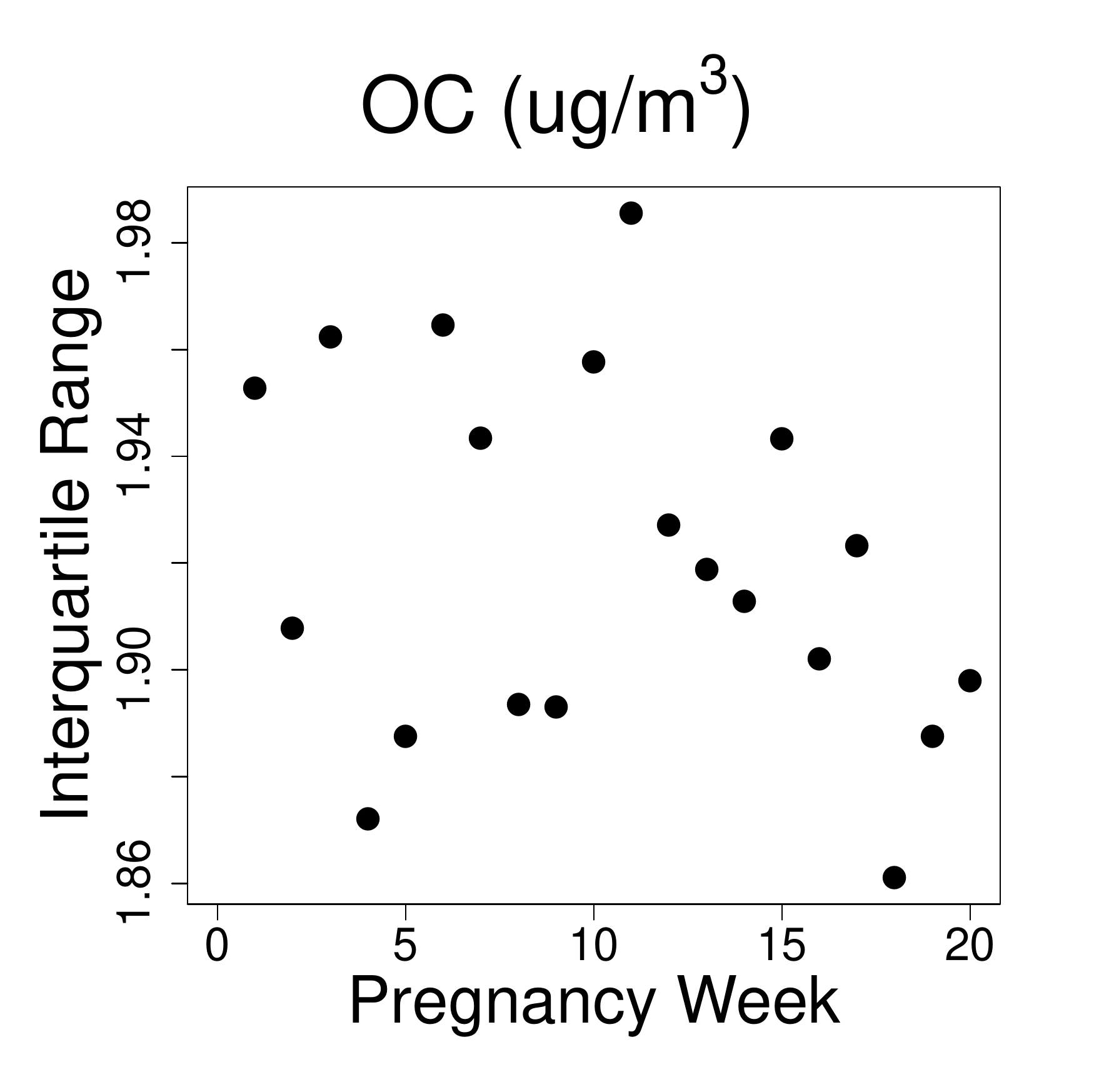}
\includegraphics[scale=0.26]{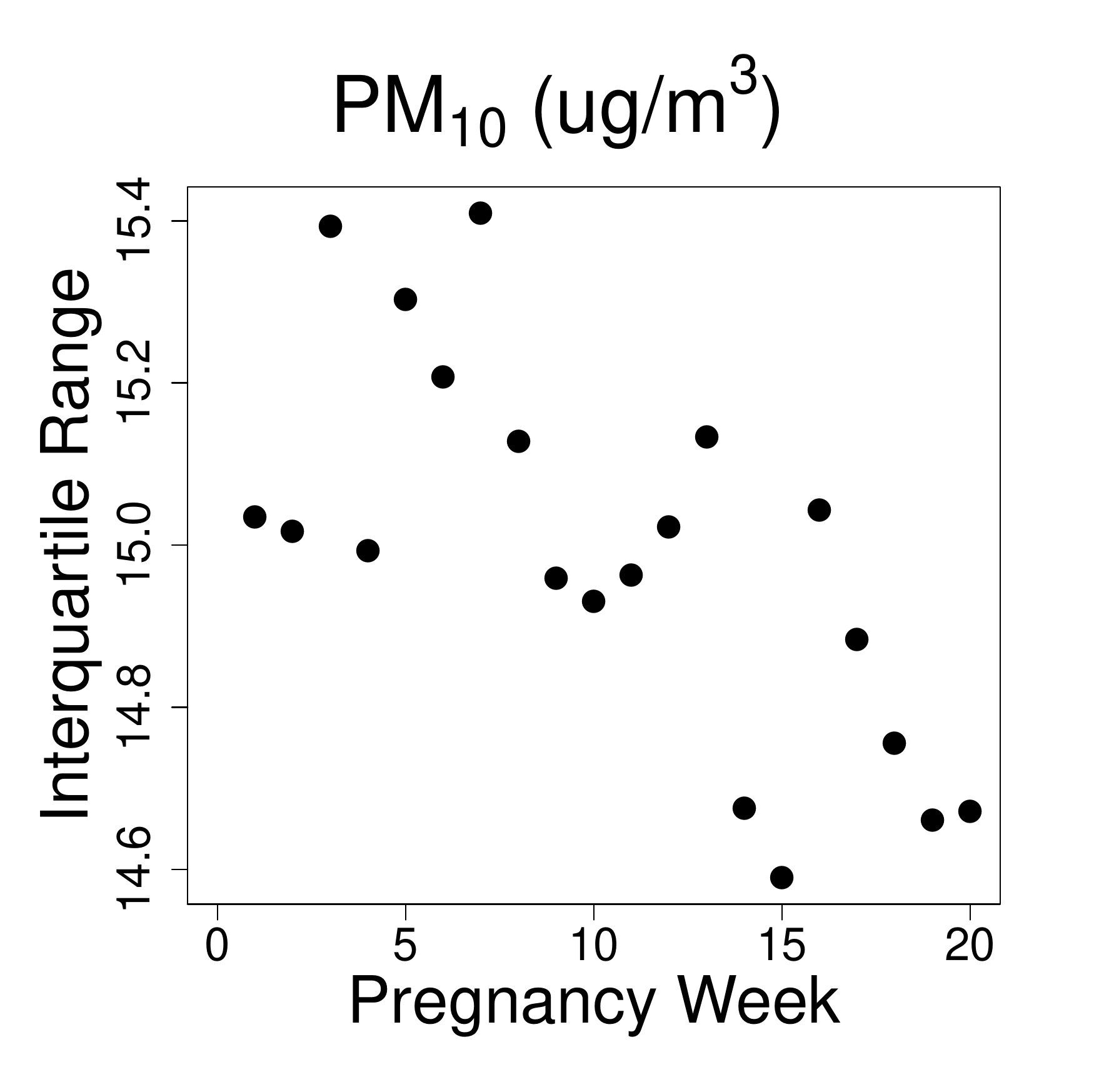}\\
\includegraphics[scale=0.26]{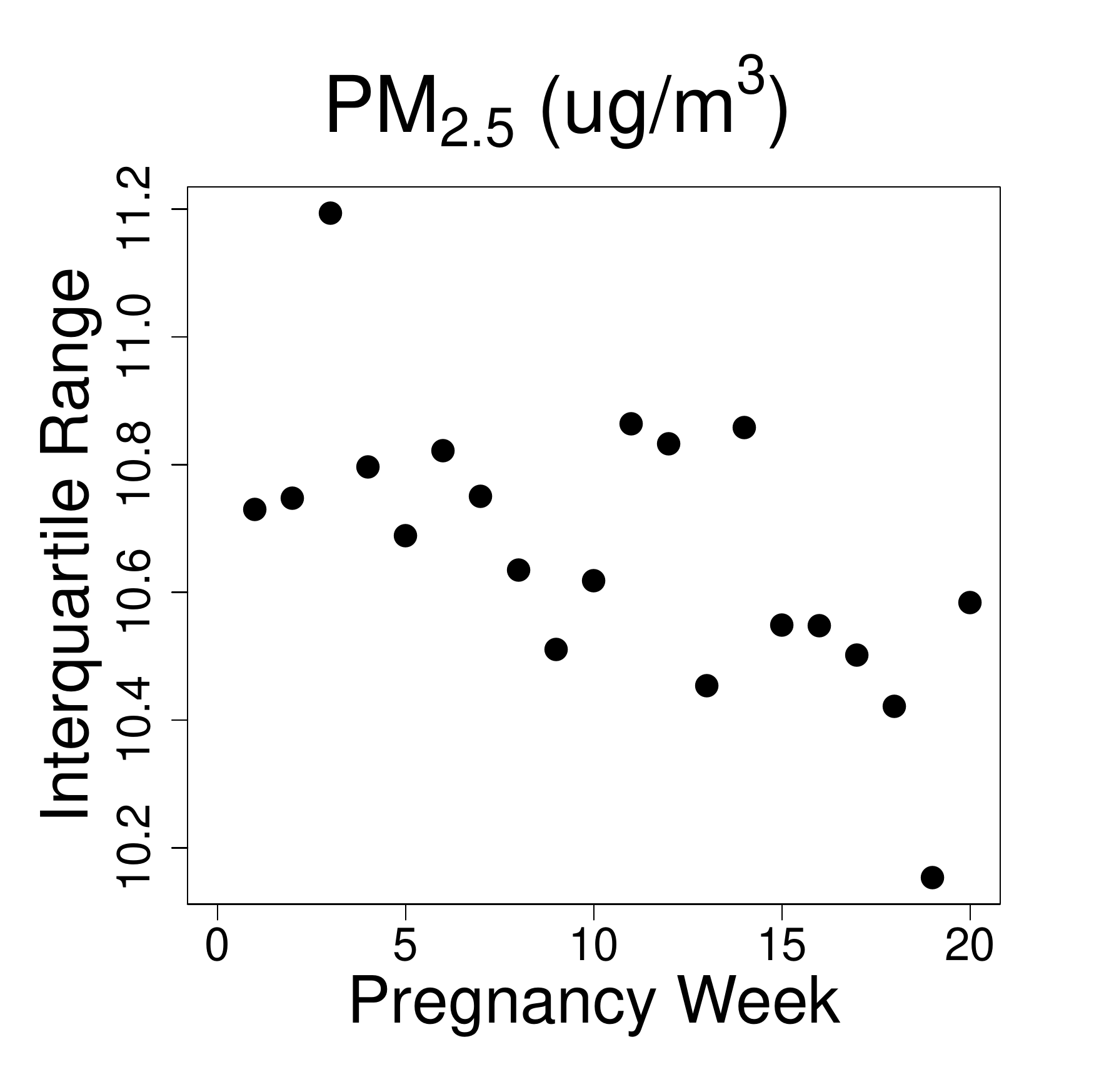}
\includegraphics[scale=0.26]{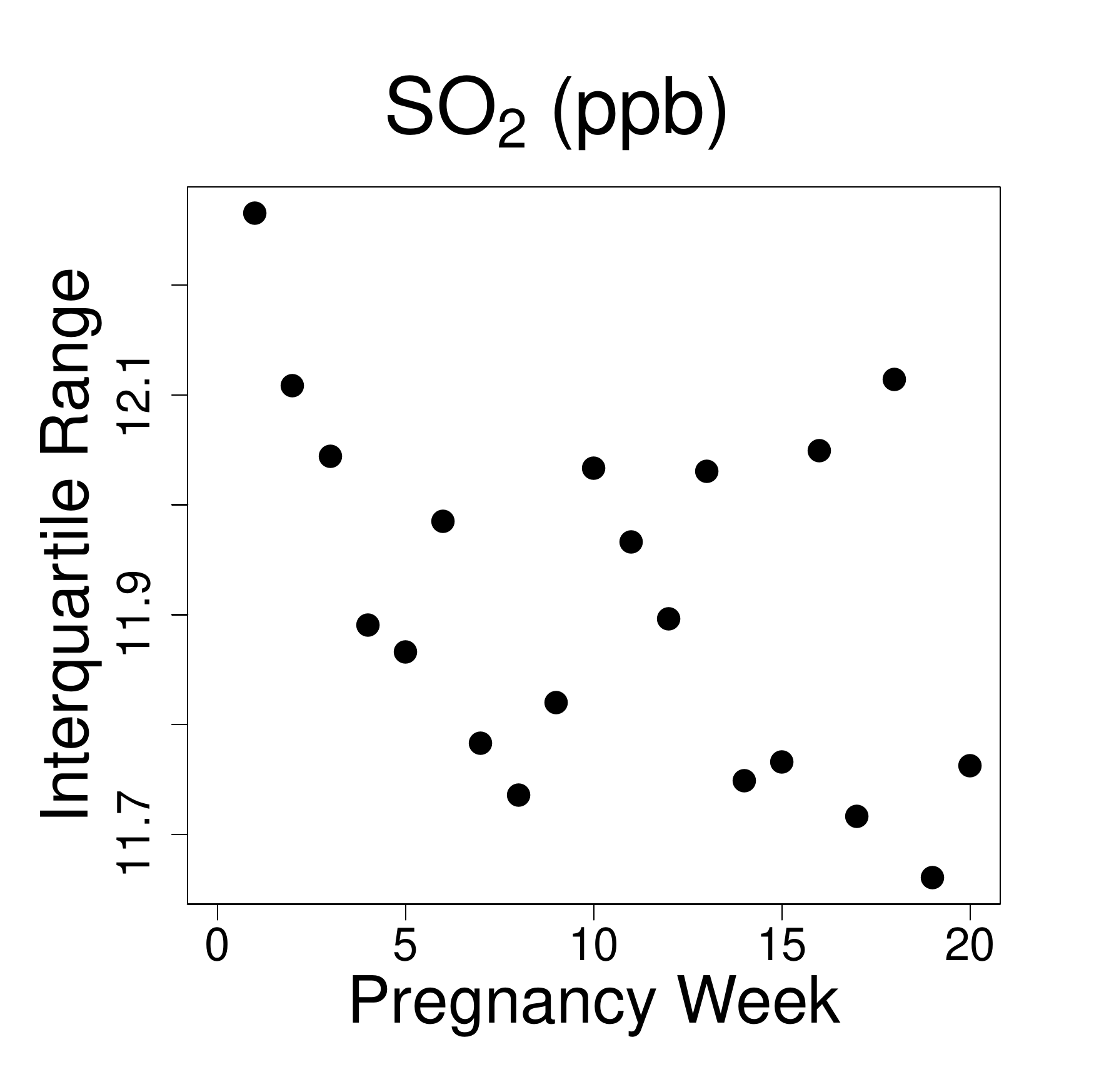}
\includegraphics[scale=0.26]{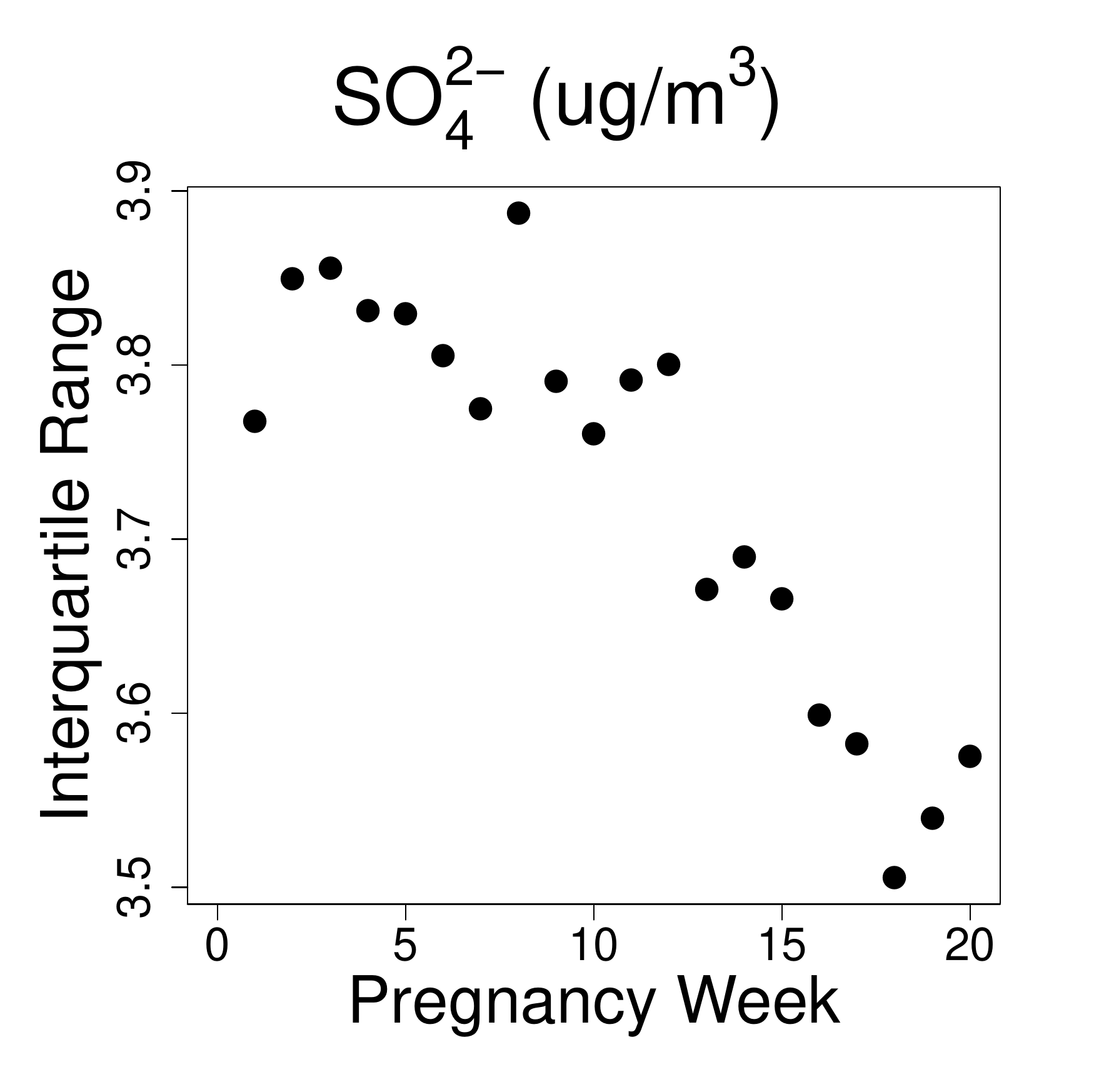}
\caption{Interquartile range for each pollutant across all gestational weeks for the \textbf{non-Hispanic White} study population in New Jersey, 2005-2014.}
\end{center}
\end{figure}
\clearpage

\begin{figure}[h]
\begin{center}
\includegraphics[scale=0.26]{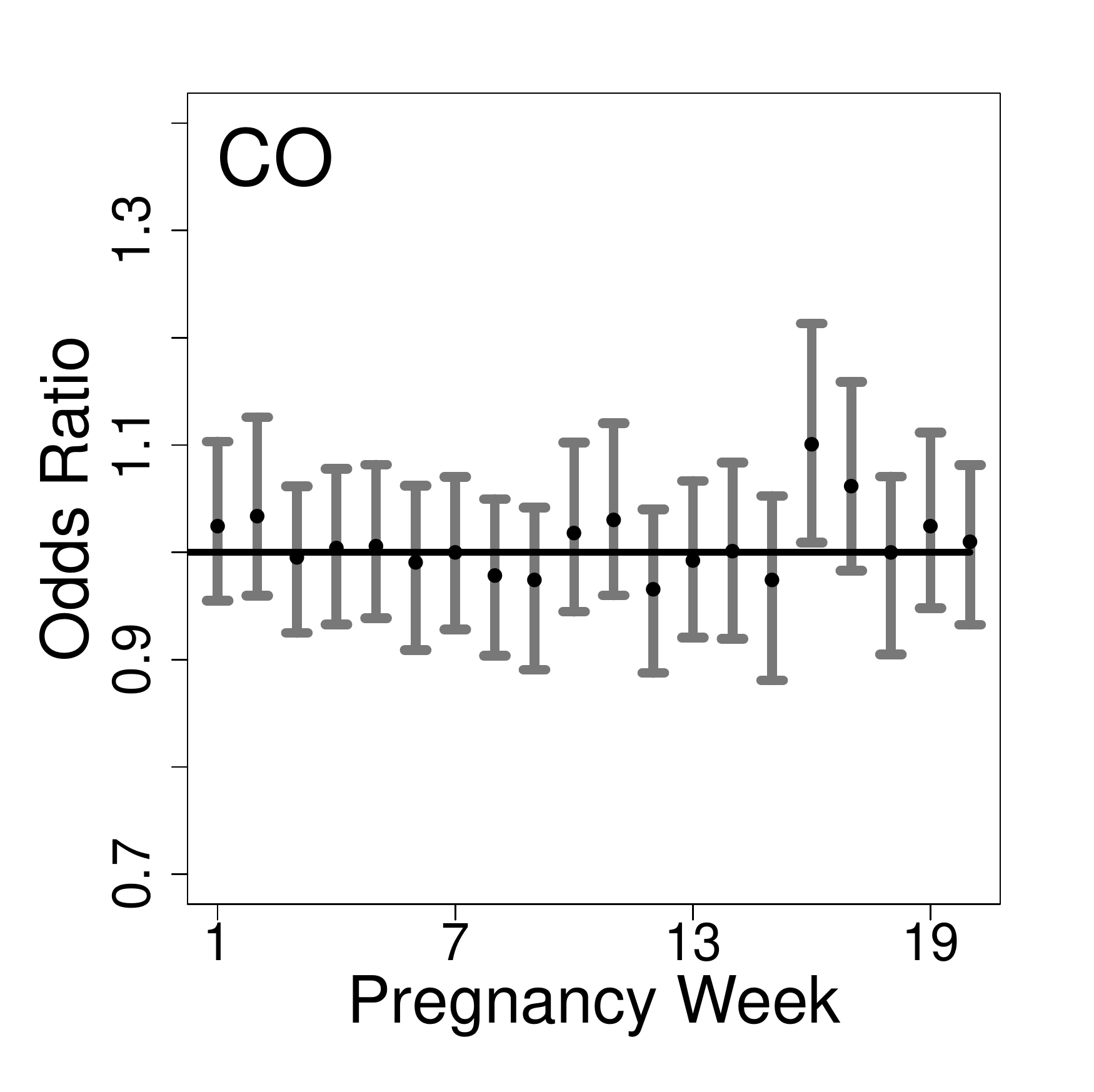}
\includegraphics[scale=0.26]{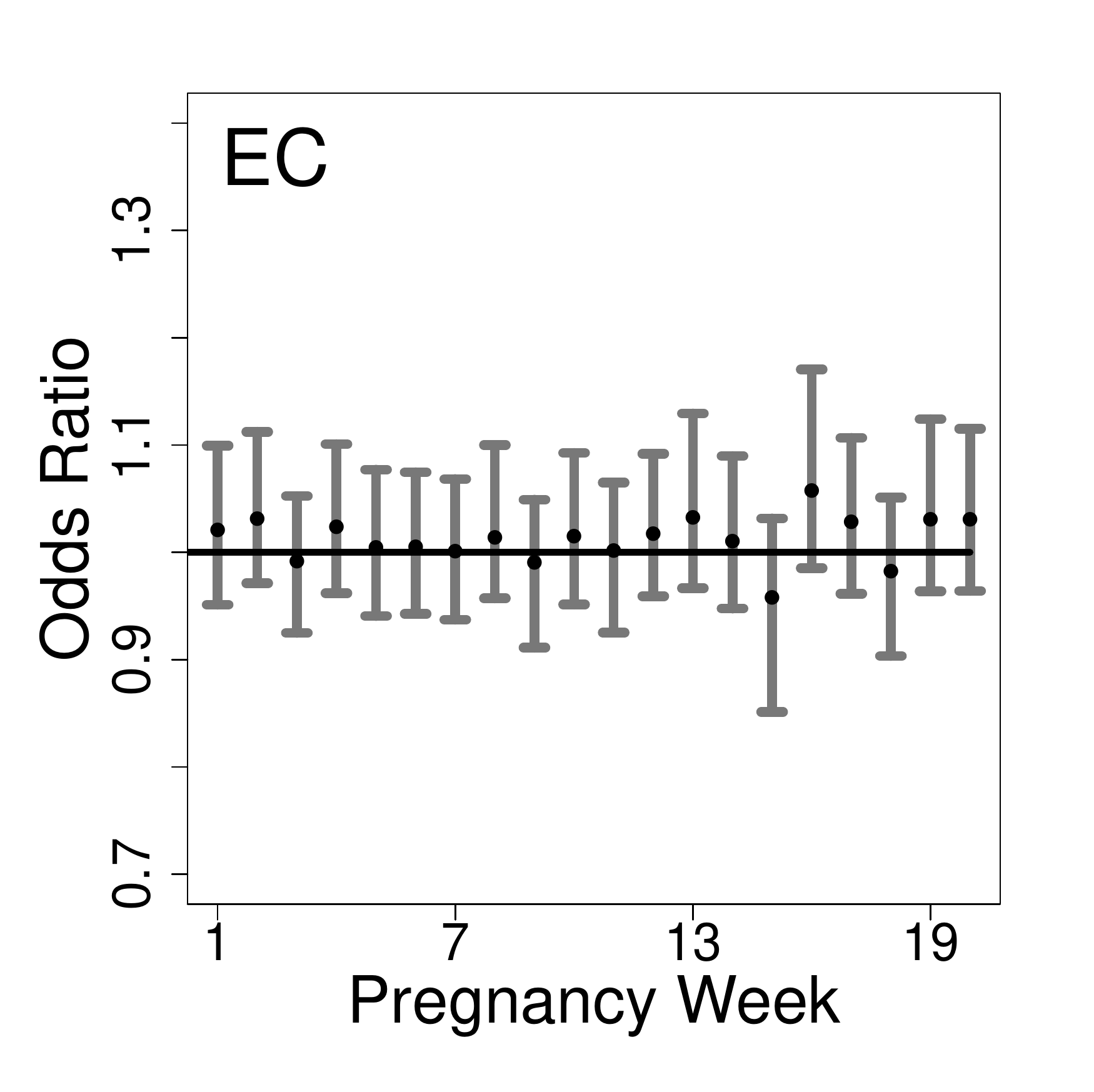}
\includegraphics[scale=0.26]{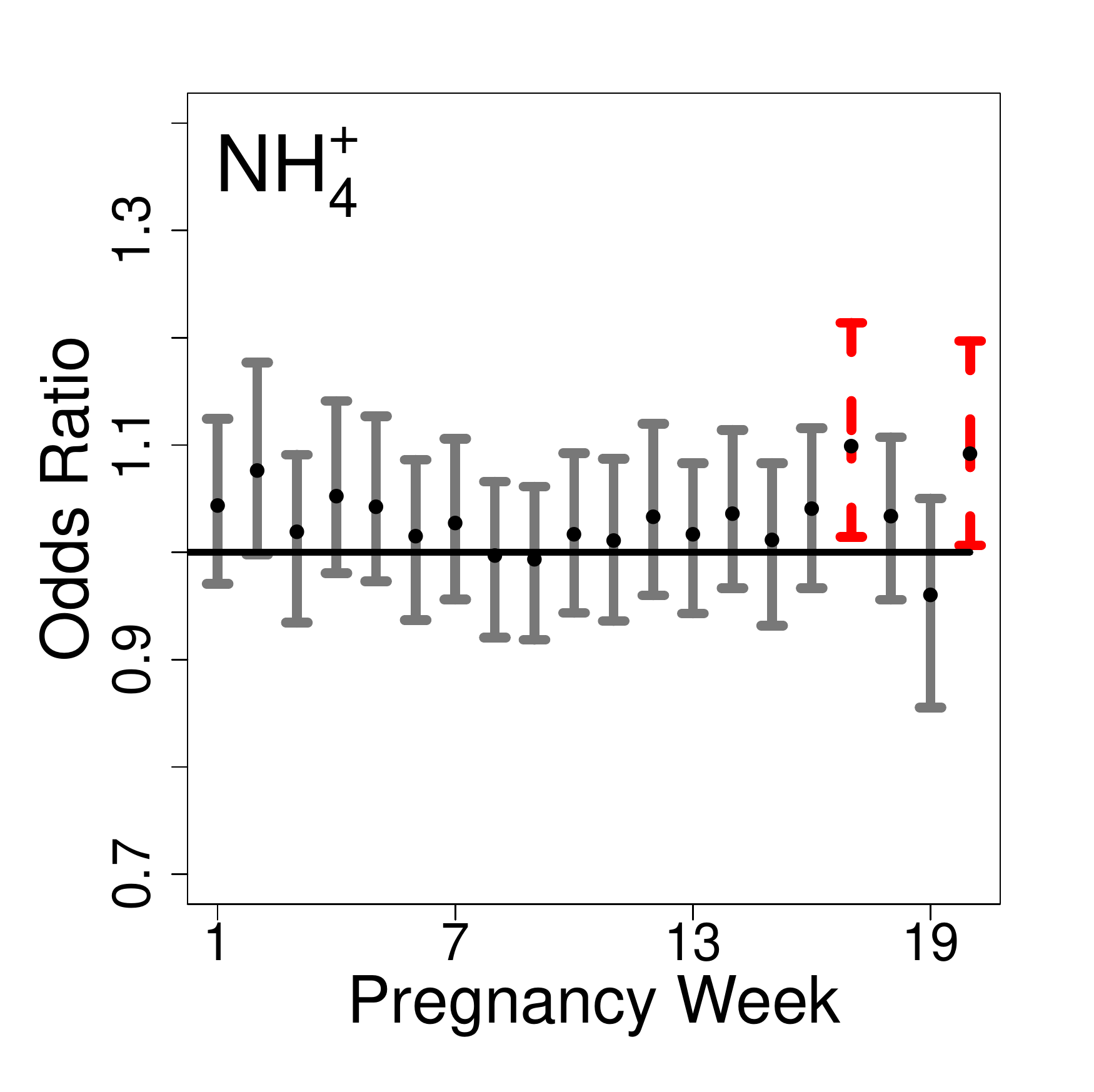}\\
\includegraphics[scale=0.26]{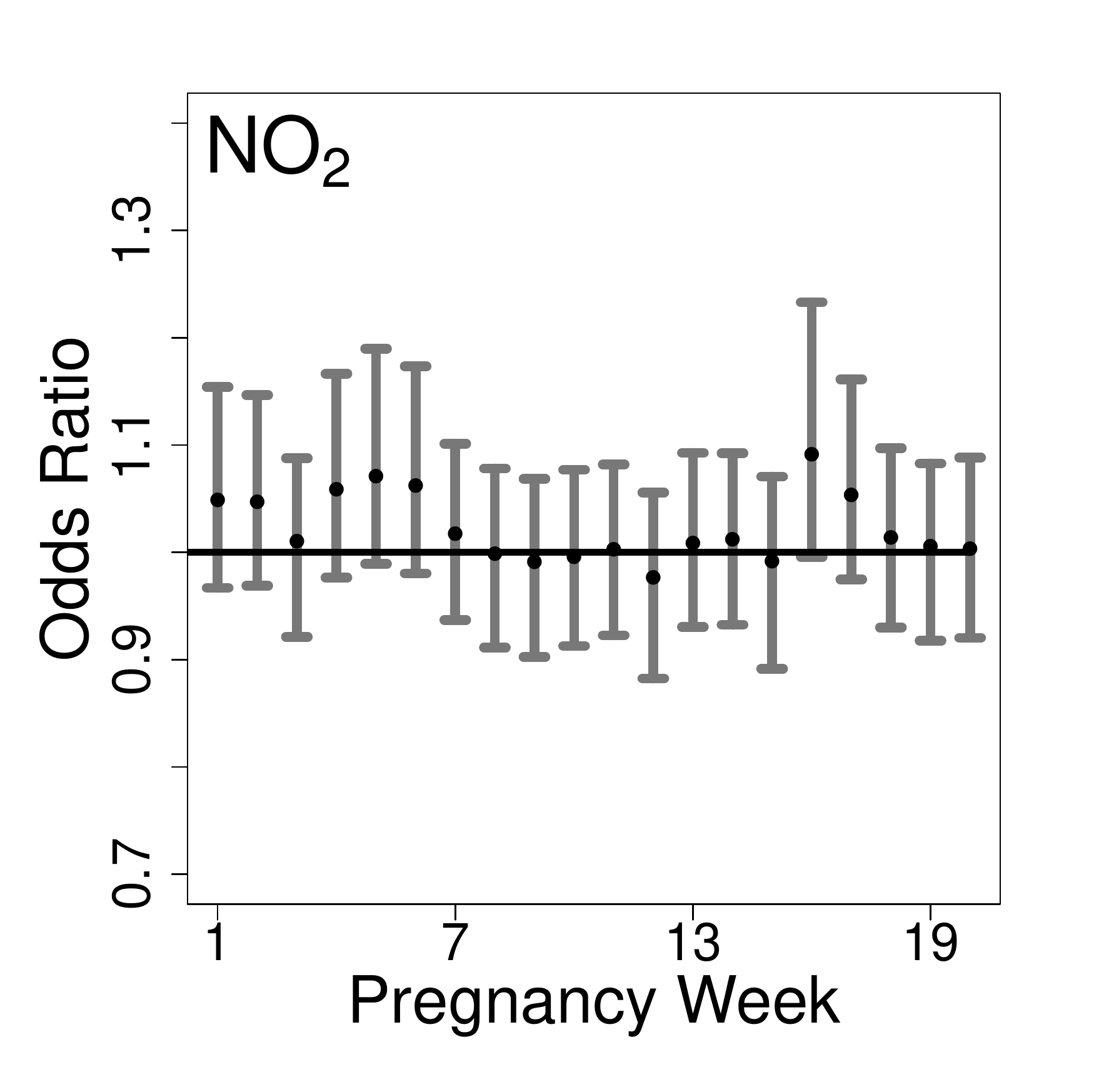}
\includegraphics[scale=0.26]{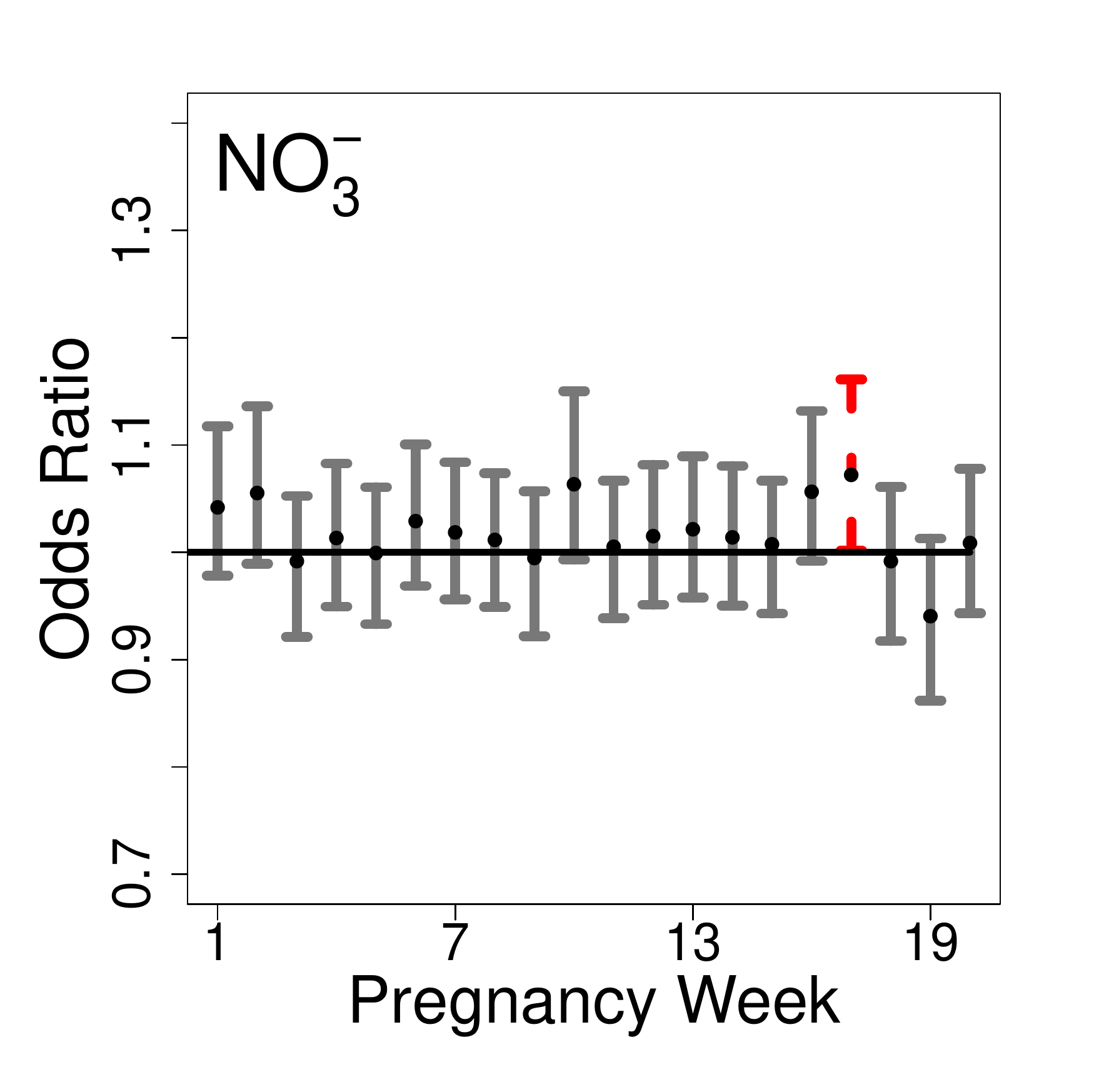}
\includegraphics[scale=0.26]{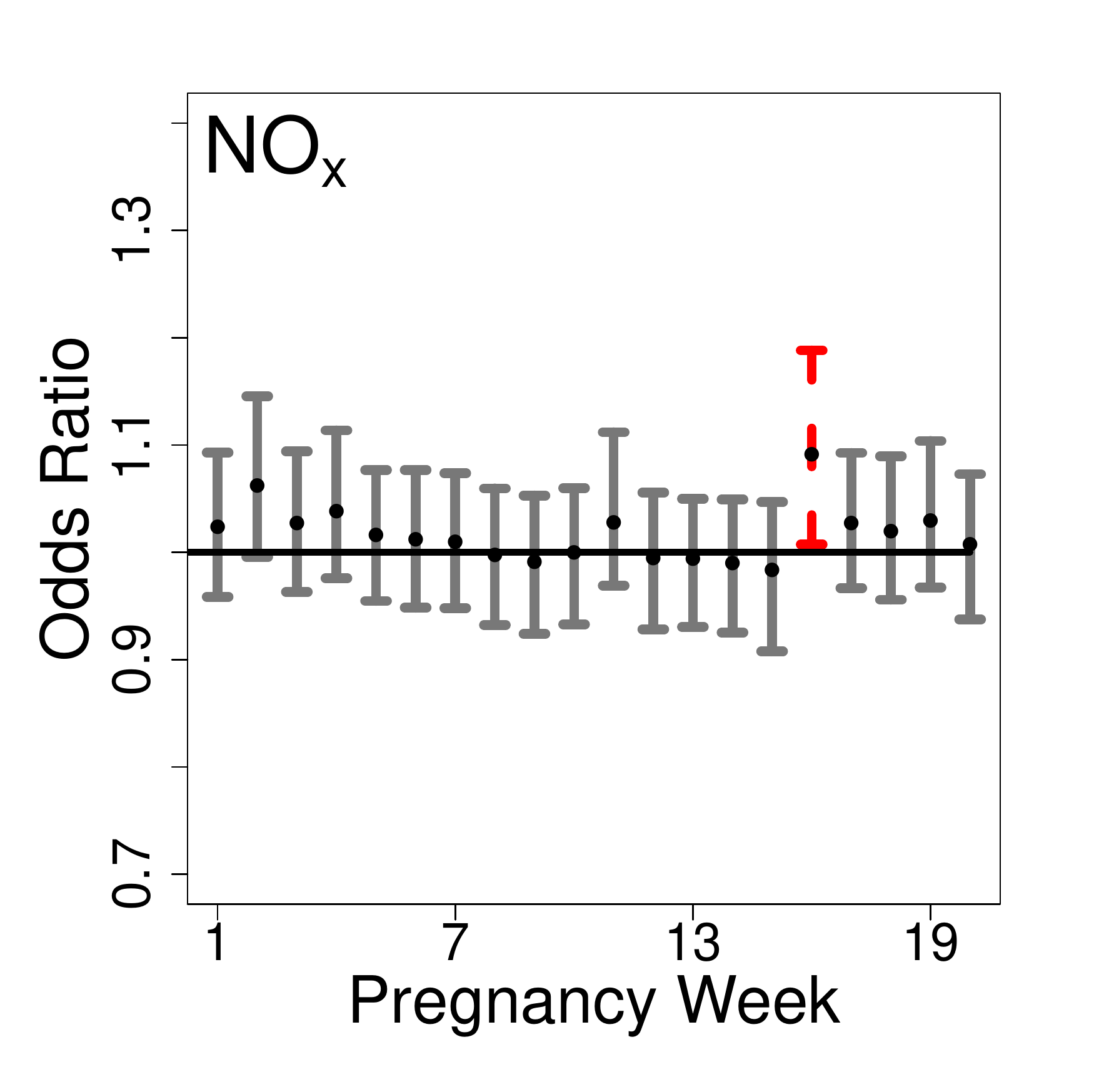}\\
\includegraphics[scale=0.26]{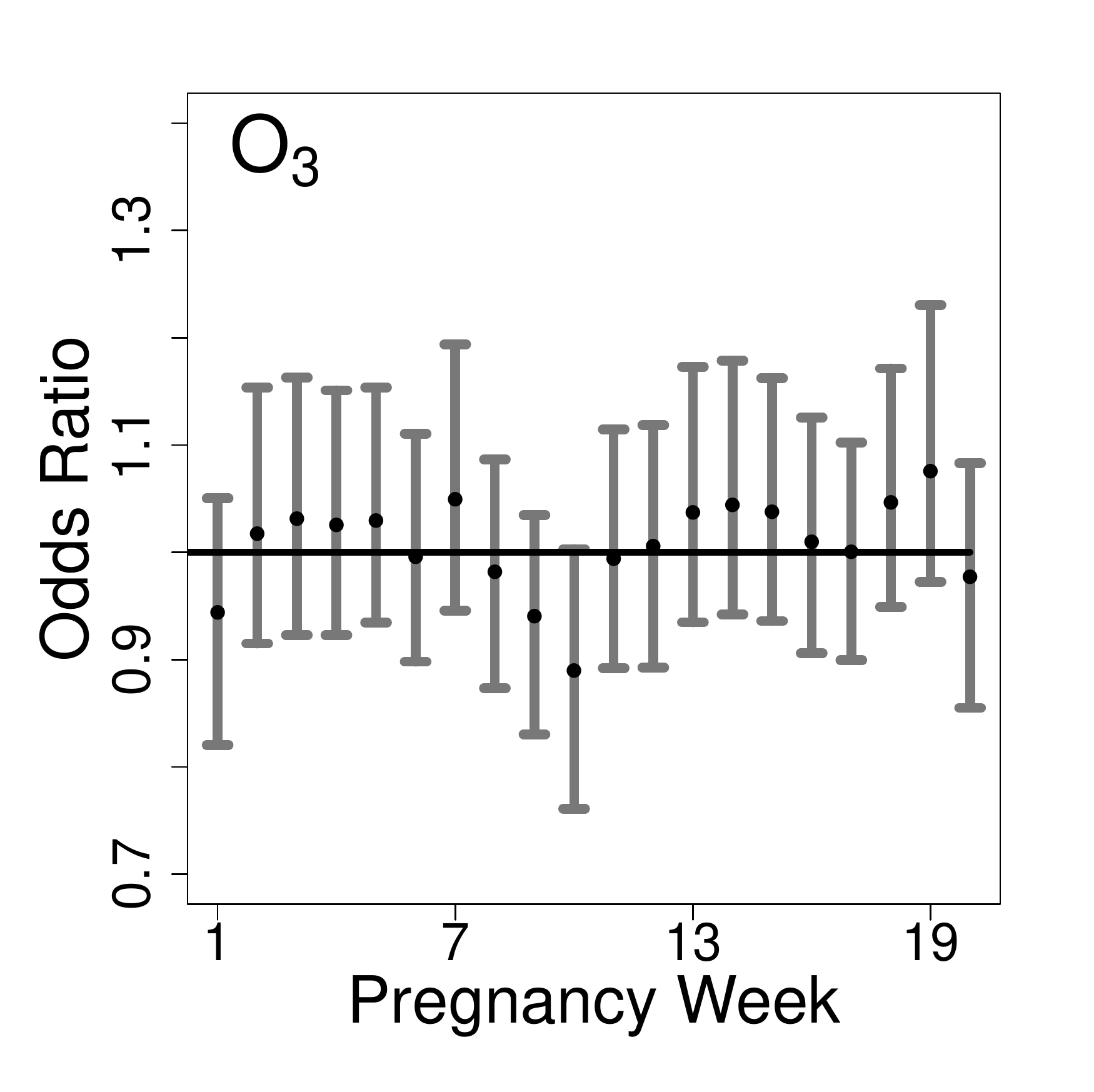}
\includegraphics[scale=0.26]{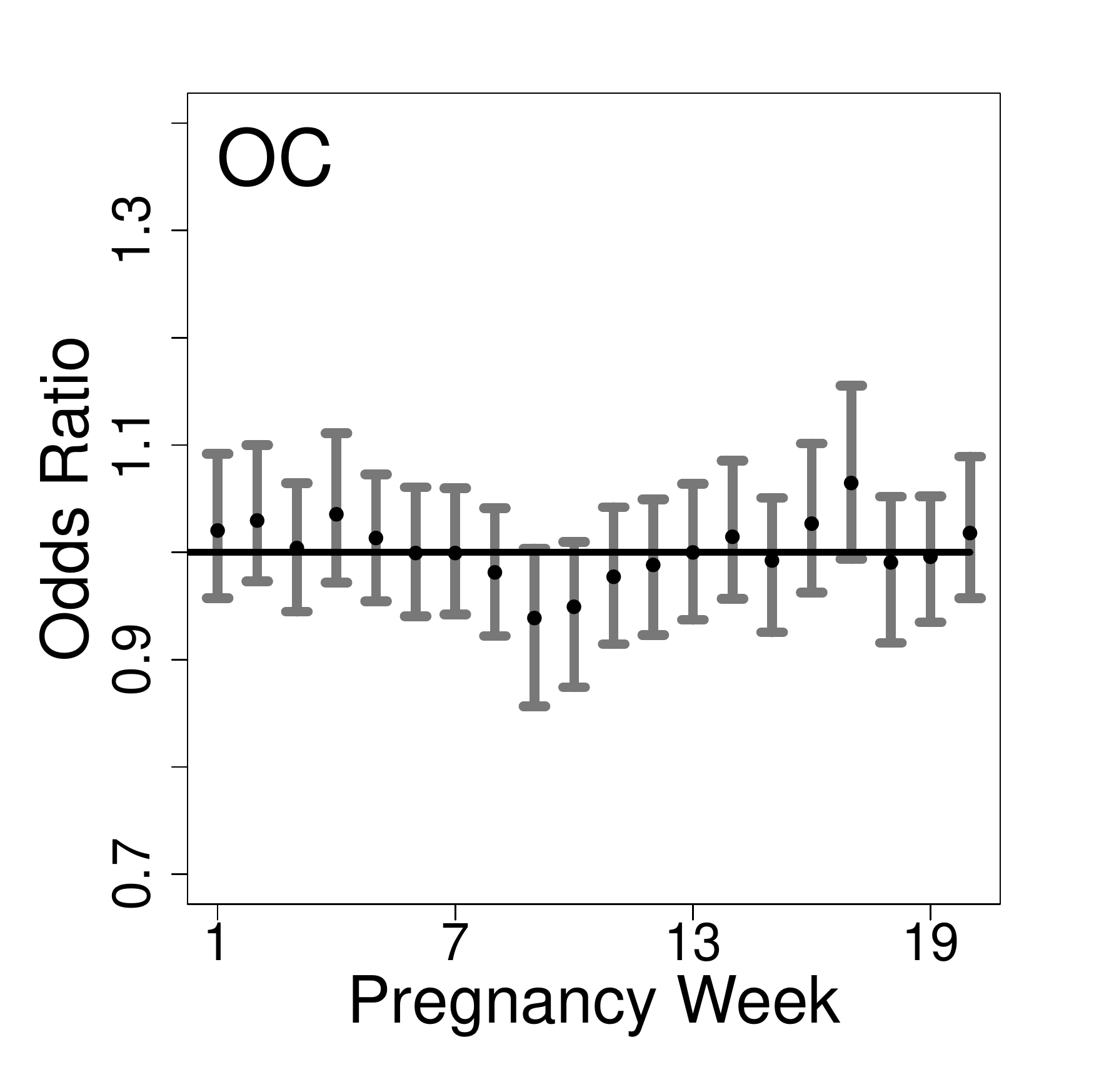}
\includegraphics[scale=0.26]{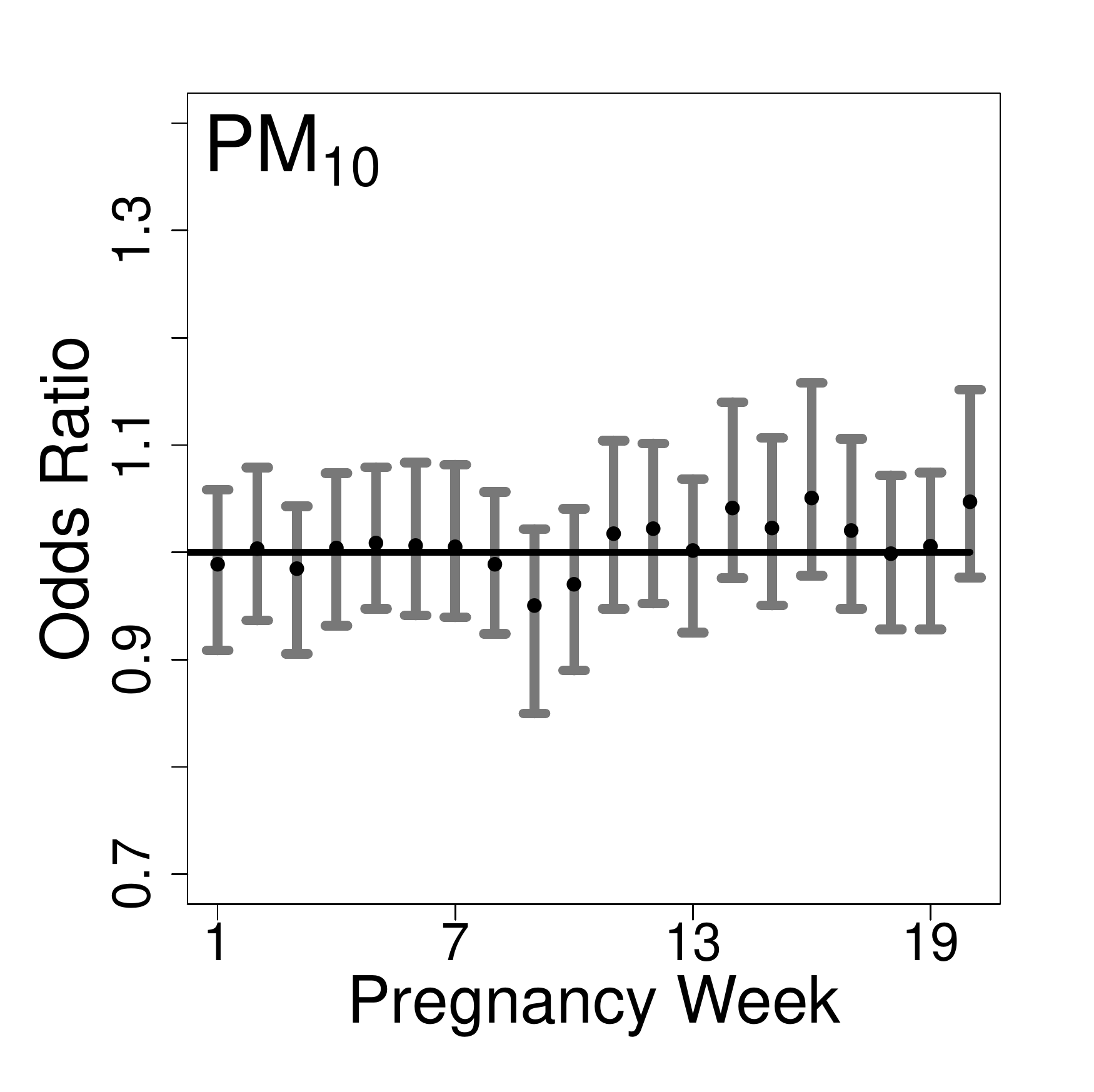}\\
\includegraphics[scale=0.26]{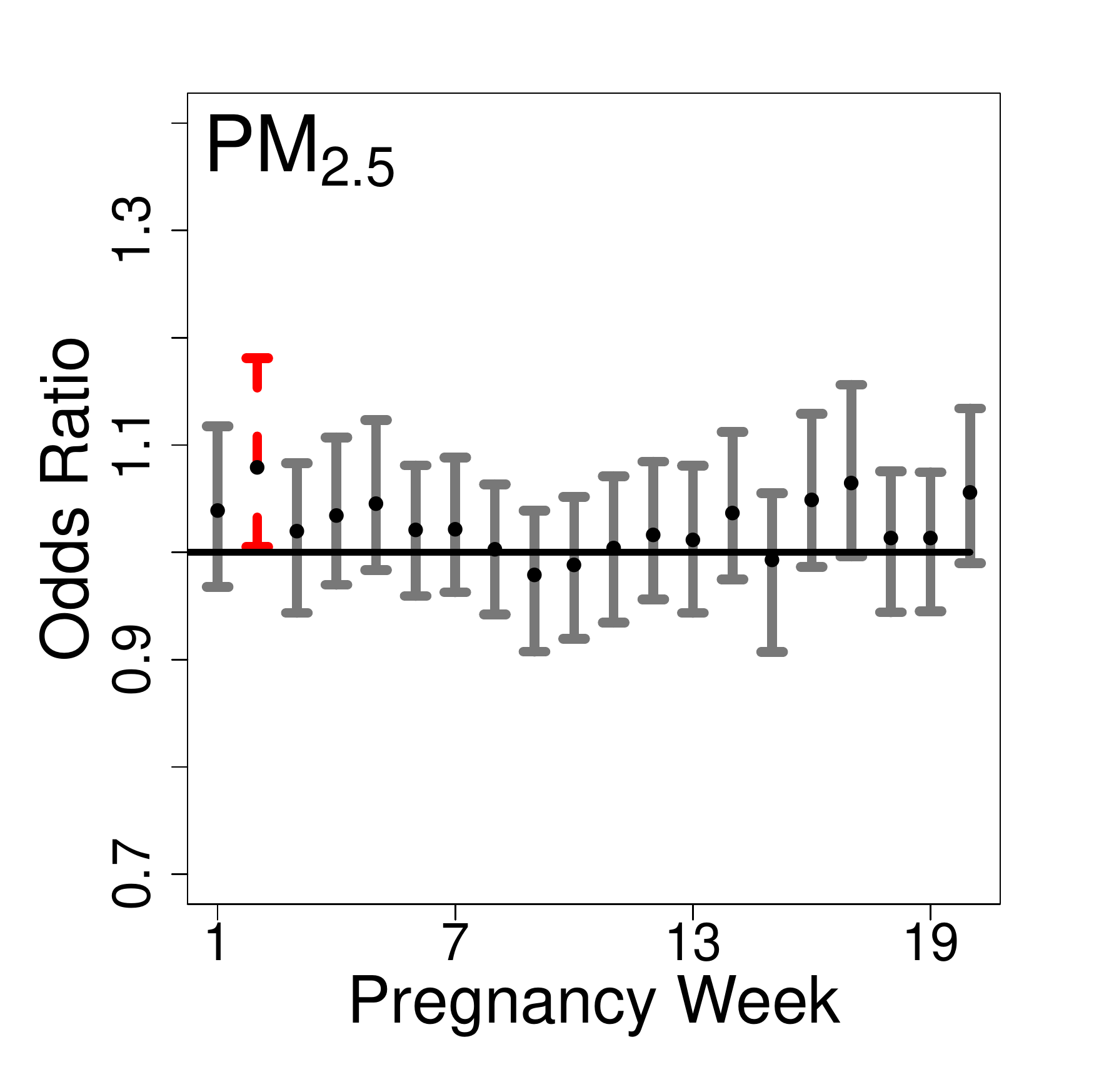}
\includegraphics[scale=0.26]{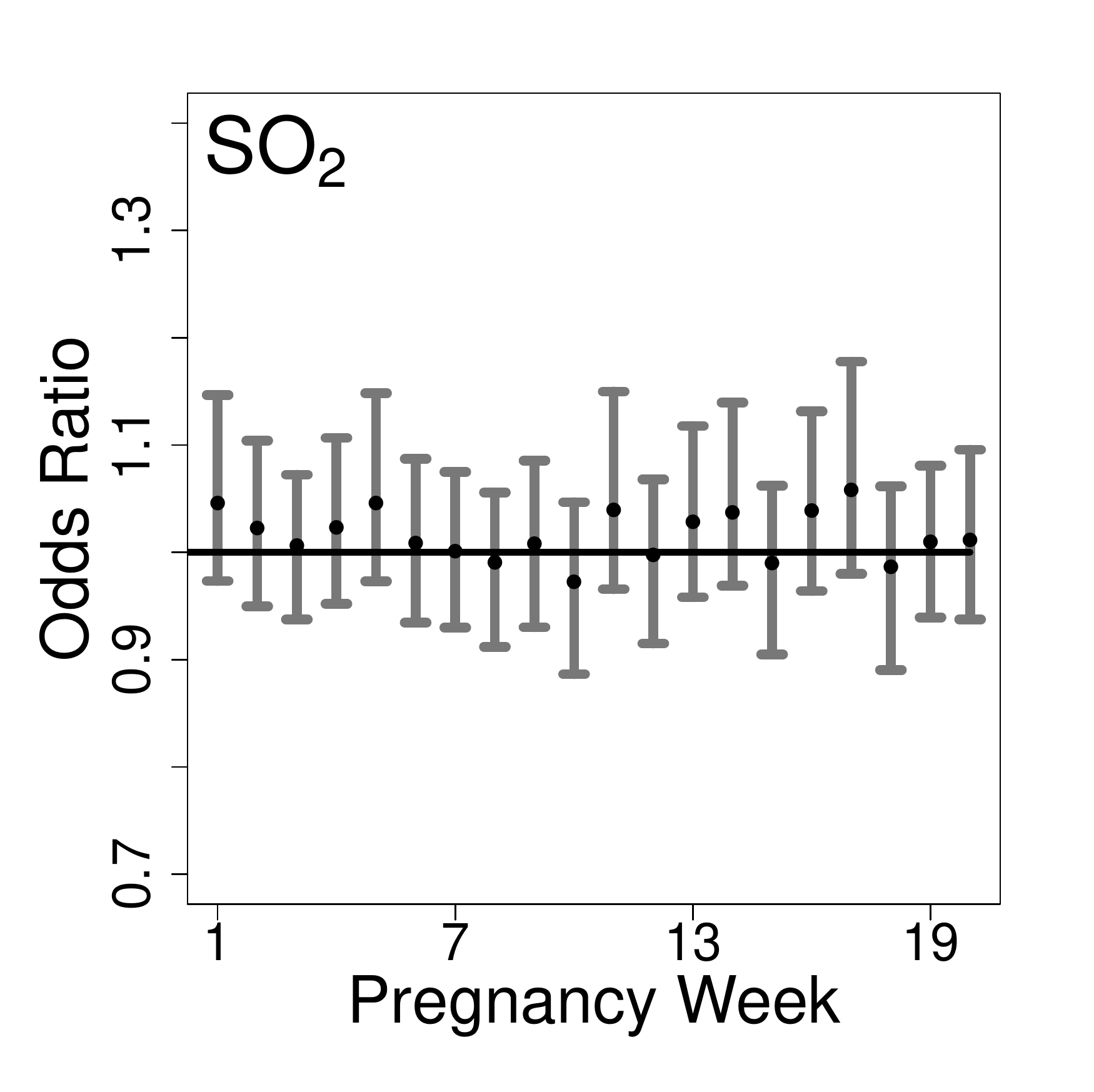}
\includegraphics[scale=0.26]{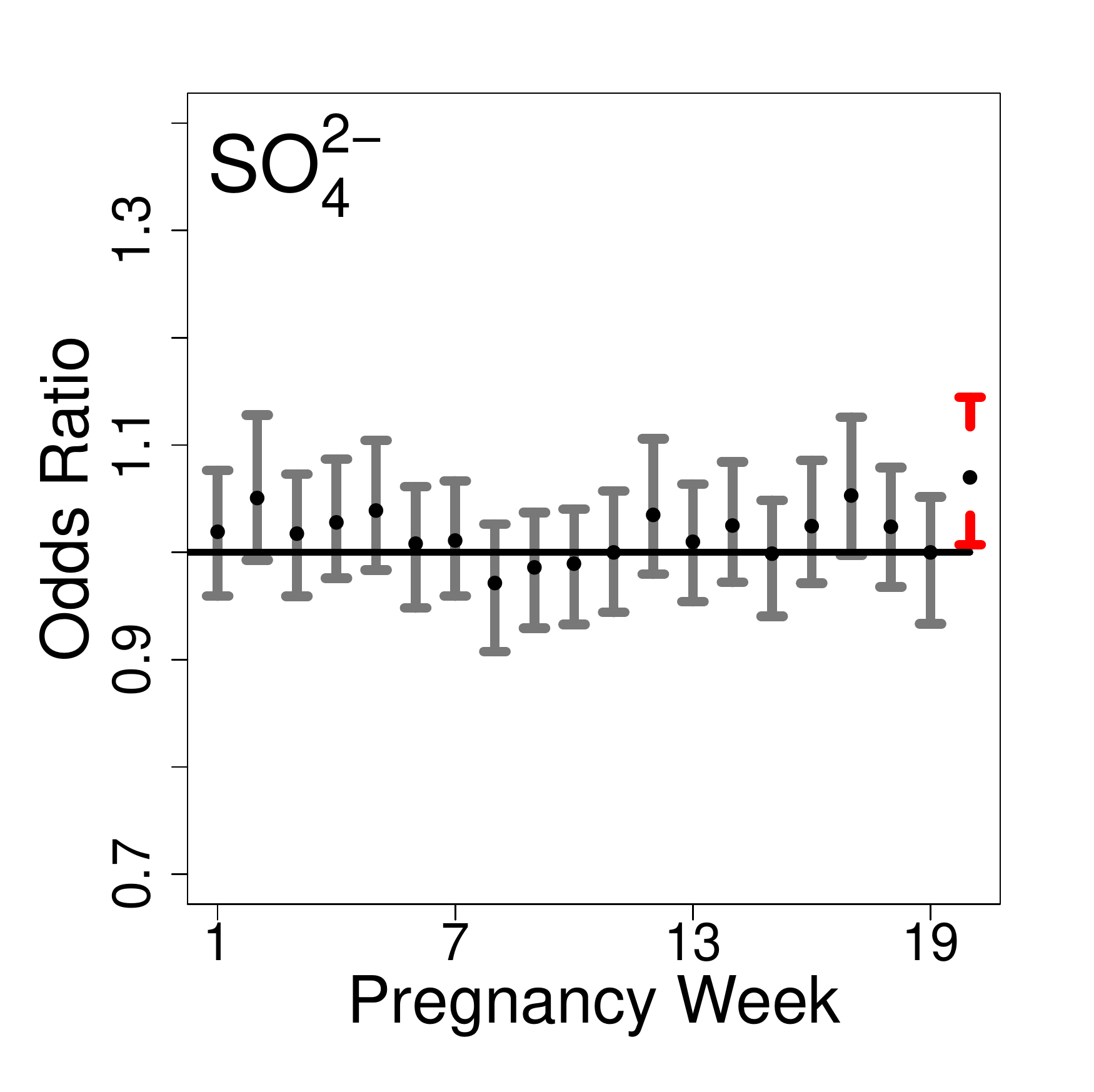}
\caption{Posterior mean and 95\% credible interval results from the \textbf{non-Hispanic Black} stillbirth and single exposure Critical Window Variable Selection (CWVS) analyses in New Jersey, 2005-2014. Results based on an interquartile range increase in weekly exposure. Weeks identified as part of the critical window set are shown in red/dashed (harmful) and blue/dashed (protective).  These definitions depend partly on the posterior inclusion probabilities in Figure S5 of the Supporting Information for the variable selection methods.}
\end{center}
\end{figure}
\clearpage

\begin{figure}[h]
\begin{center}
\includegraphics[scale=0.26]{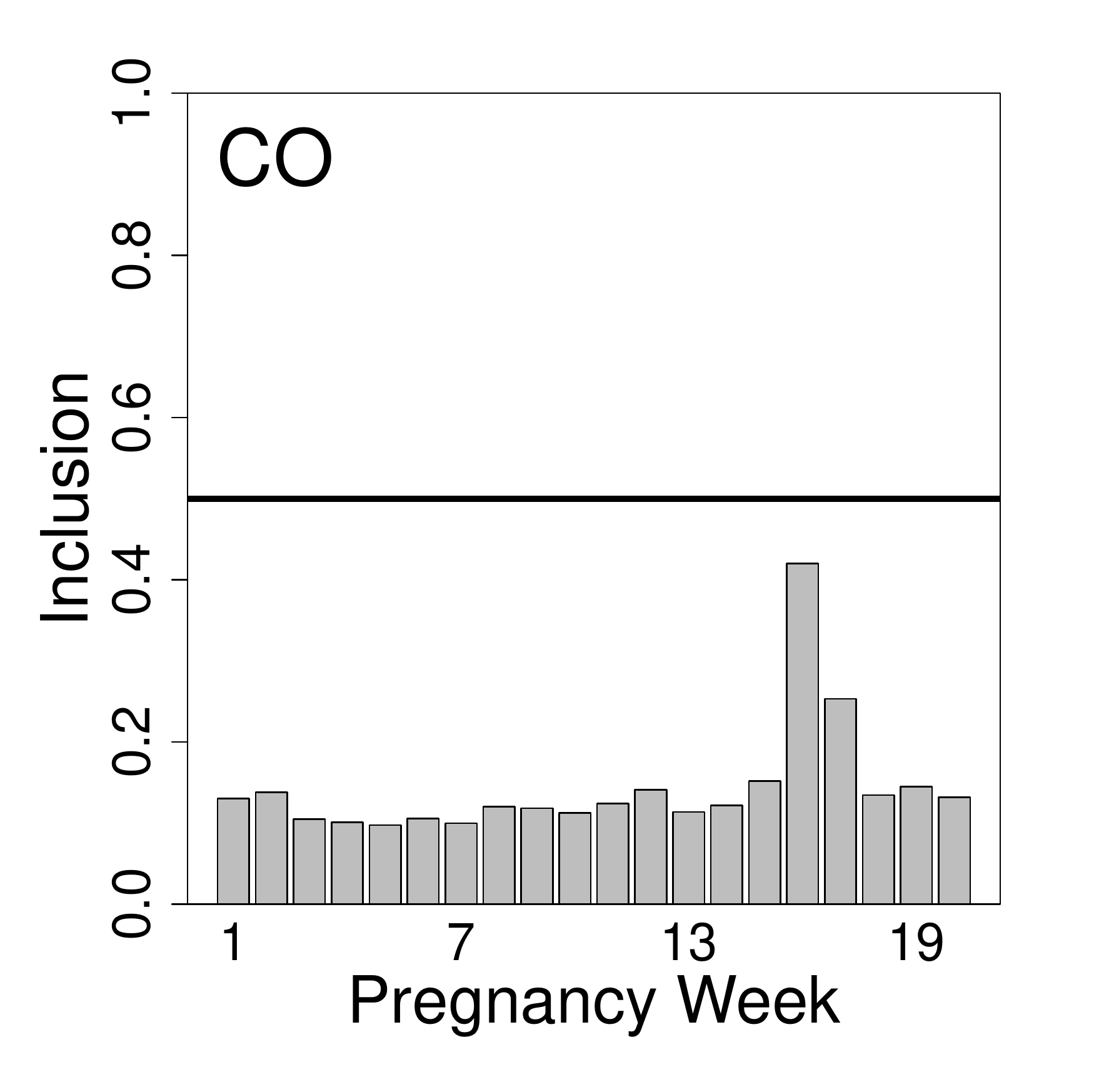}
\includegraphics[scale=0.26]{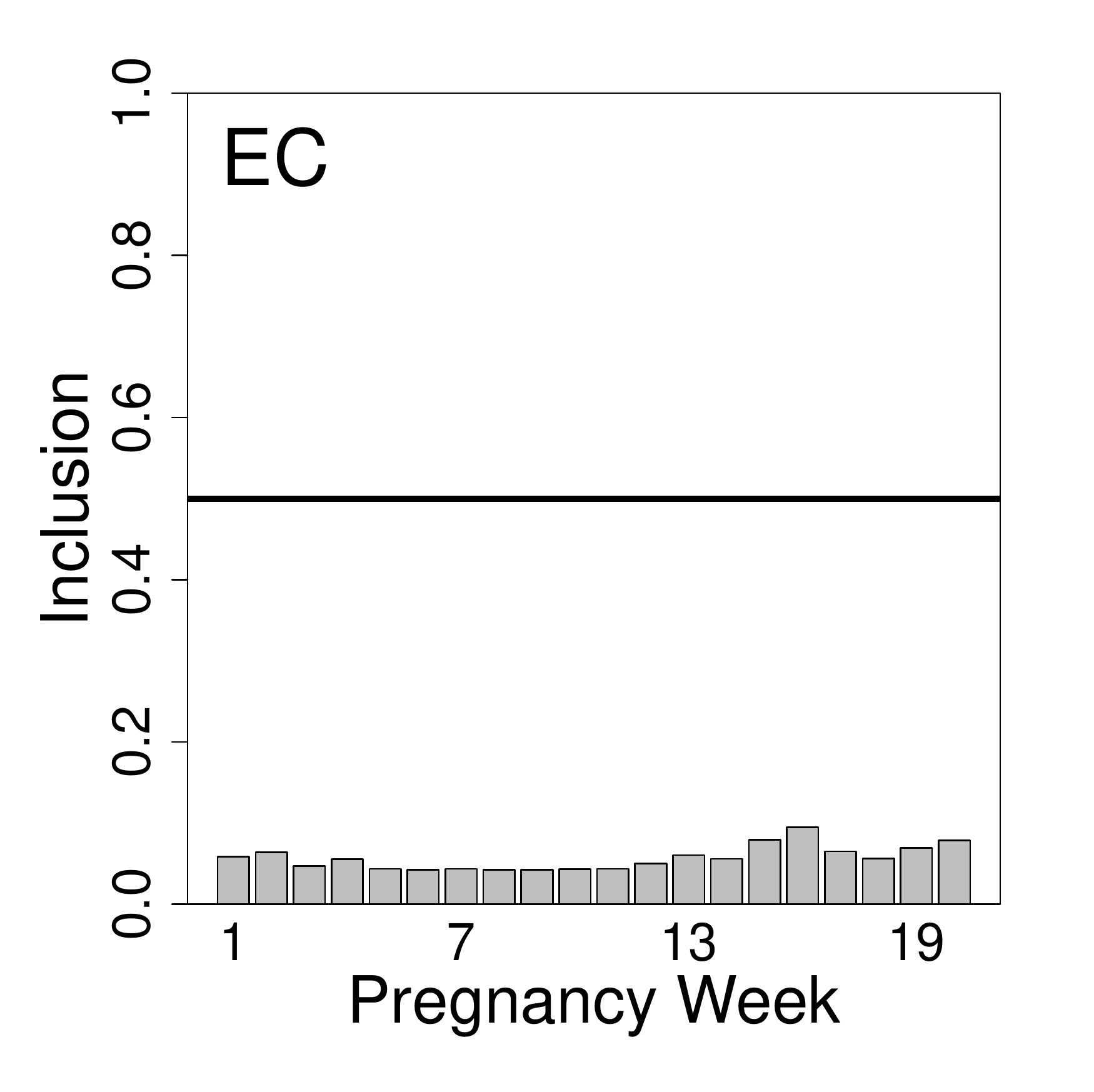}
\includegraphics[scale=0.26]{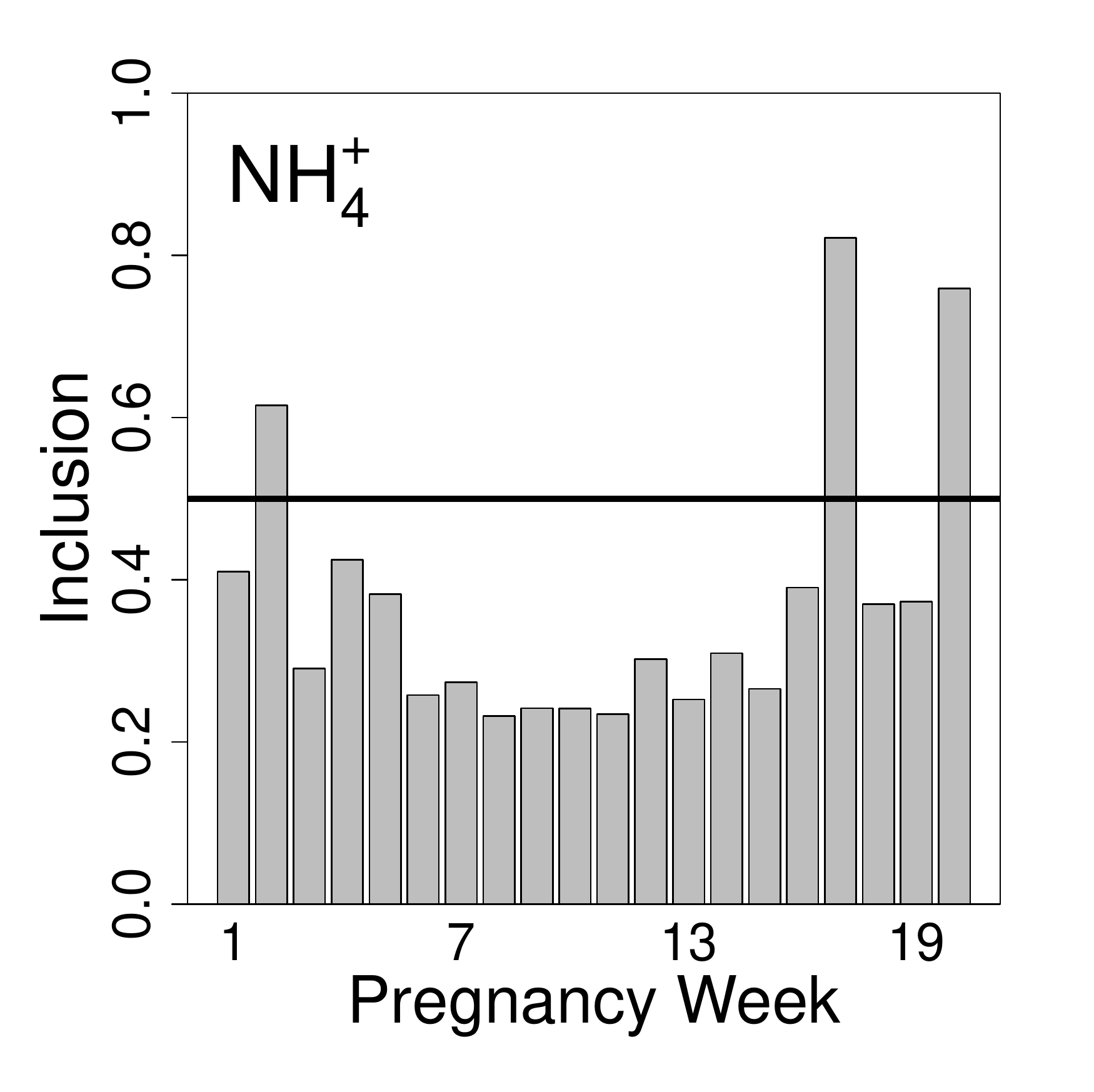}\\
\includegraphics[scale=0.26]{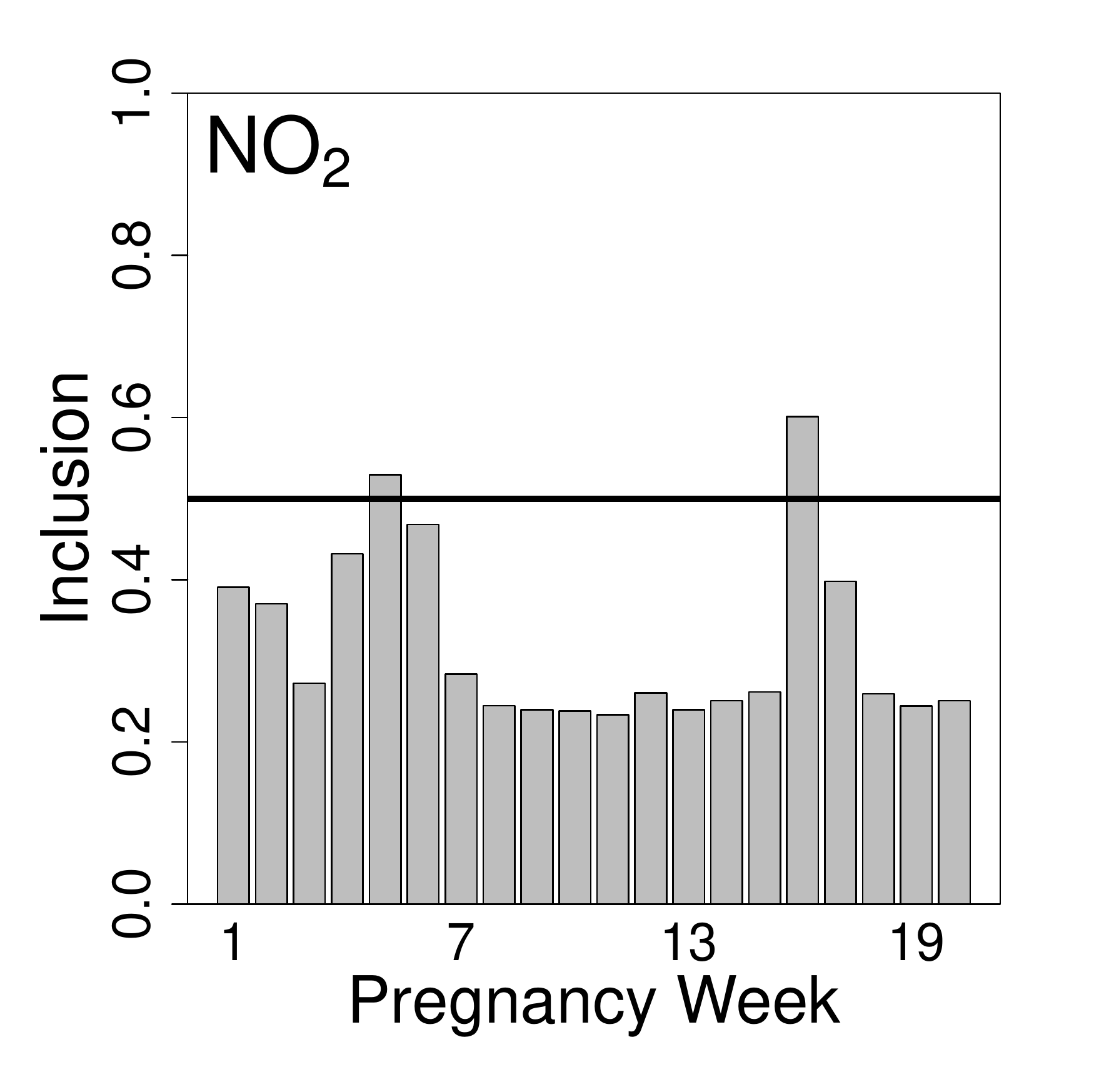}
\includegraphics[scale=0.26]{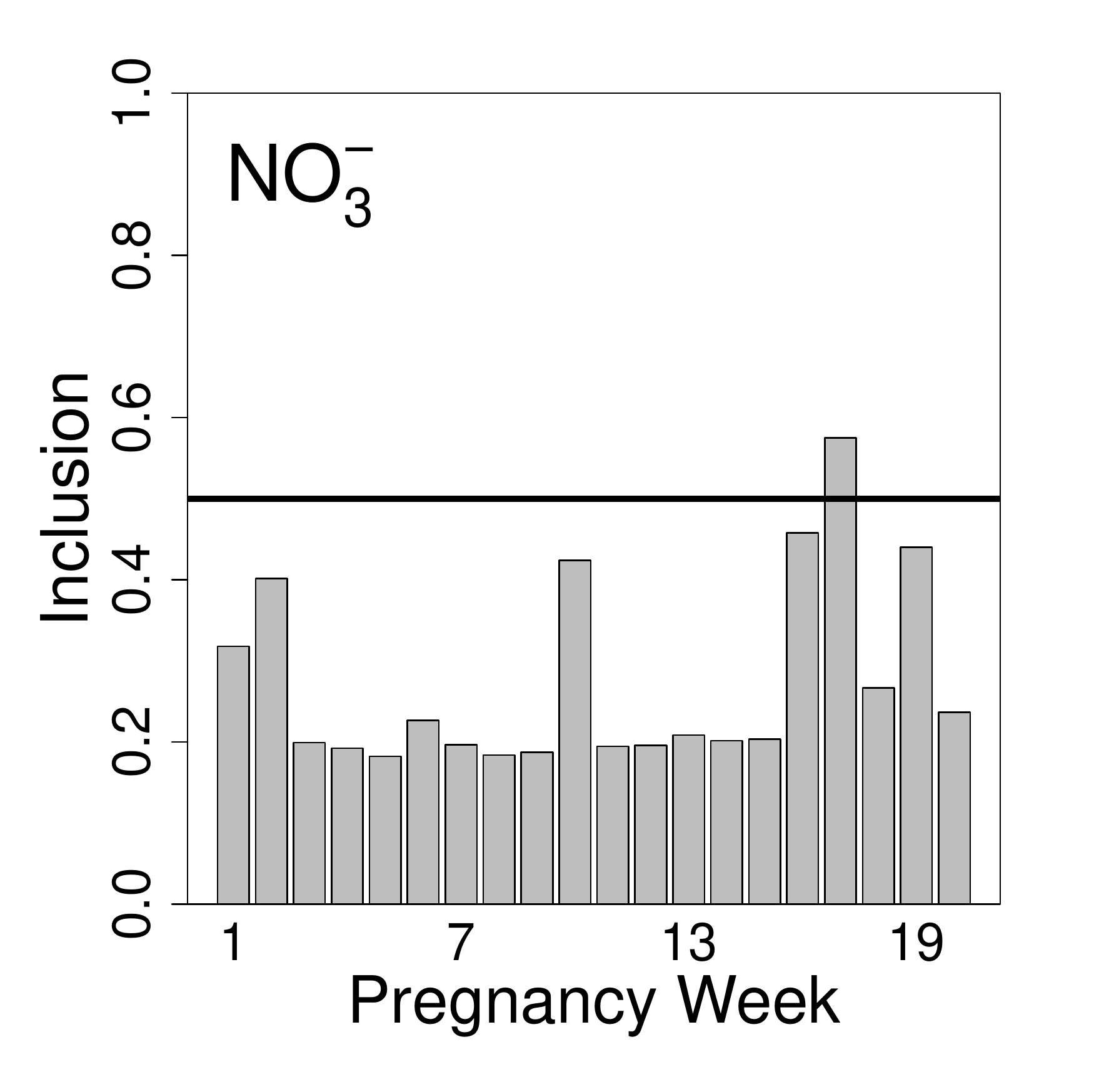}
\includegraphics[scale=0.26]{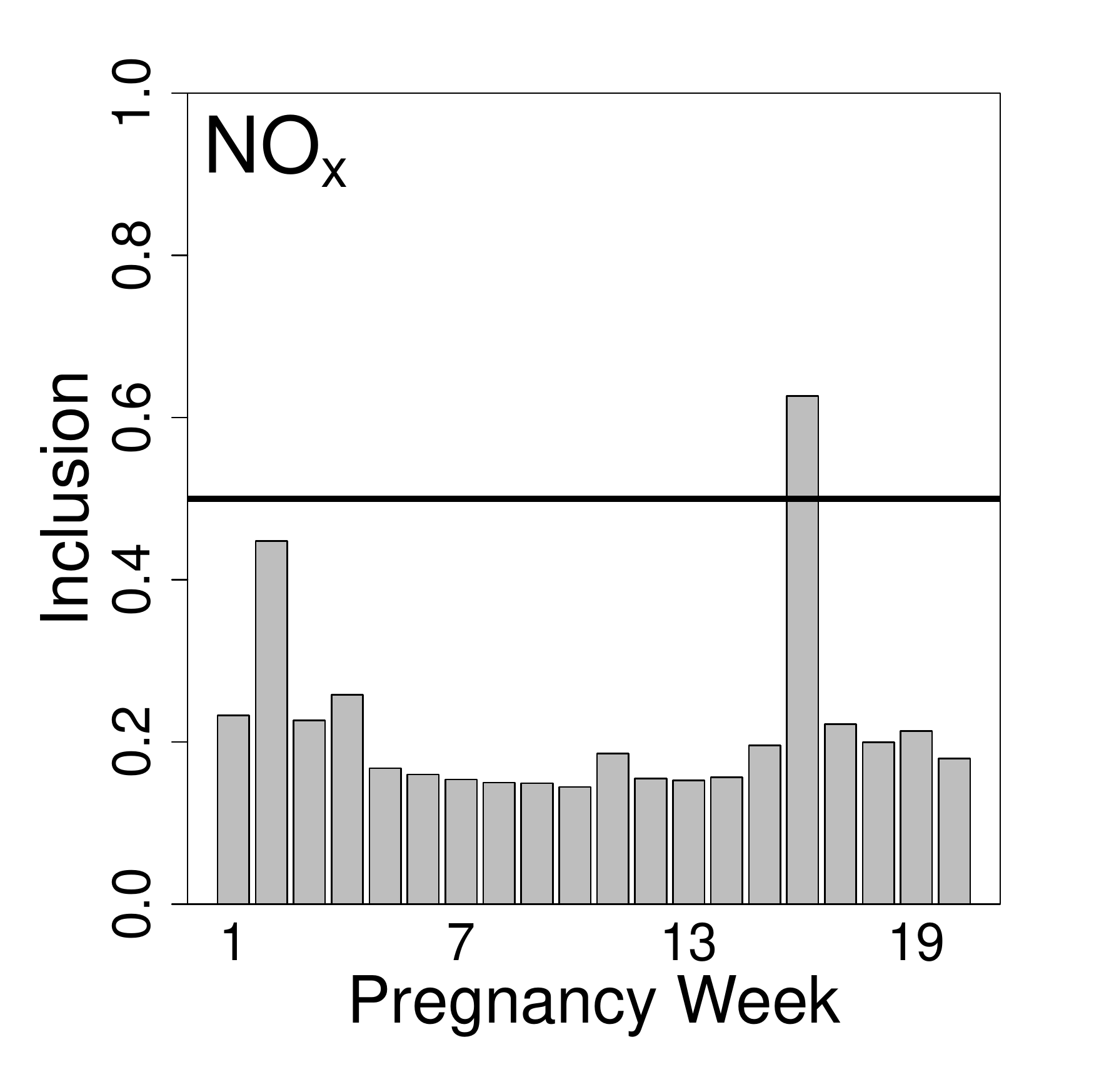}\\
\includegraphics[scale=0.26]{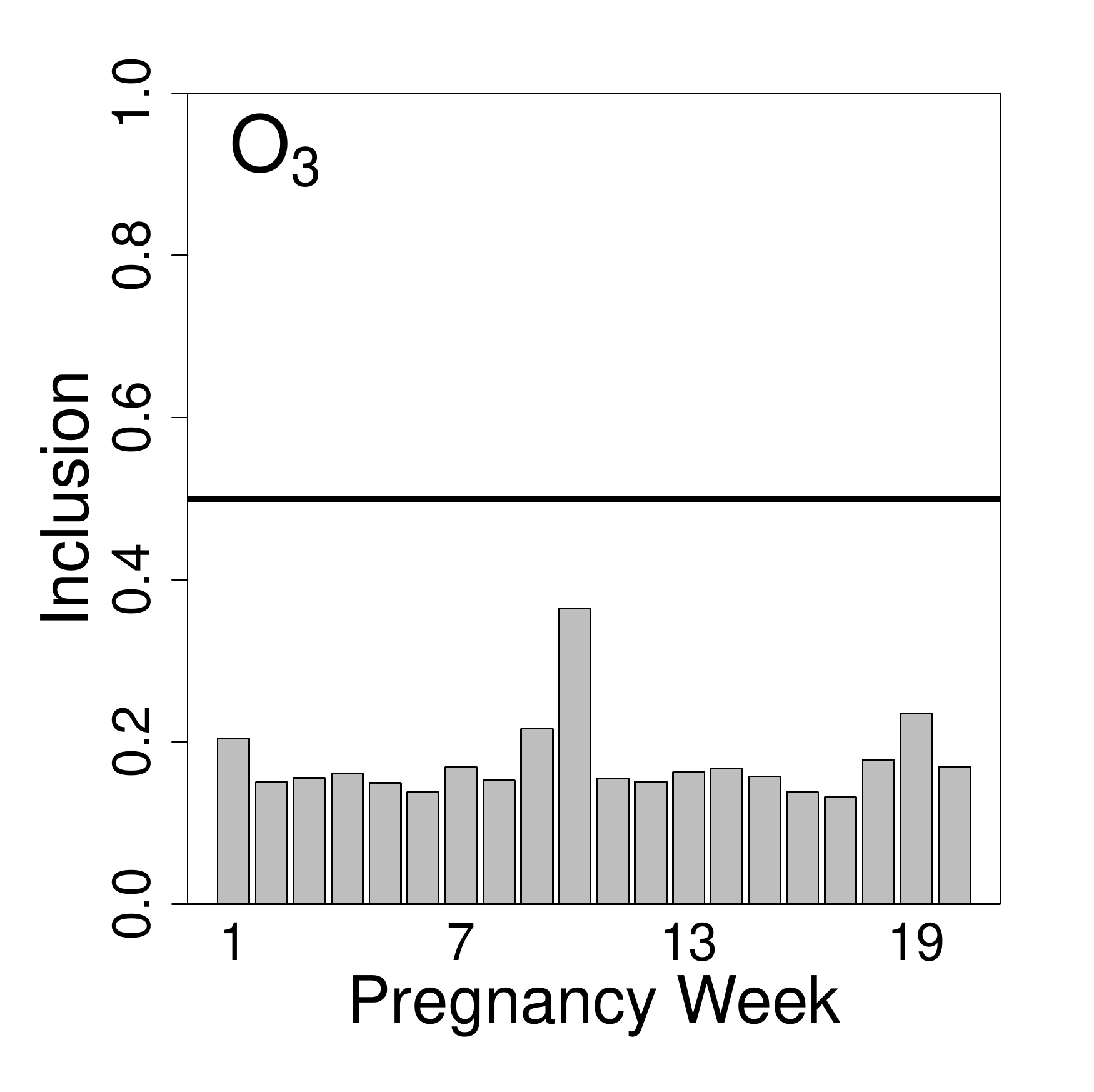}
\includegraphics[scale=0.26]{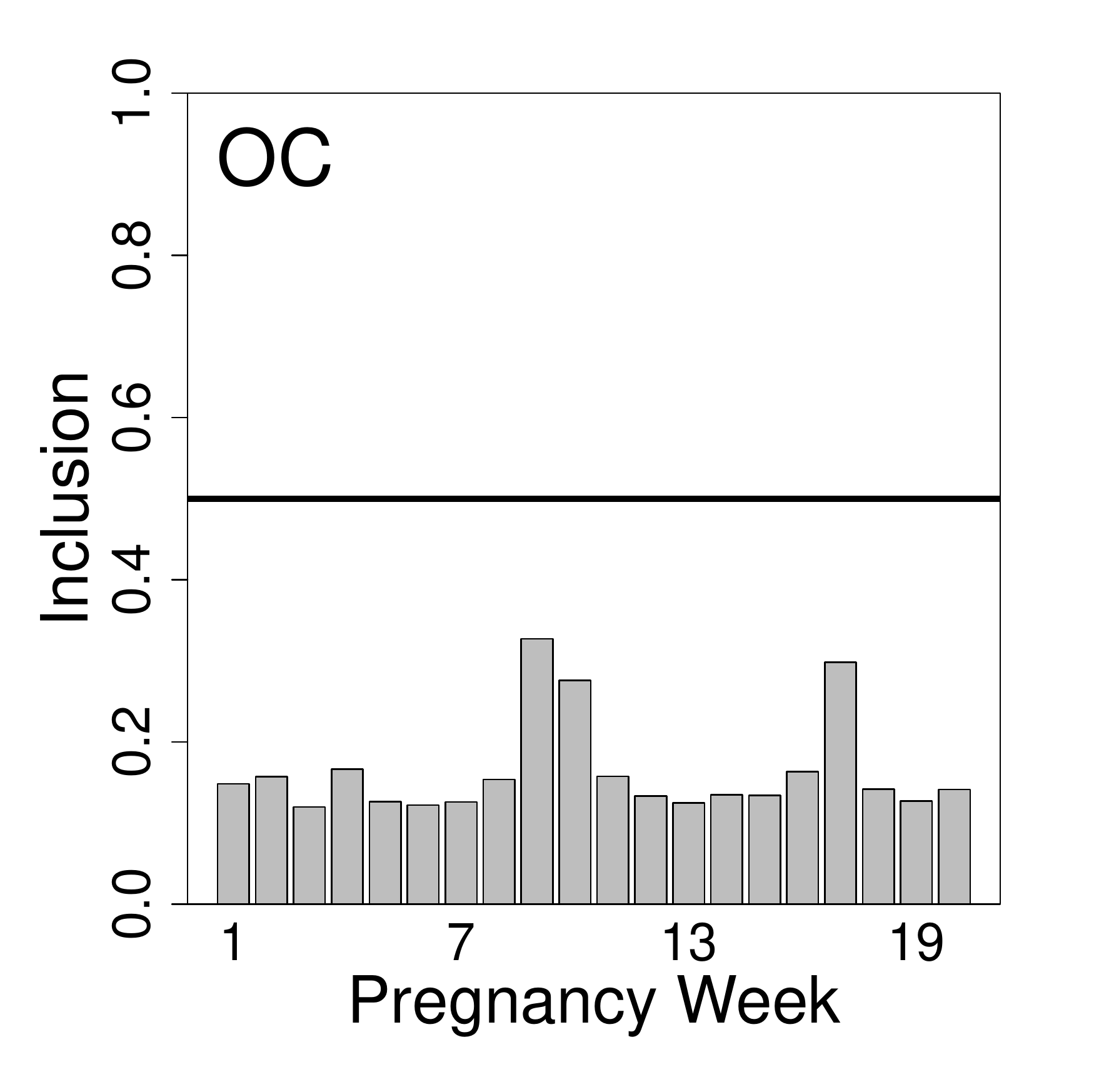}
\includegraphics[scale=0.26]{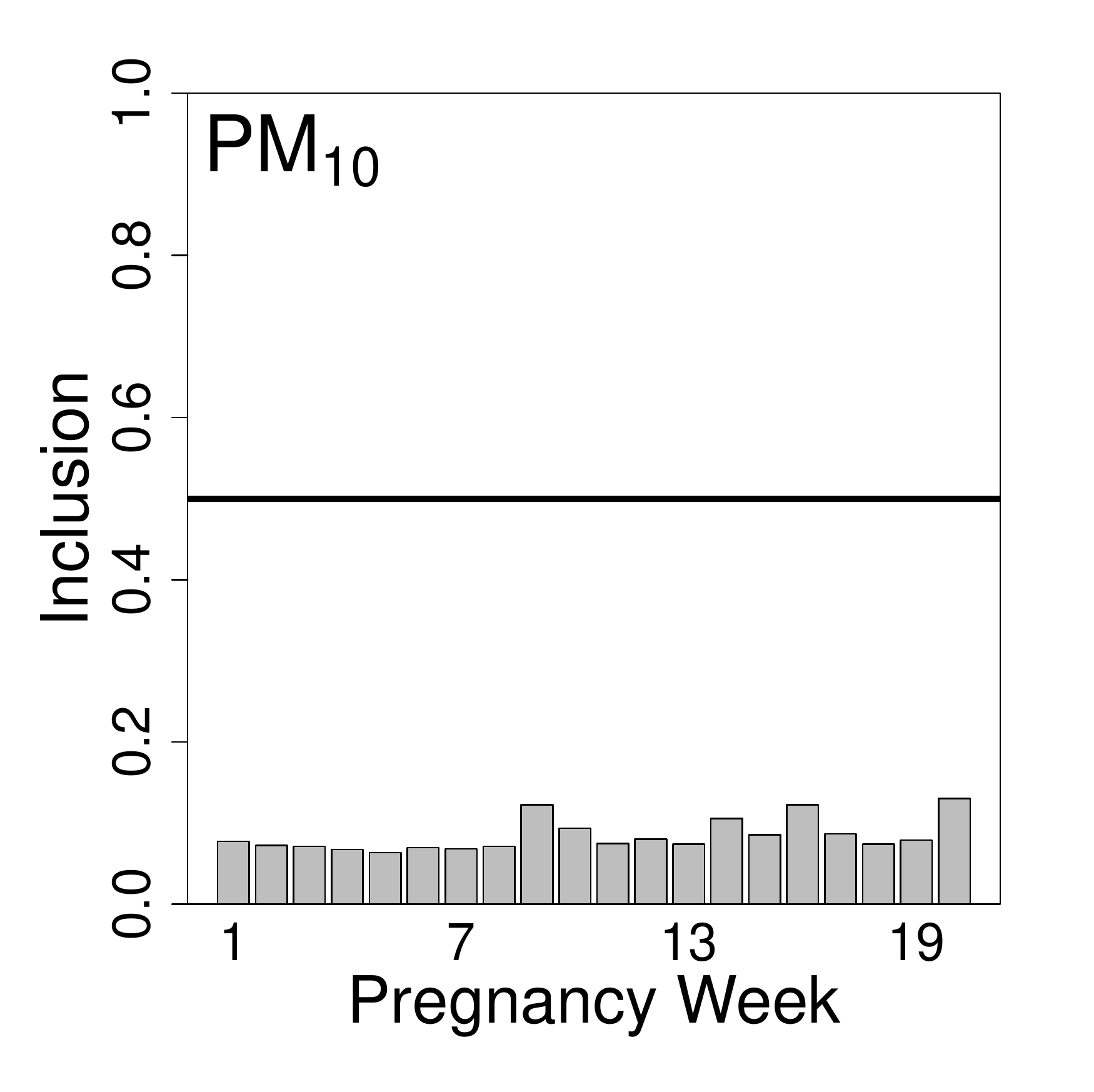}\\
\includegraphics[scale=0.26]{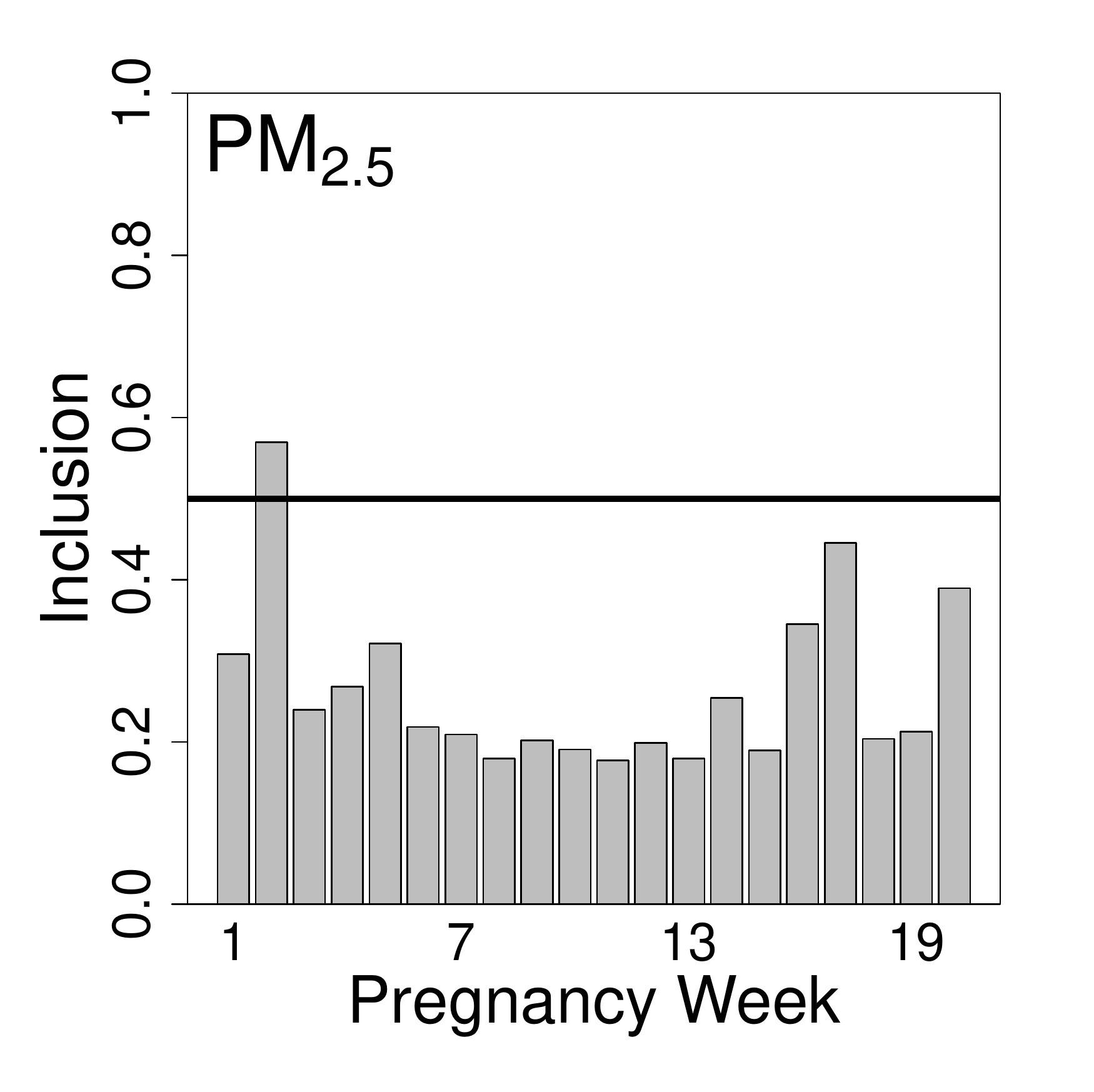}
\includegraphics[scale=0.26]{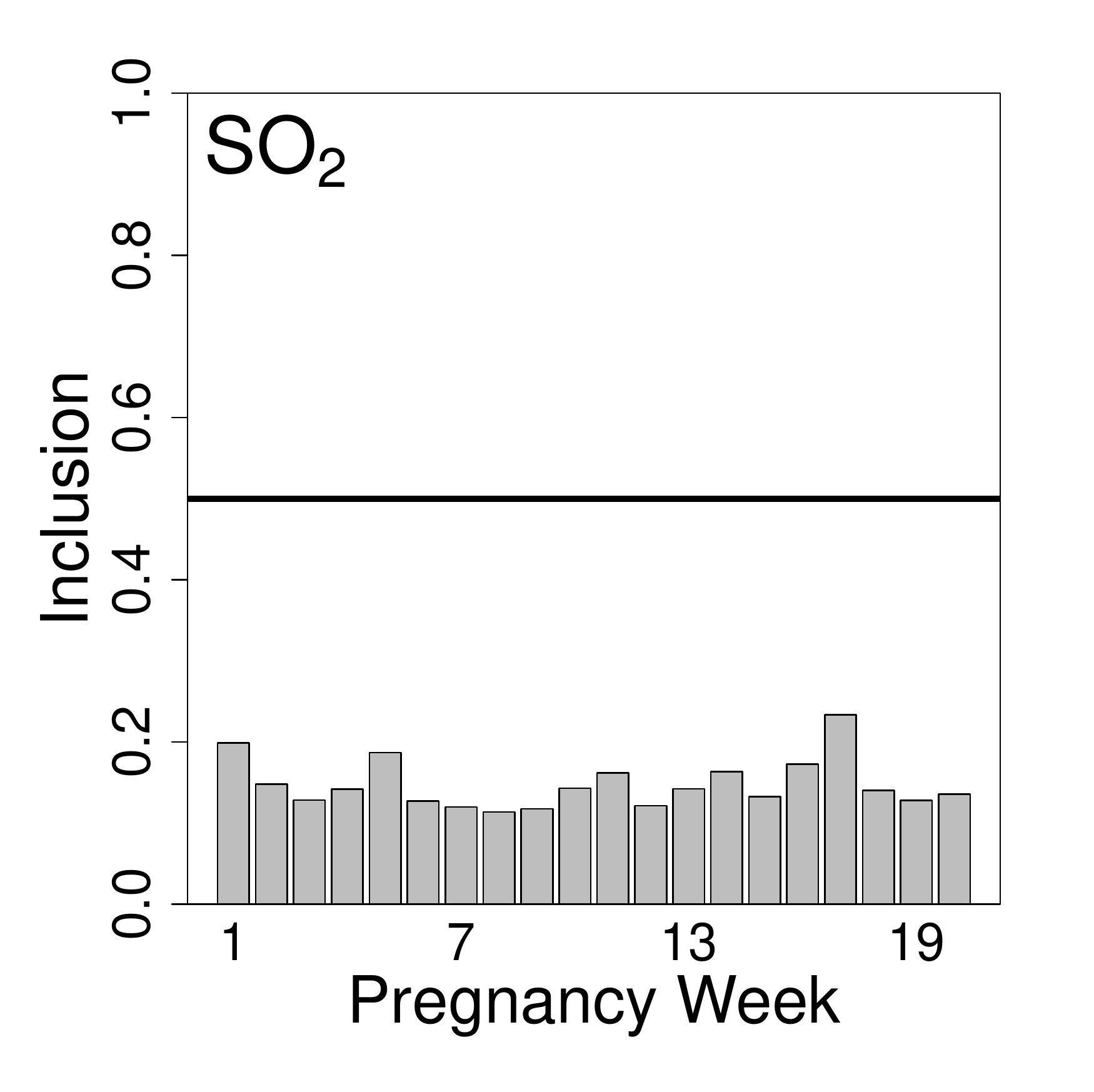}
\includegraphics[scale=0.26]{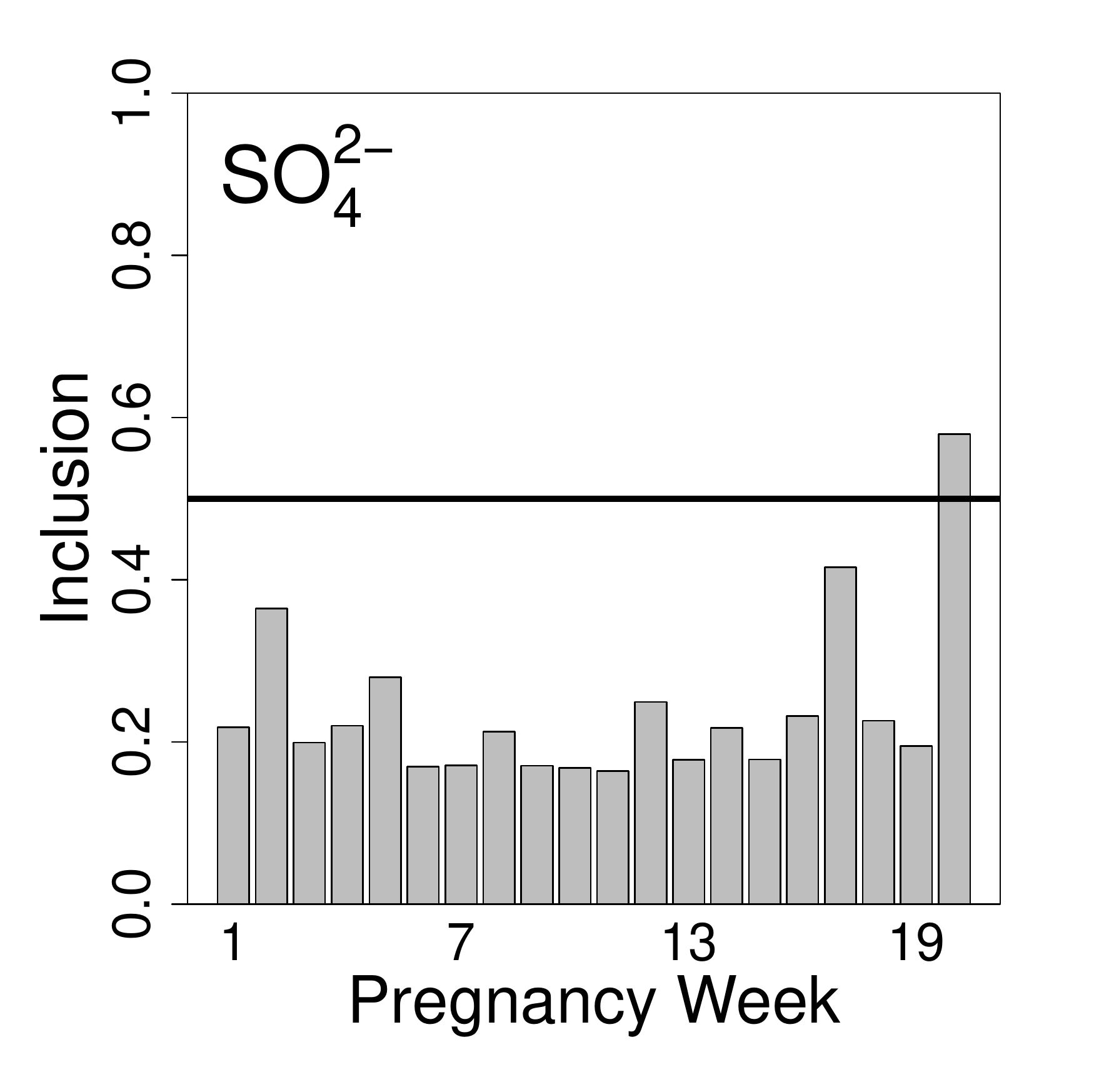}
\caption{Posterior inclusion probability results from the \textbf{non-Hispanic Black} stillbirth and single exposure Critical Window Variable Selection (CWVS) analyses in New Jersey, 2005-2014.}
\end{center}
\end{figure}
\clearpage

\begin{figure}[h]
\begin{center}
\includegraphics[scale=0.26]{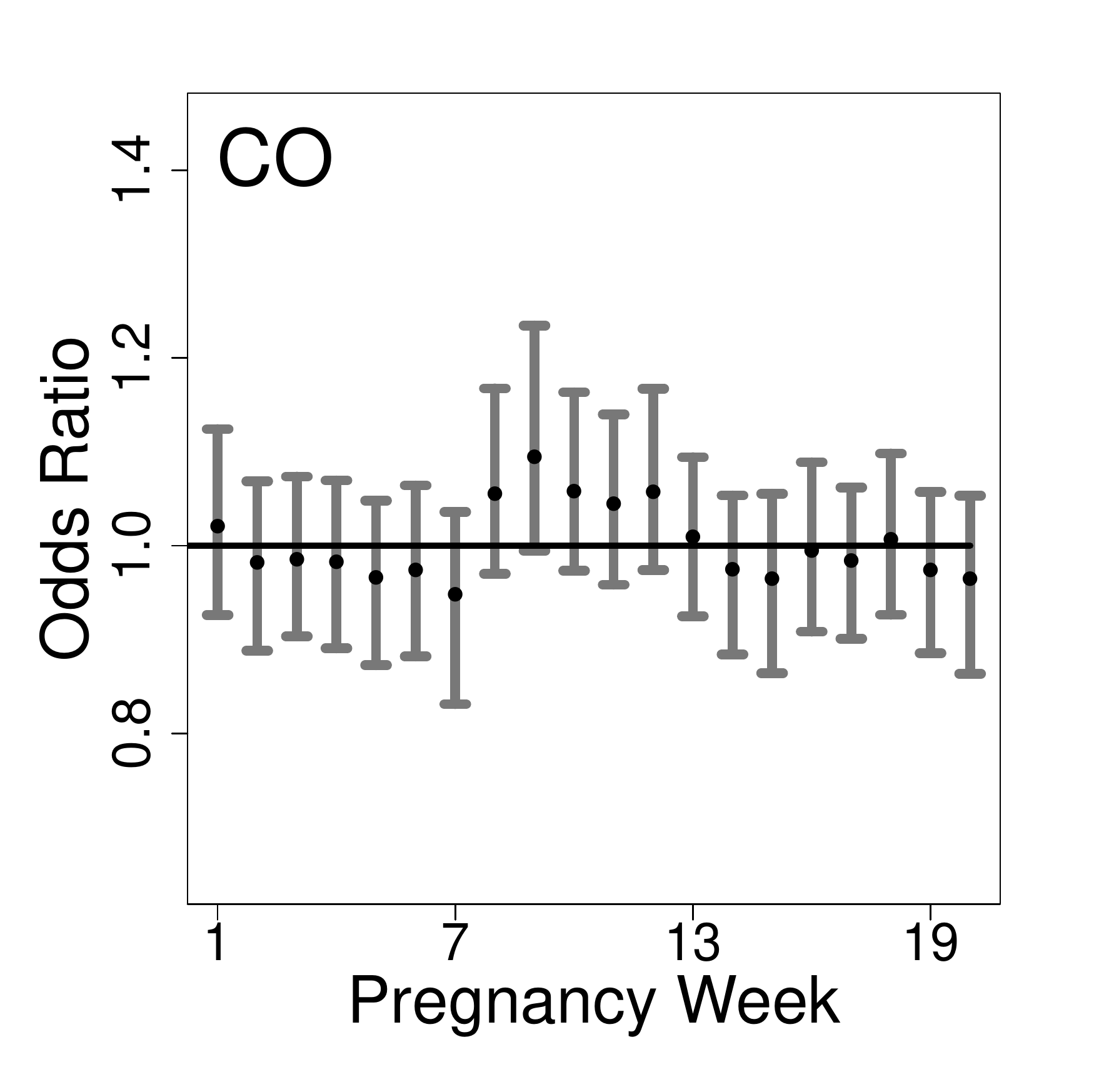}
\includegraphics[scale=0.26]{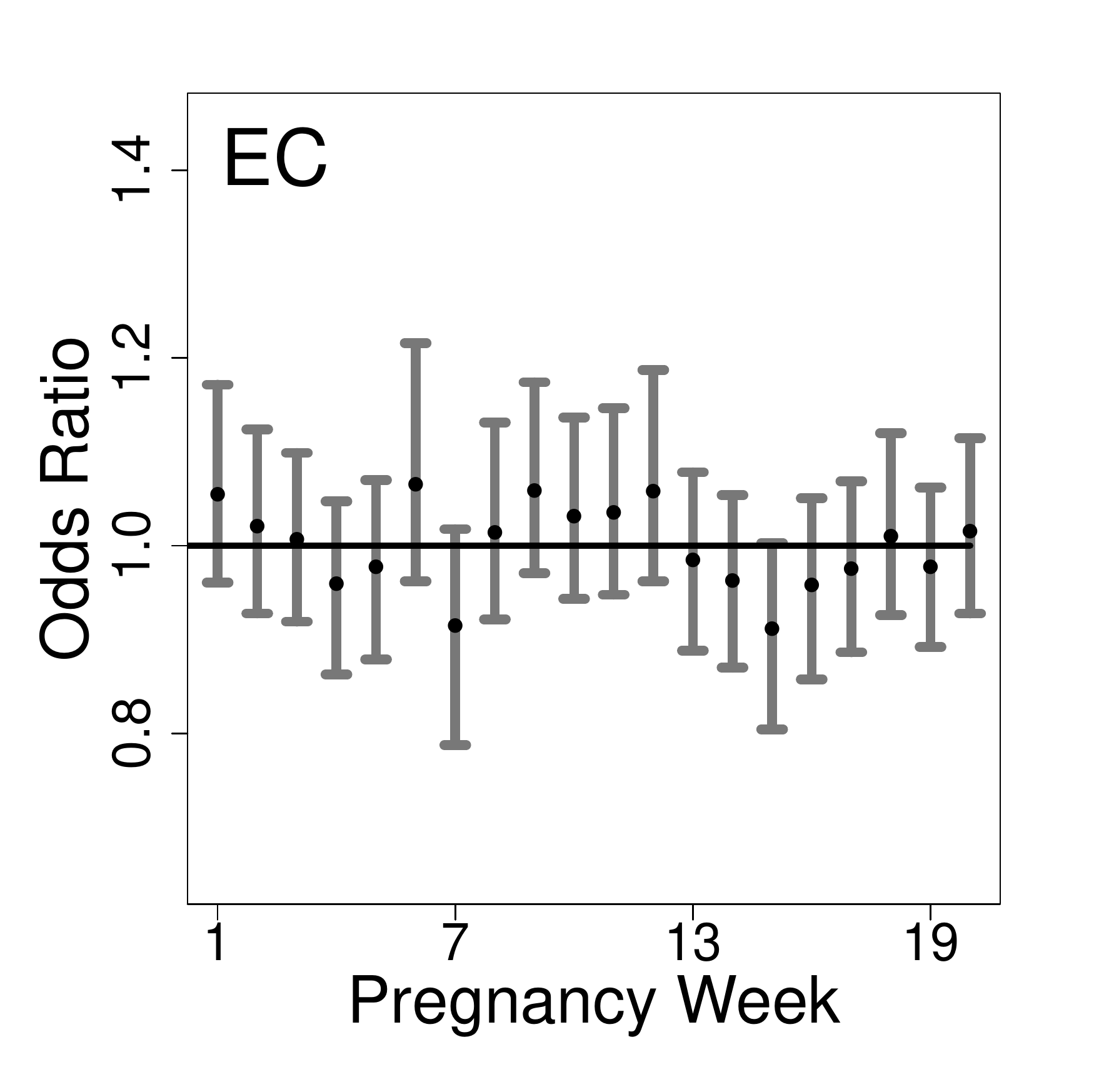}
\includegraphics[scale=0.26]{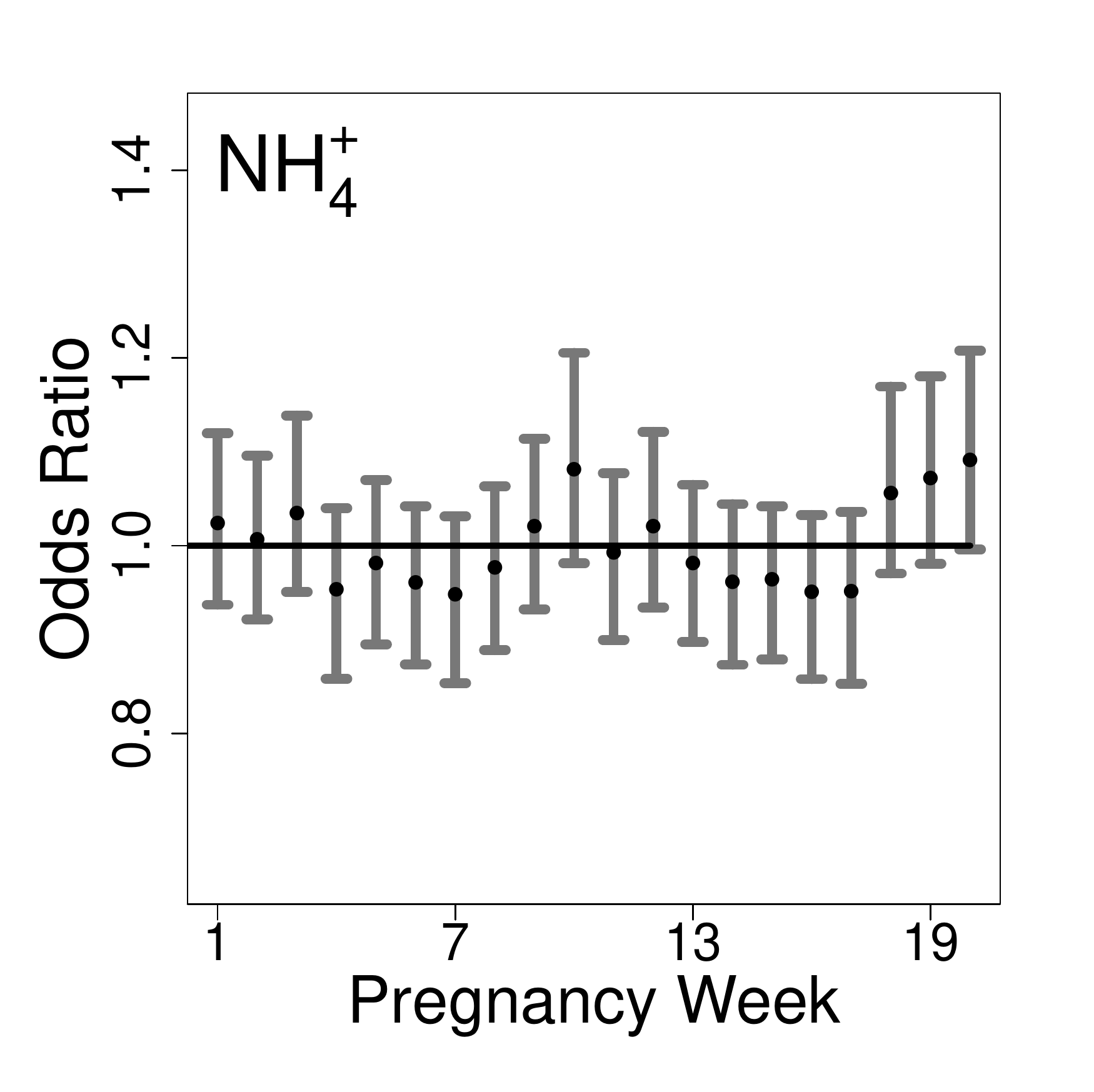}\\
\includegraphics[scale=0.26]{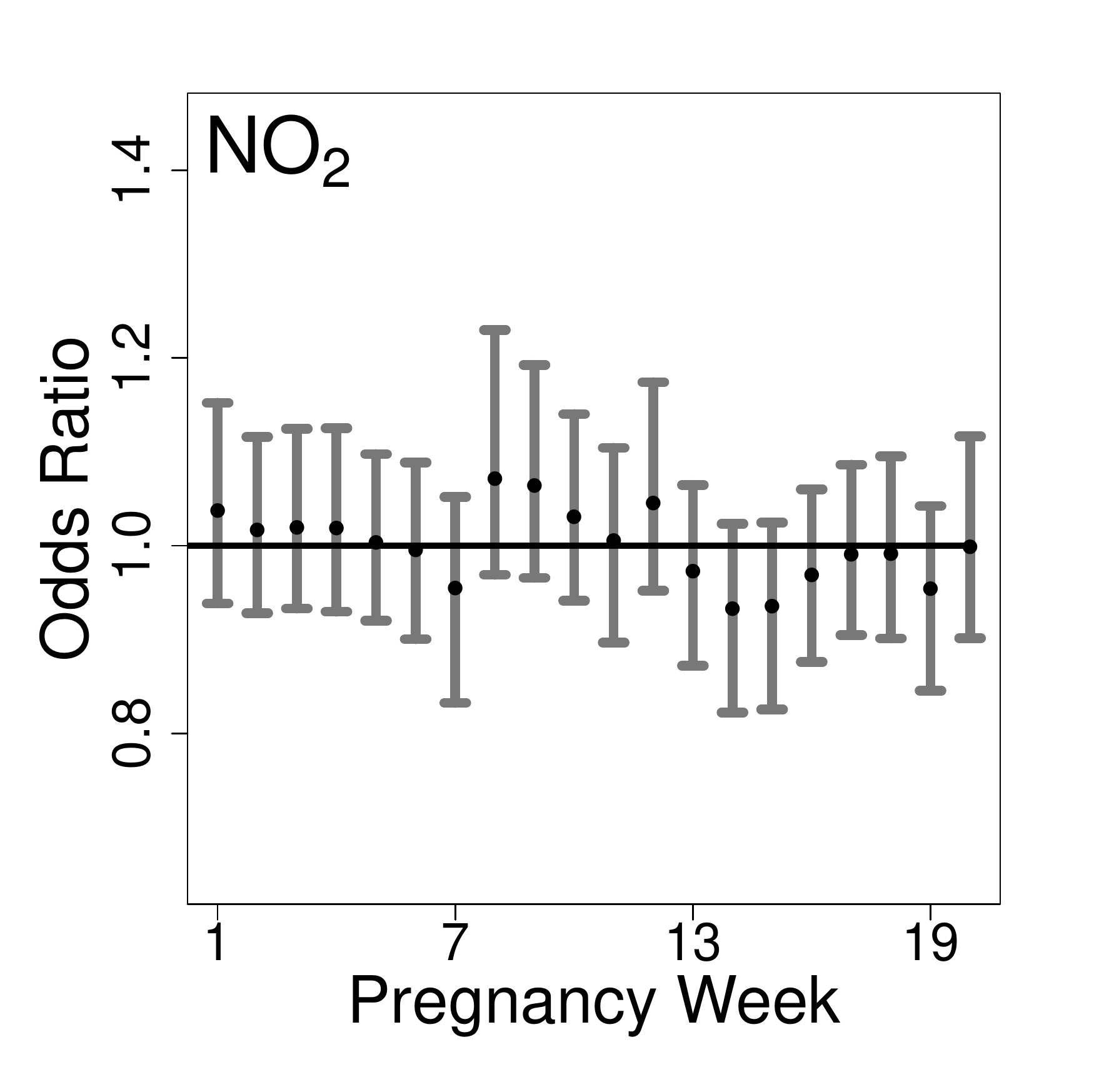}
\includegraphics[scale=0.26]{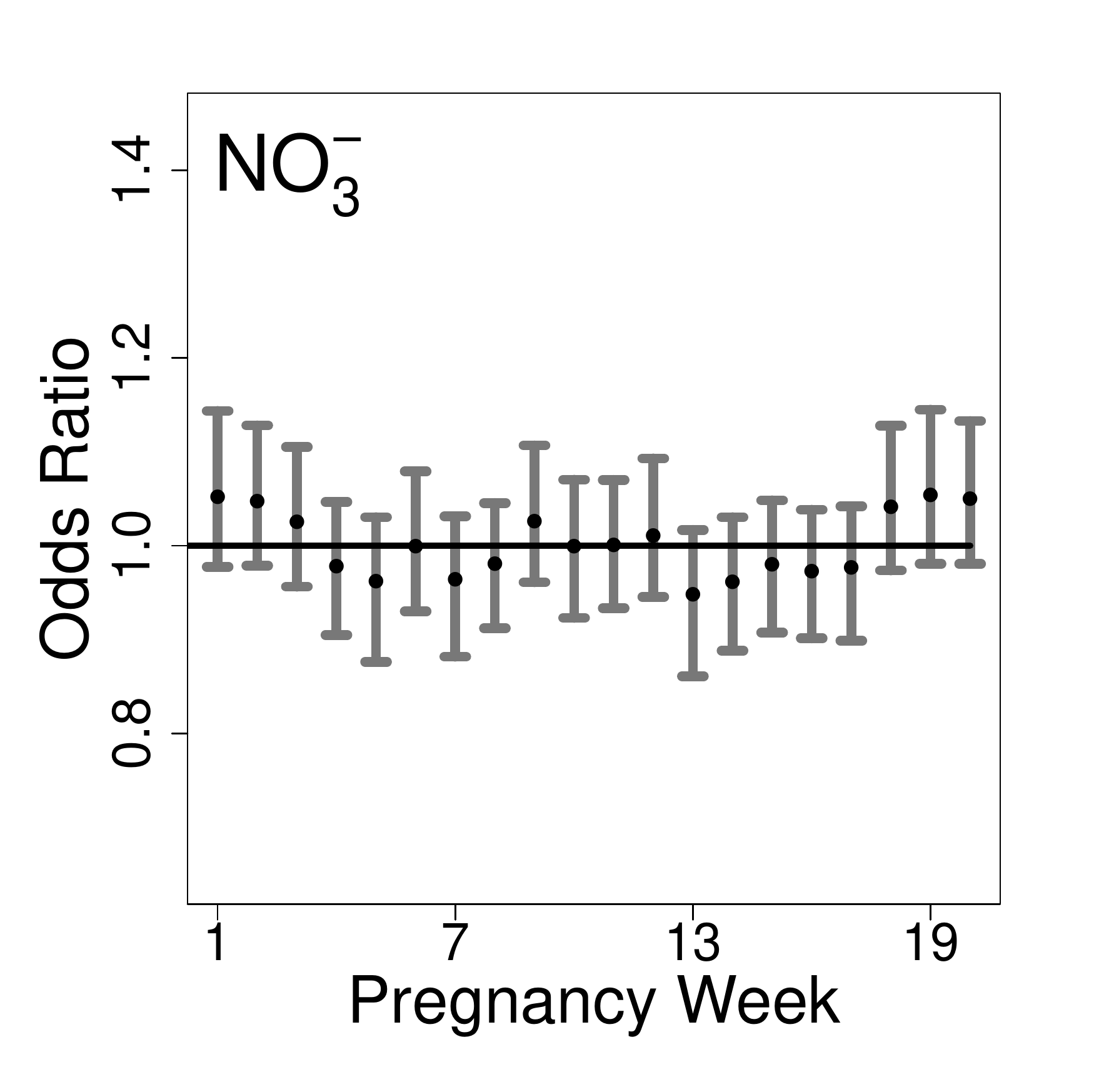}
\includegraphics[scale=0.26]{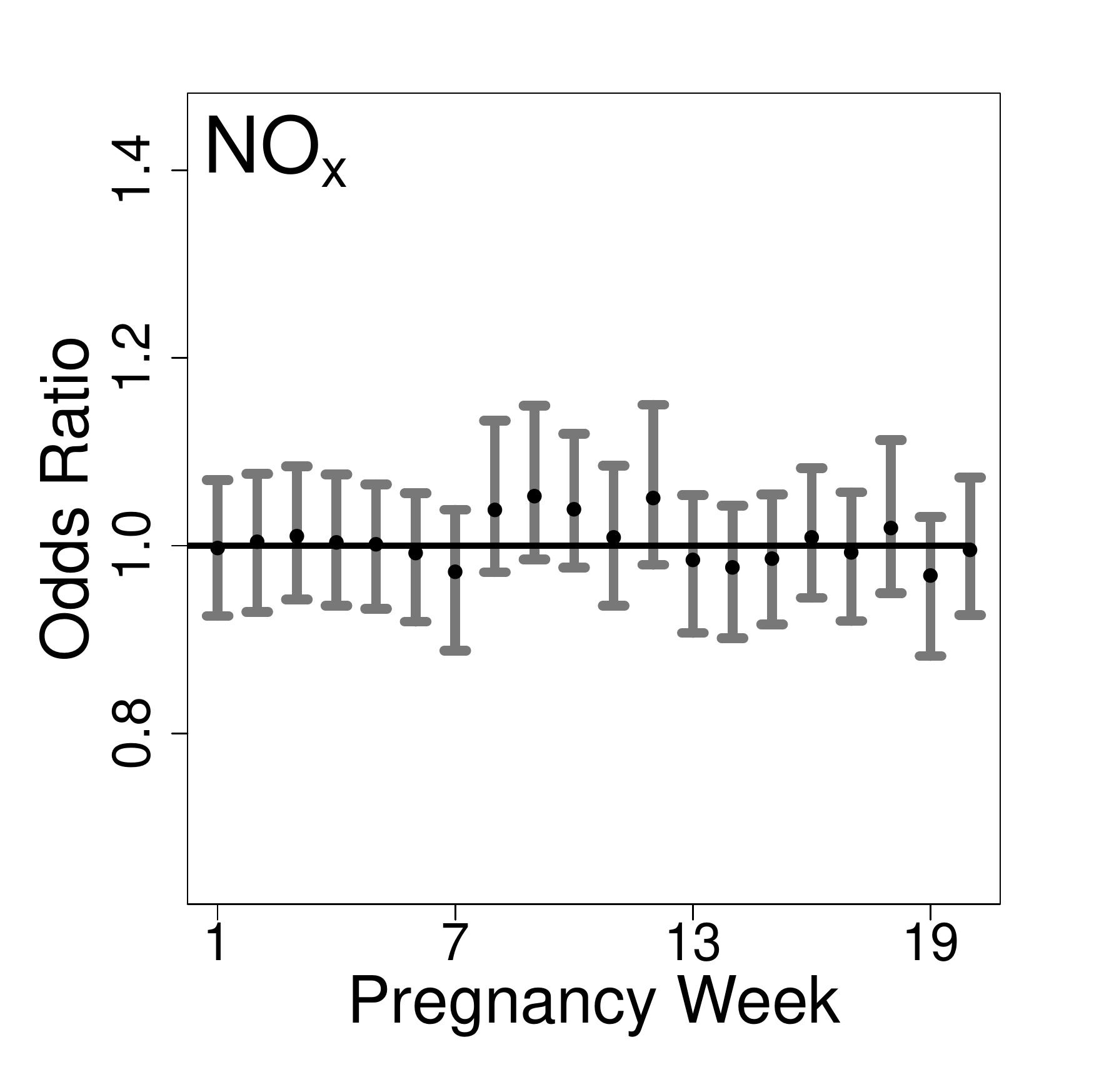}\\
\includegraphics[scale=0.26]{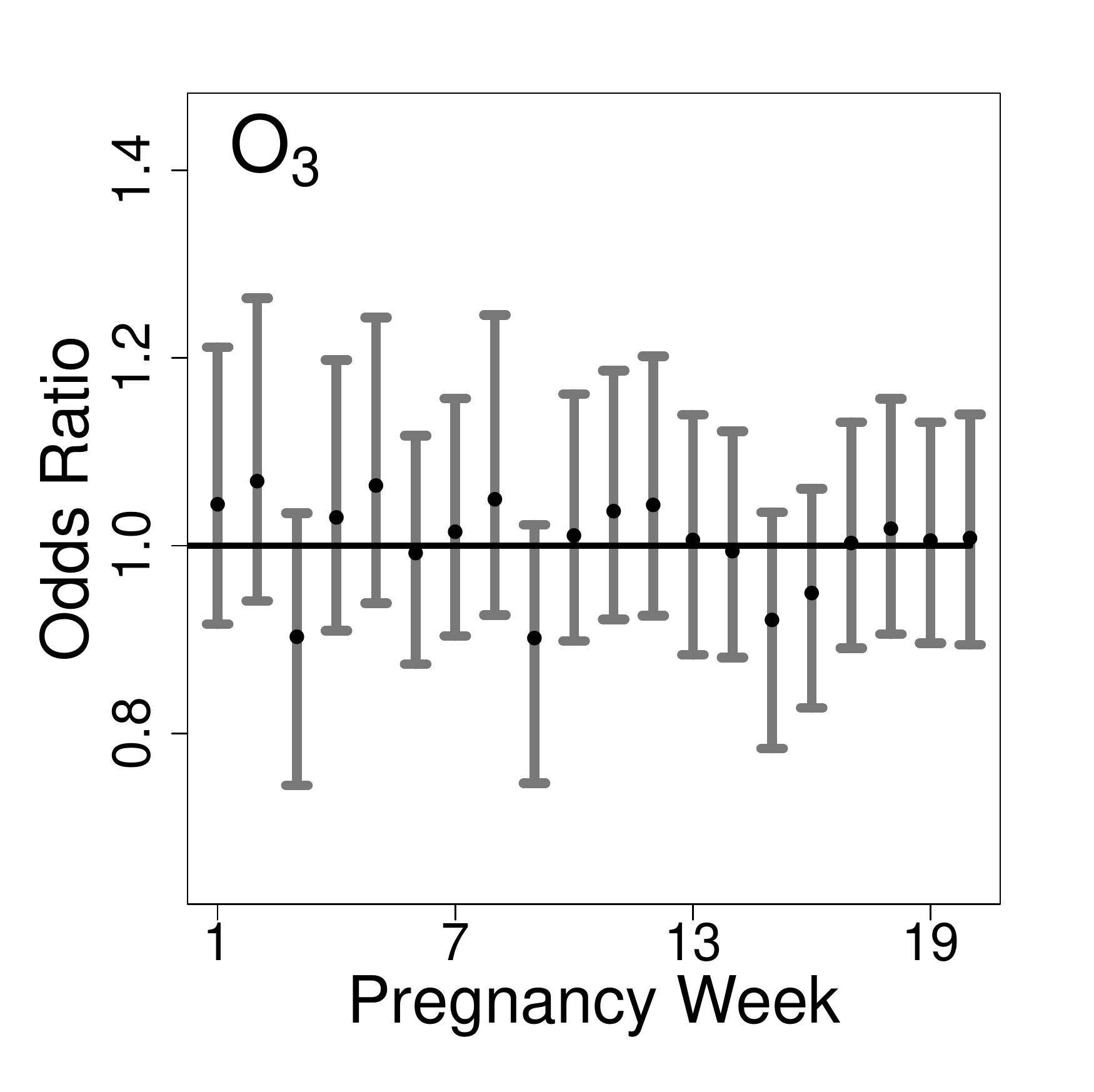}
\includegraphics[scale=0.26]{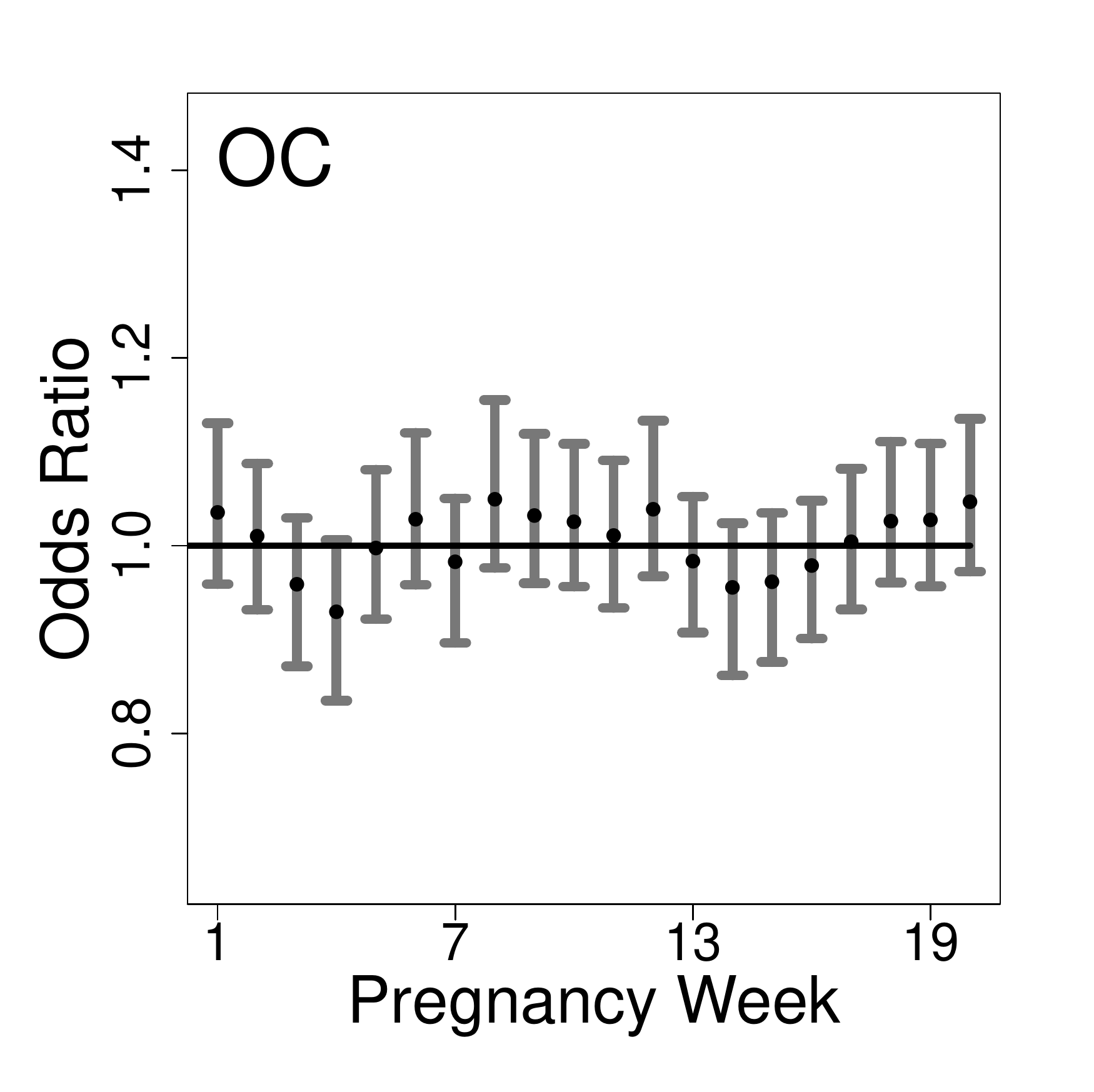}
\includegraphics[scale=0.26]{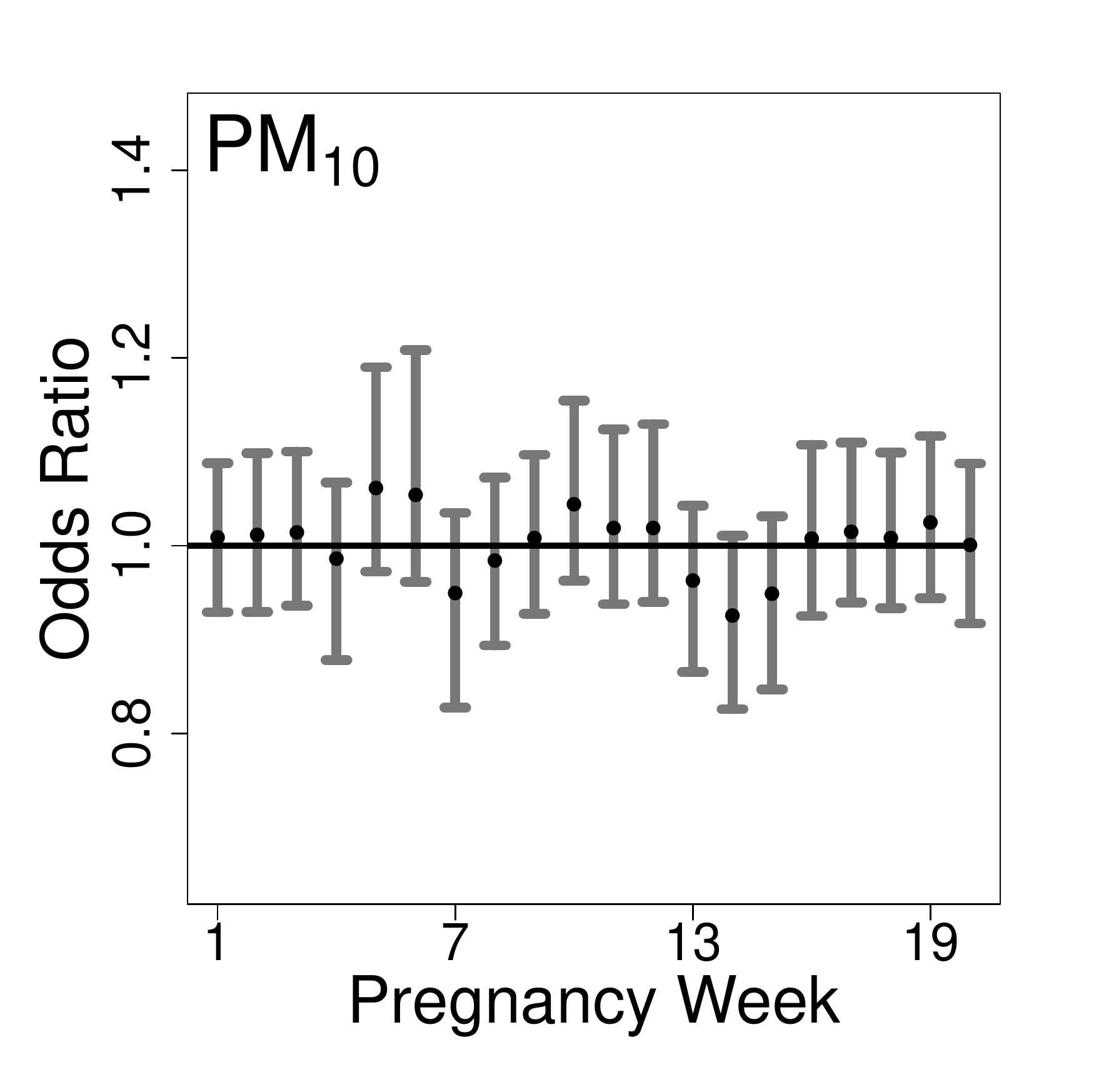}\\
\includegraphics[scale=0.26]{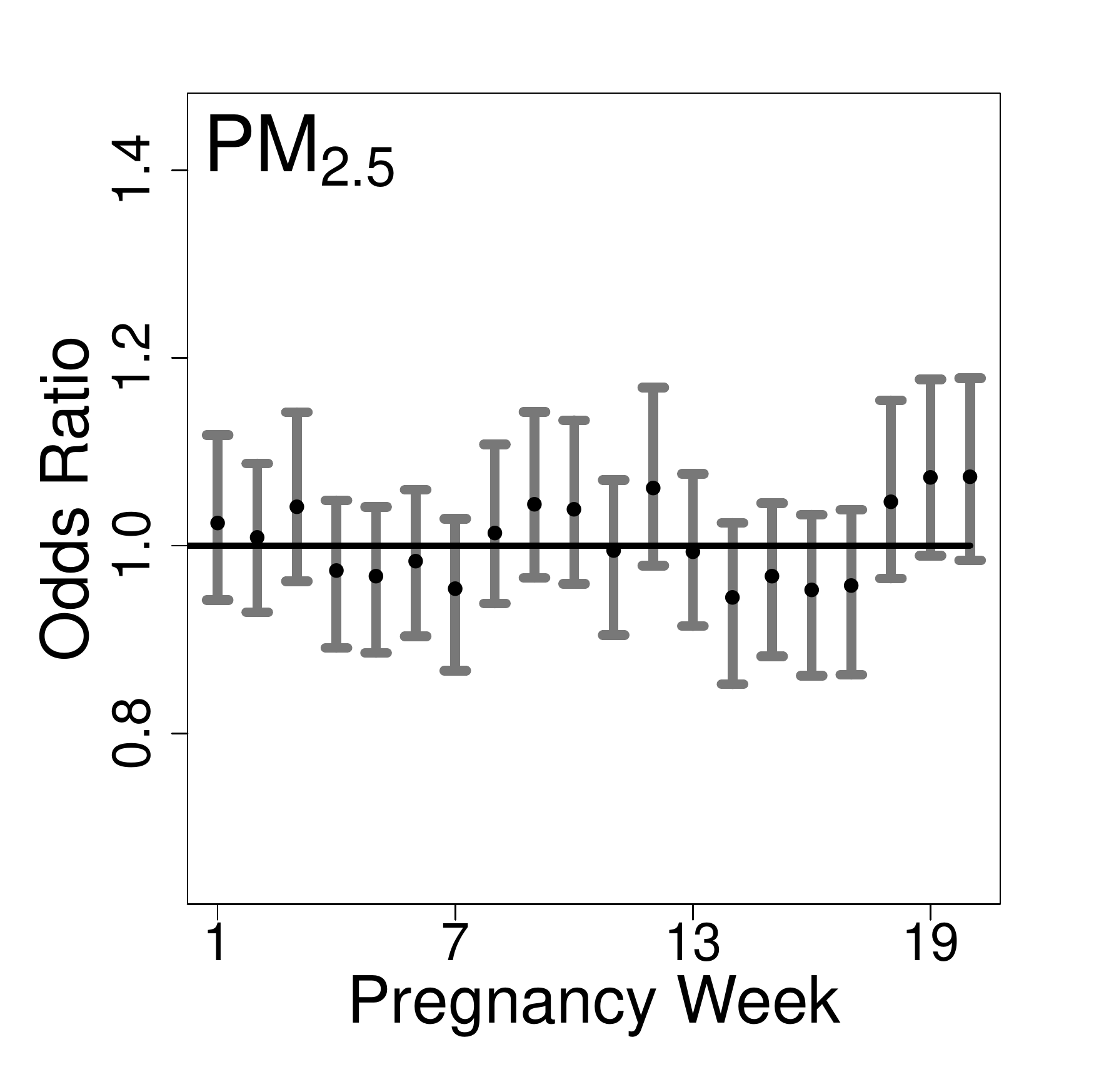}
\includegraphics[scale=0.26]{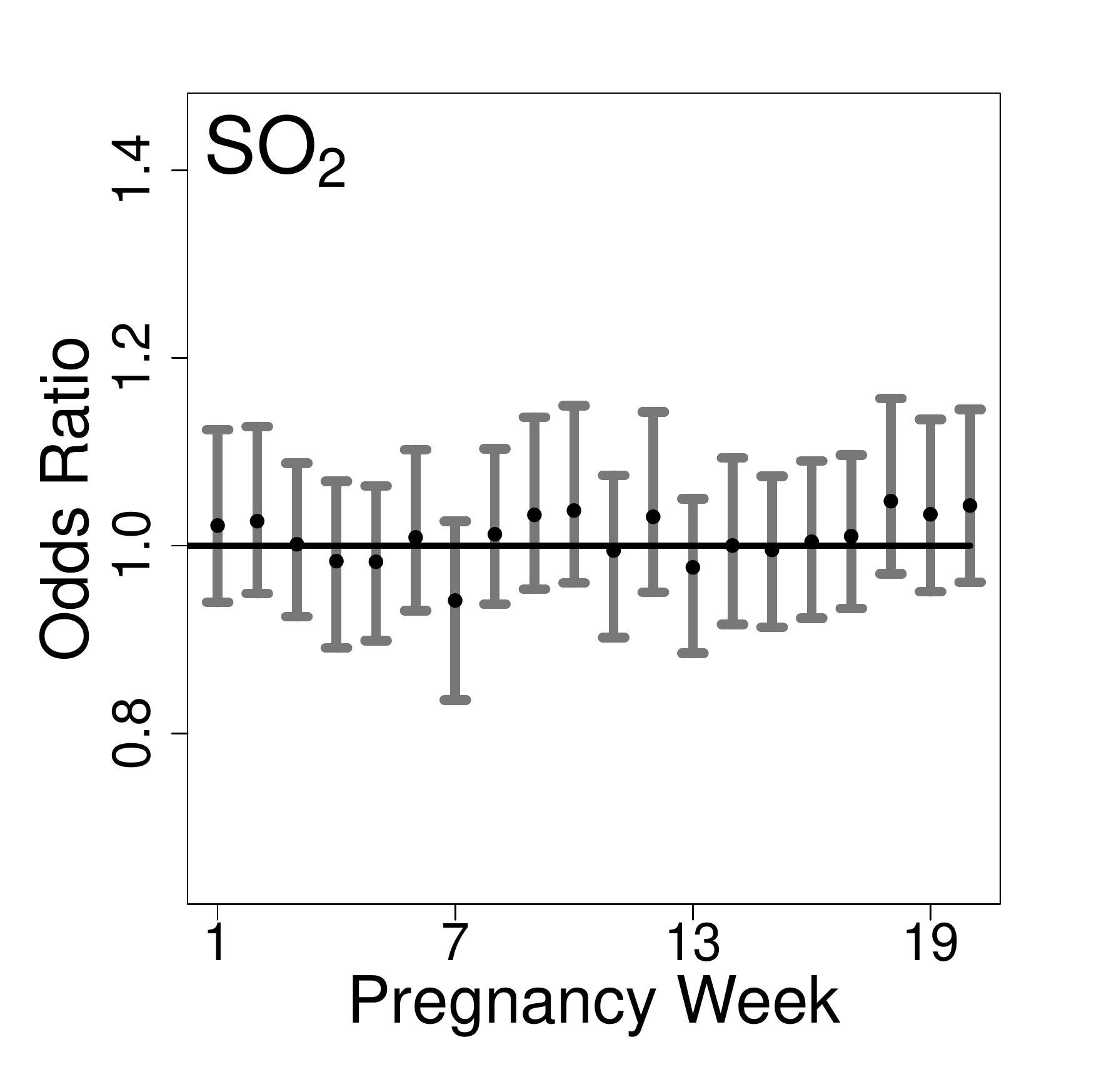}
\includegraphics[scale=0.26]{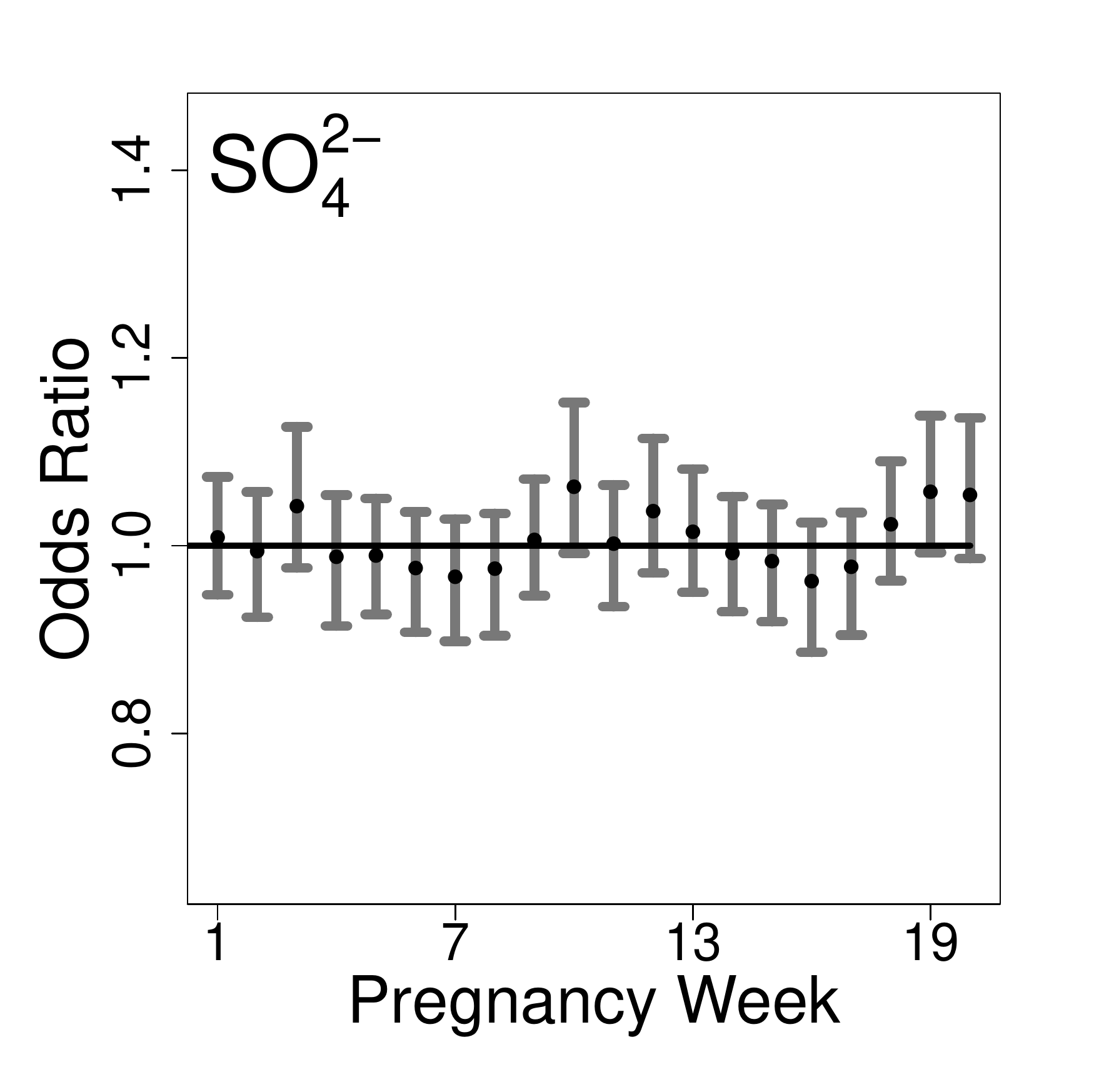}
\caption{Posterior mean and 95\% credible interval results from the \textbf{Hispanic} stillbirth and single exposure Critical Window Variable Selection (CWVS) analyses in New Jersey, 2005-2014. Results based on an interquartile range increase in weekly exposure. Weeks identified as part of the critical window set are shown in red/dashed (harmful) and blue/dashed (protective).  These definitions depend partly on the posterior inclusion probabilities in Figure S7 of the Supporting Information for the variable selection methods.}
\end{center}
\end{figure}
\clearpage 

\begin{figure}[h]
\begin{center}
\includegraphics[scale=0.26]{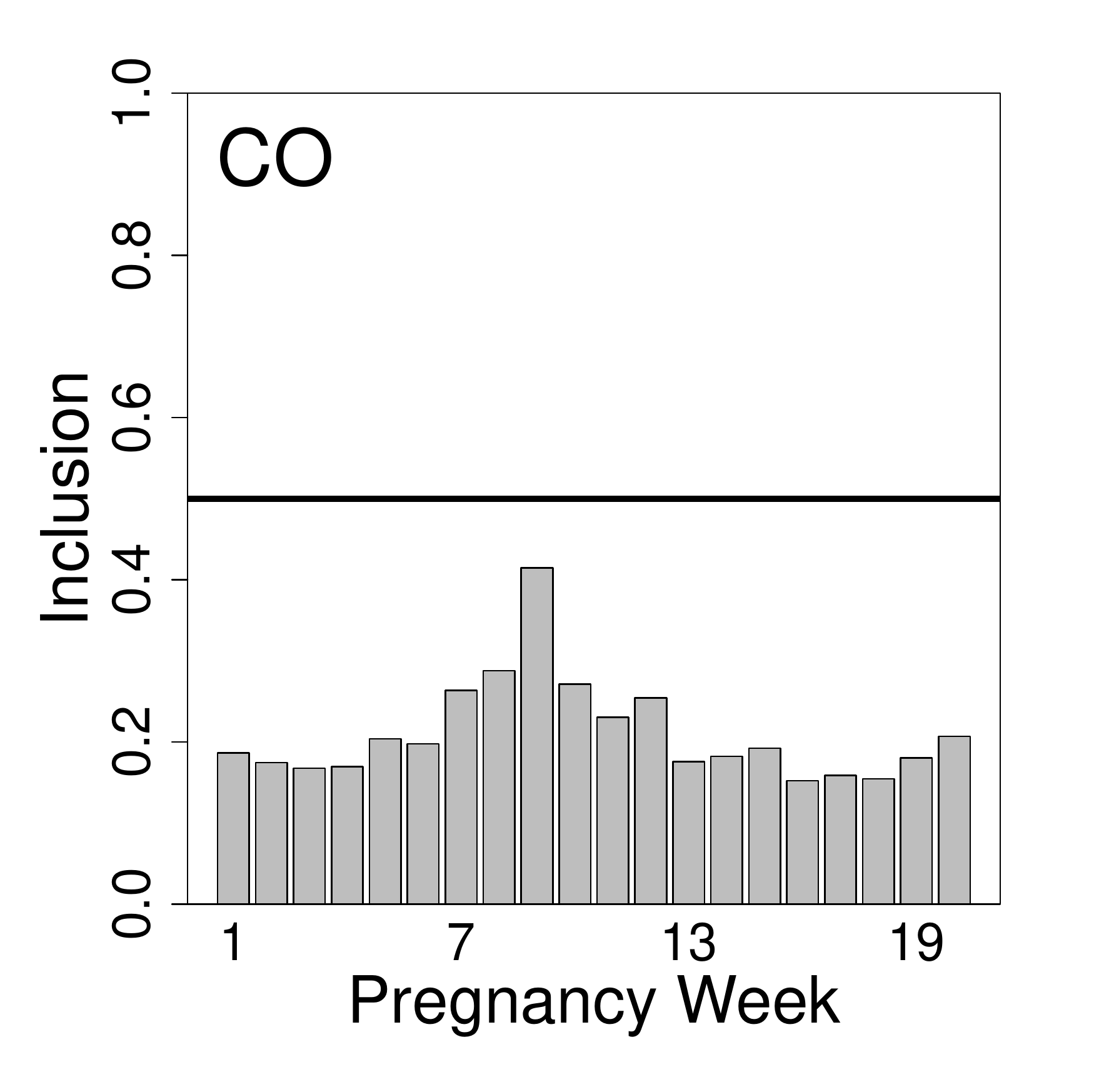}
\includegraphics[scale=0.26]{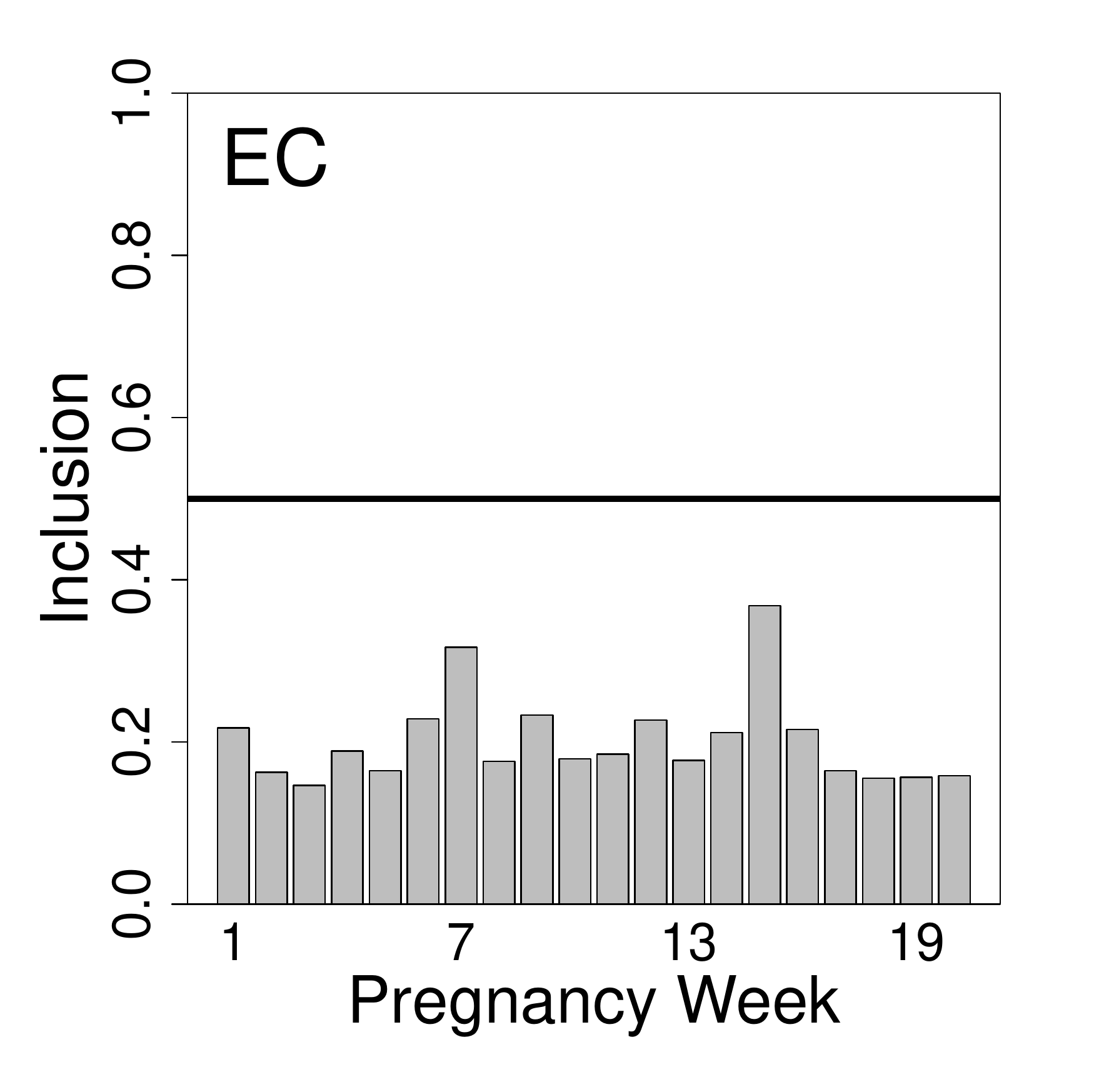}
\includegraphics[scale=0.26]{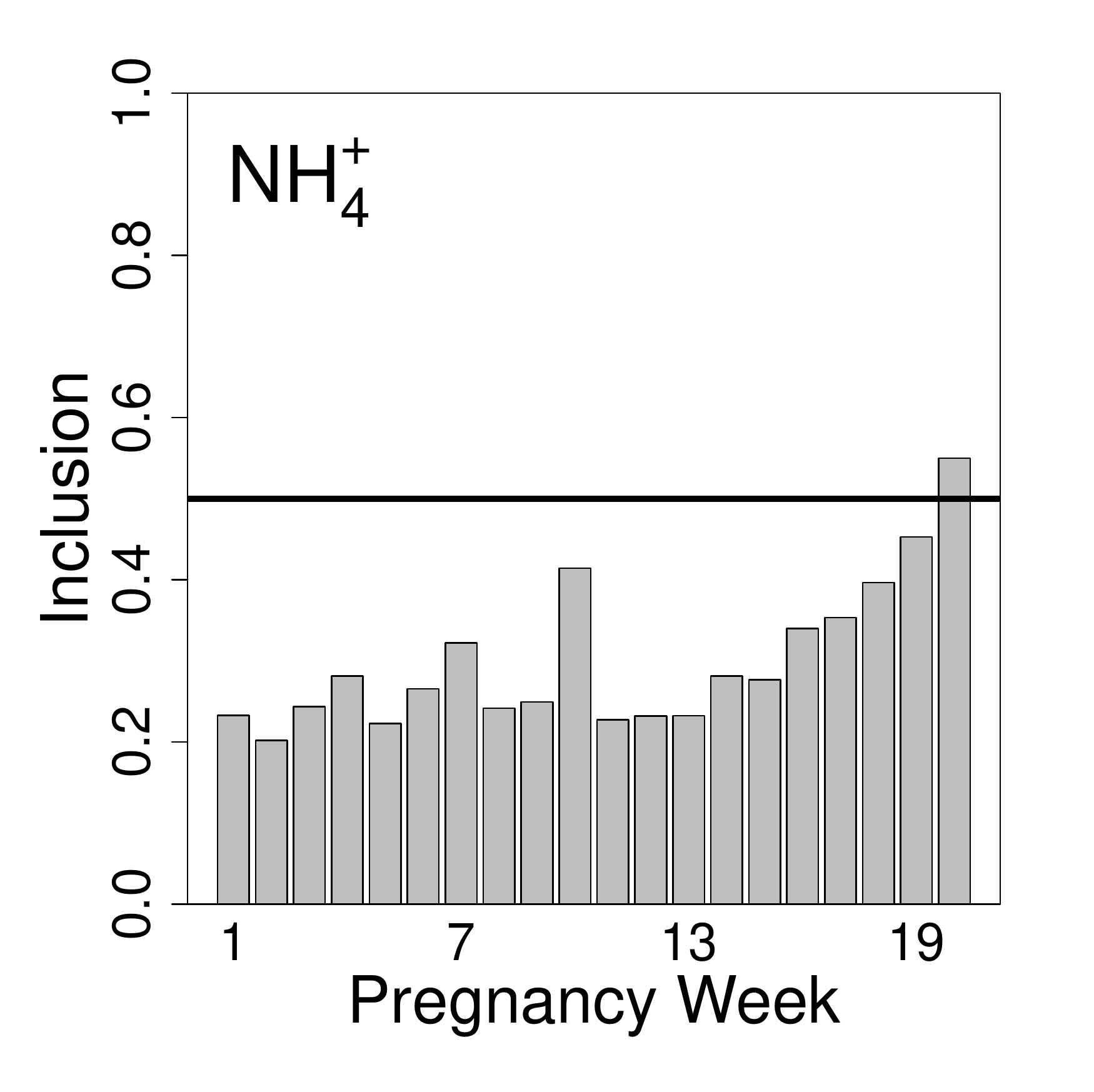}\\
\includegraphics[scale=0.26]{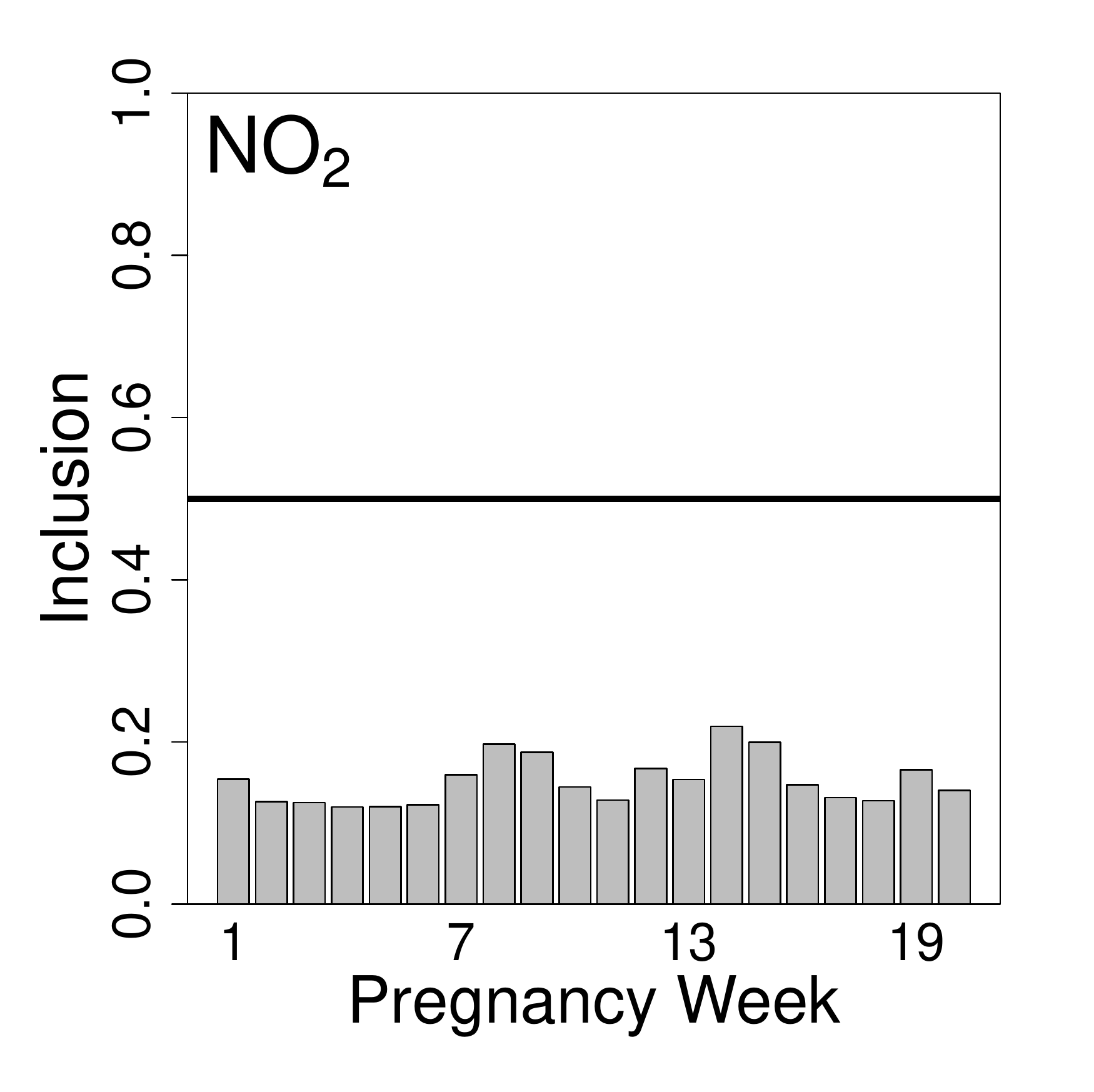}
\includegraphics[scale=0.26]{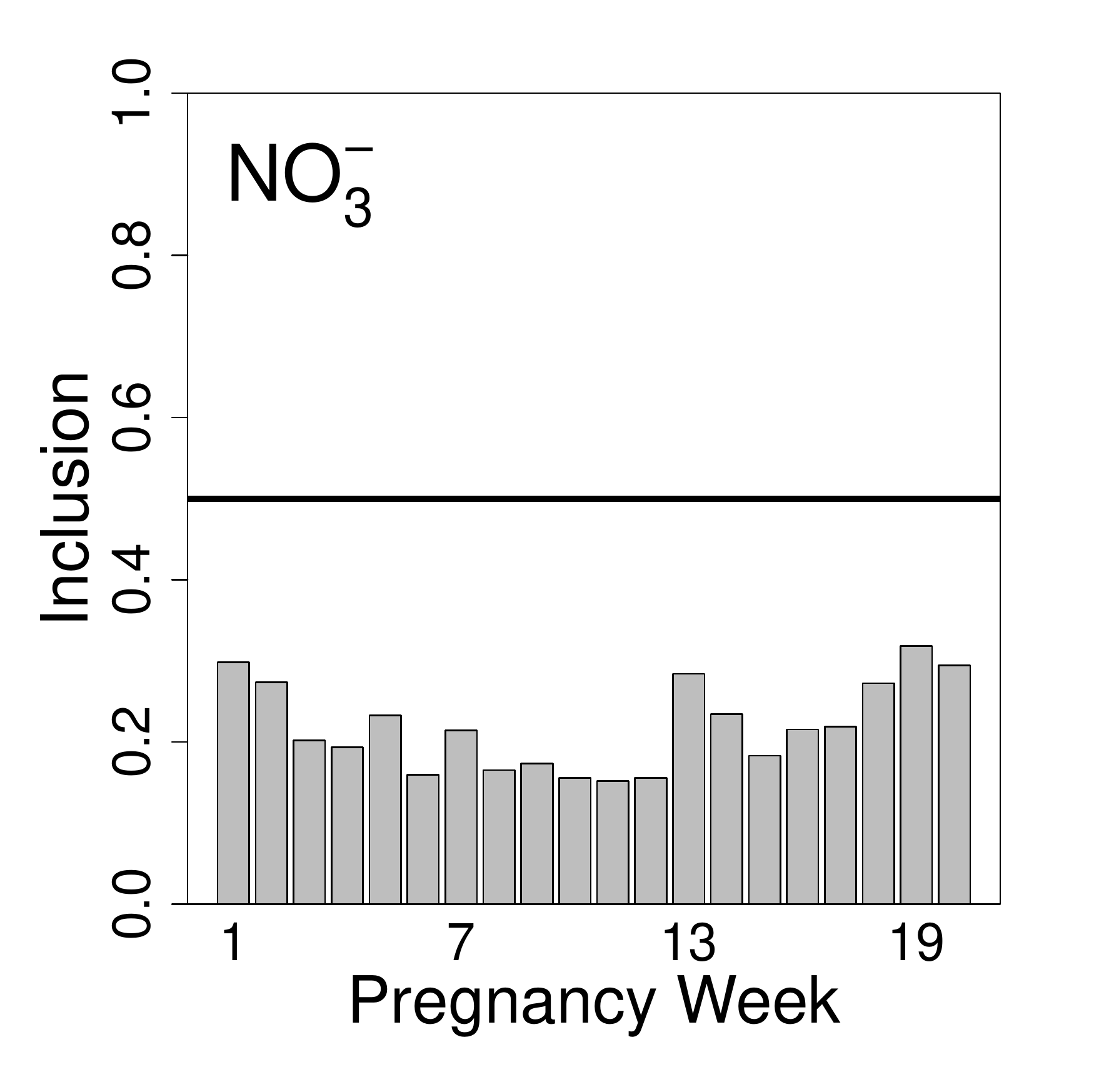}
\includegraphics[scale=0.26]{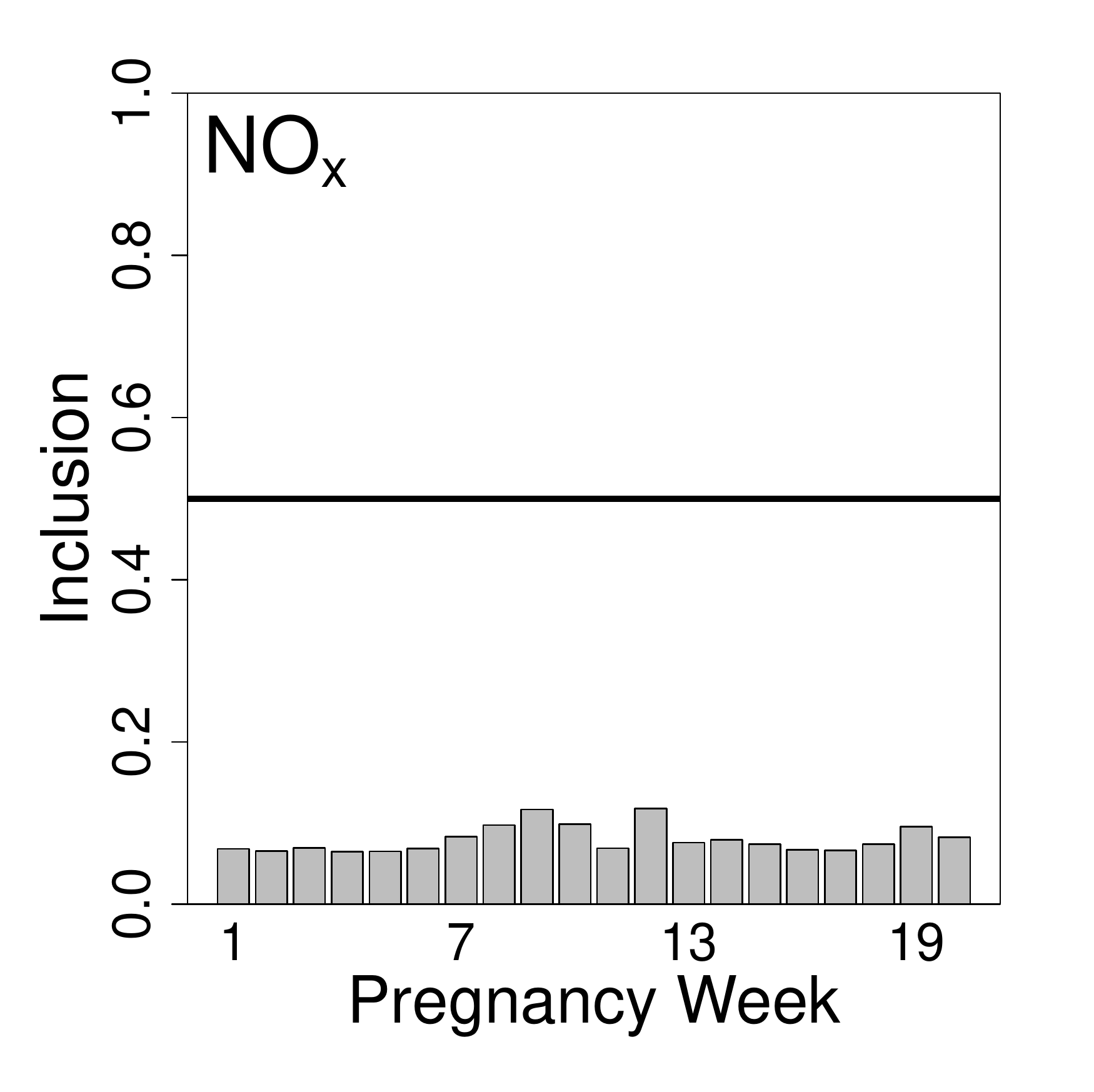}\\
\includegraphics[scale=0.26]{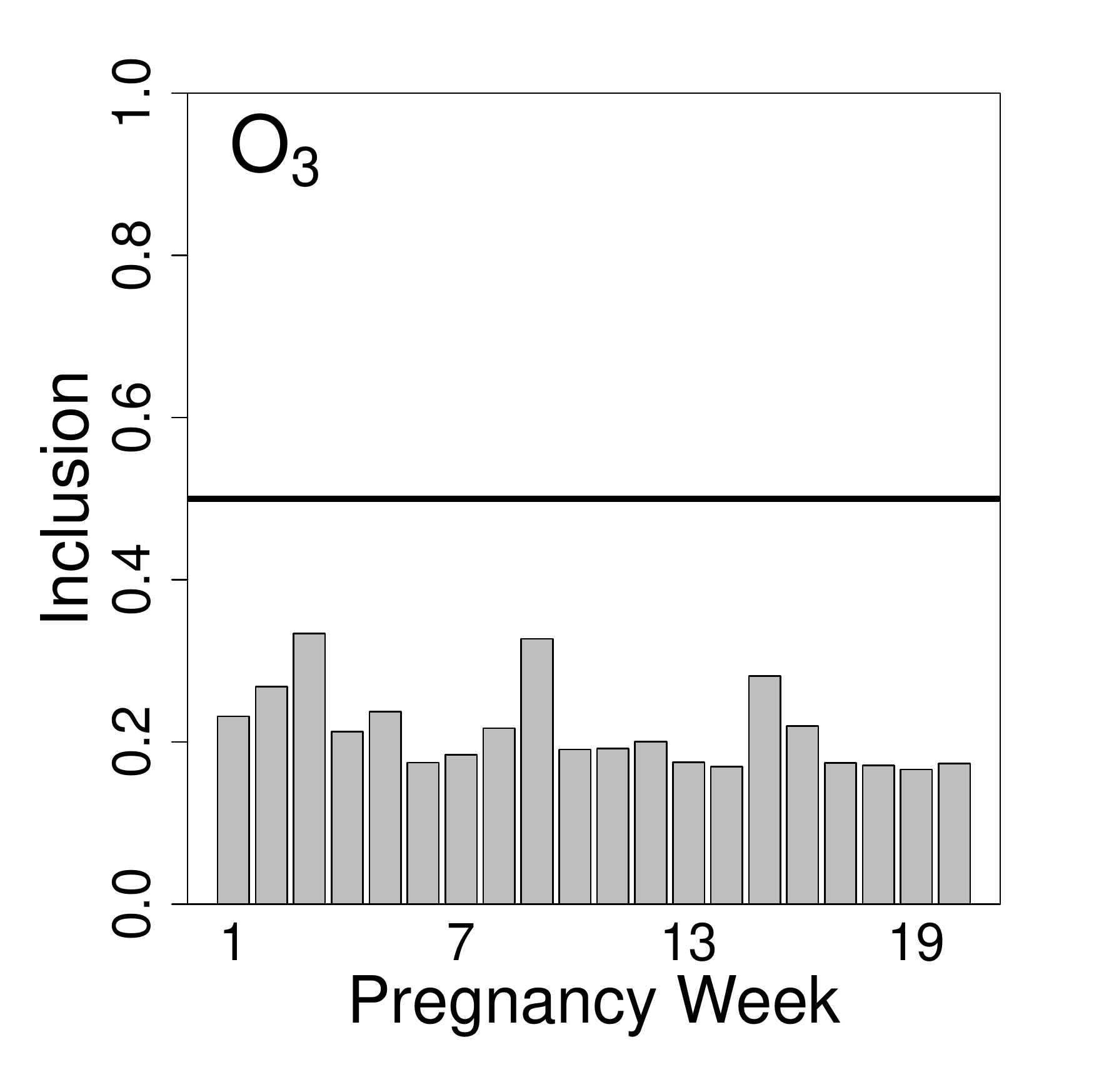}
\includegraphics[scale=0.26]{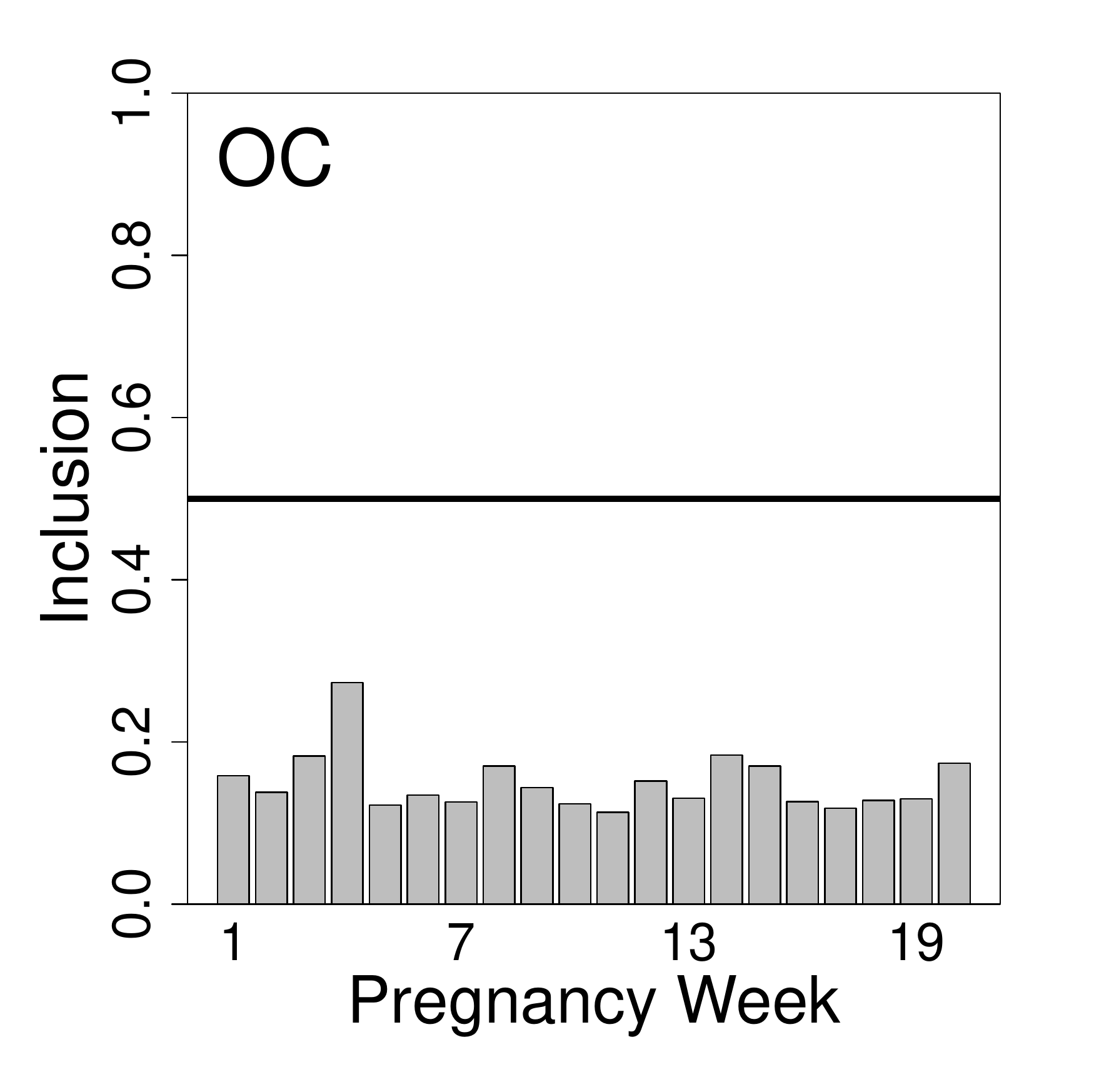}
\includegraphics[scale=0.26]{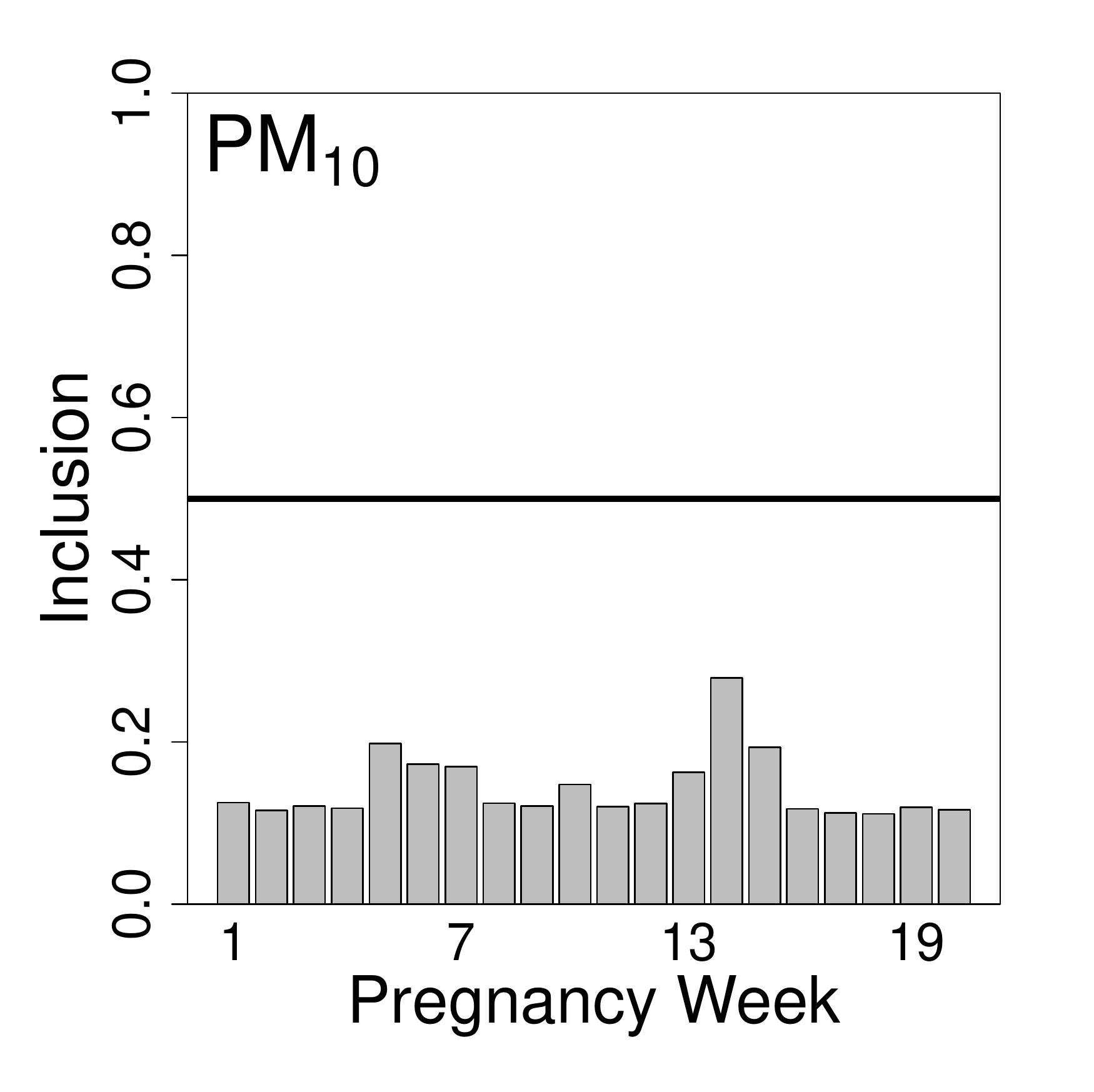}\\
\includegraphics[scale=0.26]{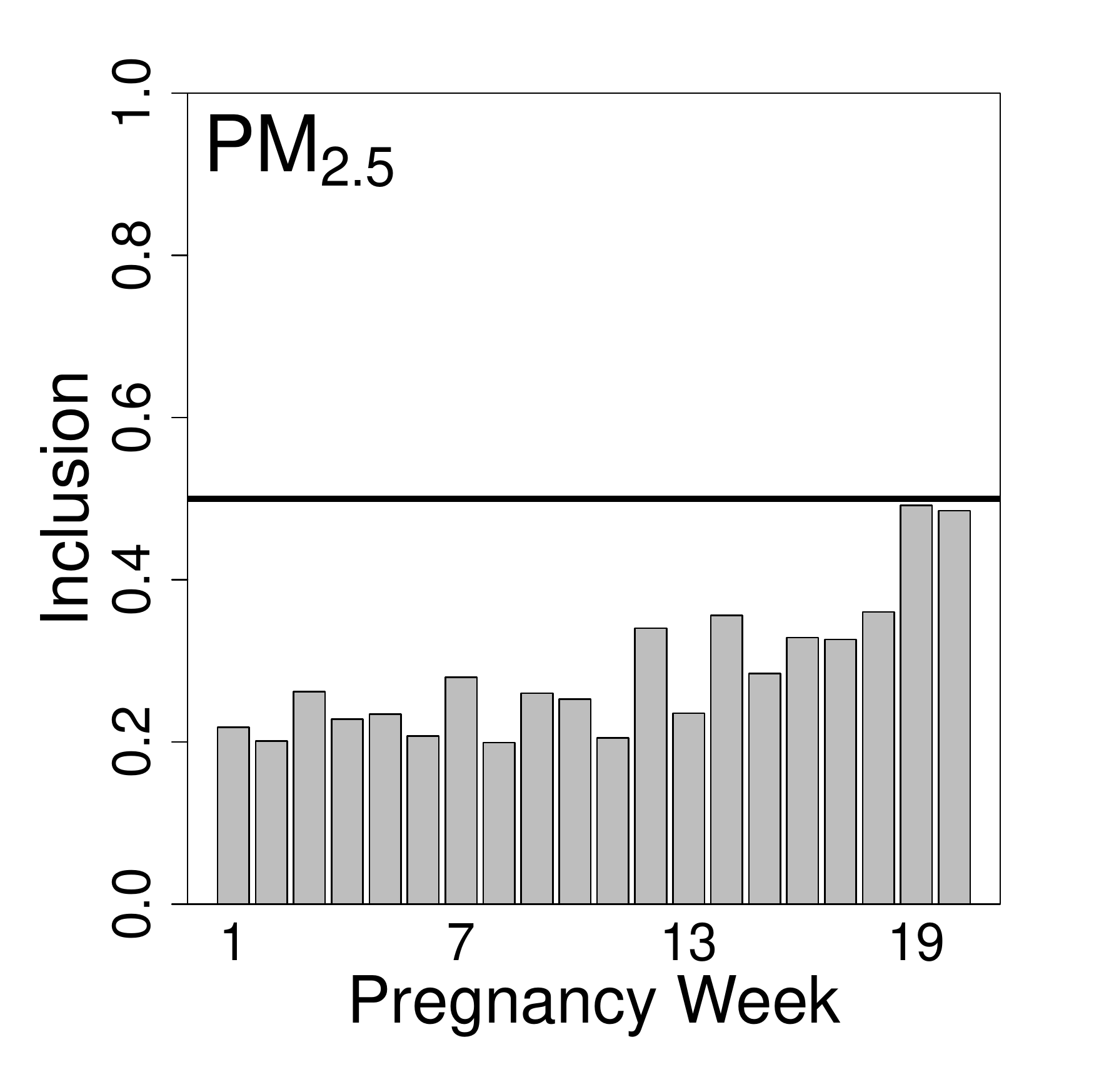}
\includegraphics[scale=0.26]{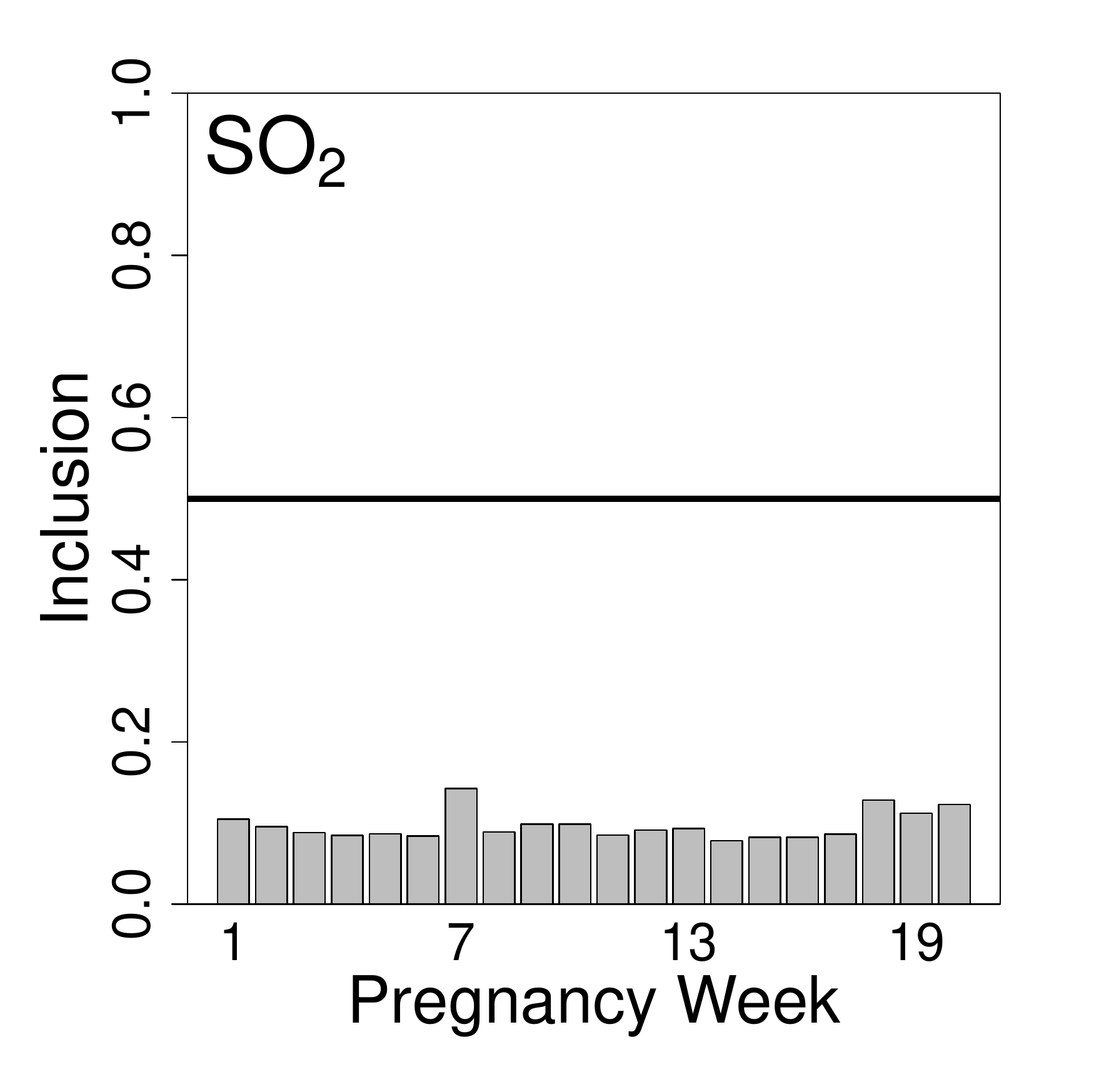}
\includegraphics[scale=0.26]{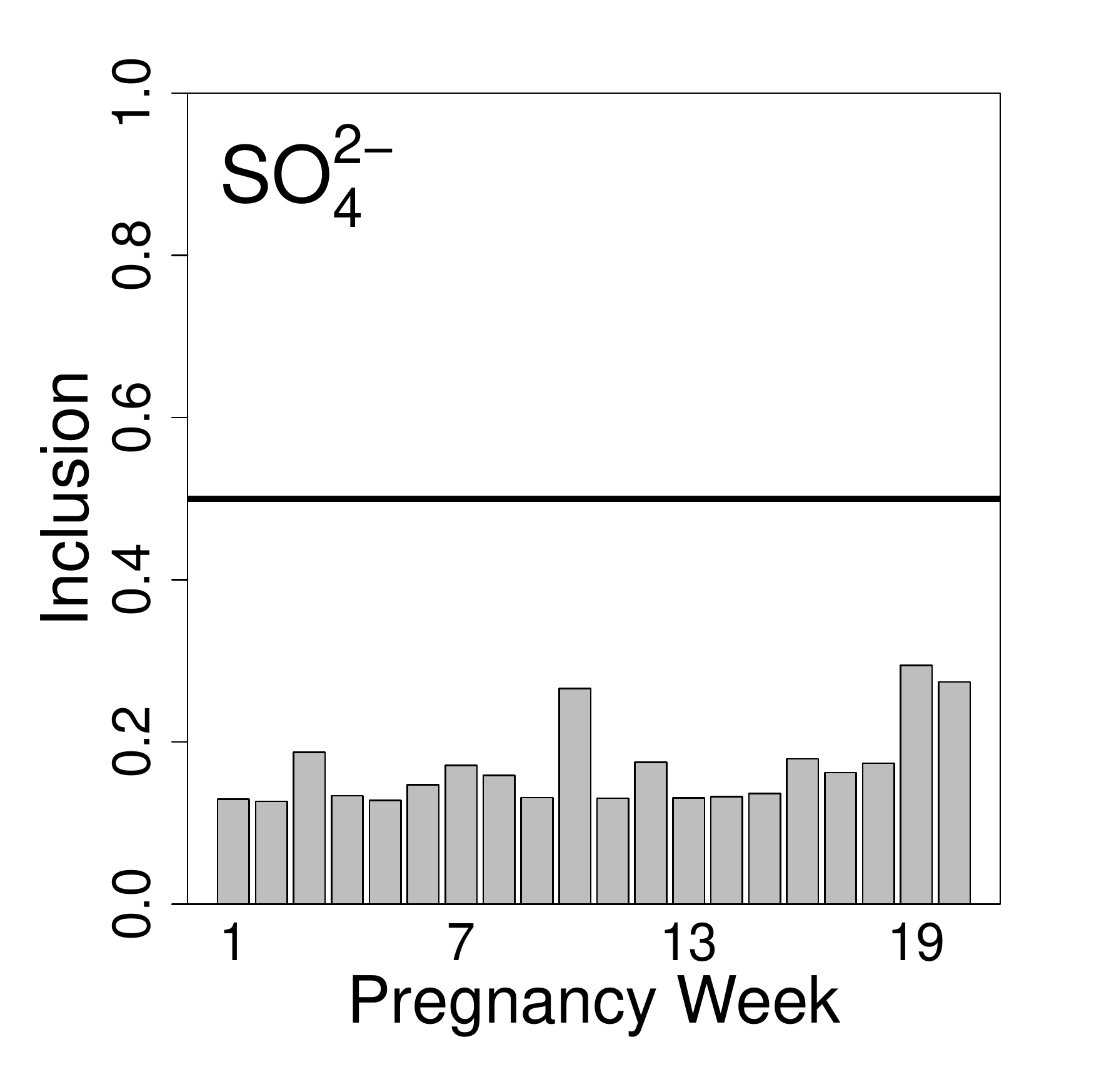}
\caption{Posterior inclusion probability results from the \textbf{Hispanic} stillbirth and single exposure Critical Window Variable Selection (CWVS) analyses in New Jersey, 2005-2014.}
\end{center}
\end{figure}
\clearpage 

\begin{figure}[h]
\begin{center}
\includegraphics[scale=0.26]{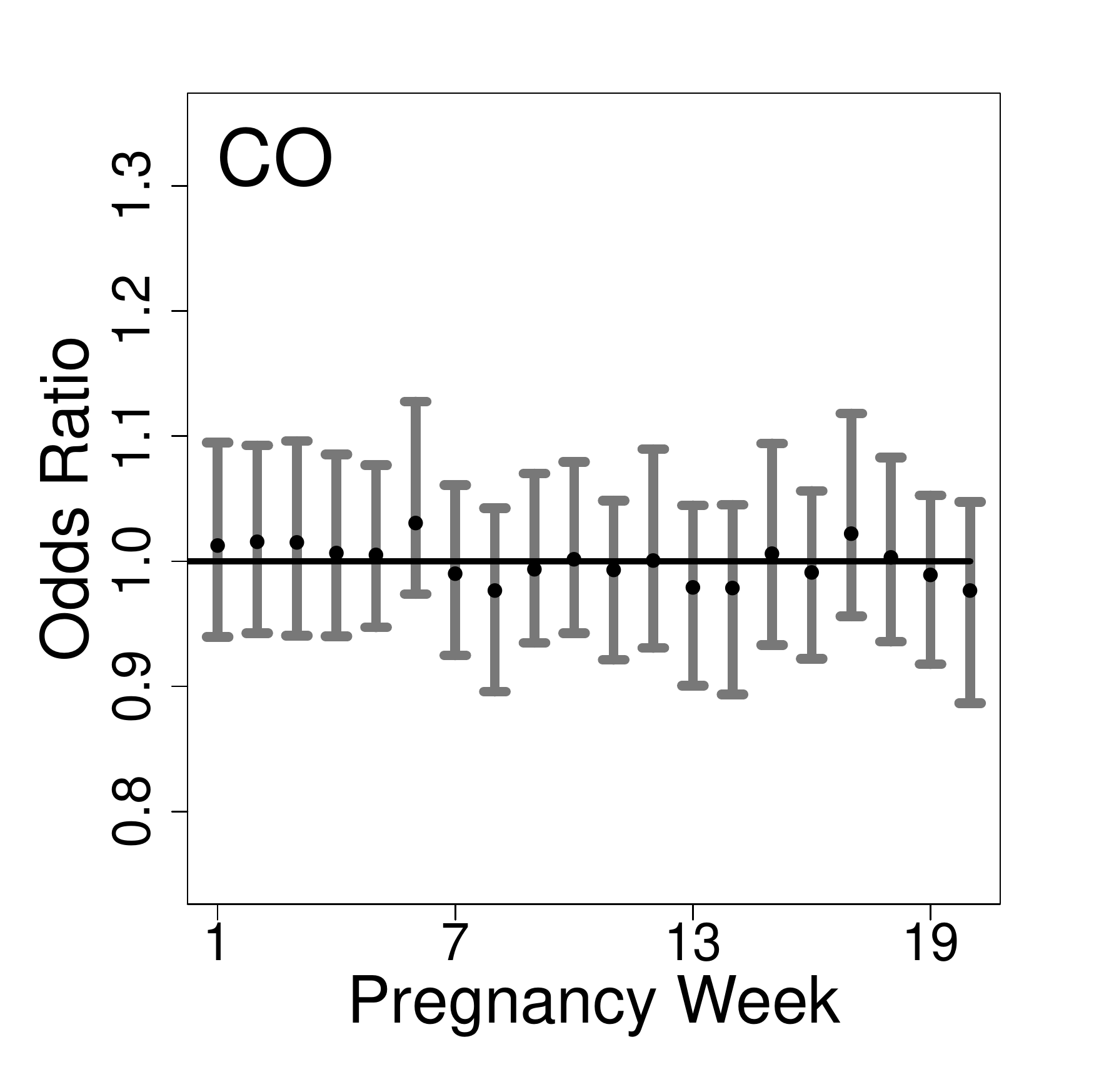}
\includegraphics[scale=0.26]{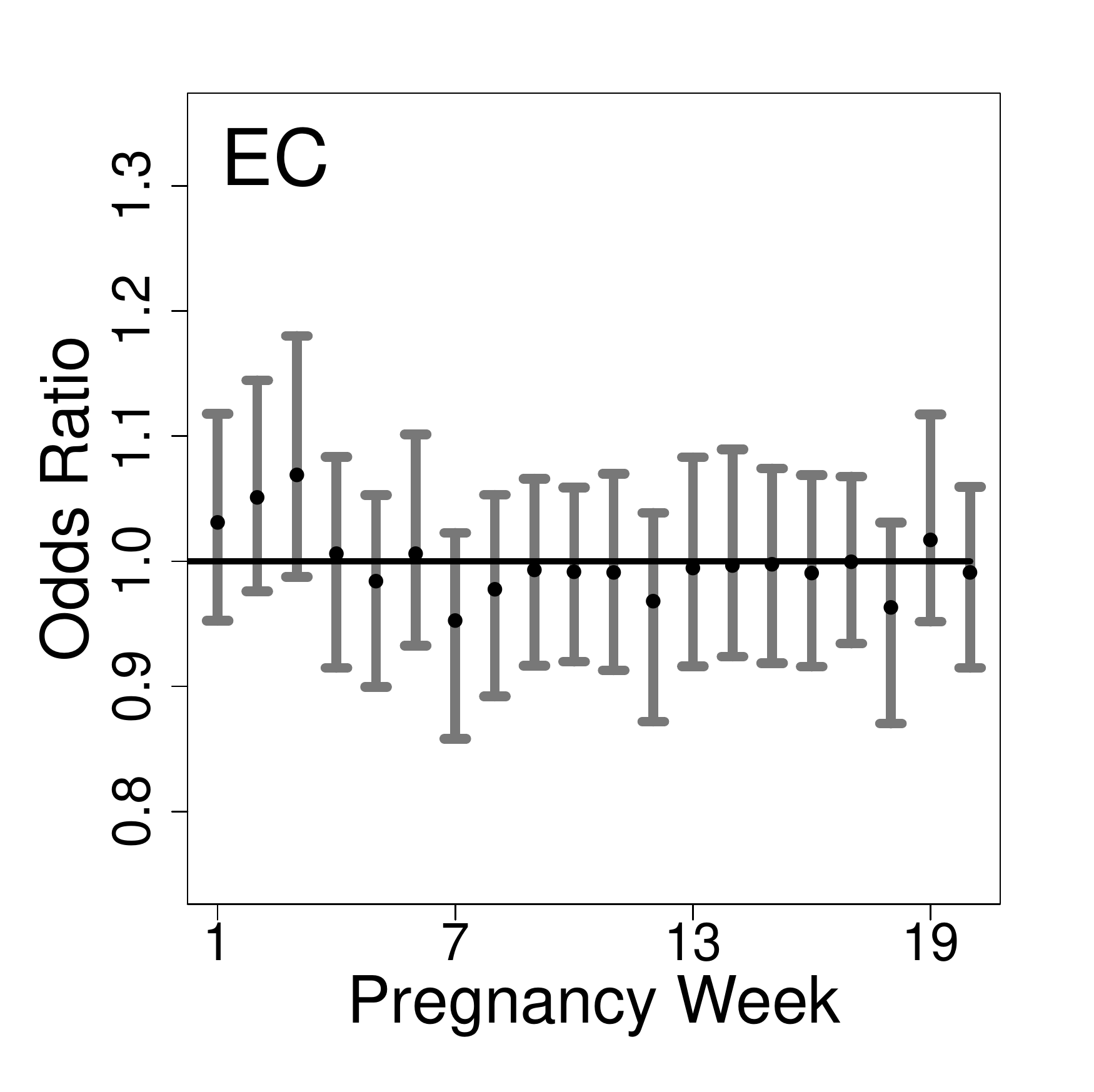}
\includegraphics[scale=0.26]{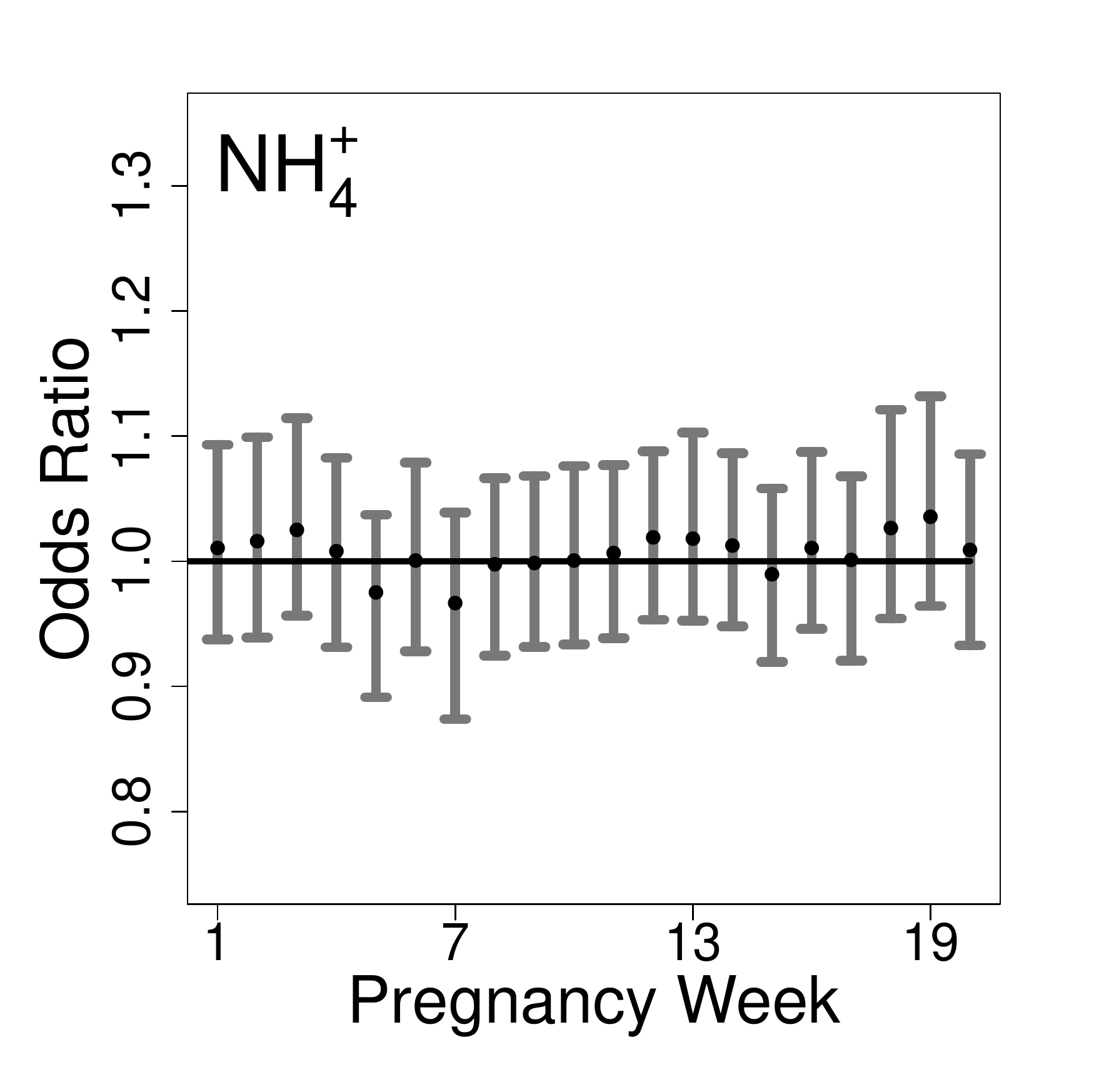}\\
\includegraphics[scale=0.26]{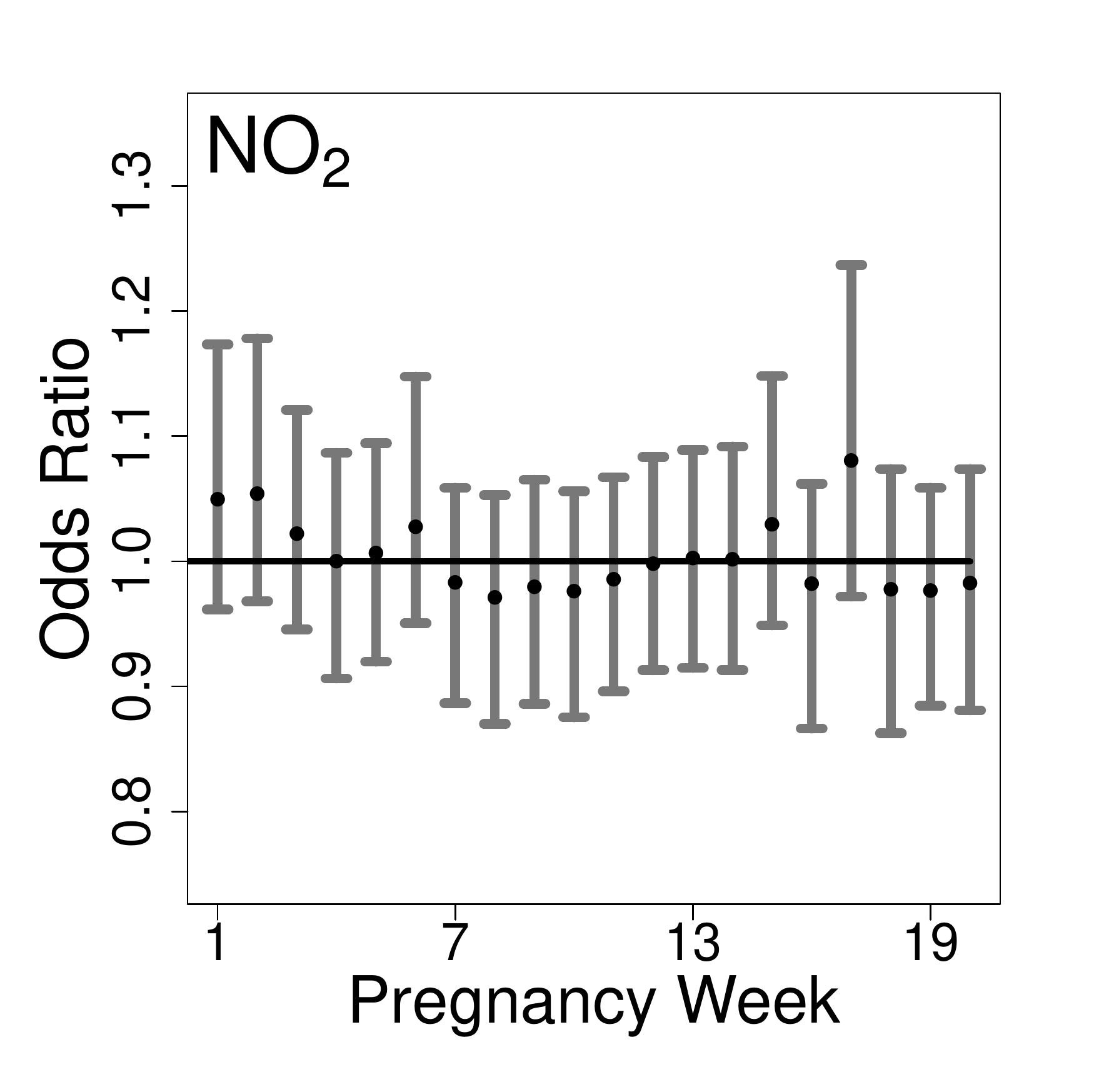}
\includegraphics[scale=0.26]{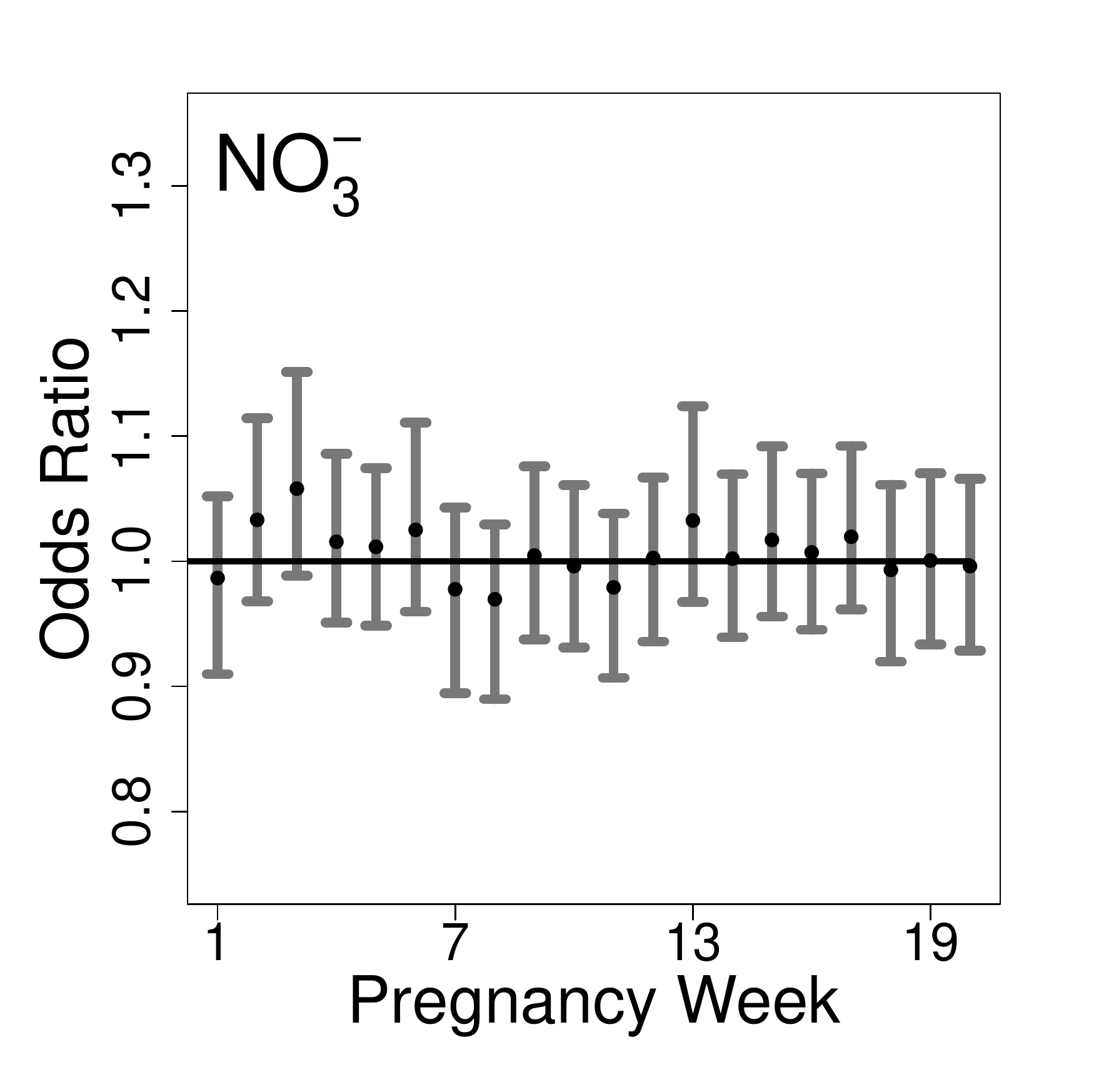}
\includegraphics[scale=0.26]{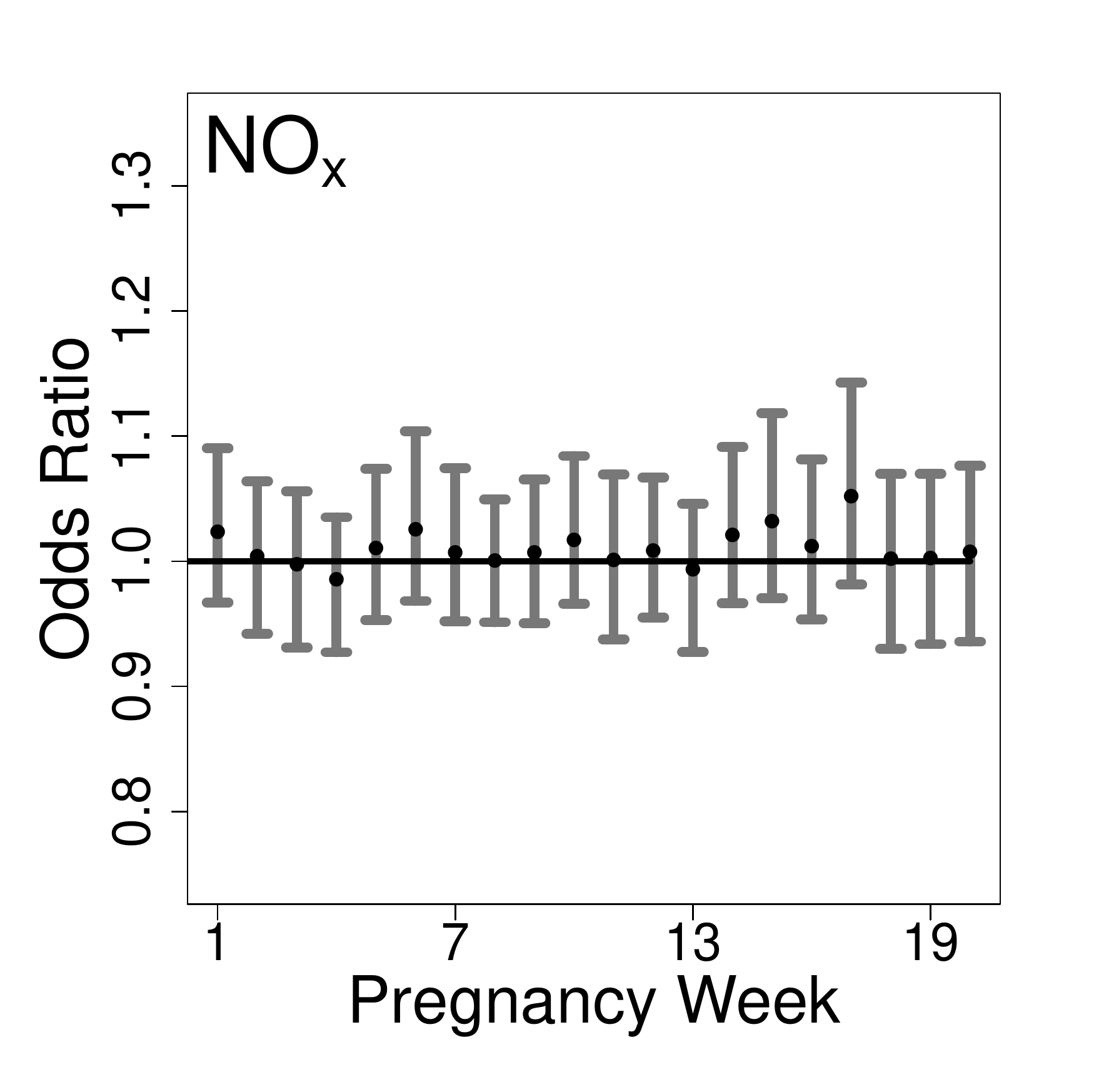}\\
\includegraphics[scale=0.26]{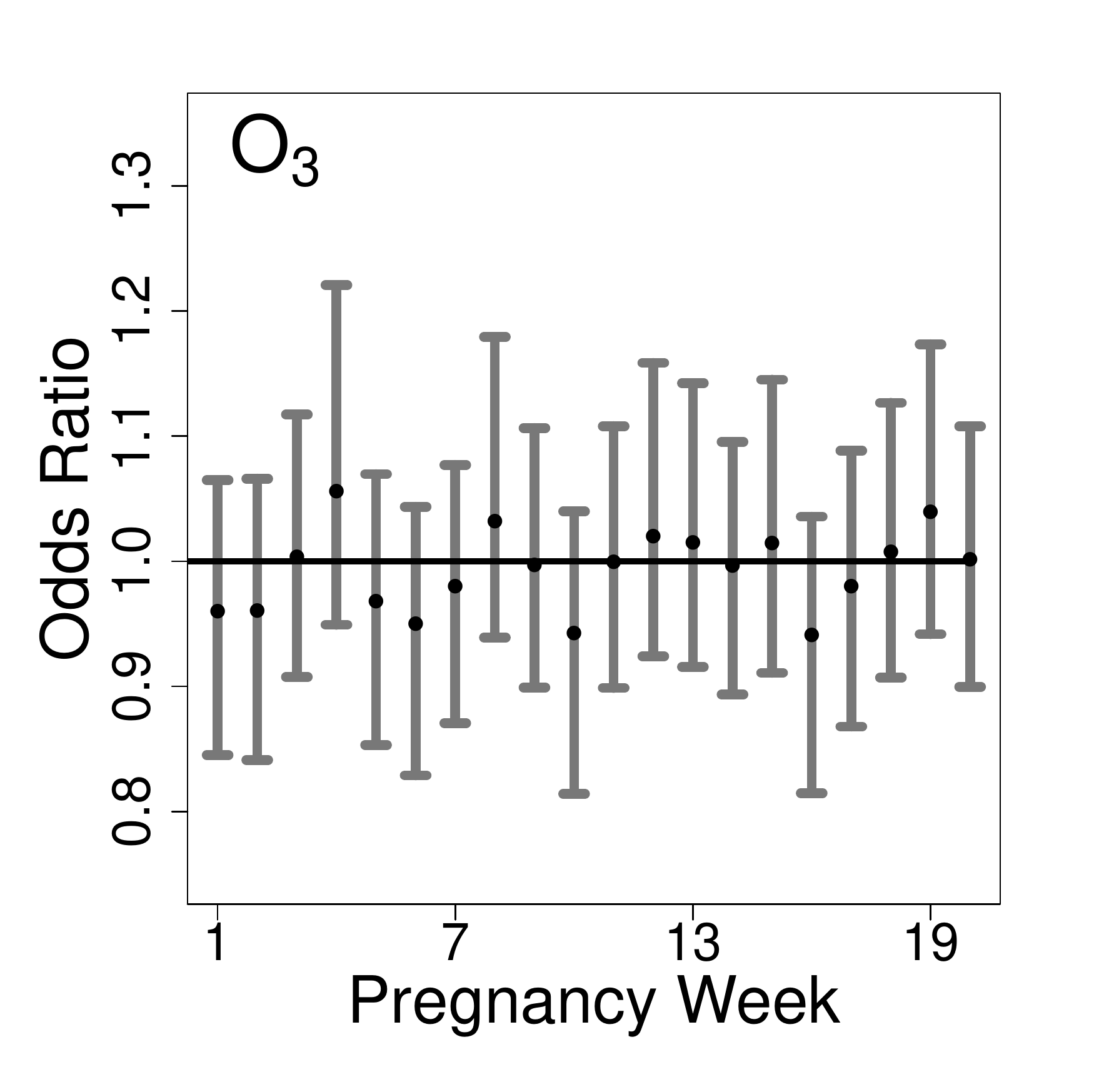}
\includegraphics[scale=0.26]{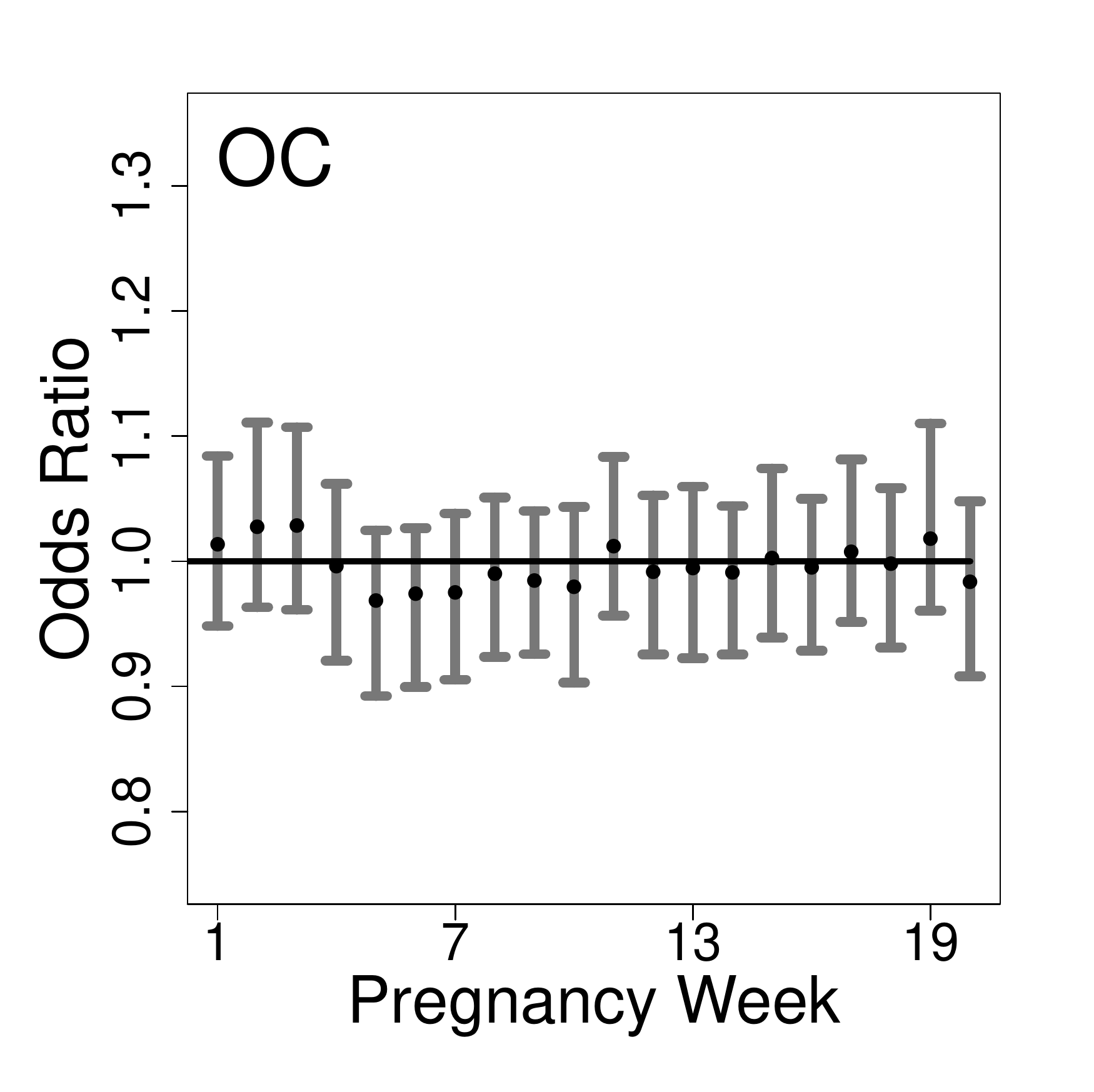}
\includegraphics[scale=0.26]{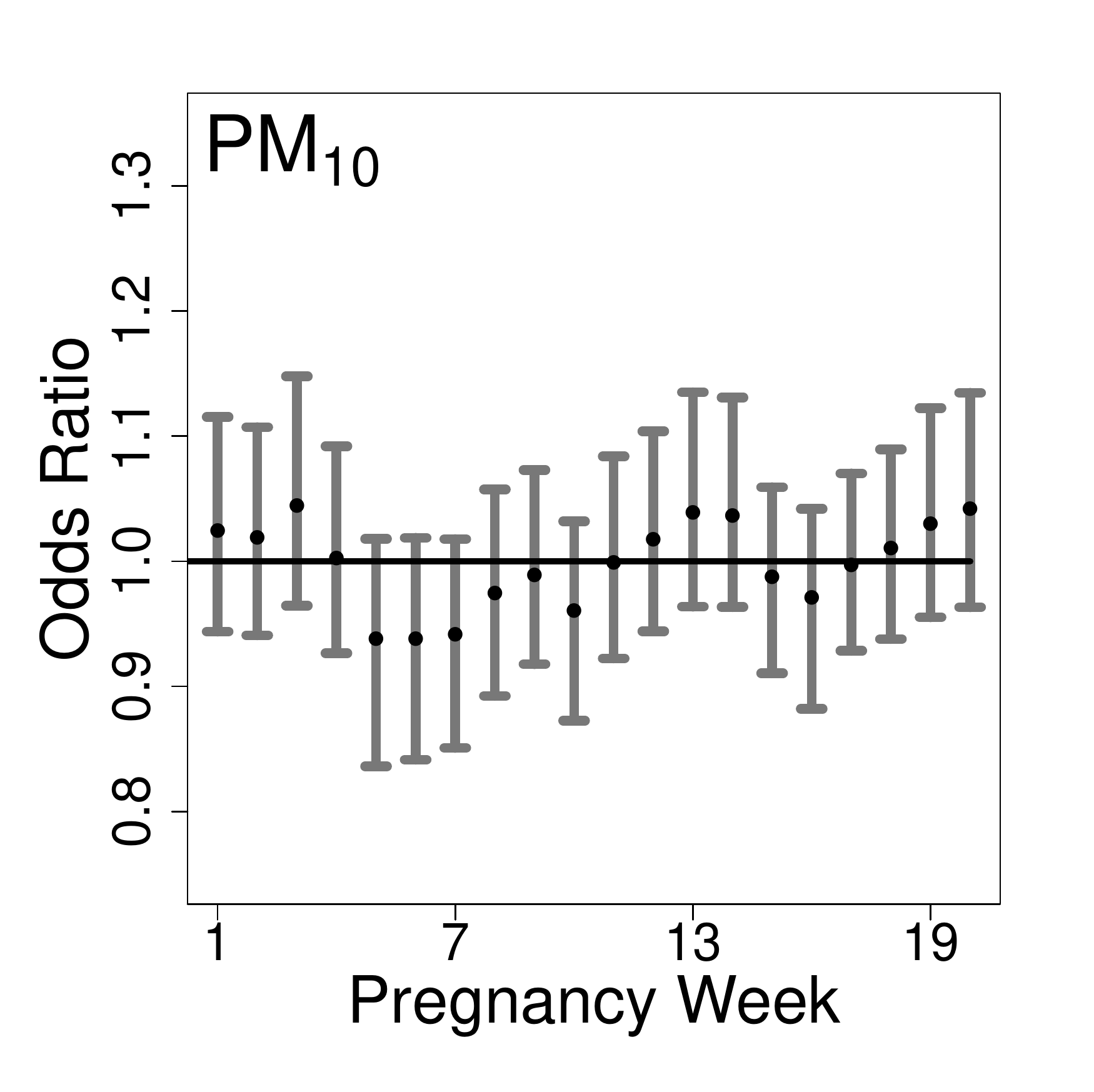}\\
\includegraphics[scale=0.26]{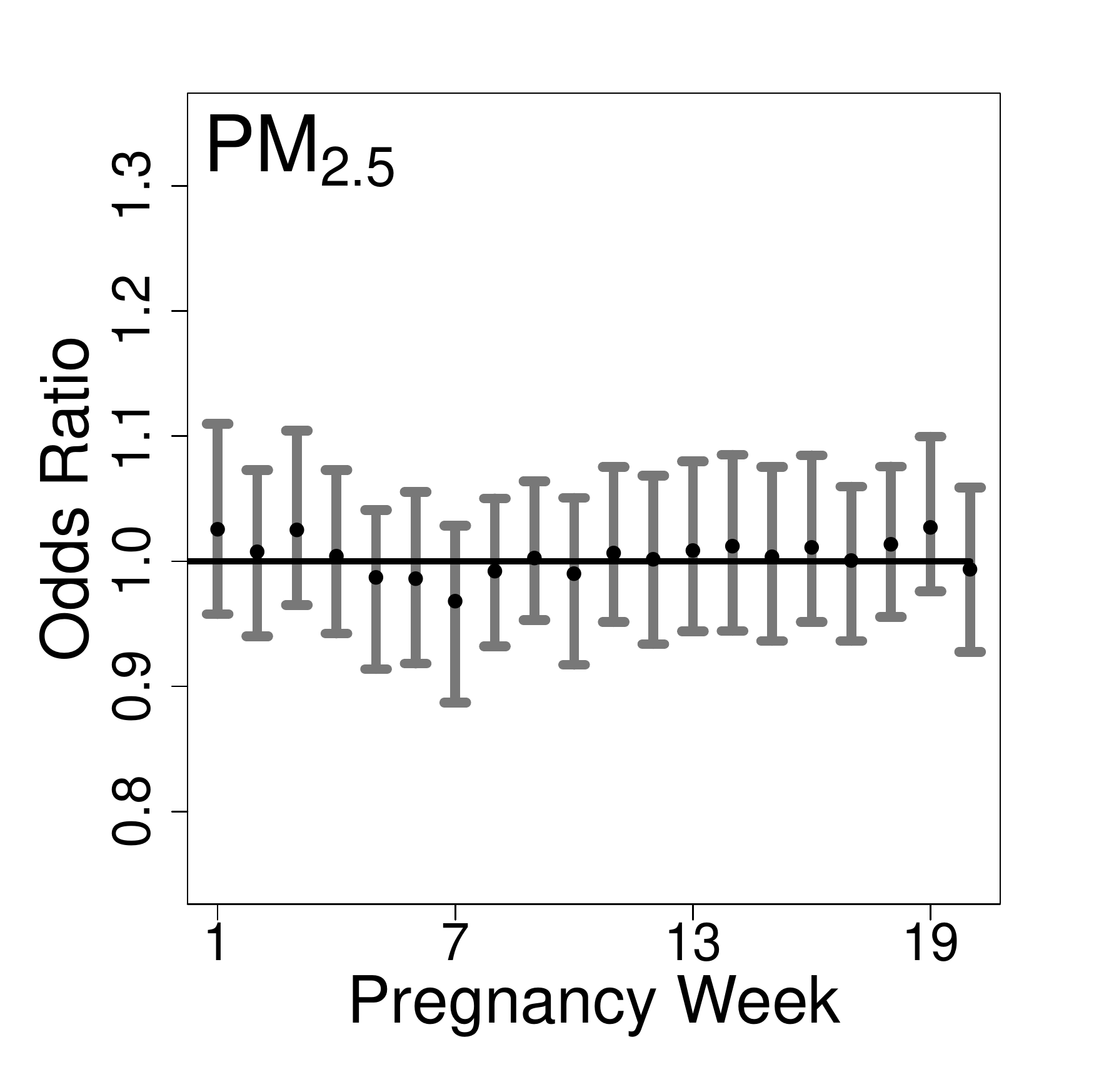}
\includegraphics[scale=0.26]{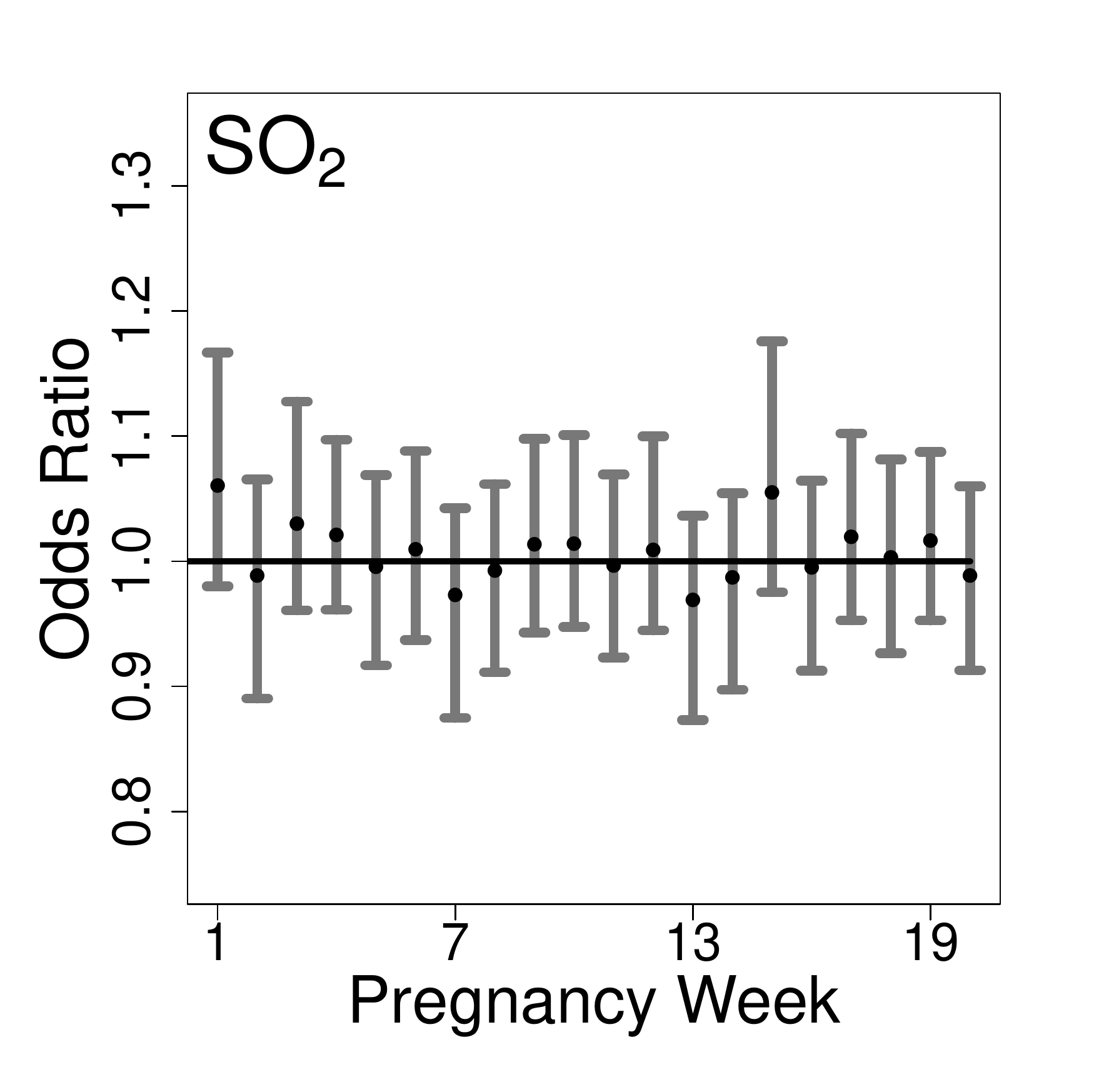}
\includegraphics[scale=0.26]{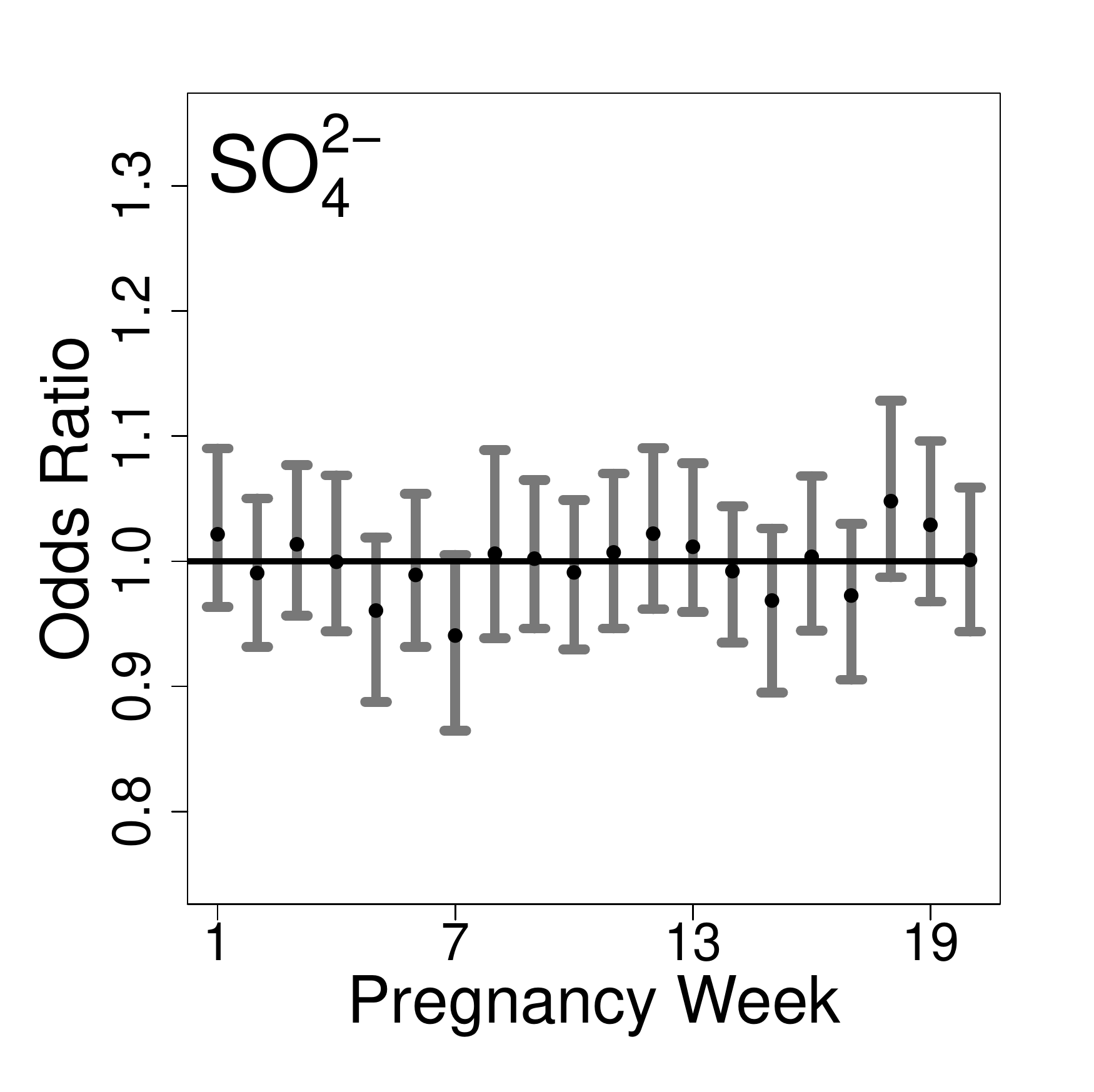}
\caption{Posterior mean and 95\% credible interval results from the \textbf{non-Hispanic White} stillbirth and single exposure Critical Window Variable Selection (CWVS) analyses in New Jersey, 2005-2014. Results based on an interquartile range increase in weekly exposure. Weeks identified as part of the critical window set are shown in red/dashed (harmful) and blue/dashed (protective).  These definitions depend partly on the posterior inclusion probabilities in Figure S9 of the Supporting Information for the variable selection methods.}
\end{center}
\end{figure}
\clearpage

\begin{figure}[h]
\begin{center}
\includegraphics[scale=0.26]{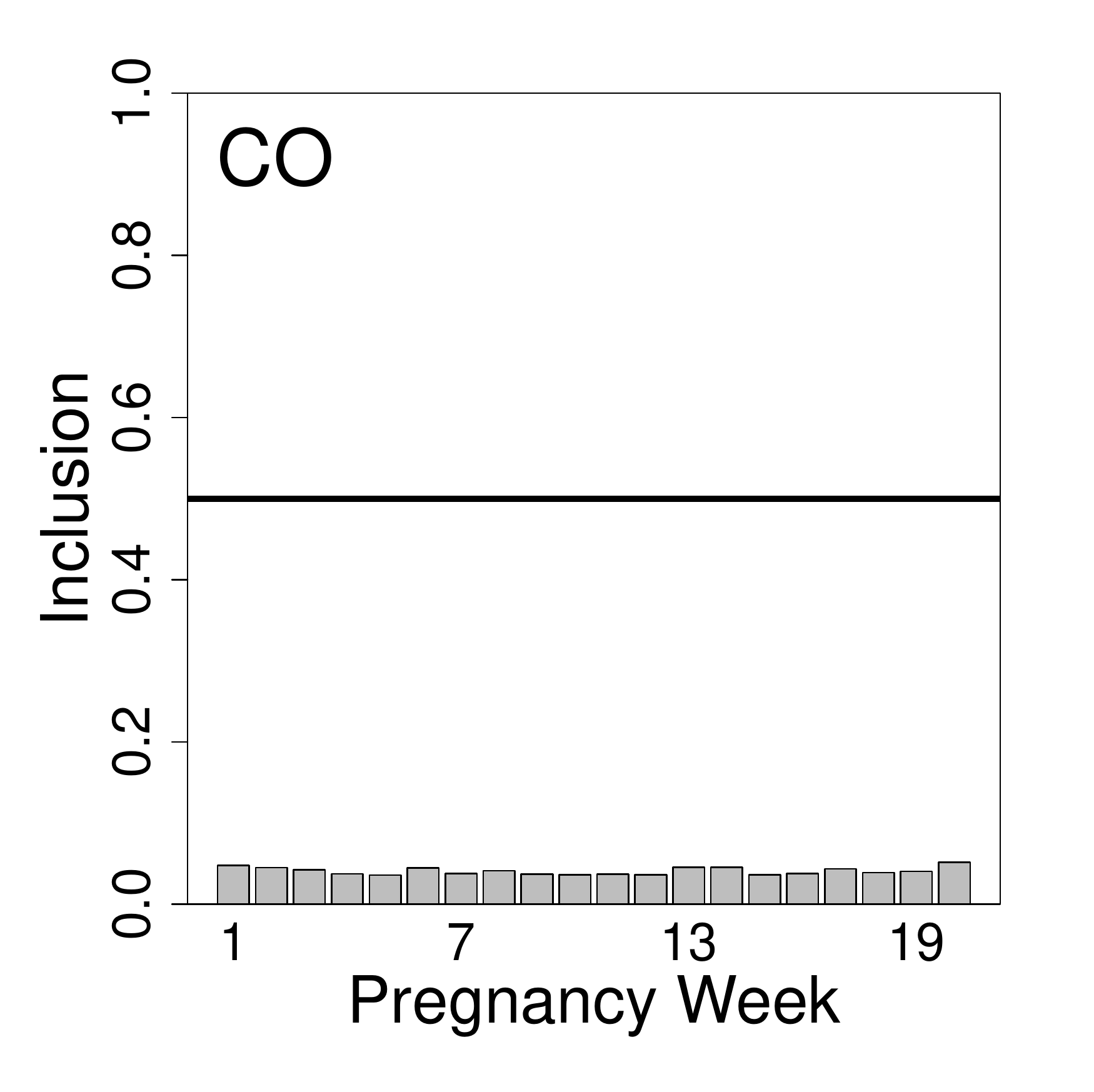}
\includegraphics[scale=0.26]{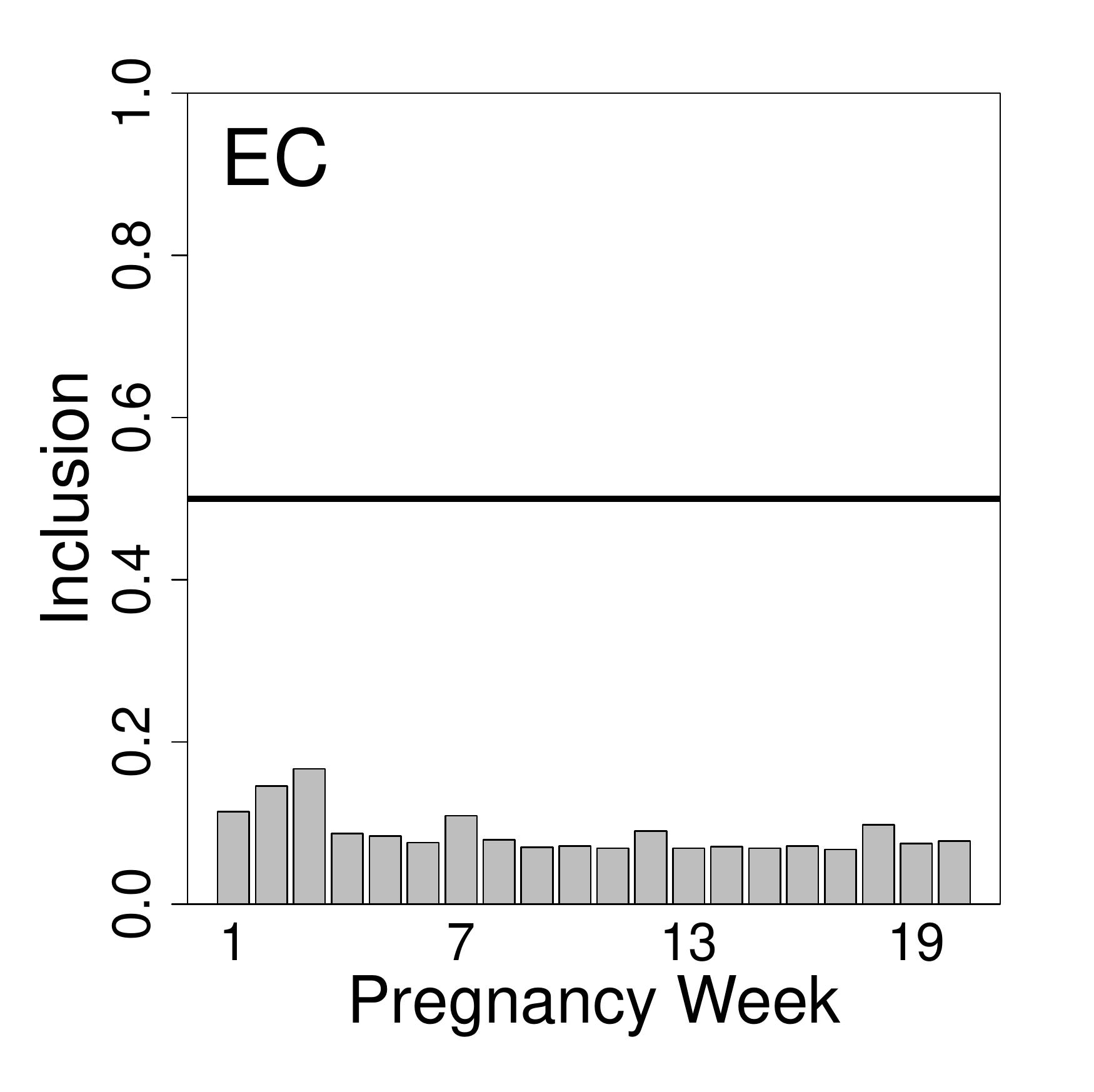}
\includegraphics[scale=0.26]{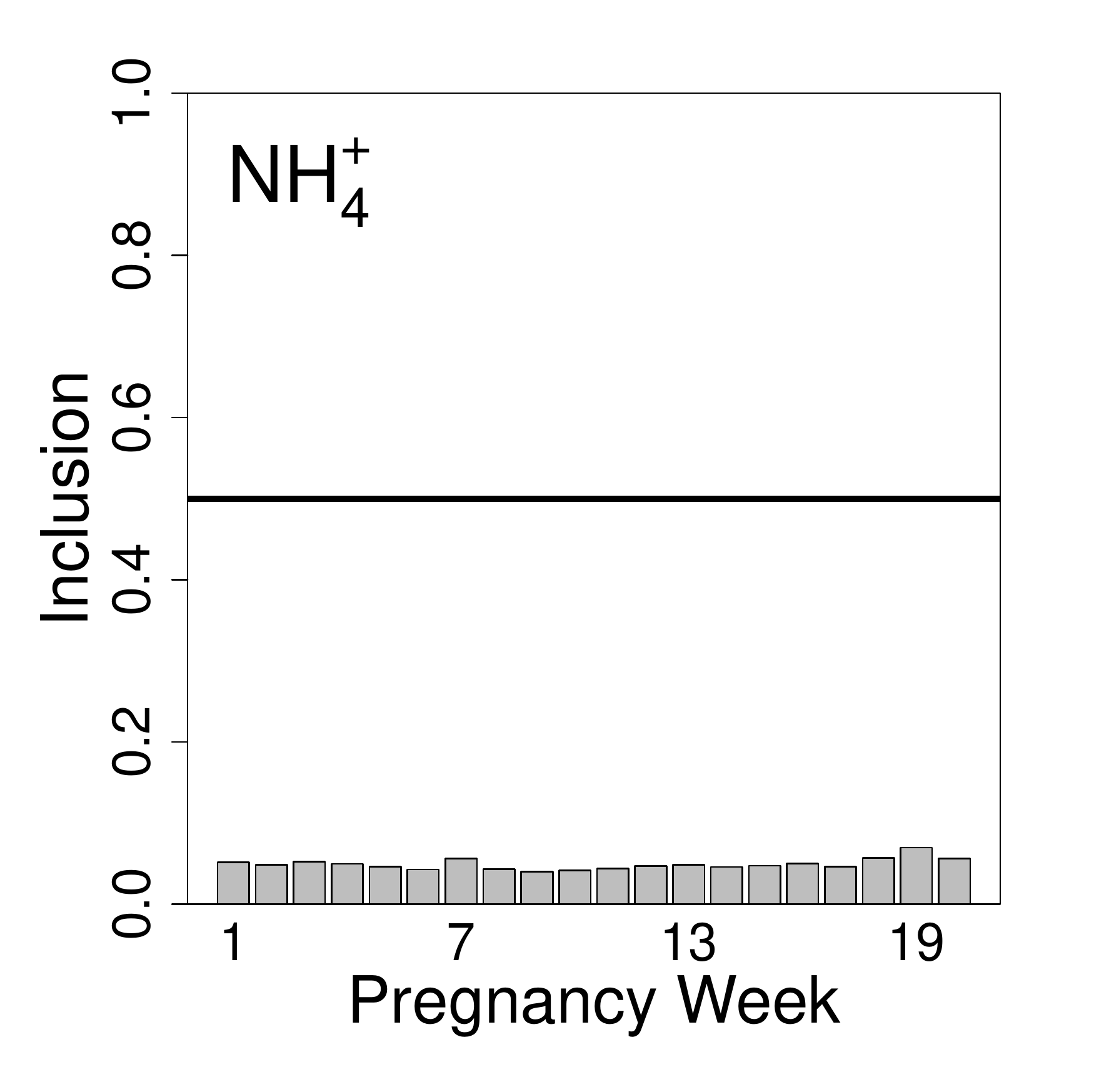}\\
\includegraphics[scale=0.26]{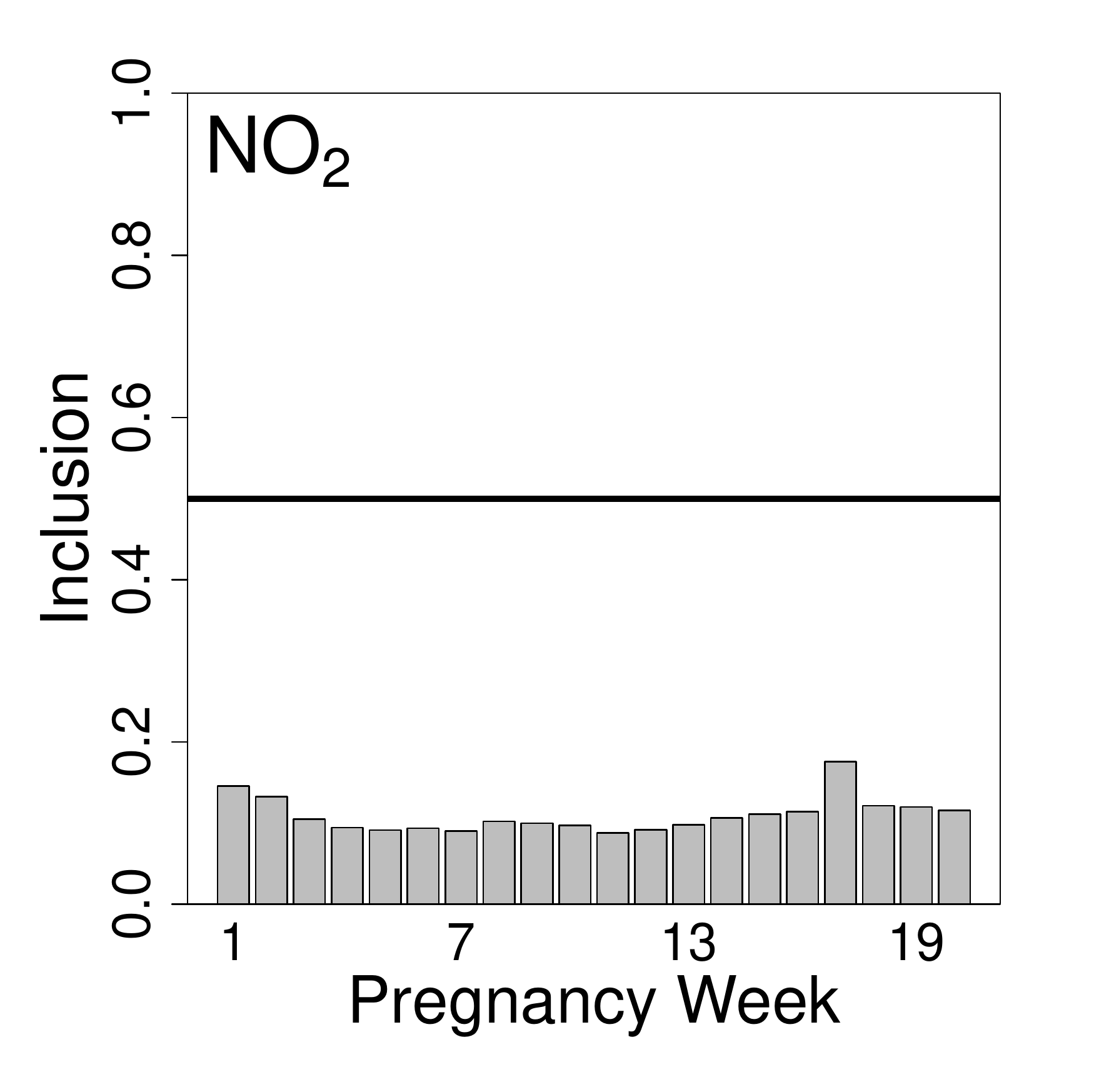}
\includegraphics[scale=0.26]{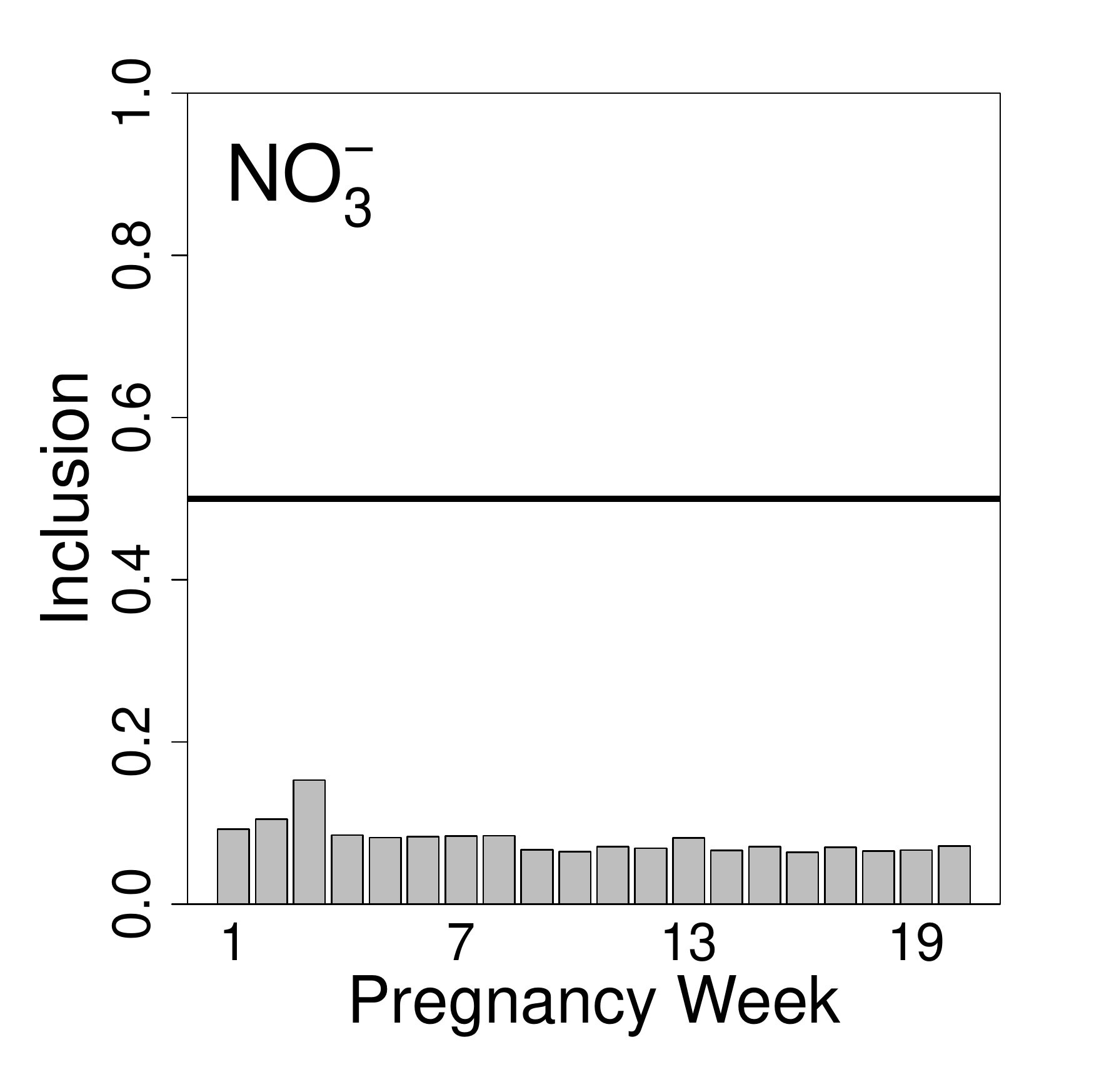}
\includegraphics[scale=0.26]{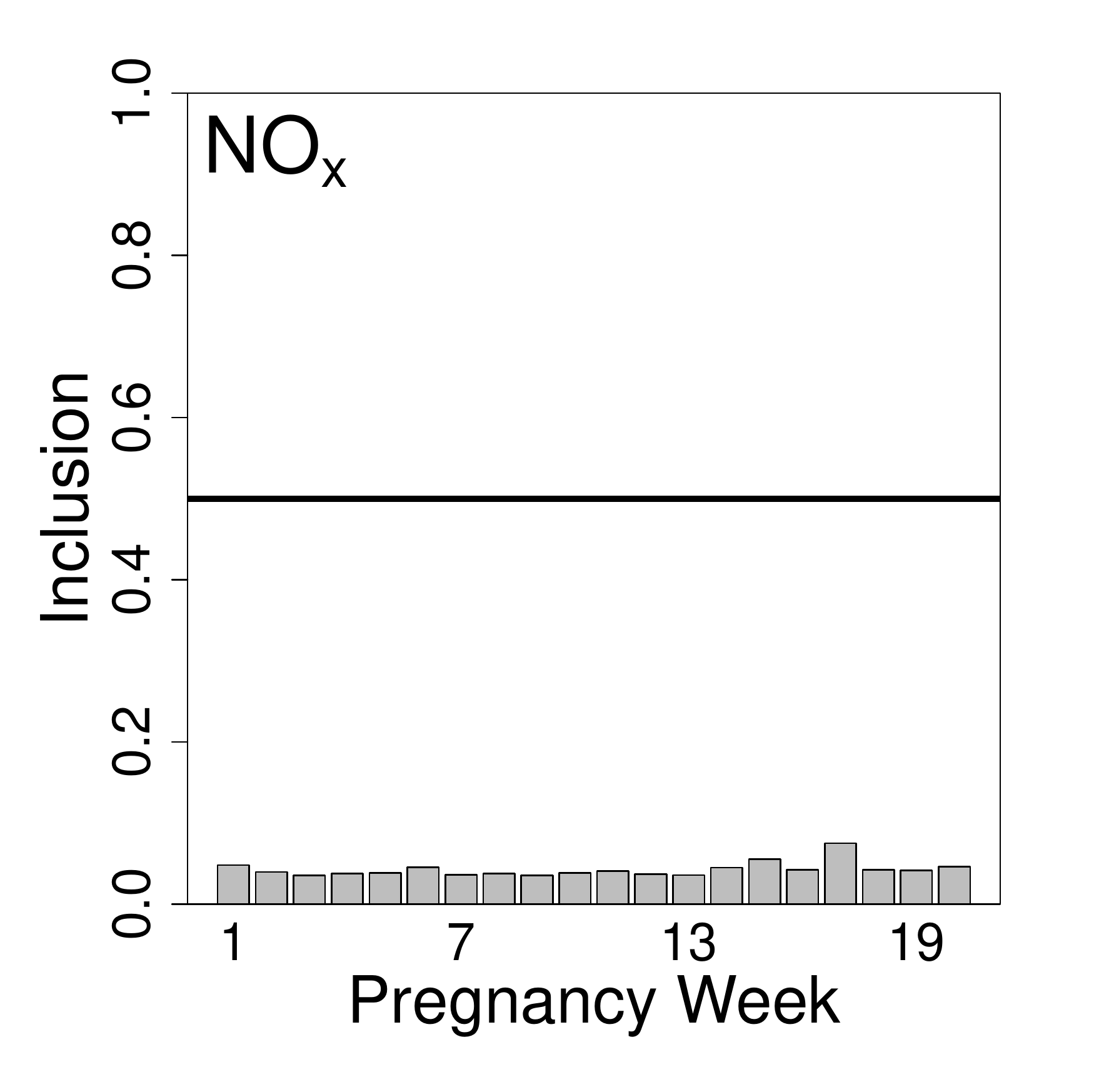}\\
\includegraphics[scale=0.26]{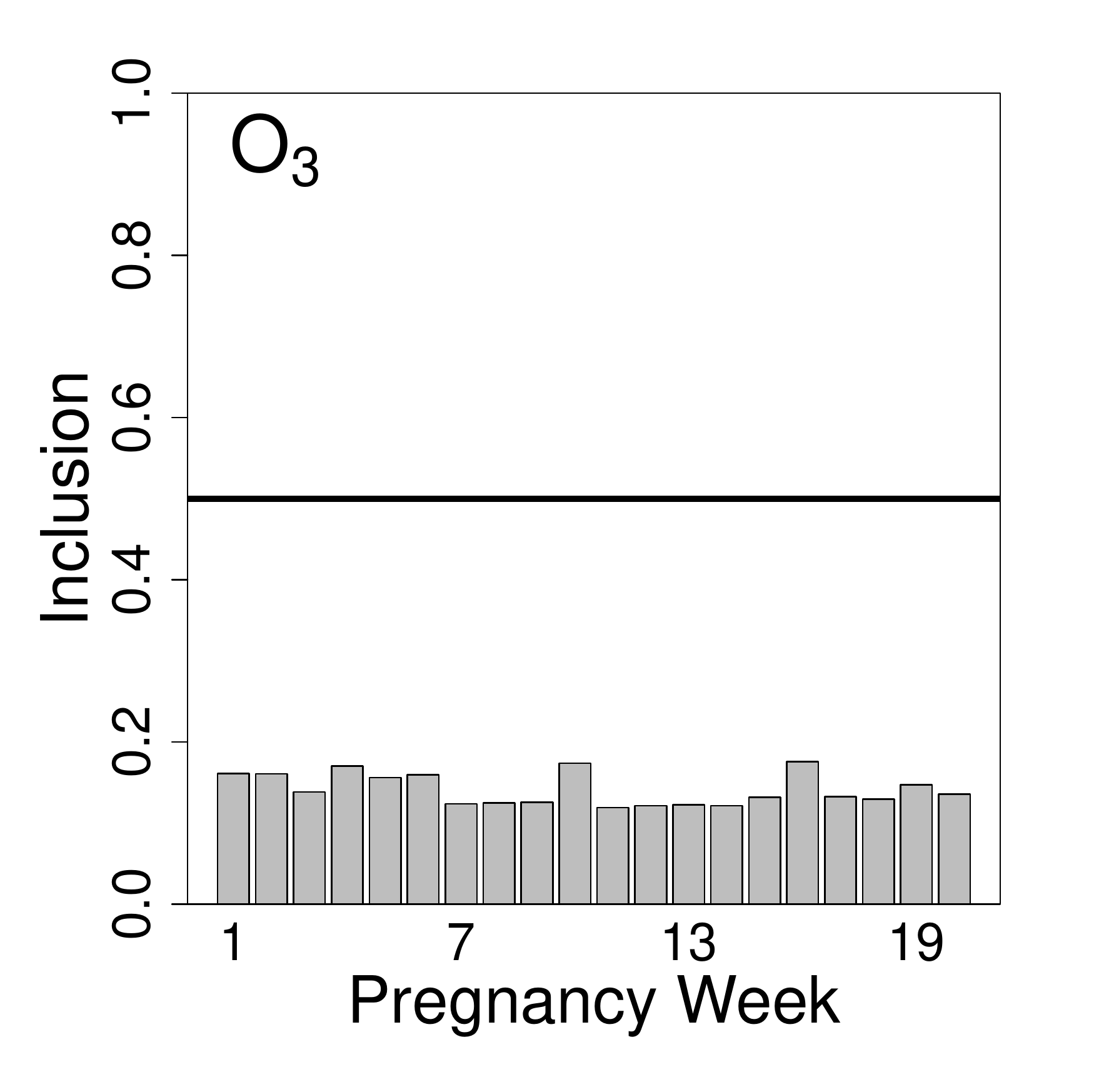}
\includegraphics[scale=0.26]{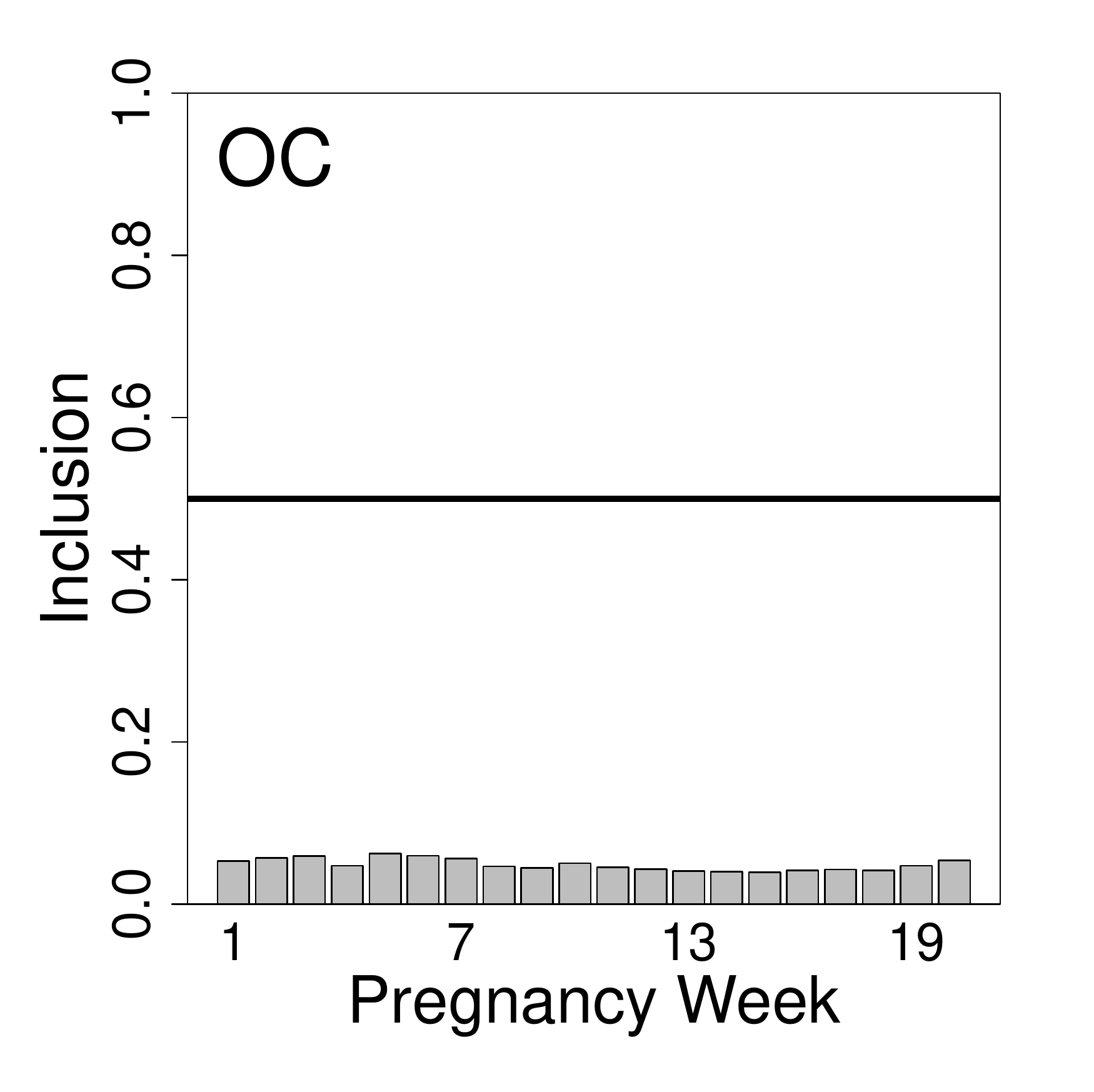}
\includegraphics[scale=0.26]{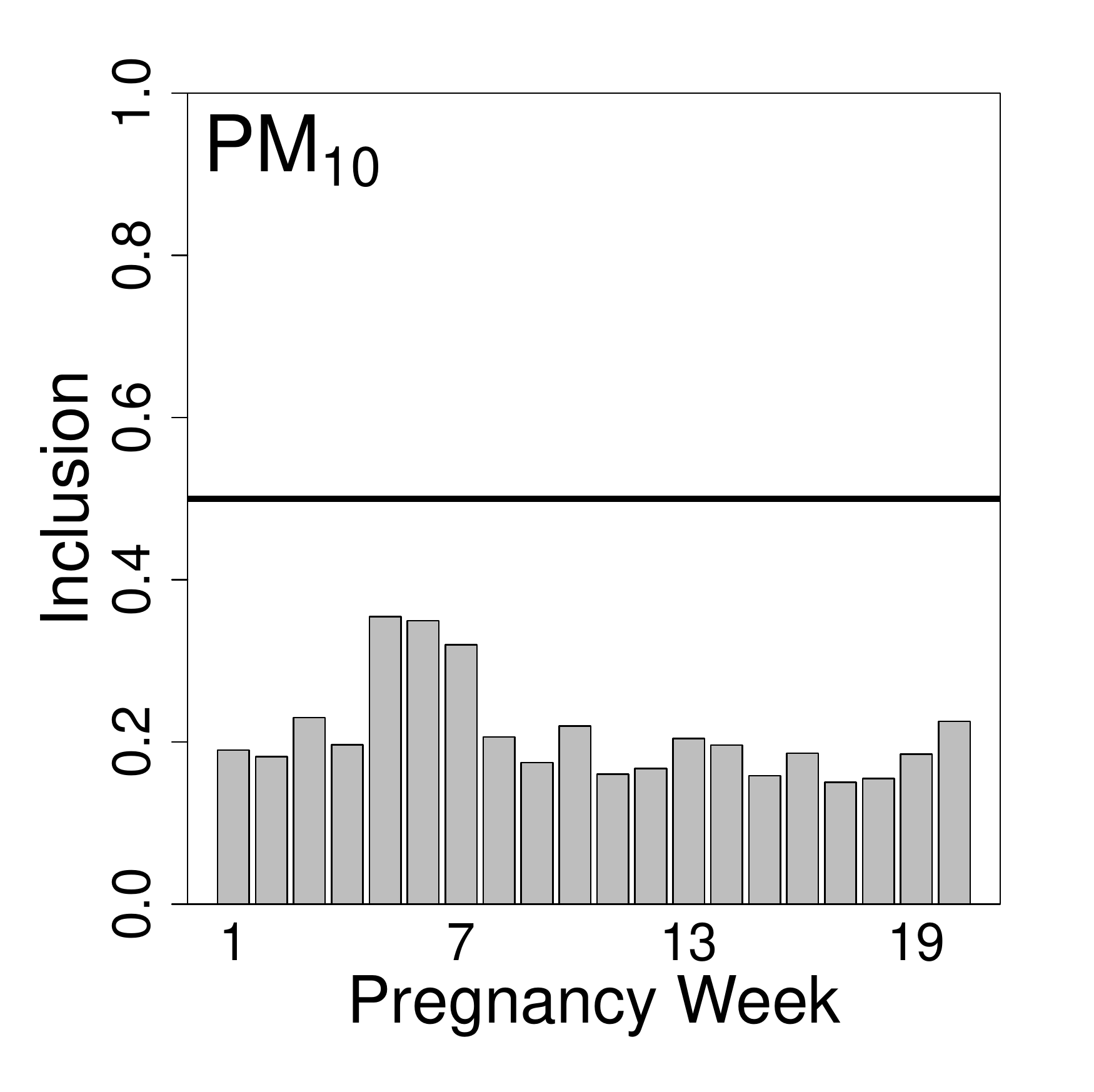}\\
\includegraphics[scale=0.26]{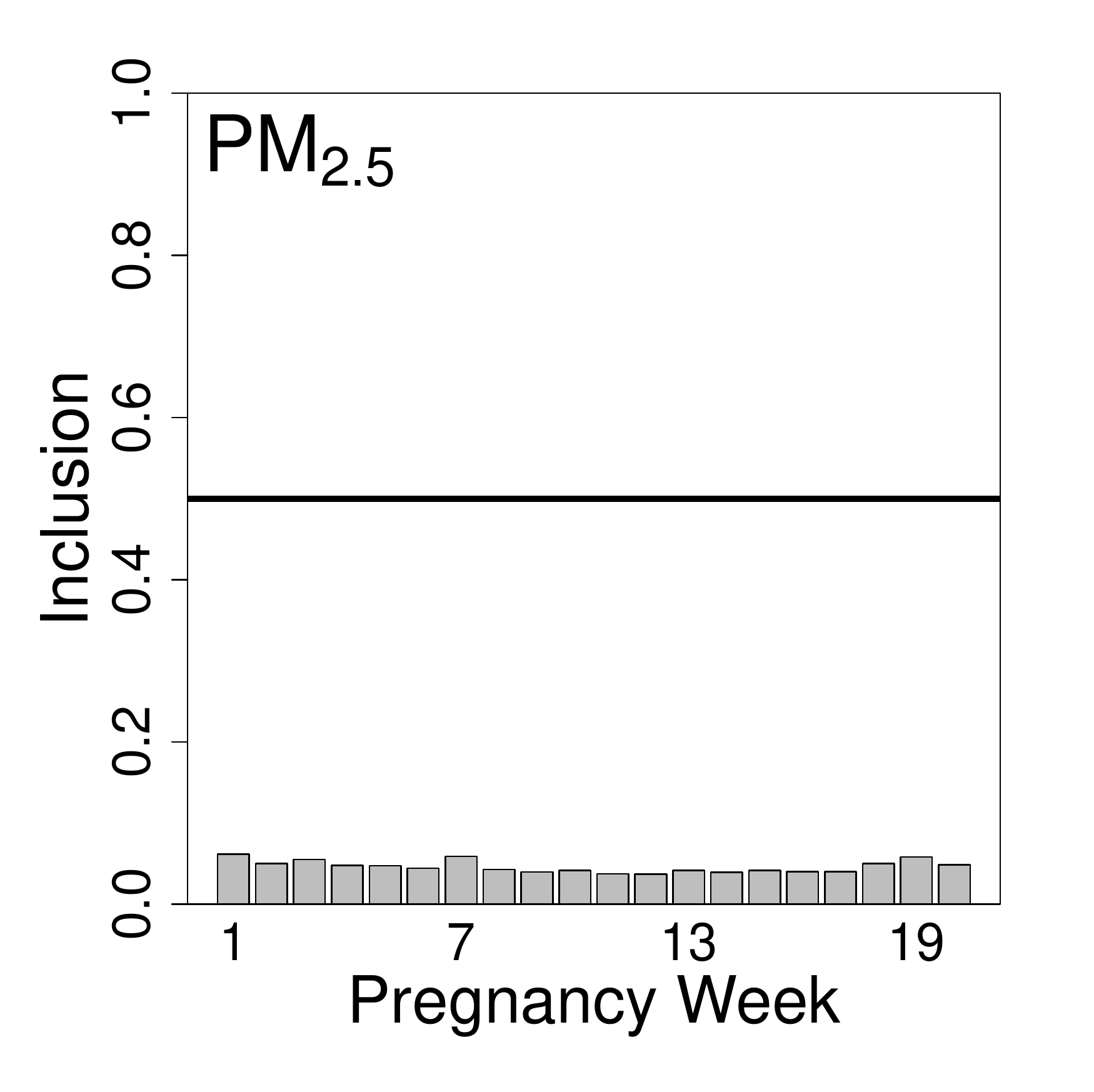}
\includegraphics[scale=0.26]{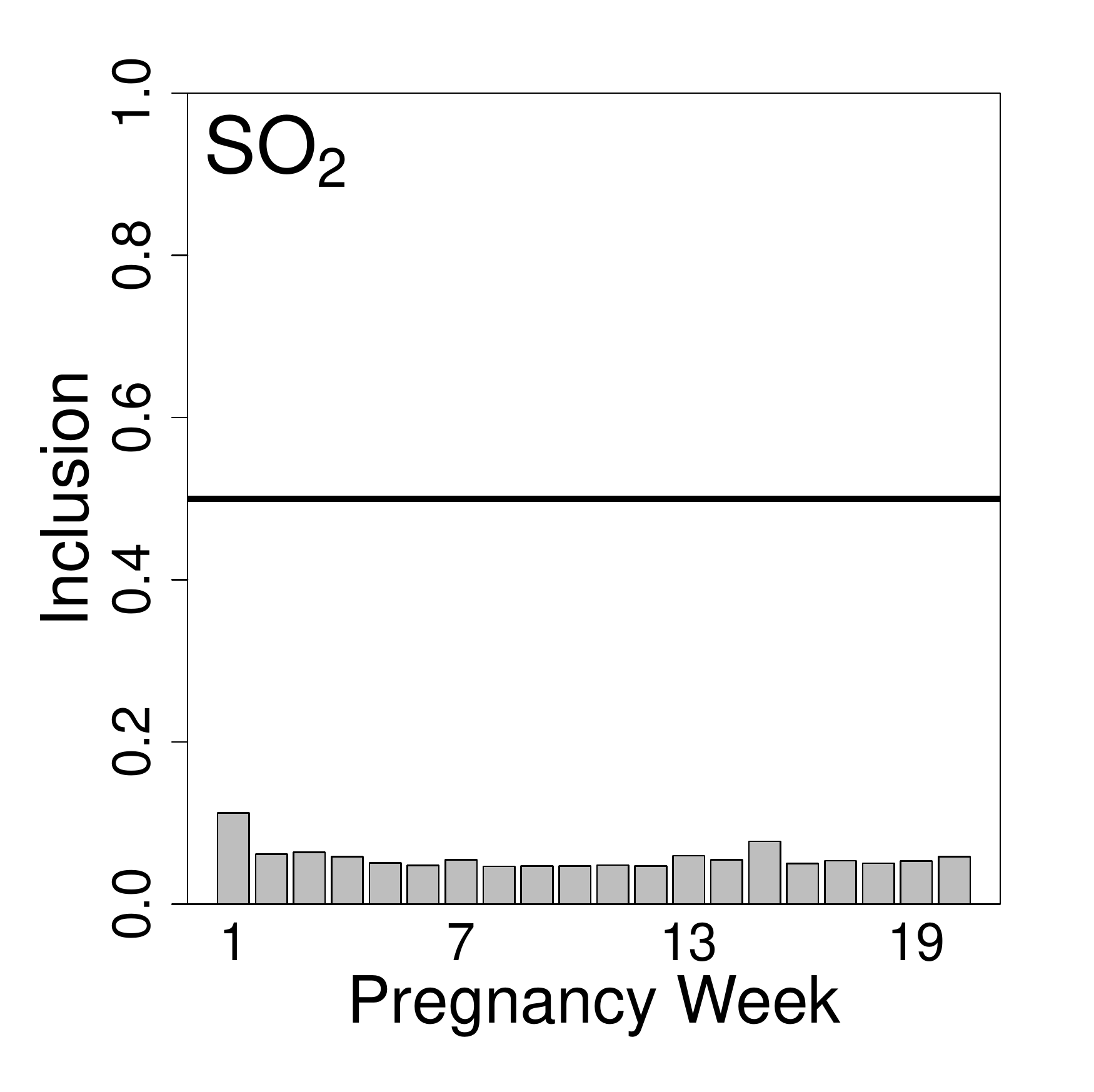}
\includegraphics[scale=0.26]{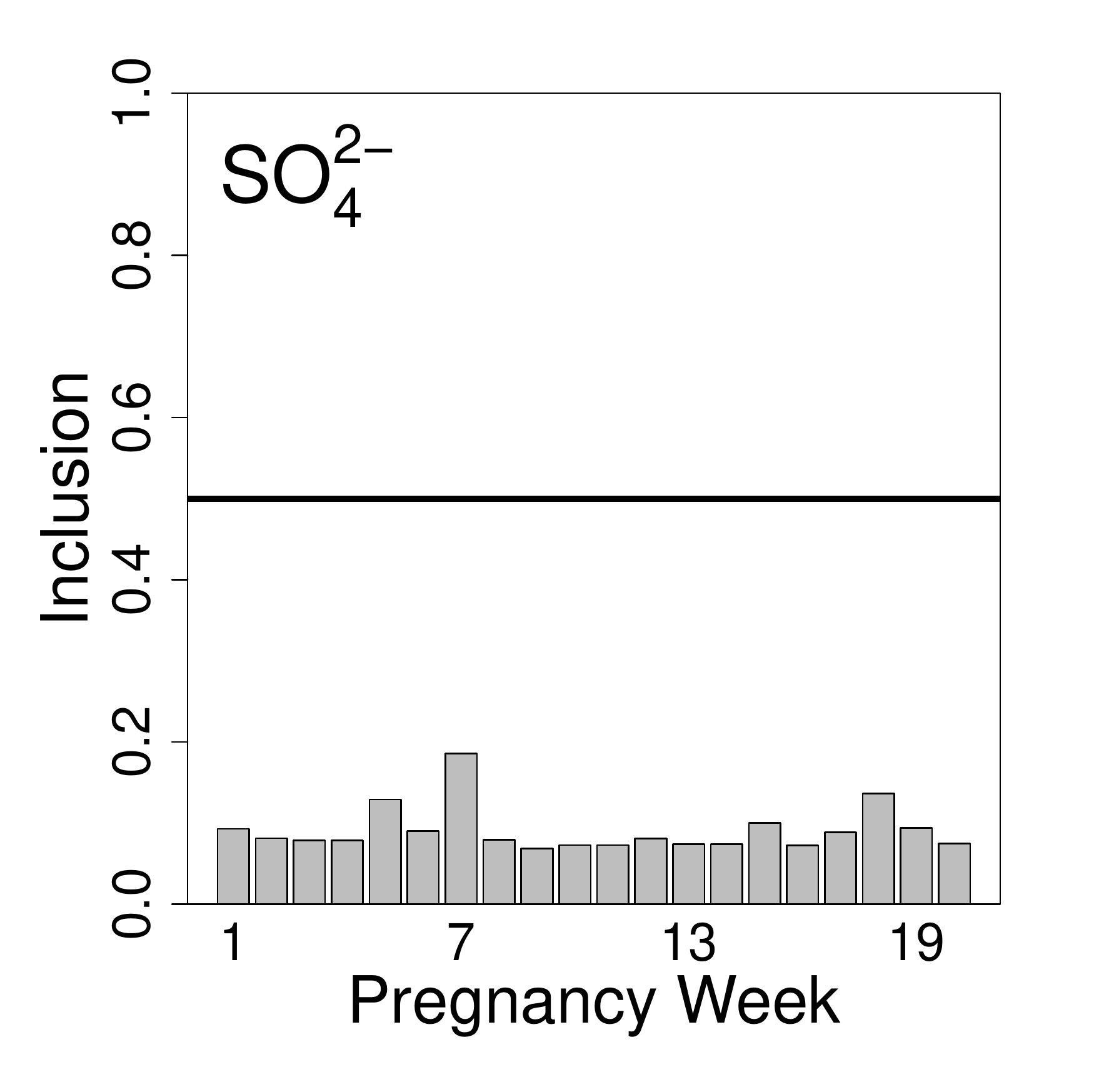}
\caption{Posterior inclusion probability results from the \textbf{non-Hispanic White} stillbirth and single exposure Critical Window Variable Selection (CWVS) analyses in New Jersey, 2005-2014.}
\end{center}
\end{figure}
\clearpage

\begin{figure}
\begin{center}
\includegraphics[trim={0.5cm 0.5cm 1cm 0.5cm}, clip, scale=0.21]{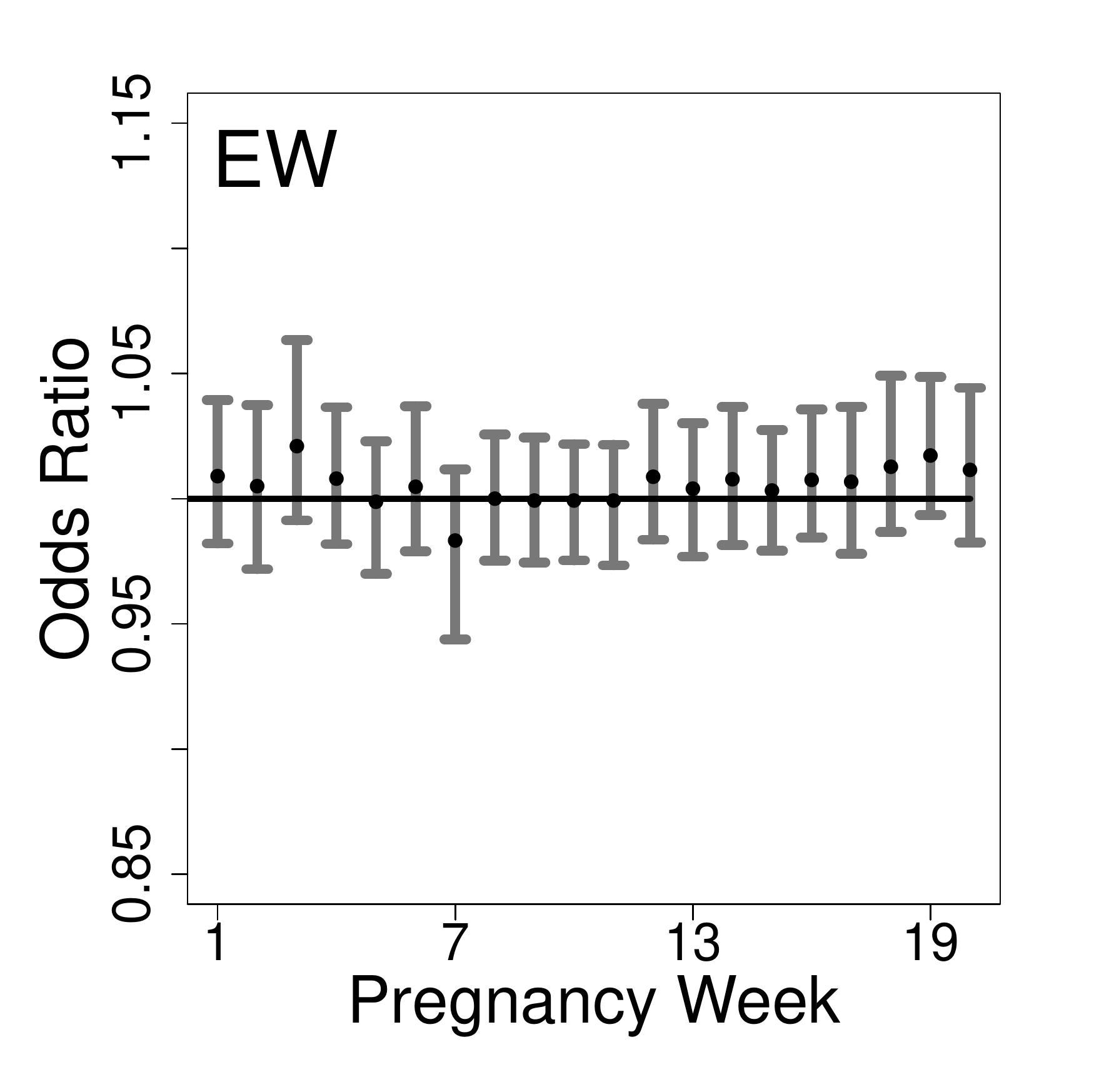}
\includegraphics[trim={0.5cm 0.5cm 1cm 0.5cm}, clip, scale=0.21]{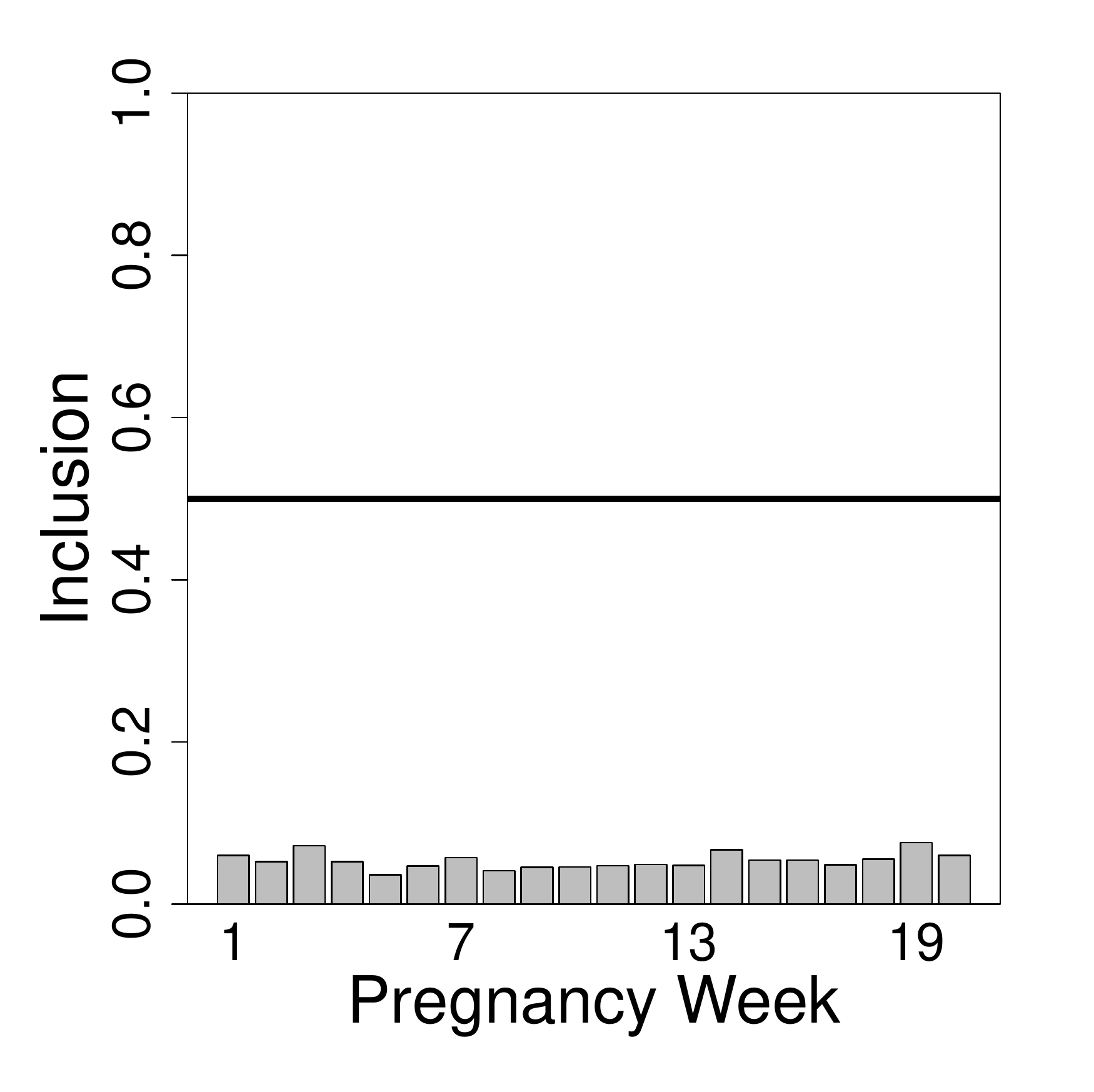}
\includegraphics[trim={0.25cm 0.5cm 1cm 0.5cm}, clip, scale=0.21]{Figures/color_key.pdf}
\includegraphics[trim={0cm 0.5cm 1cm 0.5cm}, clip, scale=0.21]{Figures/color_key-interactions.pdf}\\
\includegraphics[trim={0.5cm 0.5cm 1cm 0.5cm}, clip, scale=0.21]{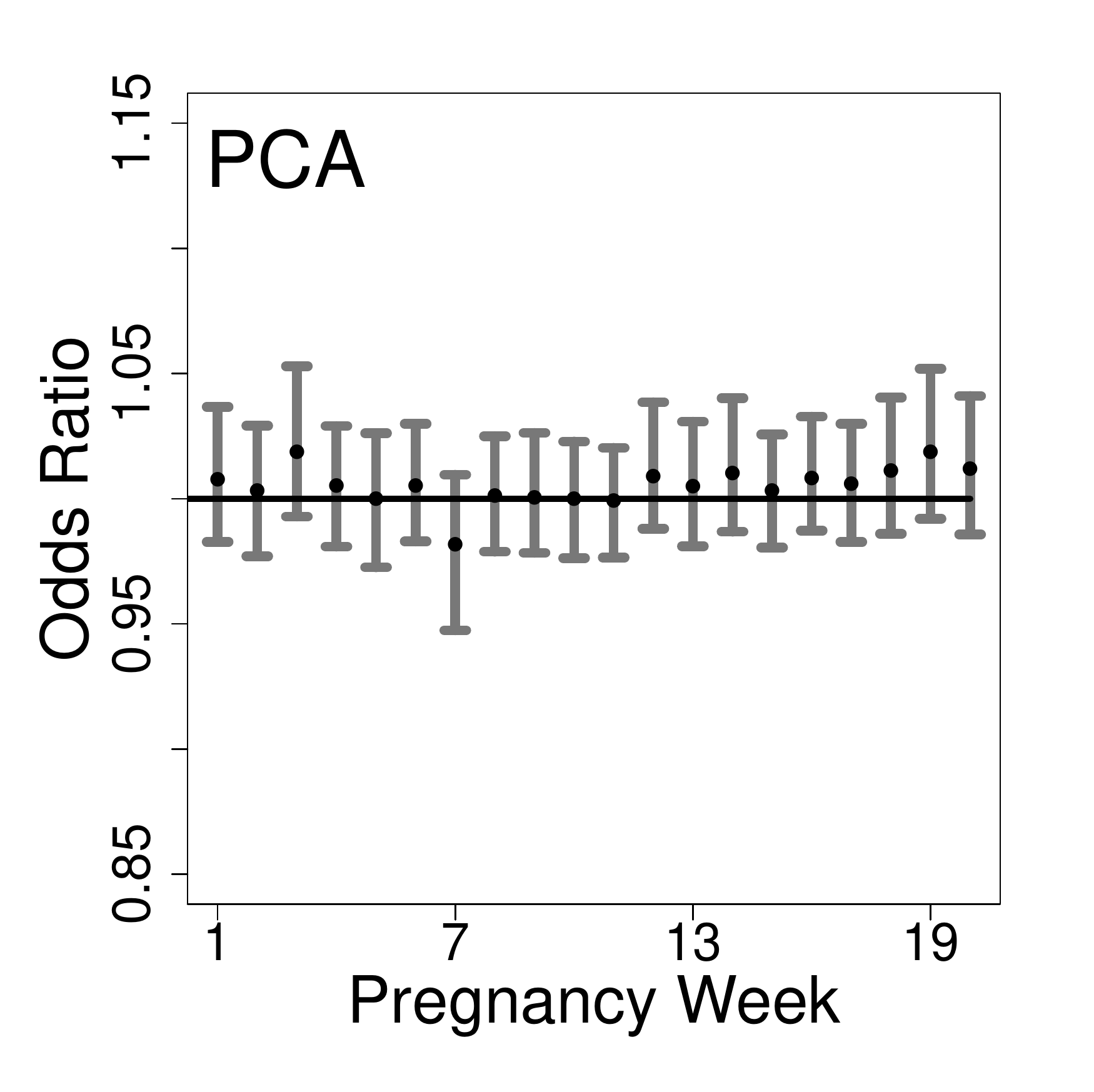}
\includegraphics[trim={0.5cm 0.5cm 1cm 0.5cm}, clip, scale=0.21]{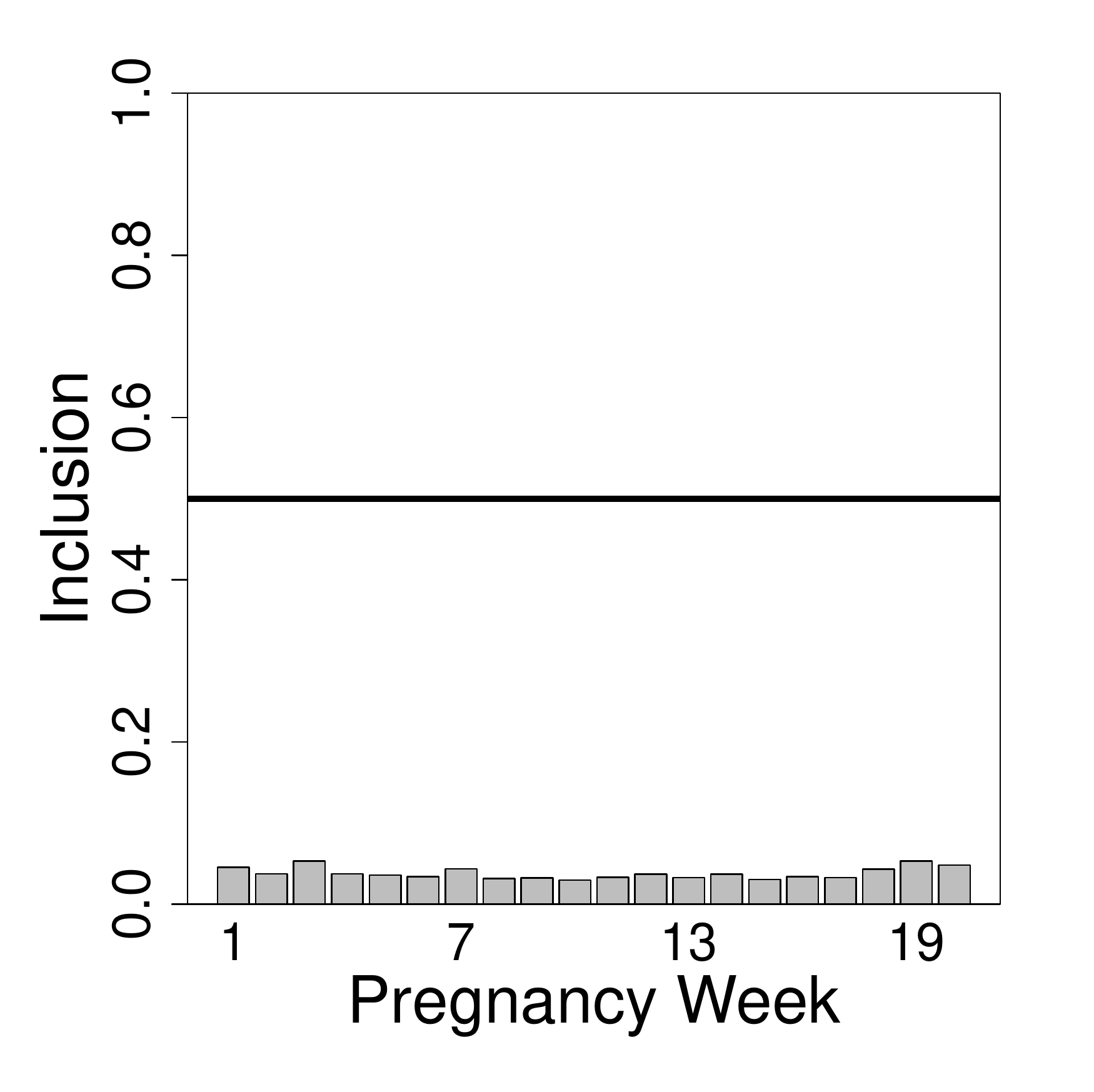}
\includegraphics[trim={0.25cm 0.5cm 1cm 0.5cm}, clip, scale=0.21]{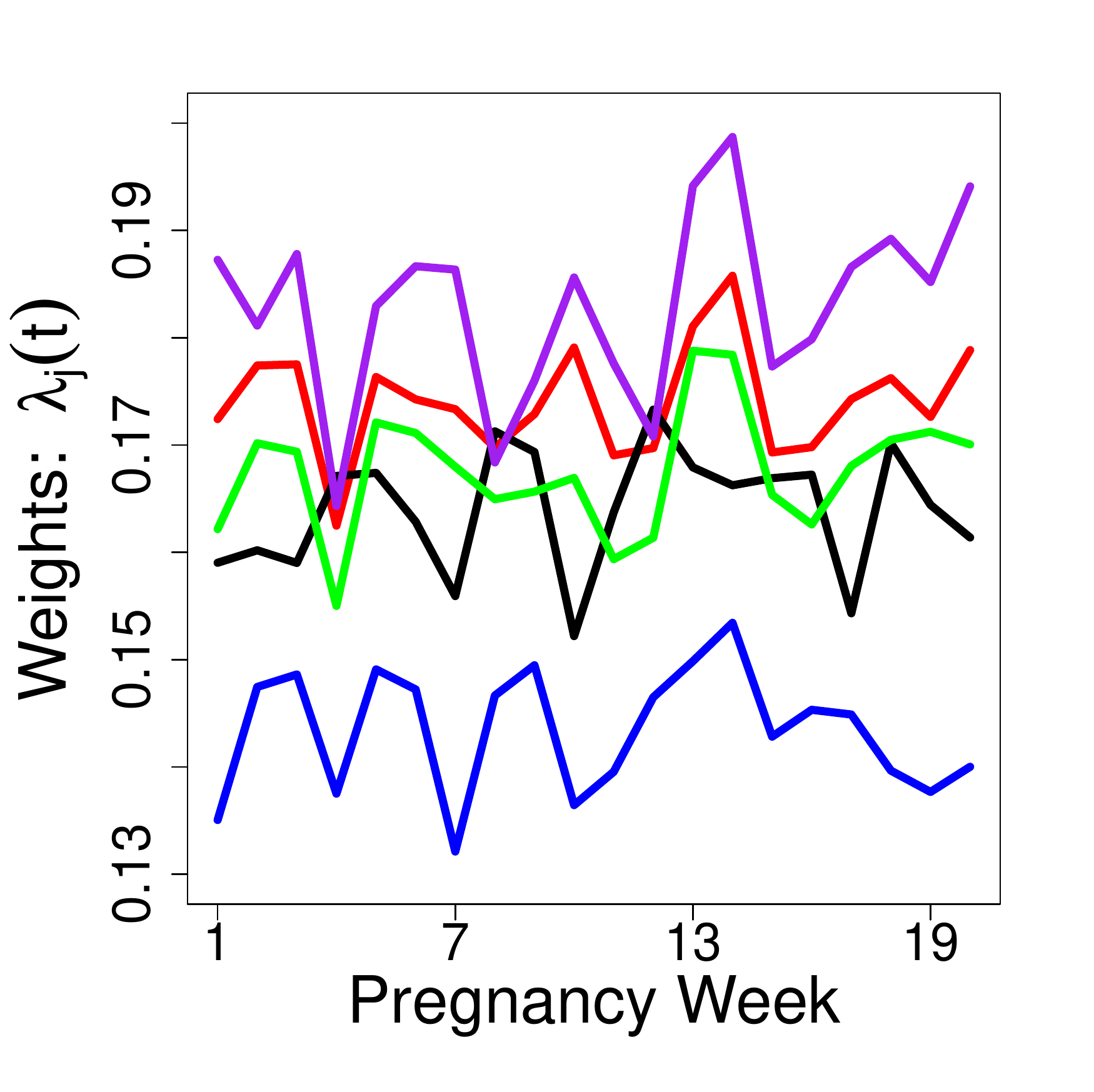}
\includegraphics[trim={0cm 0.5cm 1cm 0.5cm}, clip, scale=0.21]{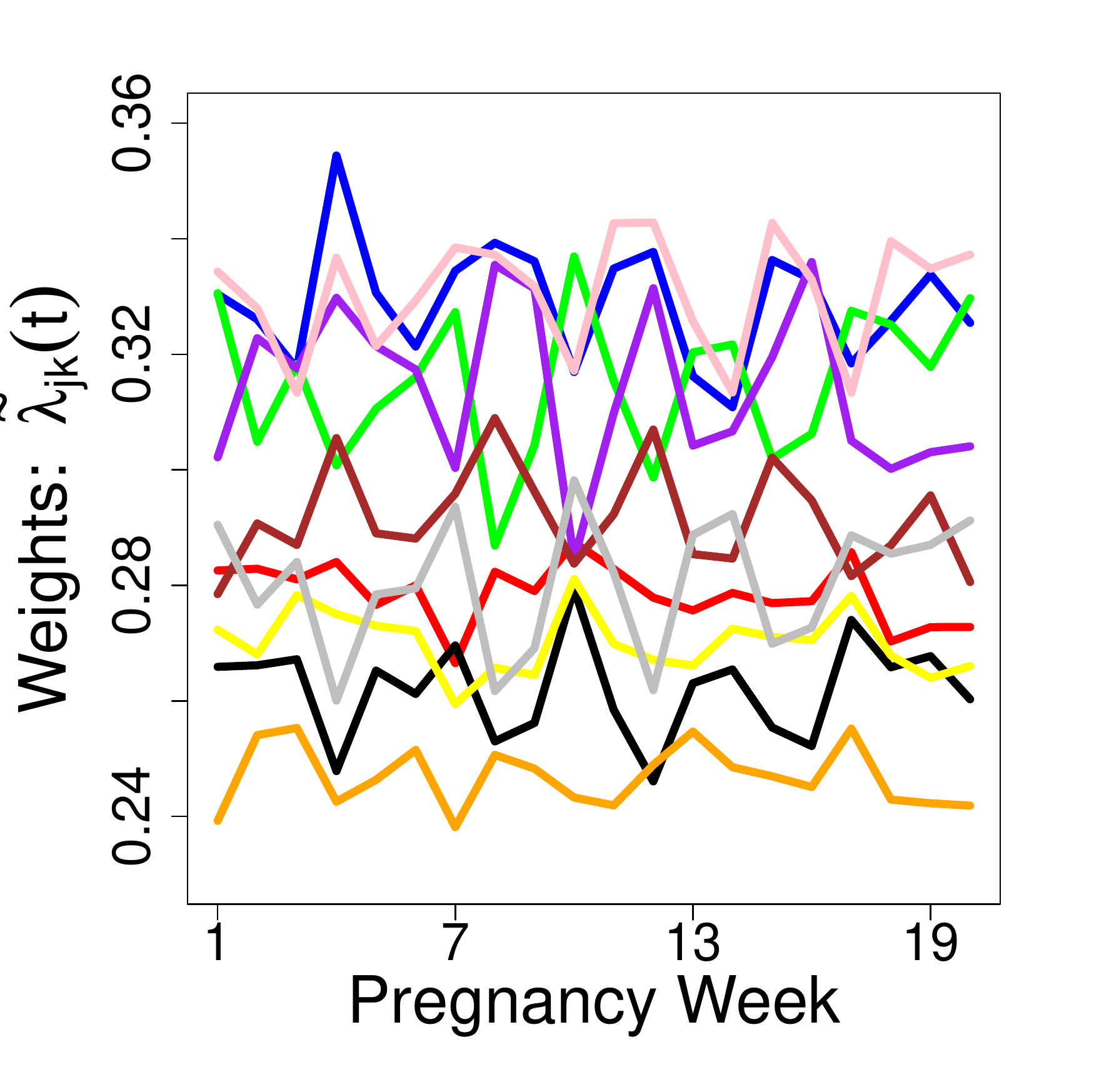}\\
\includegraphics[trim={0.5cm 0.5cm 1cm 0.5cm}, clip, scale=0.21]{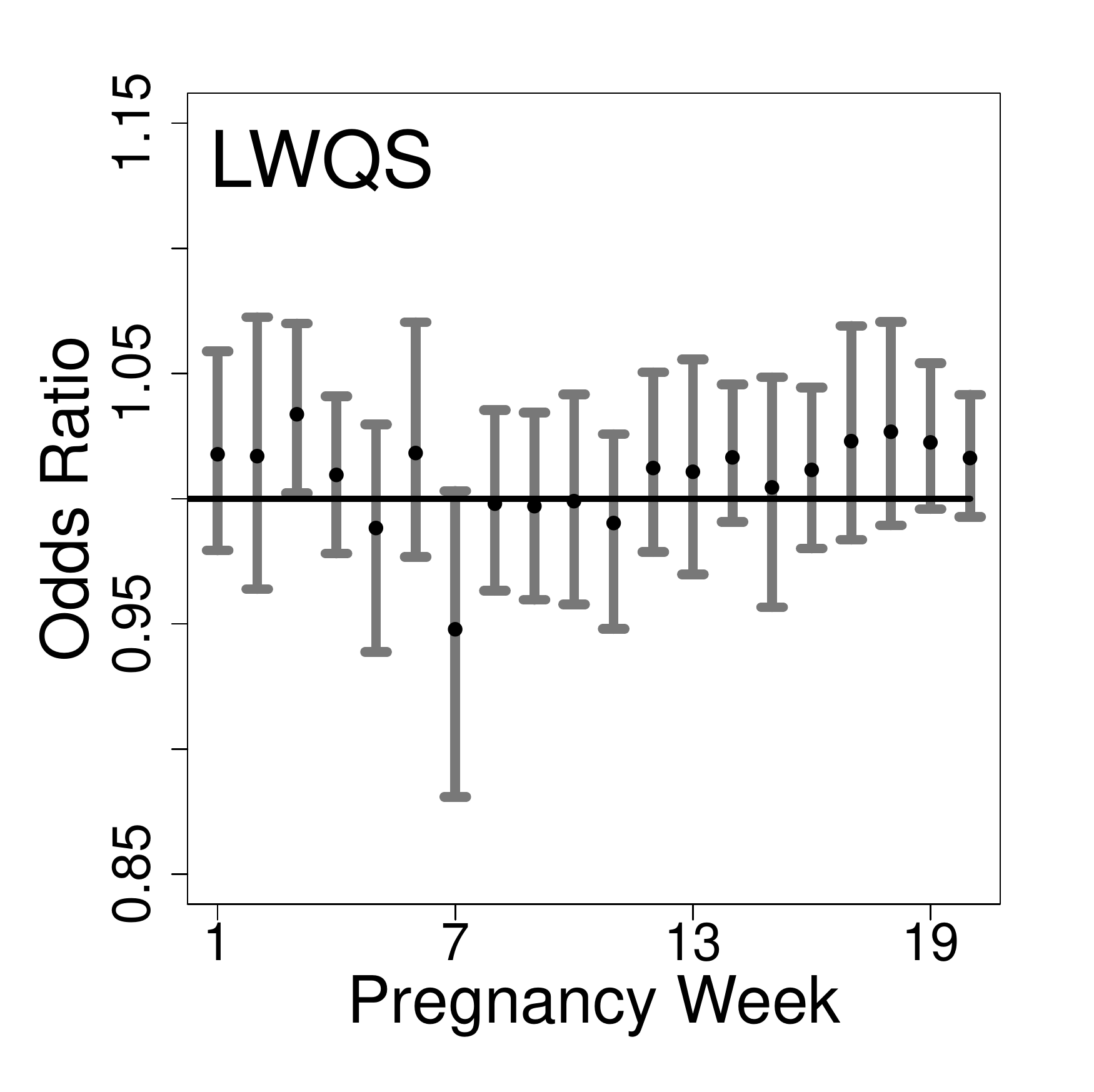}
\includegraphics[trim={0.5cm 0.5cm 1cm 0.5cm}, clip, scale=0.21]{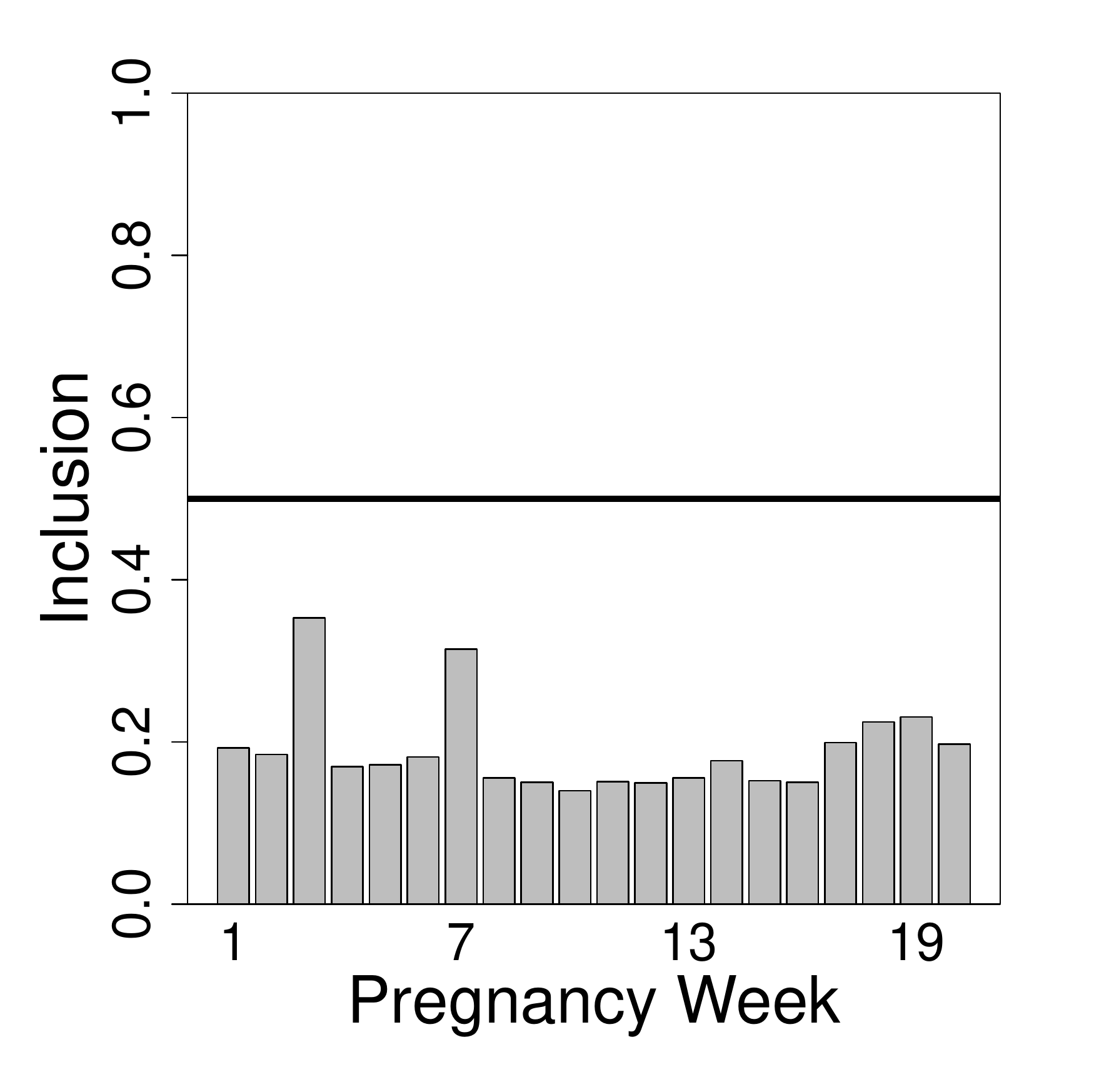}
\includegraphics[trim={0.25cm 0.5cm 1cm 0.5cm}, clip, scale=0.21]{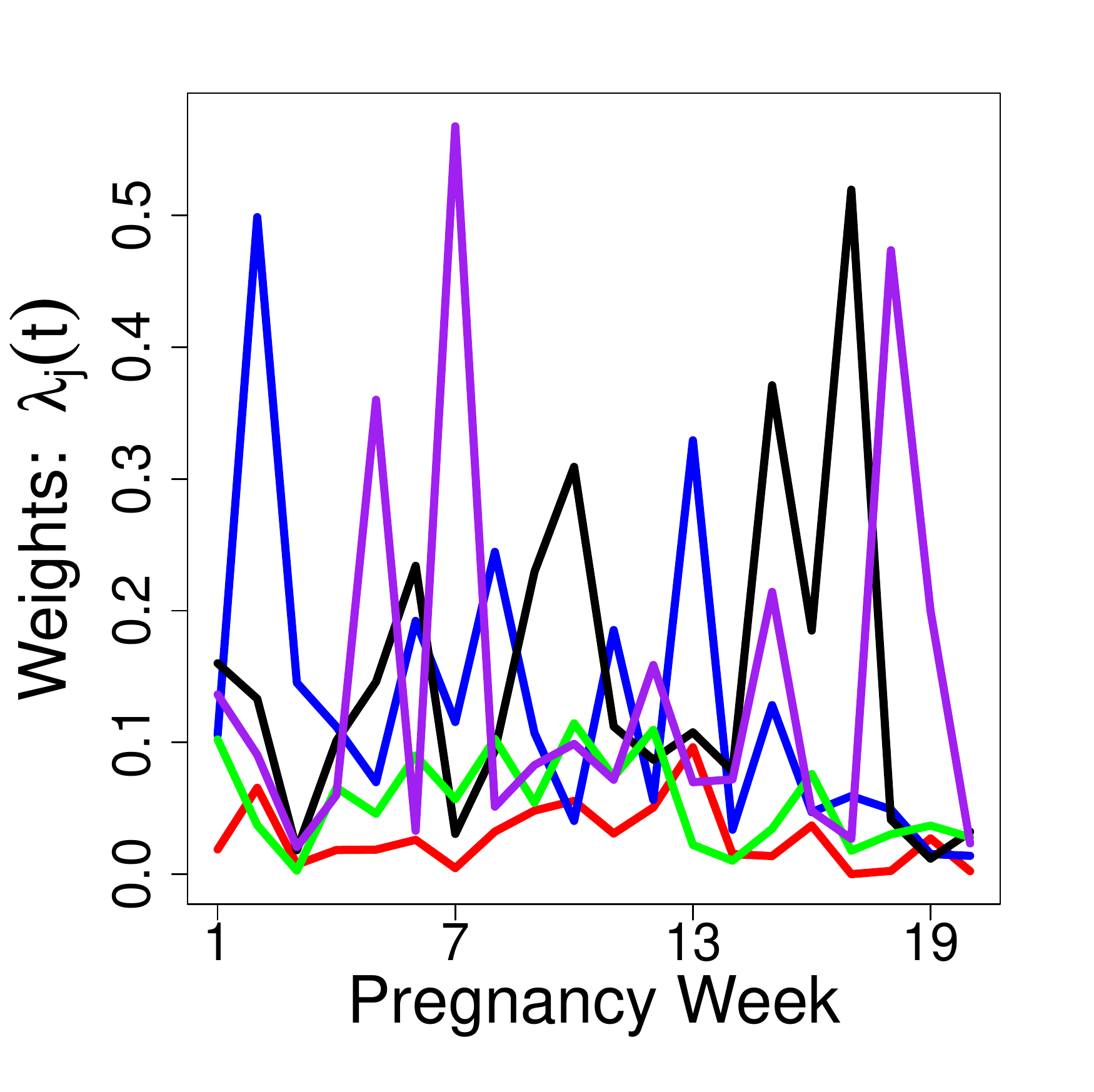}
\includegraphics[trim={0cm 0.5cm 1cm 0.5cm}, clip, scale=0.21]{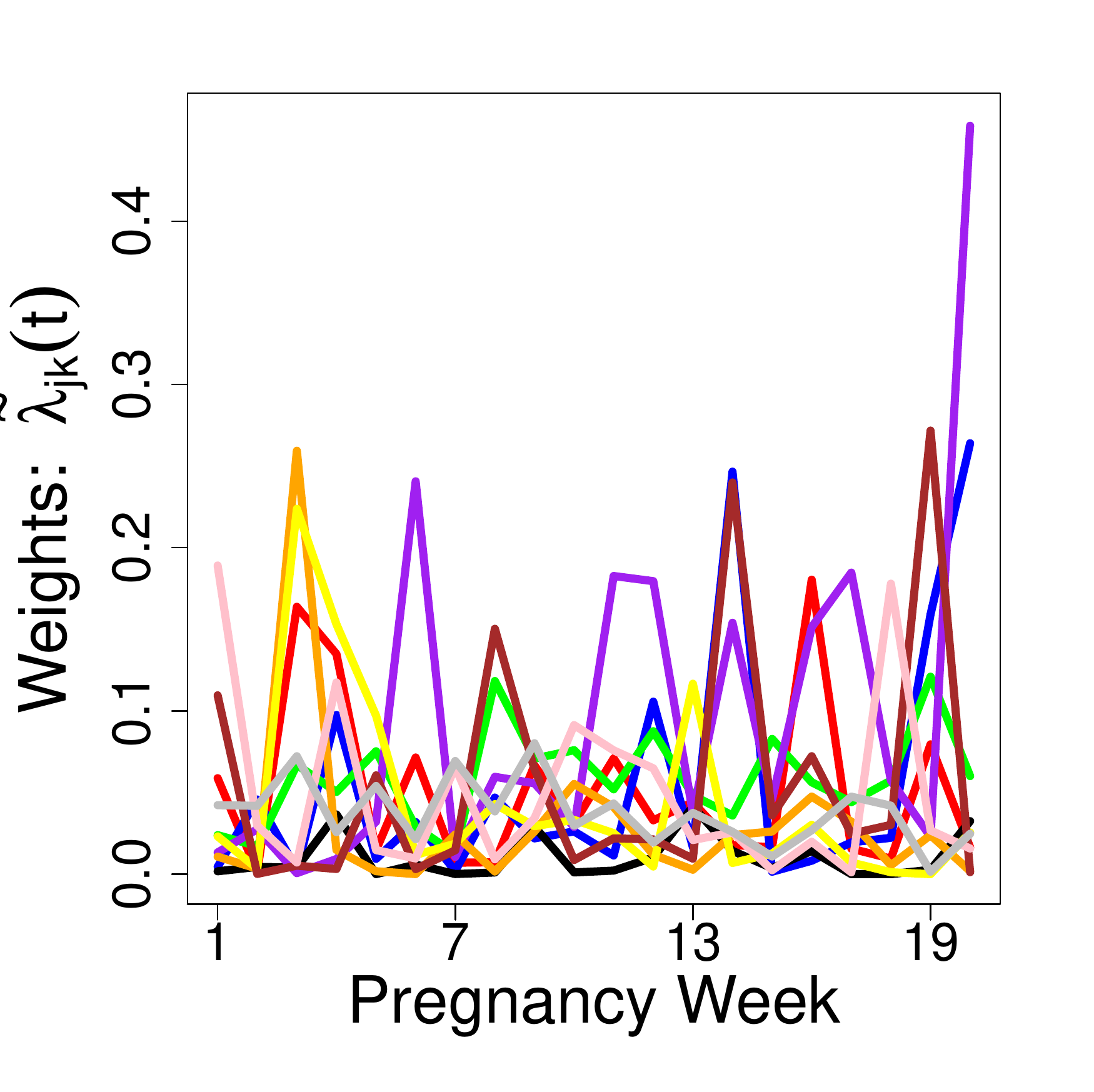}\\
\includegraphics[trim={0.5cm 0.5cm 1cm 0.5cm}, clip, scale=0.21]{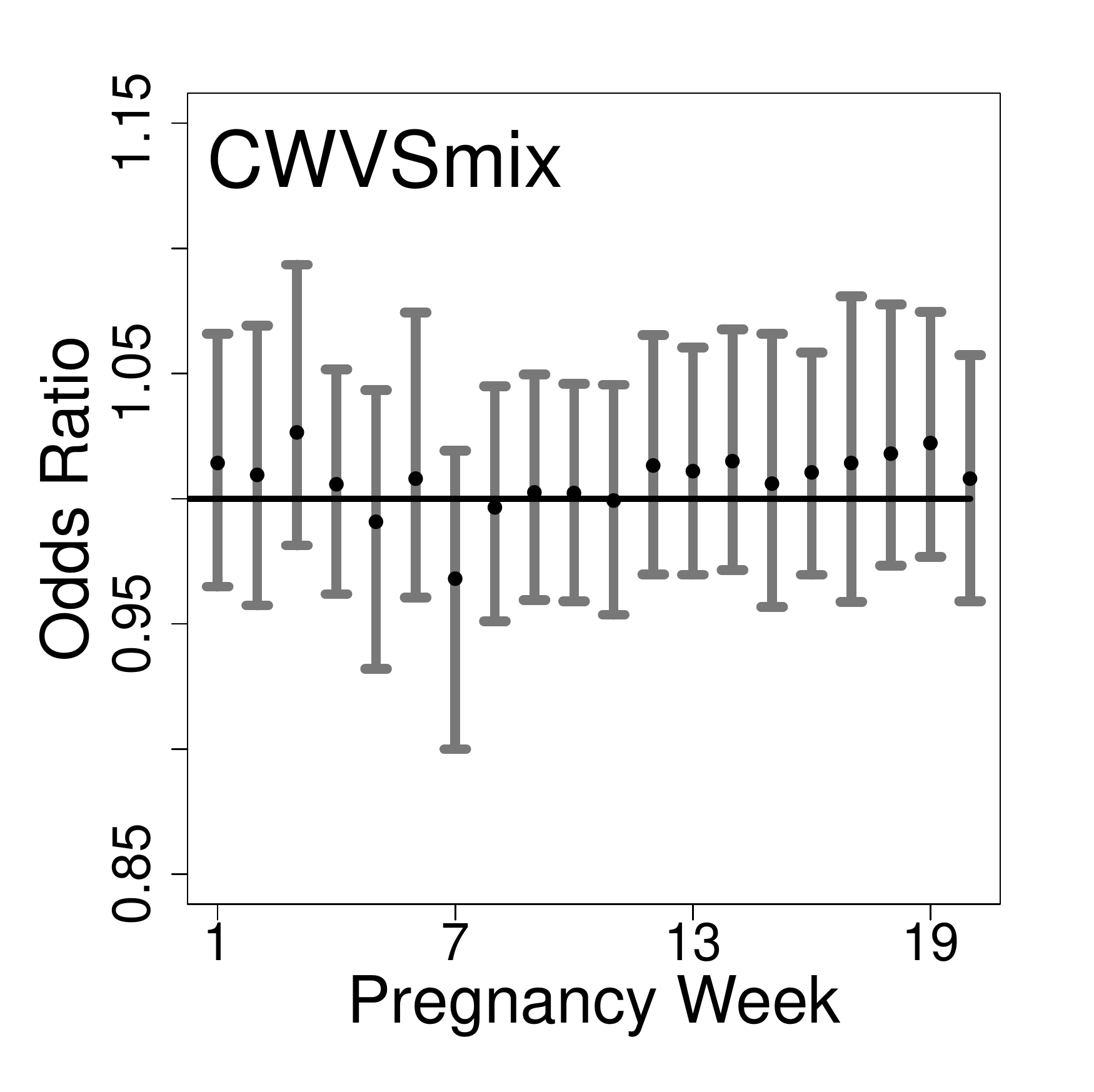}
\includegraphics[trim={0.5cm 0.5cm 1cm 0.5cm}, clip, scale=0.21]{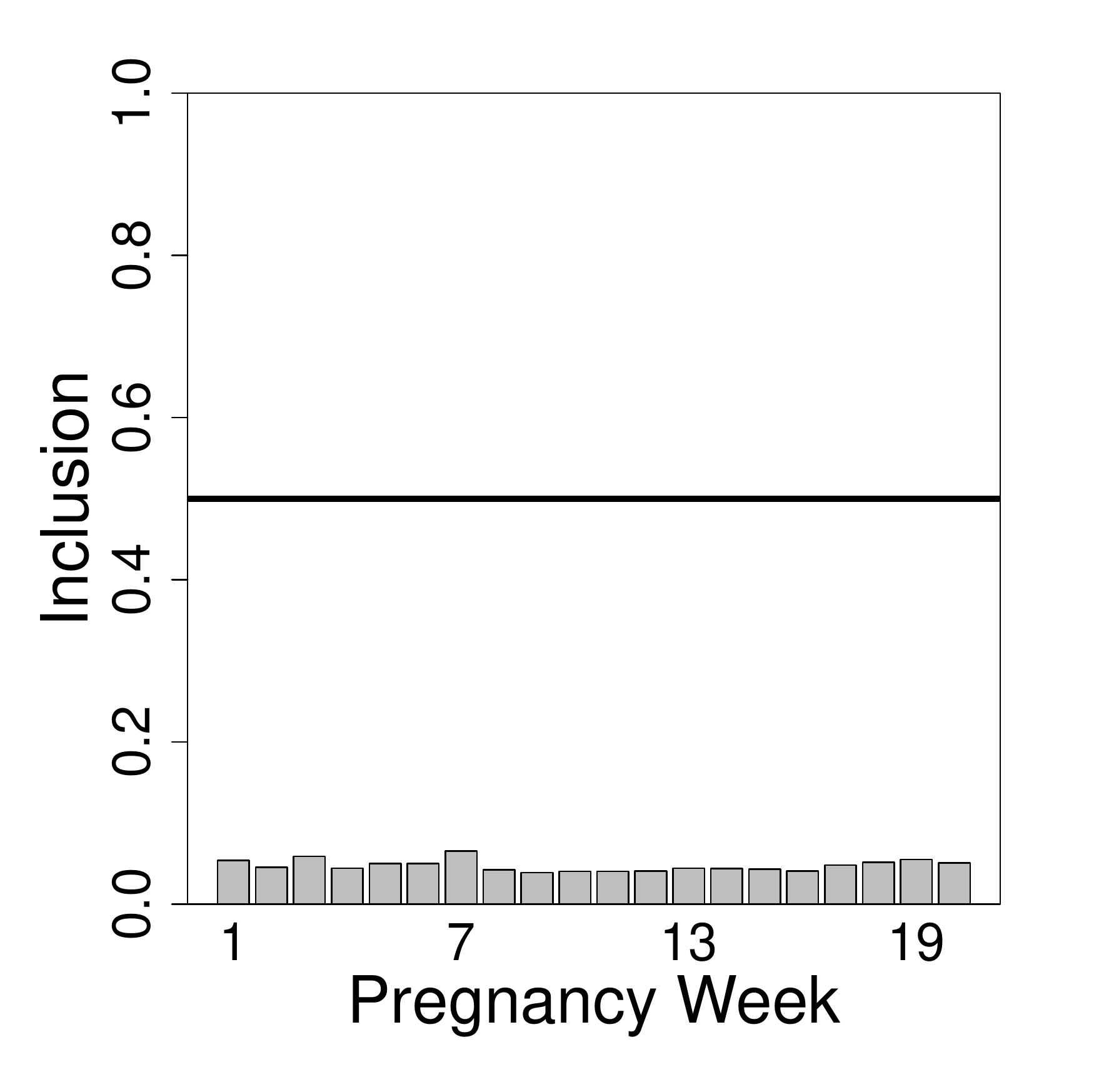}
\includegraphics[trim={0.25cm 0.5cm 1cm 0.5cm}, clip, scale=0.21]{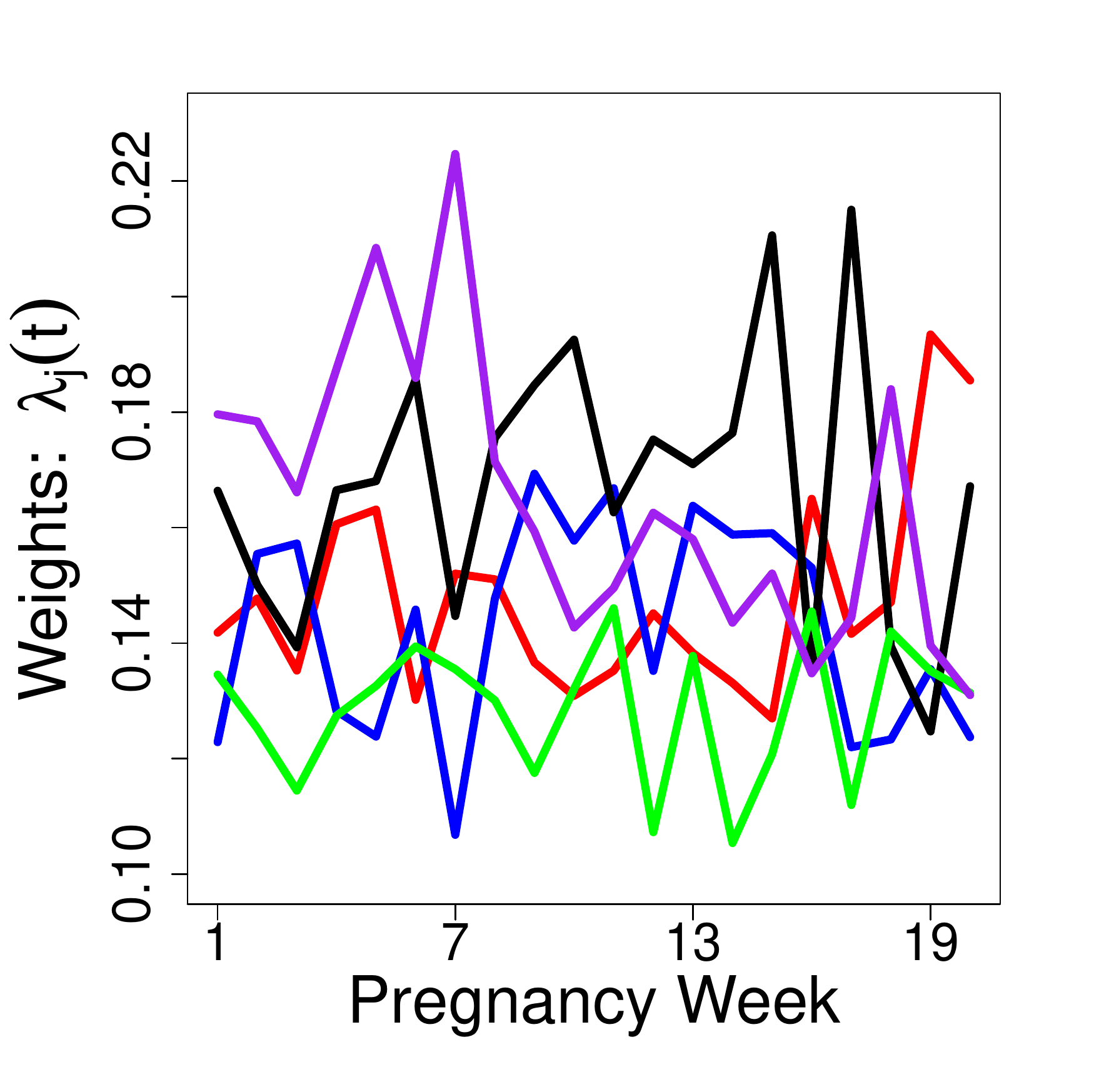}
\includegraphics[trim={0cm 0.5cm 1cm 0.5cm}, clip, scale=0.21]{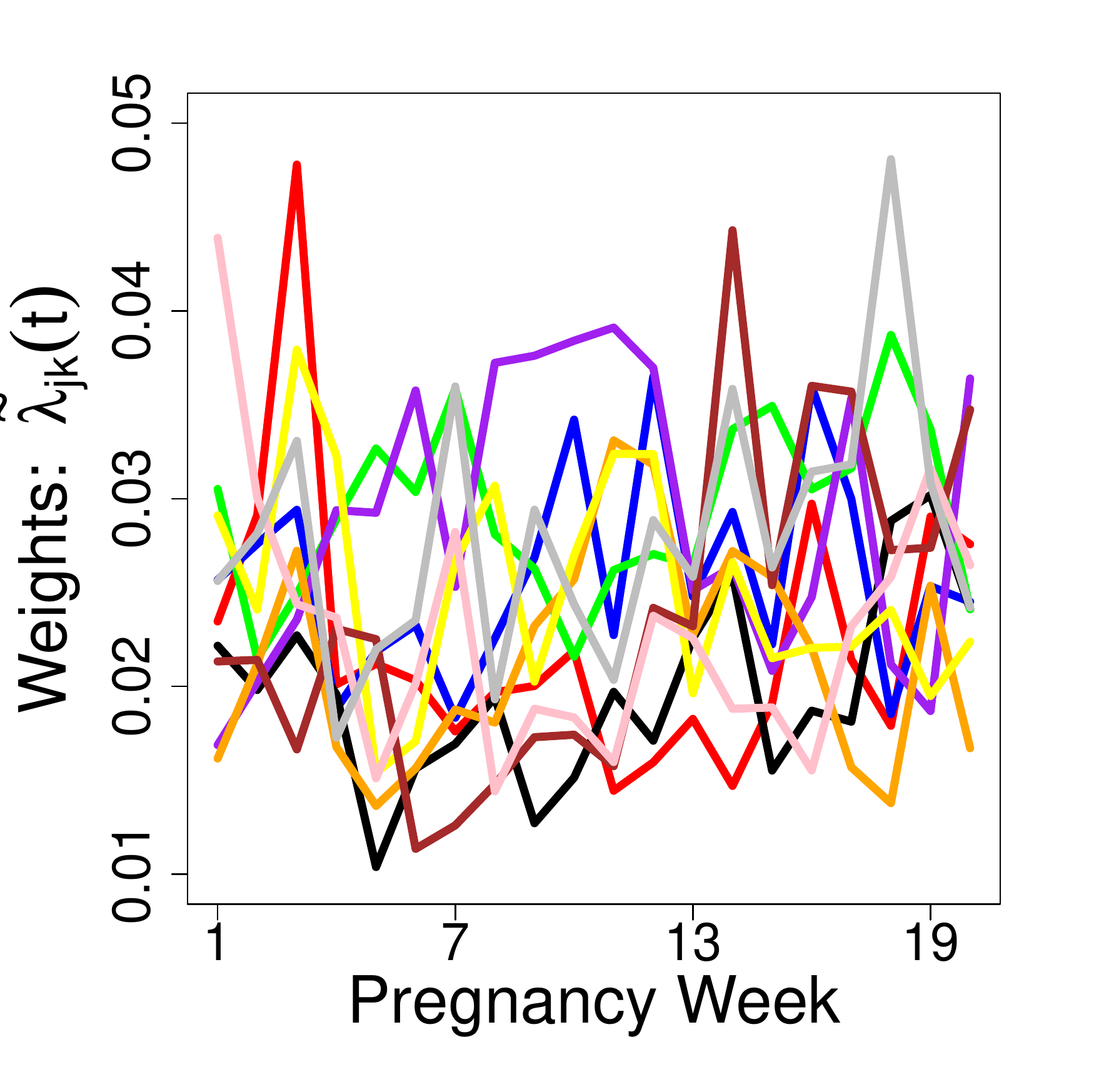}\\
\caption{Posterior means and 90\% credible intervals for the risk parameters (first column), posterior inclusion probabilities (second column), posterior means for the main effect weight parameters (third column), and posterior means for the interaction effect weight parameters (fourth column) for the \textbf{non-Hispanic White} stillbirth and multiple exposures analyses in New Jersey, 2005-2014. Results based on an interquartile range increase in weekly exposure. Weeks identified as part of the critical window set are shown in red/dashed (harmful) and blue/dashed (protective).  The average posterior standard deviation for the weight parameters from CWVSmix is 0.13 (range: 0.05-0.32).}
\end{center}
\end{figure}
\clearpage

\bibliographystyle{imsart-nameyear}
\bibliography{Supplement}